\documentclass[12pt,epsfig,english,twoside,onecolumn,titlepage,a4paper]{book}
\pdfoutput=1

\usepackage[english]{babel}
\usepackage[bf]{caption}
\usepackage{color}

\usepackage{graphics}
\usepackage{graphicx}
\usepackage{epsfig}
\usepackage{subfigure}

\usepackage{tabularx}
\usepackage{amssymb}
\usepackage{multirow}

\usepackage{fancyheadings}

\begin{document}

%
%
\thispagestyle{empty}
\begin{center}
\it \large
Warsaw University \\
Faculty of Physics \\
Institute of Experimental Physics \\
\end{center}

\begin{center}
\vspace{2.5cm}
\begin{Large}
{\bf
{Measurements and simulations of MAPS (Monolithic Active Pixel Sensors) response to charged particles  - a study towards a vertex detector at the ILC}}
\end{Large}
\end{center}

\begin{center}
\vspace{1.5cm}
{\bf \large {\L}ukasz Janusz M\c{a}czewski}
\end{center}

\begin{center}
\vspace{1.0cm}
{\large
{ PhD thesis written under supervision\\
of Prof. Jacek Ciborowski}}
\end{center}
\vspace{1.cm}
\begin{figure}[h]
  \setlength{\unitlength}{1cm}
  \begin{center}
      \mbox{\epsfxsize 3.5cm\epsfysize 3.5cm\epsffile{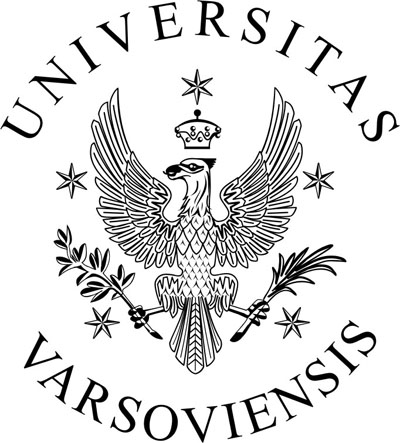}}
  \end{center}
\end{figure}
\vspace{1.2cm}
\begin{center}
 Warsaw, 2010
\end{center}
\newpage
\thispagestyle{empty}
$~~~~~~~~~~~~~~~~~~$

\thispagestyle{empty}
\vspace*{20cm}

\newpage
\thispagestyle{empty}
\vspace*{17cm}
\begin{flushright}
{\it \large  To my Wife, Ma{\l}gorzata}
\end{flushright}
\newpage
\thispagestyle{empty}
\vspace*{20cm}

\newpage
\thispagestyle{empty}
\vspace*{4cm}

\begin{center}
{\bf Abstract}
\end{center}

\vspace*{1cm}

\noindent The International Linear Collider (ILC) is a project of an electron-positron ($e^{+}e^{-}$) linear collider with the centre-of-mass energy of 200-500~GeV. Monolithic Active Pixel Sensors (MAPS) are one of the proposed silicon pixel detector concepts for the ILC vertex detector (VTX). Basic characteristics of two MAPS pixel matrices MIMOSA-5 (17~$\mu$m pixel pitch) and MIMOSA-18 (10~$\mu$m pixel pitch) are studied and compared (pedestals, noises, calibration of the ADC-to-electron conversion gain, detector efficiency and charge collection properties). The $e^{+}e^{-}$ collisions at the ILC will be accompanied by intense beamsstrahlung background of electrons and positrons hitting inner planes of the vertex detector. Tracks of this origin leave elongated clusters contrary to those of secondary hadrons. Cluster characteristics and orientation with respect to the pixels netting are studied for perpendicular and inclined tracks. Elongation and precision of determining the cluster orientation as a function of the angle of incidence were measured. A simple model of signal formation (based on charge diffusion) is proposed and tested using the collected data. 

\newpage
\thispagestyle{empty}
\vspace*{20cm}

\pagestyle{plain}
\pagenumbering{roman}
\tableofcontents
\listoffigures
\listoftables
\newpage
$~~~$ \\
\newpage

\pagestyle{headings}
\pagenumbering{arabic}

\chapter{Introduction}
\label{ch:intro}

\begin{figure}[!h]
        \begin{center}
                \resizebox{0.5\textwidth}{!}{
                        \includegraphics[]{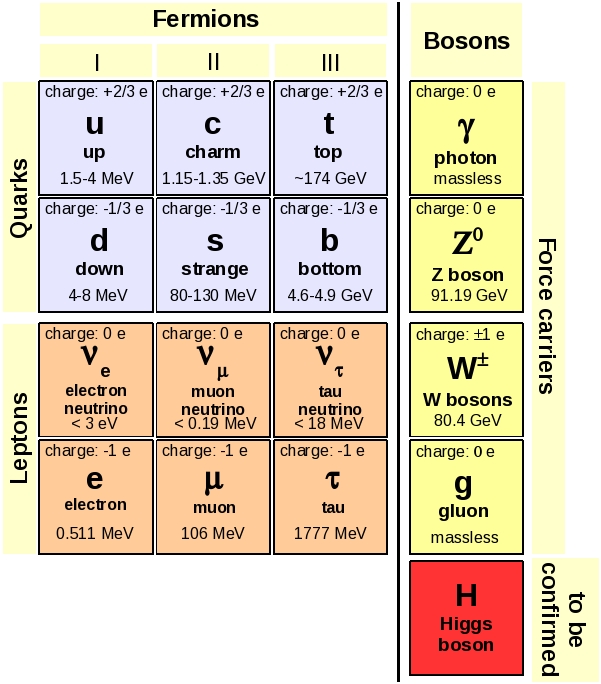}
                }
                \caption[Overview of the elementary particles according to the Standard Model]{Overview of the elementary particles according to the Standard Model.}
                \label{fig:intro:StandardModel}
        \end{center}
\end{figure}

The Standard Model (SM) \cite{ilc:StandardModel} provides a description of the basic properties of strong and electroweak interactions of leptons and hadrons. According to it, the fundamental constituents of matter are 6 spin--$\frac{1}{2}$ quarks (and 6 antiquarks) and 6 spin--$\frac{1}{2}$ leptons (and 6 antileptons), as shown in fig.~\ref{fig:intro:StandardModel}. 
\\
The fundamental forces are mediated by spin--1 vector bosons (gauge bosons). These are: the massless photon~$\gamma$, massive $W^{+}$, $W^{-} $and $Z^{0}$~bosons, and massless gluons~$g$ occurring in 8 different colour states (fig.~\ref{fig:intro:StandardModel}). The photon mediates electromagnetic forces between all charged particles while the $W^{+}$, $W^{-}$ and $Z^{0}$ bosons are carriers of weak interactions. Gluons mediate strong interactions between particles which carry the colour charge, i.e. quarks and gluons themselves. The Standard Model does not include gravitation.\\
The theoretical description of the electromagnetic, weak and strong interactions is based on symmetries. The electromagnetic interactions are described within the conserved $U(1)$ gauge symmetry and thus the photon is massless. The strong interactions are based on the conserved $SU(3)_c$ gauge symmetry ($c$ stands for colour) and thus gluons are massless. The weak interactions are described within the spontaneously broken $SU(2)$ symmetry and thus the $W^{+}$, $W^{-}$ and $Z^{0}$ bosons are massive. Predictions of the SM have been confronted with experiments and a great number of precision measurements confirmed their excellent agreement \cite{ilc:SM_LEP}.\\
The mechanism of spontaneous $SU(2)$ symmetry breaking requires existence of a scalar field with an associated massive particle, called the higgs boson, H. Despite numerous searches the higgs particle has not yet been discovered. The current lower limit for the higgs mass from direct searches at LEP is $m_{H}>~$114.4~GeV at 95\% confidence level \cite{ilc:Higgs_LEP}. Thus the higgs search is the prime objective of the current high energy experiments. The two running experiments D$\O{}$ and CDF at the Fermilab Tevatron continue to collect data of $p\overline{p}$ collisions at c.m.s energy of 1.96~TeV. Analysis of the entire available event statistics, corresponding to the delivered luminosity of 7.6~fb$^{-1}$, allowed to exclude the higgs boson in the mass region 163~GeV$~<m_{H}<~$166~GeV \cite{ilc:Higgs_Tevatron}. It is planned that both experiments will be running through 2011 and collect 10~fb$^{-1}$ of data. The oncoming experiments at CERN are ATLAS, CMS, ALICE and LHCb at the LHC (Large Hadron Collider) with the first two having the objective to discover the higgs boson. The LHC started to collide proton beams in November 2009. Wherever (and if) the higgs boson and other particles will be discovered, their appearance will happen in hadronic collisions i.e. in the presence of high background. Measurements of all their properties (masses, widths, decay branching ratios, couplings, asymmetries etc.) will be of limited precision for that reason. It would be necessary to remeasure all these quantities in low background collisions and these are offered by an $e^{+}e^{-}$ collider. Such a machine is actually under consideration. This project is called ILC (International Linear Collider).\\
Despite the excellent agreement of the SM with measurements, it has been accepted ever since that it requires an extension in order to improve certain inevitable shortcomings. One of them is the problem of the higgs self energy correction that requires for example introducing a new energy scale or, in other words, new particles beyond the SM. On the other hand, the SM does not answer all questions regarding the elementary particles and their interactions, also including many issues in the field of cosmology. For example, the masses of neutrinos in the SM are assumed zero while neutrino oscillations have been firmly established. The SM does not explain the issue of particle masses, in particular the hierarchy problem. The energy scale of the SM is of the order of $10^{2}$~GeV while the scale of the expected unification of strong end electroweak interactions is of the order of $10^{16}$~GeV and the Planck energy scale is $10^{19}$~GeV. The SM does not predict any phenomena above its energy scale. The energy gap still remains to be understood within physics beyond the SM. Apart from the issue of energy scales, there exist unexplained cosmological phenomena. Recent observations in astrophysics and cosmology \cite{ilc:WMAP} show that only around 5\% of the energy density in the universe is due to the matter that we know, i.e. hydrogen gas, heavier elements, photons and neutrinos. Around 25\% is assigned to an unknown form of matter that is not described by the Standard Model (so called Dark Matter) and the rest 70\% of the energy density in the universe is of unknown origin, the so-called Dark Energy.\\
The main concepts for an extension of the Standard Model are Super Symmetry (SUSY) (e.g. Minimal Supersymmetric Model (MSSM) \cite{ilc:SUSY}) or models assuming existence of extra dimensions \cite{ilc:ExtraDim}.\\
The SUSY, which is presumably the most popular extension of the Standard Model, postulates an additional fundamental symmetry between fermions and bosons, thereby introducing a bosonic partner to each fermion and vice versa. According to theoretical arguments, the supersymmetric energy scale should not be too high -- it is expected that at last some of the supersymmetric particles are lighter than 1~TeV thus they are in a range of planned experiments at the Large Hadron Collider (LHC) and the International Linear Collider.\\ 
At the LHC, where protons will be colliding at the centre-of-mass energy of 14~TeV, the higgs boson and signals of a new physics beyond the Standard Model will be searched. It will however take several years from now to detect such particles if they exist. The ILC, colliding electron-positron beams at the centre-of-mass energy of 200-500~GeV (possibly up to 1~TeV), will provide complementary, high precision measurements of the LHC discoveries. An electron-positron collider provides clean experimental environment in the sense that background rates of physics processes are orders of magnitude lower than those at hadron colliders. However $e^{+}e^{-}$ collisions are accompanied by machine induced backgrounds with a major contribution due to $e^{+}e^{-}$ pairs of low transverse momentum. Such pairs are created from bremsstrahlung photons that are radiated by beam particles deflected in the electromagnetic field of the colliding beams -- for that reason this effect is called beamsstrahlung. The vertex detector, which is located closest to the primary interaction point, suffers the most from this source of background. The pile up of hits leads to high detector occupancy. Moreover, the beam related background induces also radiation damages in the silicon sensors of the vertex detector, leading to deterioration of its overall tracking performances.\\\\
The present thesis is devoted to studies regarding a vertex detector for the ILC.\\\\
The vertex detector is a key component for precise heavy flavour identification ($c$, $b$ quarks) which is based on reconstruction of secondary decay vertices. Heavy flavour identification is critical at the ILC e.g. for measuring the higgs branching ratios with high precision as well as in search for new physics phenomena (for example charm identification in conjunction with observation of missing energy could be a signature of a stop decay into charm and a neutralino). The need for maximising the efficiency and purity of heavy flavour identification pushes the required vertex detector efficiency, angular coverage and impact parameter resolution beyond the current state of the art.\\
The ILC vertex detector will be built of thin, highly segmented pixel detectors arranged in quasi-cylindrical layers surrounding the beam pipe. The basic characteristics of a vertex detector to be installed in the future at the ILC are the following:
\begin{itemize}
 \item High granularity satisfying the demand for precision measurements of physics processes
 \item Lowest possible thickness to minimise multiple Coulomb scattering
 \item High efficiency of track detection (exceeding 99\%)
 \item Low noise and high signal-to-noise ratio
 \item Sufficient radiation tolerance
 \item Satisfactory mechanical rigidity
 \item Sufficient readout speed ($\sim$40~MHz)
 \item Low power consumption and dissipation
\end{itemize}
Several pixel detector technologies are presently under investigation (chapter~\ref{ch:silicon}). A good trade-off between the above features is offered by Monolithic Active Pixel Sensors (MAPS). Granularity is understood in this context as the pixel size or the distance between charge collecting diodes -- the pitch. It might seem that the most suitable pixel matrix should have the smallest possible pitch to allow sufficient accuracy of track reconstruction. This is only partially true since decreasing the pitch implies a quadratic increase of the number of readout channels. This, in turn, causes an increase of power needed for running the detector which results in more power dissipation in the matrices. The decision regarding the pixel size for the ILC vertex detector will be taken in the future based on results of tests of numerous devices and supplemented by results of simulations.\\
The MAPS detectors are relatively cheap since they are fabricated in the commercially available CMOS process. Charge carriers, generated by ionising particles traversing the active volume, diffuse isotropically in the active volume of the MAPS detector. One of the major advantages of the MAPS devices is the possibility of integrating the readout and signal processing electronics on the same substrate as the sensor. Existing MAPS prototypes of the MIMOSA series are reviewed in chapter~\ref{ch:tests:exp}.\\
Beamsstrahlung $e^{+}e^{-}$ tracks impact the vertex detector planes at angles generally larger than those of the secondary hadrons (from beam-beam interactions). One thus expects that the beamsstrahlung related pixel clusters should be elongated w.r.t the clusters of the latter origin. Moreover, such clusters, when sufficiently elongated, are expected to have also different orientations w.r.t. the pixel netting. These two pieces of information regarding clusters -- elongation and orientation -- could be exploited in order to distinguish between beamsstrahlung and hadronic clusters.\\
Studies and measurements described in the present thesis regarded two MAPS detectors: MIMOSA-5 and MIMOSA-18 which have pixels of 17~$\mu$m and 10~$\mu$m pitch, respectively. These particular matrices were chosen for the purpose of comparing various characteristics of devices with two significantly different pixels sizes.\\
The first aim of the present thesis was to measure and compare several basic properties of the MIMOSA-5 and MIMOSA-18 pixel matrices, including: ($i$) pedestals and noises, ($ii$) calibration of the ADC-to-electron conversion gain, ($iii$) detector efficiency, ($iv$) charge collection properties.\\
The second aim was to perform a detailed study of cluster characteristics for different angles of incident tracks. A method of determining the cluster shape (elongation and its orientation with respect to the pixel netting) was proposed and tested.\\
The third aim was to develop a simple, effective model of signal formation in a MAPS detector and compare it with the above experimental results. This simple model has been implemented in detailed Monte Carlo simulations of the ILC vertex detector.\\
The second and third aims are particularly important for investigations of the readout strategy as well as possible rejection of beamsstrahlung clusters. The experimental setup, data analysis procedure and obtained results are presented in chapters~\ref{ch:tests:exp} and~\ref{ch:tests:exp_results}, while the signal formation model for the MAPS detectors and its tests are discussed in chapter~\ref{ch:digi}.

\chapter{The International Linear Collider Project}
\label{ch:ilc}

The International Linear Collider \cite{ilc:GDE_Accelerator} is a project of a linear accelerator, in which positron-electron ($e^{+}e^{-}$) beams will collide at the centre-of-mass (cms) energy of 200-500~GeV (possible extension of the maximal cms energy up to 1~TeV). The ILC has been conceived to be the next large experimental facility in the high energy physics, following the Large Hadron Collider \cite{ilc:lhc} which is presently starting operation. The LHC with high energy of colliding proton beams of 14~TeV at the cms and with high peak luminosity of $10^{34}$~cm$^{-2}$s$^{-1}$ is a perfect tool for discovering new heavy particles. The ILC, with its much lower experimental background, polarised beams and tunable collision energies will provide precision measurements regarding the LHC discoveries.\\
The ILC machine will have variable cms energy in a range between 200~GeV and 500~GeV. The beam energies will be measured with a precision better than 0.1\%. The ILC design should allow an upgrade to reach 1~TeV cms energy. The planned integrated luminosity of 500~$fb^{-}1$ in the first four years requires the running efficiency of 75\% assuming an annual physics run of 9 months with the peak luminosity of $2\times10^{34}$~cm$^{-2}$s$^{-1}$ at 500~GeV cms energy. The polarisation of the electron beam is expected to exceed 80\% while the positron beam could be optionally polarised in $\sim$60\%.

\section{Collider and beam parameters}
\label{ch:ilc:collider_beam}

\begin{figure}[!htbp]
        \begin{center}
                \resizebox{\textwidth}{!}{
                        \includegraphics[]{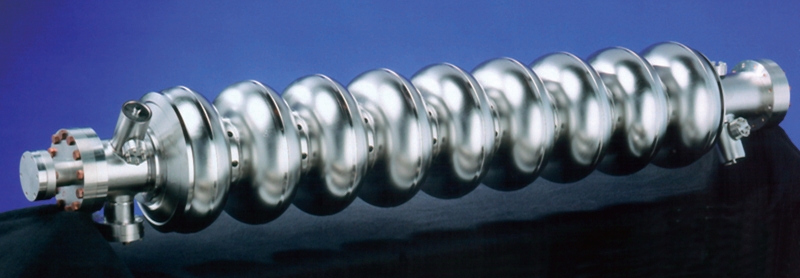}
                }
                \caption[A nine-cell 1.3~GHz superconducting niobium cavity]{A nine-cell 1.3~GHz superconducting niobium cavity (from~\cite{ilc:GDE_Accelerator}).}
                \label{fig:ilc:RFCavity}
        \end{center}
\end{figure}

\begin{figure}[!htbp]
        \begin{center}
                \resizebox{\textwidth}{!}{
                        \includegraphics[]{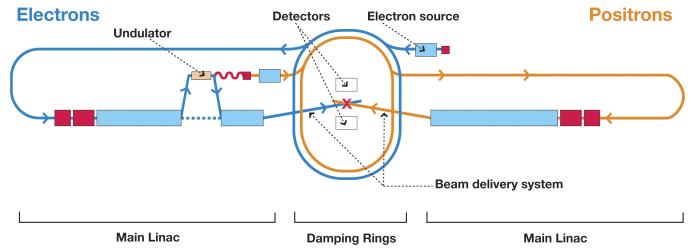}
                }
                \caption[Schematic layout of the ILC accelerator]{Schematic layout of the ILC accelerator.}
                \label{fig:ilc:beam_delivery_system}
        \end{center}
\end{figure}

The ILC will be based on the 1.3~GHz superconducting RF niobium cavities operating at 2~K. The 1~m long cavity is composed of 8 or 9 cells, as shown in fig.~\ref{fig:ilc:RFCavity}. The ILC will need approximately 17000 superconducting niobium cavities, each having the gradient of 31.5~MV/m. Up till now 160 cavities have been produced as part of the ongoing R\&D program at DESY. Several cavities have already achieved the desired and higher gradients however it is still a challenge to achieve the required production yield of 80\% for nine-cell cavities.
\\
Preparation and assembly of cavities are performed in ultra high clean-room environment, after which the cavities are submitted to electropolishing in order to provide an ultra smooth inner surface.\\
Since the value of the gradient is the key parameter for the ILC, intensive studies are under way all over the world on alternative cavity shapes and materials. One promising technique is the use of `large-grain' niobium \cite{ilc:large-grain}, as opposed to the small-grain material that has been used in the past. Use of large grain material may remove the need for electropolishing, since the same surface finish can potentially be achieved with Buffered Chemical Polishing - a possible cost saving.\\
In order to reach high peak luminosity of $2\times10^{34}$~cm$^{-2}$s$^{-1}$ at 500~GeV cms, beams of high power and low emittance are required. The choice of 1.3 GHz superconducting RF is well suited to the requirements, primarily because the very low power loss in superconducting RF cavity walls allows the use of long RF pulses, relaxing the requirements on the peak-power generation, and ultimately leading to high wall-plug to beam transfer efficiency.
\begin{figure}[!htbp]
        \begin{center}
                \resizebox{\textwidth}{!}{
                        \includegraphics[]{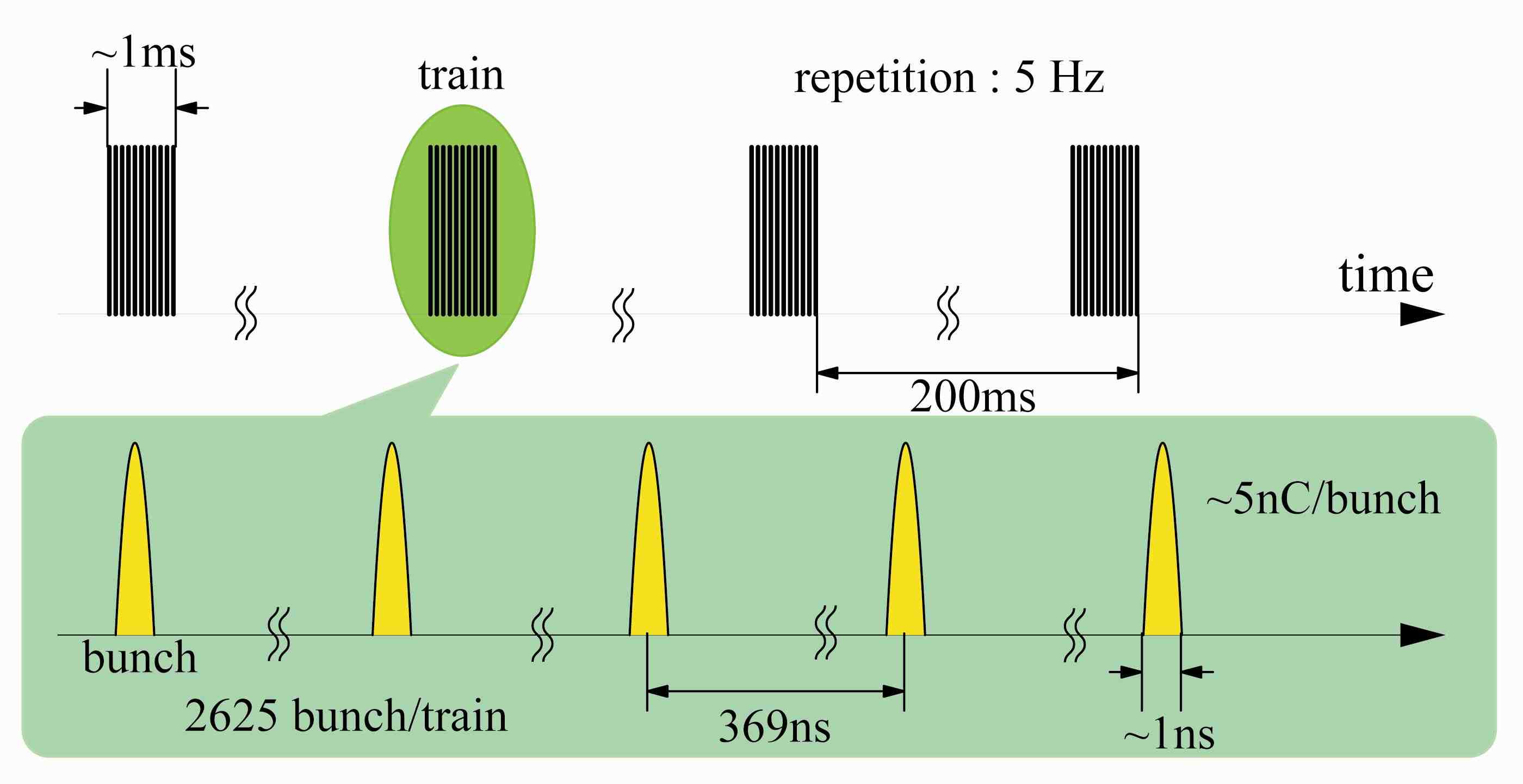}
                }
                \caption[ILC beams structure]{ILC beams structure.}
                \label{fig:ilc:beam_structure}
        \end{center}
\end{figure}\\
The beam will be divided into pulses (so called trains) as shown in fig.~\ref{fig:ilc:beam_structure}. In the nominal design the beam pulse of $\sim$1~ms length is composed of 2625 bunches. Each bunch of $\sim$1~ns length contains 2$\cdot$10$^{10}$ particles. The bunches included in a single train are separated by $\sim$369~ns. The trains are delivered with a repetition rate of 5~Hz.\\
The required luminosity of $2\times10^{34}$~cm$^{-2}$s$^{-1}$ at 500~GeV cms can be provided with different values for the specific beam parameters. Apart from nominal set of parameters three additional sets of parameters have been proposed labelled `Low N`, `Large Y` and `Low P`, also summarized in table~\ref{tab:ilc:set_of_beam_para}.
\begin{itemize}
 \item \textbf{Low N}\\
 In the Low N mode the number of particles in a single bunch is reduced in order to avoid problems such as microwave instabilities in the damping rings, single bunch wakefield emittance dilutions, or a large disruption parameter at the interaction point (IP). The assumed luminosity of $2\times10^{34}$~cm$^{-2}$s$^{-1}$ in the Low N mode is reached due to the increase of the bunch number in the train, decrease of the pulse duration and reduction of the beam emittance. With this set of parameters the lower beamsstrahlung and possibly lower background in the detectors closest to the IP is expected. 
 \item \textbf{Large Y}\\
 If the vertical emittance at the IP of 4$\times$10$^{-8}$~m$\cdot$rad could not be obtained due to the tuning difficulties in the damping rings and beam delivery system or wakefield effects in the linac, the assumed luminosity of $2\times10^{34}$~cm$^{-2}$s$^{-1}$ can be achieved by reducing horizontal emittance and increasing the length of the bunch. In the Large Y mode a similar level of beamsstrahlung is expected as in the nominal case. 
 \item \textbf{Low P}\\
 The set of parameters for the Low P mode is devoted to the situation in which limitations of the beam current or the beam power occurs. These may arise in the injector systems, damping rings, main linacs or beam delivery system. In order to keep the same luminosity as in the nominal mode the horizontal emittance is decreased by increased focusing at the IP. Since the vertical emittance is very low it is impossible to reduce beamsstrahlung background by increasing the bunch length, resulting in a roughly double rate of beamsstrahlung than in the case of nominal parameters.
\end{itemize}
\begin{table}[!htbp]

	\begin{tabularx}{1.1\textwidth}{@{\extracolsep{\fill}} |>{\small}c>{\small}c|>{\small}c|>{\small}c>{\small}c>{\small}c|} \hline

	Parameter & Symbol/Unit & Nominal & Low N & Large Y & Low P \\ \hline \hline
	Repetition rate & $f_{rep}(Hz)$ & 5 & 5 & 5 & 5 \\ \hline
	Number of particles per bunch & N(10$^{10}$) & 2 & 1 & 2 & 2 \\ \hline
	Number of bunches per pulse & $n_{b}$ & 2625 & 5120 & 2625 & 1320 \\ \hline
	Bunch interval in the Main Linac & $t_{b}$ (ns) & 369.2 & 189.2 & 369.2 & 480.0 \\ \hline
	Average beam current in pulse & $I_{ave}$ (mA) & 9.0 & 9.0 & 9.0 & 6.8 \\ \hline
	Normalized emittance at IP & $\gamma\epsilon_{x}^{*}$ (mm$\cdot$mrad) & 10 & 10 & 10 & 10 \\ \hline
	Normalized emittance at IP & $\gamma\epsilon_{y}^{*}$ (mm$\cdot$mrad) & 0.04 & 0.03 & 0.08 & 0.036 \\ \hline
	R.m.s beam size at IP & $\sigma_{x}^{*}$ (nm) & 639 & 474 & 474 & 474 \\ \hline
	R.m.s beam size at IP & $\sigma_{y}^{*}$ (nm) & 5.7 & 3.5 & 9.9 & 3.8 \\ \hline
	R.m.s bunch length & $\sigma_{z}$ ($\mu$m) & 300 & 200 & 500 & 200 \\ \hline
	Energy loss by beamsstrahlung & $\delta_{BS}$ & 0.024 & 0.017 & 0.027 & 0.055 \\ \hline
	Luminosity & $\mathcal{L}$ 10$^{34}$/cm$^{2}$/s & 2 & 2 & 2 & 2 \\ \hline

	\end{tabularx}

\caption[Beam parameters at the interaction point for 500 GeV cms]{Beam parameters at the interaction point for 500 GeV cms \cite{ilc:large-grain}.}
\label{tab:ilc:set_of_beam_para}

\end{table}
According to the current design, the 31~km long ILC will consist of two arms as presented in fig.~\ref{fig:ilc:beam_delivery_system}. The electron source based on a photocathode DC gun, the undulator-based positron source, driven by the 150 GeV main electron beam and the damping rings of 6.7~km circumference will be located around the interaction region (IR). After damping the $e^{+}$ and $e^{-}$ beams to the desired emittances the beams will be transported with the RTML (Ring To Main Linac) system of 15~km length to the Main Linacs where they will be accelerated. The high energy main linacs of 11~km will be followed with the Beam Delivery System (BDS). The BDS of 4.5~km length, used for focusing the beams to the sizes required to meet the ILC luminosity goals, will bring them into collision and then transport the spent beams to the main beam dumps. The beams in the ILC will collide at 14~mrad crossing angle. Two detectors will be alternately moved into the beam position with a ``push-pull'' scheme. 

\section{ILC detector concepts}
\label{ch:ilc:detectors}

The International Linear Collider will be a perfect tool for precision studies of particle production and decays. In contrast to hadronic interactions, like those to be realised at the LHC, the e$^{+}$e$^{-}$ collisions are characterised by much lower background. The experimental conditions in the ILC provide an accurate knowledge of the initial conditions like centre-of-mass energy, initial state helicity and charge. A unique feature of electron beams is their possible polarisation allowing to study specific processes requiring knowledge of the spin states.\\
The ILC detectors will not need to cope with extreme data rates and high radiation doses. In fact events may be record events without preselection. The challenge for the subdetectors of the ILC is the unprecedented precision of measurements.

\subsection{Challenges for detector design and technology}
\label{ch:ilc:detectors:challanges}

\subsubsection{Challenges for the calorimetry}
\label{ch:ilc:detectors:challanges:cal}

Numerous processes of interest at the ILC are characterised by multi-jet final states, often accompanied by charged leptons or missing energy, e.g. the measurement of the higgs self-coupling using the following sequence$e^{+}e^{-} \rightarrow ZHH \rightarrow qqbbbb$ \cite{ilc:HihggsSelf-Coupling1,ilc:HihggsSelf-Coupling2}, higgs mass measurement in 4-jet channel $e^{+}e^{-} \rightarrow ZH \rightarrow qqbb$ \cite{ilc:HiggsMass}, branching fraction for $e^{+}e^{-} \rightarrow ZH \rightarrow ZWW^{*}$ \cite{ilc:Branchingfraction} or measurement of the cross section for $e^{+}e^{-} \rightarrow \nu \overline{\nu}W^{+}W^{-}$ \cite{ilc:CrossSection}. Reconstruction of the invariant mass of two or more jets in the final state will provide identification of the W, Z and H bosons, top quarks and new states or decay modes. The physics program of the ILC requires the di-jet mass resolution to be comparable with the natural decay width of the parent particle, around few GeV or less. The demanded di-jet mass resolution can be achieved with the jet energy resolution of $\Delta E/\sqrt{E} = 30\%$, which is about factor of two better than that achieved at LEP (Large Electron--Positron Collider).\\
Jet energy resolution of $\Delta E/\sqrt{E} = 30\%$ can be reached in the detector equipped with highly efficient and nearly hermetic tracking system and a calorimeter of a very fine transverse and longitudinal segmentation. The charged particle tracks will be associated with the energy deposits in the calorimeter. By excluding the energy deposited in the calorimeter by charged particles but including that from neutral hadrons and photons a significant improvement in the overall jet resolution is possible. Simulations of the so called ``particle flow'' approach have shown that the calorimeter should posses the following features:
\begin{itemize}
 \item 
 very fine granularity in the transverse and longitudinal directions -- segmentation of the order of 1$\times$1~cm$^{2}$ in the electromagnetic and hadronic calorimeter provides jet energy resolution $\Delta E/\sqrt{E} = 15\%$ and $\Delta E/\sqrt{E} = 40\%$, respectively,
 \item
 large inner radius and high magnetic field - important for good separation of the charged and neutral particles contained in jets,
 \item
 material providing compact electromagnetic shower development. 
\end{itemize}
The challenging jet energy resolution has inspired another approach as well. A transversely segmented, dual readout compensating calorimeter which also promises excellent jet energy resolution.

\subsubsection{Challenges for tracking}
\label{ch:ilc:detectors:challanges:track}

The ILC tracking system should provide a very accurate measurement of the charged particle momentum in the full solid angle for a wide range of energies. The tracker must be built with minimal material to preserve lepton identification and high performance calorimetry.\\
Precise momentum determination is needed for a number of key measurements at the ILC. As an example the higgs mass measurement $M_{h}$ in the process $e^{+}e^{-} \rightarrow ZH \rightarrow l^{+}l^{-}X$ was simulated \cite{ilc:HiggsMass}. The uncertainty of this measurement, $\Delta$M$_{h}$, depends on the momentum resolution \cite{ilc:HiggsMass}:
\begin{equation}
\label{eq:ilc:sigma_track_momentum}
	\frac{\Delta p_{t}}{p_{t}^{2}} = a \oplus \frac{b}{p_{t}\sin{\theta}},
\end{equation}
where $p_{t} = p\sin{\theta}$ is the transverse momentum and $a$, $b$ are constants. When the $Z$ boson decays in the leptonic channel it is possible to reconstruct the mass of an object recoiling against it with high precision and without any assumptions on the nature of the recoiling particles or its decays. Results for $\Delta$M$_{h}$ for various sets of parameters are summarised in tab.~\ref{tab:ilc:Higgs_Mass_Recoil}. The precise measurement of the higgs mass ($\Delta$M$_{h} < 100~MeV$) requires $a \leqslant 2\times10^{-5}$~GeV$^{-1}$ and $b \leqslant 1\times10^{-3}$.
\begin{table}[!htbp]
\begin{center}

	\begin{tabularx}{0.5\textwidth}{@{\extracolsep{\fill}} |>{\small}c|>{\small}c|>{\small}c|} \hline

	a [10$^{-5}$~GeV$^{-1}$] & b [10$^{-3}$] & $\Delta$M$_{h}$~[MeV] \\ \hline \hline
	1.0 & 1.0 & 85 \\ \hline
	2.0 & 1.0 & 103 \\ \hline
	4.0 & 1.0 & 153 \\ \hline
	8.0 & 1.0 & 273 \\ \hline \hline
	2.0 & 0.5 & 96 \\ \hline
	2.0 & 1.0 & 103 \\ \hline
	2.0 & 2.0 & 124 \\ \hline
	2.0 & 4.0 & 188 \\ \hline
	\end{tabularx}

\caption[The precision of the recoil higgs mass for the several values of parameters characterising the tracker momentum resolution of the tracker]{The precision of the recoil higgs mass for the several values of parameters characterising the tracker momentum resolution of the tracker, according to formula (\ref{eq:ilc:sigma_track_momentum}).}
\label{tab:ilc:Higgs_Mass_Recoil}

\end{center}
\end{table}\\
The recoil mass of the hypothetical higgs boson also depends on the precision of the centre-of-mass energy measurement. Precisely set centre-of-mass energy of the colliding beams, $E_{cm}$, is important for many physics studies, and a major effort will be devoted to measuring the beam energies before and after the interaction. Since the $E_{cm}$, measured upstream and downstream of the interaction point can differ from the luminosity-weighted $E_{cm}$ it is important to enable comparing such measurements with a direct detector measurement of the centre-of-mass energy based on physics events. Precise measurement of the luminosity-weighted $E_{cm}$ can be performed by studying the muon pair produced in the following processes $e^{+}e^{-} \rightarrow \mu^{+}\mu^{-}$ or $e^{+}e^{-} \rightarrow Z\gamma \rightarrow \mu^{+}\mu^{-}\gamma$ \cite{ilc:HiggsMass}. The combined measurement of the muon angles and their momenta provides $E_{cm}$ determination with an accuracy of about 20~MeV. It is also important to provide an excellent forward tracking with a minimal material budget to minimise multiple scattering.\\
The ILC tracker momentum resolution $\Delta p_{t}/p_{t}^{2}$ is expected to be below $5\cdot10^{-5}$~GeV$^{-1}$ \cite{ilc:GDE_Detector}.

\subsubsection{Challenges for vertexing}
\label{ch:ilc:detectors:challanges:vertex}

The ILC vertex detector is conceived to determine very precisely the actual e$^{+}$e$^{-}$ interaction coordinates and to reconstruct vertices of decaying secondary particles (determine their invariant mass which provide flavour tagging). The required precision of the impact parameter reconstruction must be better than 5$\mu$m$\oplus$10$\mu$m/($p\sin{^{3/2}\theta}$).\\
The vertex detector also plays a significant role in the global tracking. Excellent single point resolution of pixel detectors ($\sim1~\mu$m), which are currently used in construction of the vertex detectors, can provide the seeds for recognising tracks in forward and central trackers (global fitting). Moreover, a multi-layer vertex detector provides efficient and standalone pattern recognition and even momentum measurement, which may be essential in measuring low transverse momentum tracks.\\
The ILC vertex detector has to cope with the machine background, so called beam\-sstrah\-lung, which is the source of a large number of low momentum e$^{+}$e$^{-}$ pairs. The latter generate approx. 100~hits/mm$^{2}$/train in the innermost layer of the vertex detector (it is one order of magnitude more than pattern recognition algorithms can handle). The above considerations stimulate active development of many new technologies providing much faster readout electronics and high radiation resistance.
 
\subsection{The Silicon Detector (SiD)}
\label{ch:ilc:detectors:SiD}

\begin{figure}[!htbp]
        \begin{center}
                \resizebox{0.5\textwidth}{!}{
                        \includegraphics[]{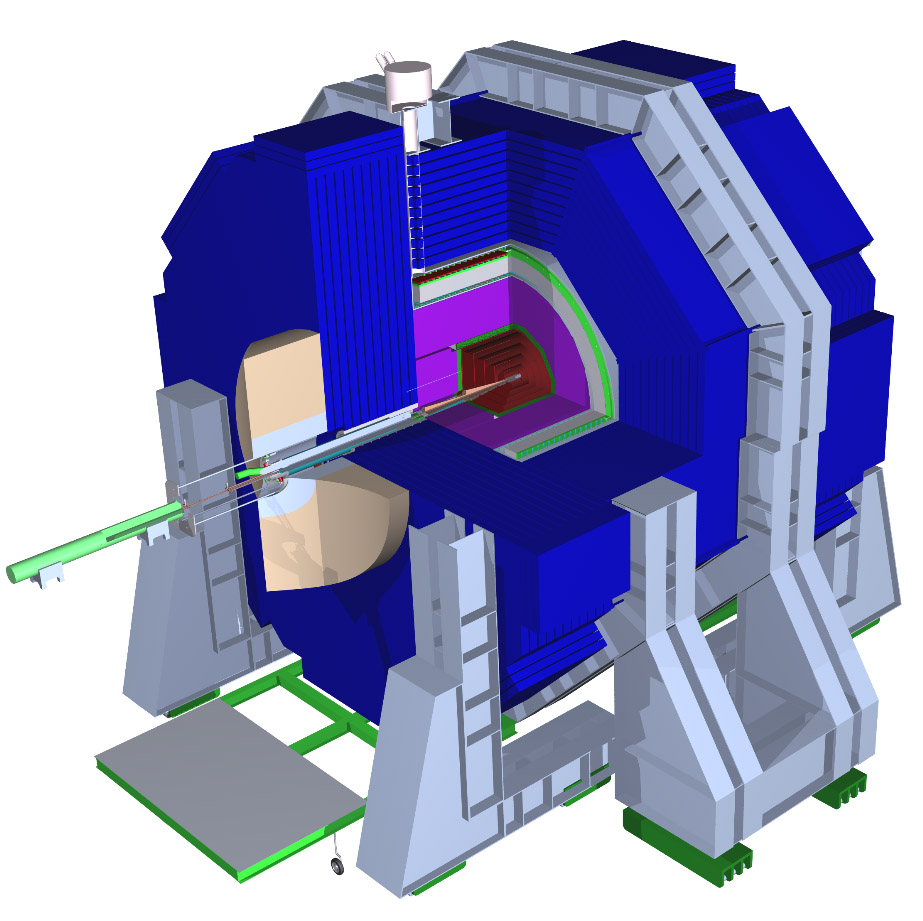}
                }
                \caption[Silicon detector concept]{Silicon detector concept (from~\cite{ilc:SiD_LoI}).}
                \label{fig:ilc:SiD_Concept}
        \end{center}
\end{figure}

The SiD (Silicon Detector) concept is based on silicon technologies~\cite{ilc:SiDConcept}, fig.~\ref{fig:ilc:SiD_Concept}. It consists of a silicon tracking system, silicon-tungsten hadronic calorimeter and a muon system. Silicon detectors are fast and robust, and they can be finely segmented. Due to the fast readout of the silicon detectors, the SiD system will only record background associated with the single bunch crossing accompanying a physics event, minimising the overlay background.\\
The SiD is optimised for the Particle Flow Approach. Thus electromagnetic and hadron calorimeters of high granularity are to be used. Achieving excellent jet energy resolution requires both calorimeters to be located within the solenoid. Since a high granularity silicon-tungsten calorimeter and a large solenoid are expensive, the cost considerations push the design to be as compact as possible, with the minimum possible radius and length. The reduced radius of the solenoid is compensated with the high magnetic field of 5~Tesla, improving the separation of charged and neutral particles in the calorimeter. The high magnetic field in conjunction with an excellent intrinsic resolution of the silicon sensors will provide superb charged particle momentum resolution, despite the limited radius of the tacking system. The high magnetic field strongly constrains beamsstrahlung e$^{+}$e$^{-}$ pairs to the small radius and allows a beam-pipe of minimal radius for high performance vertex detection.
\\
The Silicon Detector has the following components:
\begin{itemize}
 \item 
 In order to provide excellent pattern recognition and impact parameter resolution, over the full solid angle, the vertex detector based on the silicon pixel detectors has five barrel layers of 12.5~cm length ended with four endcap layers on each side. The inner radius of the vertex is 1.4~cm and the outer layer is 6.1~cm.
 \item
 The central tracker consists of five cylindrical layers of axially oriented silicon strip detectors. Both sides of the single layer are completed with the endcap disks. The latter are equipped with crossed pairs single-sided strip sensors. Individual layer are only 0.8\%~$X_{0}$ thick, including the sensor, readout ASICs (Application Specific Integrated Circuit) and cables.
 \item
 The electromagnetic calorimeter (ECAL) of 29~$X_{0}$ depth is a 3.59~m long cylinder of the 1.27~m inner radius, closed with two endcaps. It consists of 30 alternating layers of silicon pixel sensors of 14~mm pitch and tungsten absorber. The first 20 layers of the absorber feature 2.7~mm thickens while the last 10 are two times thicker. The gap between absorber layers is of the order of 1~mm in order to preserve a small Moliere radius. This design should provide jet energy resolution of $\Delta E/\sqrt{E} = 17\%$.
 \item
 The hadronic calorimeter (HCAL), following the ECAL, is composed of 40 layers of 20~mm Fe absorber slices, separated with 12~mm thick gaps housing highly pixellated detectors (1~cm$^{2}$). There are four pixel detector options under consideration: RPC (Resistive Plate Chamber), GEM (Gas Electron Multipliers), MicroMegas (MICRO-MEsh GASeous detector) and scintillators. The HCAL consists of 5.54~m long cylindrical barrel of 1.4~m inner radius, closed with two disks. The total depth of the HCAL is 4 interaction lengths.
 \item
 The 5.6~m long cylindrical 5 Tesla solenoid is based on the CMS design. With the inner radius of 2.5~m and outer radius of 3.3~m it envelopes the tracker and both calorimeters. The high field provides: high momentum resolution in the tracker, separation of particles entering the calorimeters and constraint of the beamsstrahlung e$^{+}$e$^{-}$ pairs within small radii.
 \item
 The most outer component of the SiD detector is the iron yoke which returns the magnetic flux. It consists of 23 layers of 10~cm iron plates which are separated with the gaps housing the RPCs or scintillator strips for muon detection. The iron yoke is built as a barrel with the inner radius of 3.33~m and the outer radius of 6.45~m.
 \item
 The SiD is also equipped with the forward system composed of a luminosity calorimeter and a beamcal to catch very forward produced particles.
\end{itemize}
The tracking algorithms developed for the SiD will exploit information from the vertex detector, central tracker and electromagnetic calorimeter. Since most of the tracks are found in the vertex detector it plays a key role in track pattern recognition. The tracks reconstructed in the vertex detector are extrapolated to the central tracker where additional hits are used for more accurate reconstruction of their curvature. This procedure misses roughly 5\% of tracks because they result from neutral decays outside the vertex detector. Those originating from within the second layer of the central tracker are reconstructed by the standalone central tracking algorithm. Tracks produced by decays beyond the second layer of the central tracker, but within the ECAL, are captured with a calorimeter-assisted tracking algorithm. The latter uses the track entry points to the calorimeter and their directions as seeds for extrapolation backward into the tracker. The described tracking approach provides high precision of momentum measurements for which the value of the $a$ parameter in the (\ref{eq:ilc:sigma_track_momentum}) is below $2\cdot10^{-5}$~GeV$^{-1}$. 

\subsection{The International Large Detector (ILD)}
\label{ch:ilc:detectors:ILD}

The International Large Detector (ILD) is based on two previous designs of the GLD (Gaseous Large Detector) and the LDC (Large Detector Concept) presented in~\cite{ilc:GLDConcept} and \cite{ilc:LDCConcept}, respectively.\\
The central component of the ILD tracker is a Time Projection Chamber (TPC) which provides up to 224 precise measurements along the high momentum track of a charged particle. This is supplemented by a system of silicon based tracking detectors, which provide additional measurement points inside and outside of the TPC, and extend the angular coverage down to very small angles. The $b$ and $c$ quark tagging as well as vertex reconstruction is performed with an accurate vertex detector based on silicon pixel detectors. The precise jet energy resolution is provided by the highly granular calorimeter working in conjunction with the accurate tracking system (Particle Flow Approach). The tungsten absorber based electromagnetic calorimeter (ECAL) covers the first interaction length, followed by a steel based sampling hadronic calorimeter (HCAL). Several ECAL and HCAL readout technologies are being studied. 
\begin{figure}[!htbp]
        \begin{center}
                \resizebox{0.5\textwidth}{!}{
                        \includegraphics[]{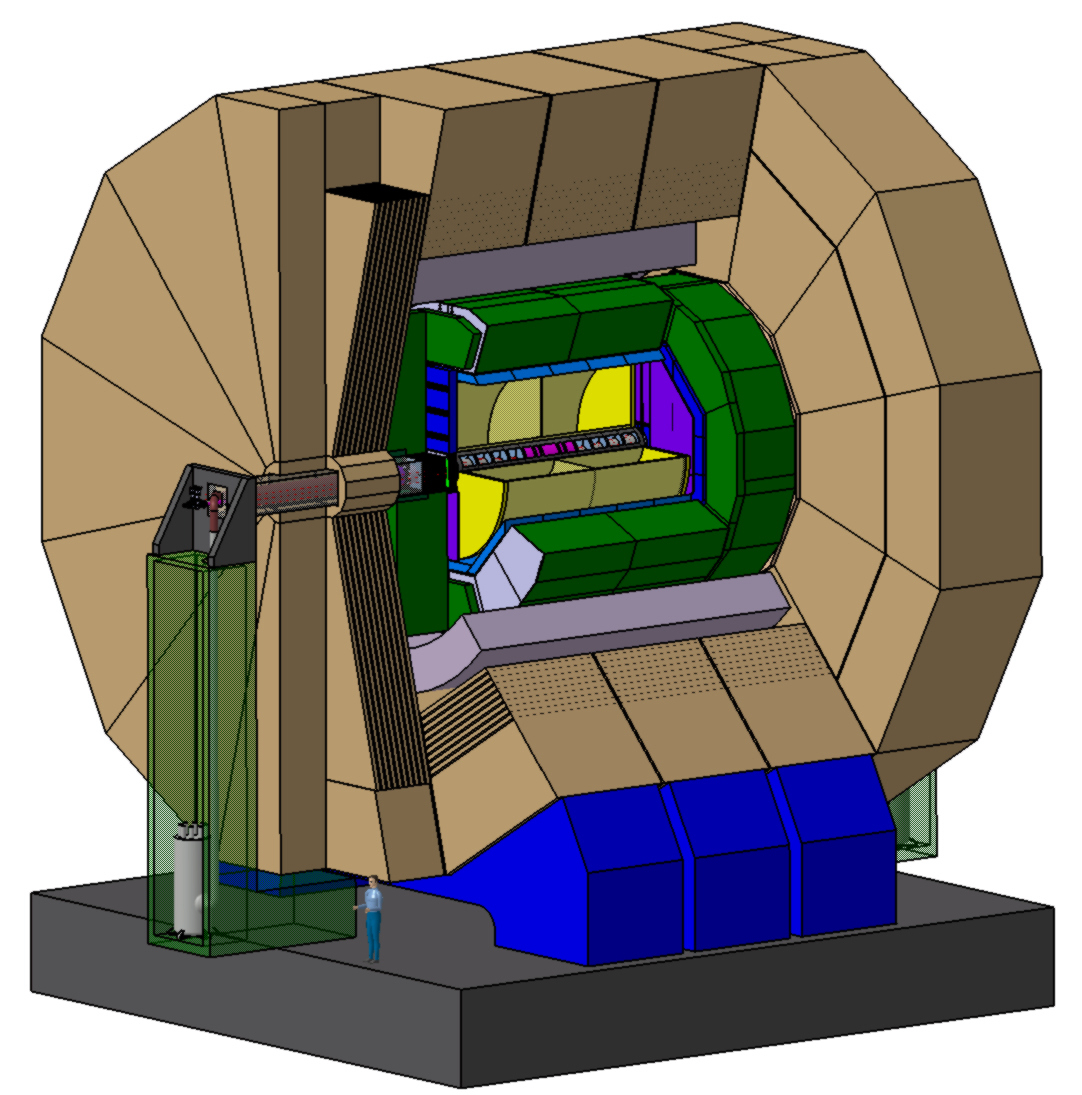}
                }
                \caption[International large detector concept]{International large detector concept (from~\cite{ilc:ILD_LoI}).}
                \label{fig:ilc:ILD_Concept}
        \end{center}
\end{figure}\\
The detailed description of the ILD (fig.~\ref{fig:ilc:ILD_Concept}) can be found in~\cite{ilc:ILD_LoI}. Below a basic layout of the ILD is presented:
\begin{itemize}
 \item
 A multi-layer pixel-vertex detector (VTX), composed of three super-layers (VTX-DL), each comprising two layers, or five single layers spaced at equal distances to the IP (VTX-SL). To minimise the occupancy from background hits, the first super-layer (single layer) is only half as long as the outer two (four). In either case the vertex detector has a purely barrel geometry of 15/16~mm inner radius and 60~mm outer radius.
 \item 
 A system of strip and pixel detectors surrounding the VTX detector. In the barrel, two layers of Si strip detectors (SIT) are arranged to bridge the gap between the VTX and the TPC. In the forward region, a system of Si-pixel and Si-strip disks (FTD) provides low angle tracking coverage.
 \item
 A large volume time projection chamber (TPC) with up to 224 points per track. It is a 4.5~m long cylinder of 1.8~m outer radius. The TPC is optimised for excellent 3-dimensional point resolution and minimum material in the field cage ($\sim$0.04~$X_{0}$) and in the end-plate ($\sim$0.15~$X_{0}$). It also provides dE/dx based particle identification capabilities. Three possible technologies are considered for reading out the ILD TPC: MPGD (Micro-Pattern Gas Detectors), GEM (Gas Electron Multipliers) and MicroMegas (MICRO-MEsh GASeous detector). 
 \item
 A system of Si-strip detectors, one behind the end-plate of the TPC (ETD) and one in between the TPC and the ECAL (SET). These provide additional high precision space points which improve the tracking measurements and provide additional redundancy in the regions between the main tracking volume and the calorimeters.
 \item
 A sampling tungsten electromagnetic calorimeter (ECAL) consisting of 4.7~m long octagonal barrel and two endcaps subdivided into 4 quadrants. The ECAL barrel is composed of 8 identical parts (so-called slaves) segmented along beam axis into 5 modules. Single module is divided approximately to 30 tungsten absorber layers, possibly with varying thickness, separated with silicon detectors or scintillators of fine segmentation 5~-~10~mm. The ECAL is 24~$X_{0}$ deep. 
 \item
 A highly segmented hadron calorimeter (HCAL) with up to 48 longitudinal samples and small transverse cell size. Two options are considered, both based on a Steel-absorber structure. One option uses scintillator tiles of 3$\times$3~cm$^{2}$, which are readout with an analogue system. The second uses a gas-based readout which allows a 1$\times$1~cm$^{2}$ cell geometry with a binary or semi-digital readout of each cell. The HCAL of 5.5 interaction length depth is build of short barrel (4.7~m long) and two large endcaps subdivided in to 4 quadrants. Depending on the design the barrel is subdivided along the $z$ direction into 2 sections of 8 octants or 5 sections of 8 identical modules.
 \item
 A system of high precision, radiation hard, calorimetric detectors in the very forward region (LumiCAL, BCAL, LHCAL). These extend the calorimetric coverage to almost 4$\pi$, measure the luminosity, and monitor the quality of the colliding beams.
 \item
 A large volume superconducting coil surrounds the calorimeters, creating an axial B field of nominally 3.5~Tesla. The iron flux return yoke houses muon system based on the Plastic Steamer Tubes (PST) or Resistive Plate Chambers (RPC). The RPCs tend to be preferred over PSTs due to their reduced cost and greater flexibility in the segmentation achievable. In a single gap between iron plates two layers of RPC with an orthogonal strip orientation could be located, providing two dimensional information on the muon track.
\end{itemize}
The expected ILD TPC momentum resolution for 3.5~Tesla magnetic field is $\Delta p_{t}/p_{t}^{2} = 9\cdot10^{-5}$~GeV$^{-1}$ while with a support from other tracking subsystems (SET, SIT, VTX) it improves and amounts $\Delta p_{t}/p_{t}^{2} = 2\cdot10^{-5}$~GeV$^{-1}$.

\subsection{The 4-th detector}
\label{ch:ilc:detectors:4-th}

The so-called 4-th detector concept differs from the previously presented SiD and ILD in several aspects. The 4-th design utilises a novel implementation of the compensating calorimeter, which balances the response to hadrons and electors and so is insensitive to fluctuations in the fraction of electromagnetic energy in shower. The second innovation introduced by the 4-th concept is a dual solenoid system with endcap coils used for returning magnetic flux and identification of the muons.\\
The compensating calorimeter of the 4-th design is equipped with dual-readout fibres enabling separation of the hadronic and electromagnetic components of hadronic showers. The fibres made of scintillator (used for measuring energy of the charged particles contained in the shower) and quartz (enabling detection of the Cerenkov light generated mainly by the relativistic electrons - electromagnetic component of the shower) will be located in the channels drilled in the absorber material (tungsten or brass). It is also considered to include a third type of fibres sensitive to the low energy neutrons produced in the shower. The signals referring to the electromagnetic and hadronic shower components are readout with the photo detectors and afterwards compensated with the dedicated software. It is expected that the dual-readout compensating calorimeter can provide energy resolution of $\Delta E/\sqrt{E} = 20\%$. Obtained resolution is comparable with the one assumed for Particle Flow Calorimeters featuring much higher granularity.\\
In front of the dual-readout calorimeter an electromagnetic section is placed. It consists of crystals sensitive to both scintillation and Cerenkov light. With the electromagnetic calorimeter a better energy and spatial resolution for photons and electrons is provided than in the fibre calorimeter.\\
The dual-solenoid system surrounds the detector and provides 3.5~Tesla magnetic field inside the detector volume. The flux from the inner solenoid is returned by the outer one which is oppositely driven with a smaller turn density. The dual-solenoid field is terminated by a novel ``wall of coils``. In the gap between solenoids a magnetic field of 1.5~Tesla is present. With an addition tracking system located in between solenoids it is possible to measure momentum of muons which have penetrated calorimeter. This solution introduces a low material budget to the muon system, in contrast to the conventional systems exploiting the iron yokes, resulting in a high momentum resolution of $\Delta p_{\mu}/p_{\mu}^{2} \approx 10^{-4}$~GeV$^{-1}$. 
\begin{figure}[!htbp]
        \begin{center}
                \resizebox{0.8\textwidth}{!}{
                        \includegraphics[]{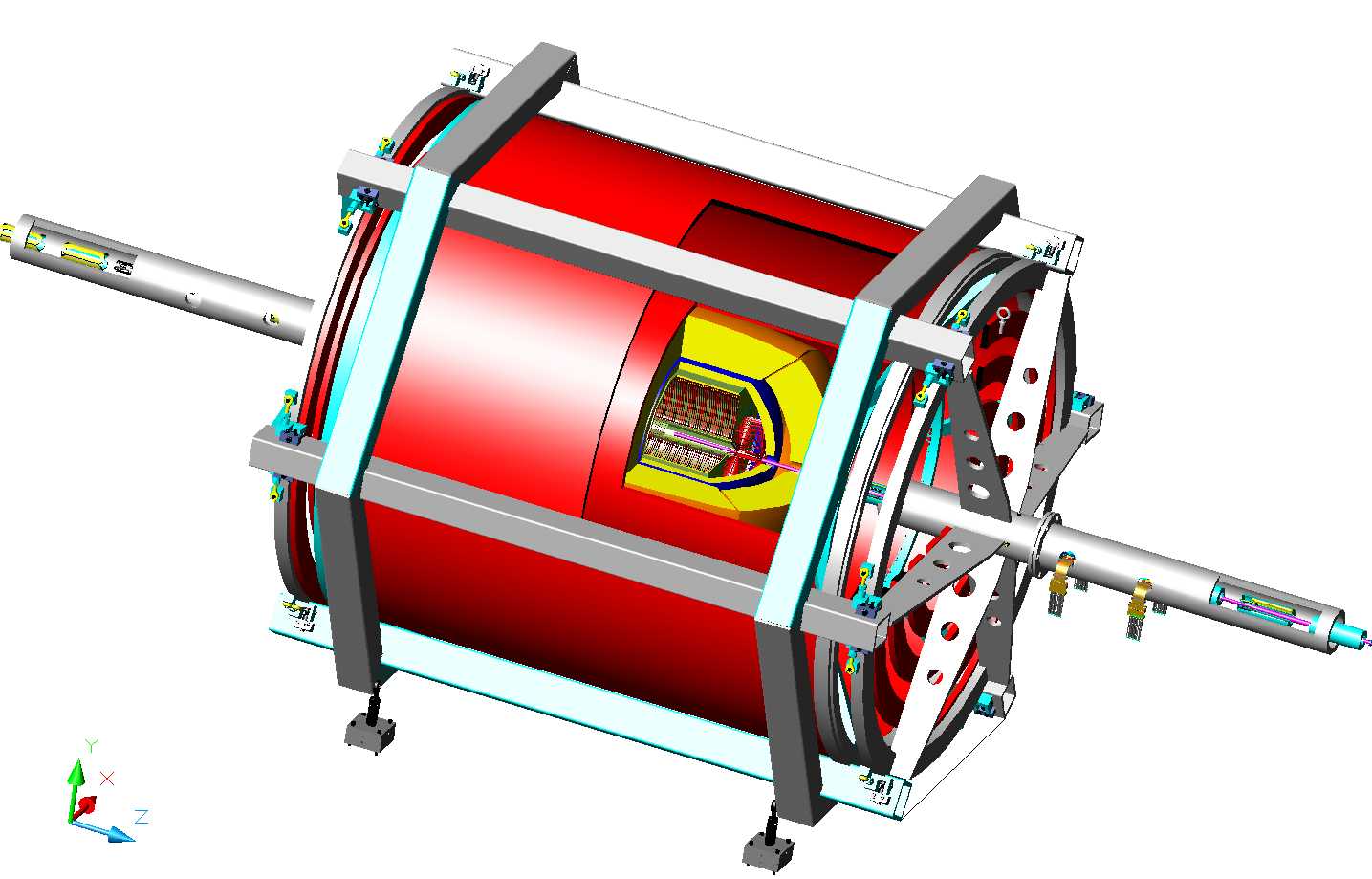}
                }
                \caption[4-th detector concept]{4-th detector concept (from~\cite{ilc:4thConcept}).}
                \label{fig:ilc:4th_Concept}
        \end{center}
\end{figure}\\
The 4-th detector (fig.~\ref{fig:ilc:4th_Concept}) contains following key elements~\cite{ilc:4thConcept} :
\begin{itemize}
 \item
 The pixel vertex detector for $b$ and $c$ quark tagging and accurate vertex reconstruction. The 4-th concept vertex detector features the same design as the one developed for SiD, with the inner and outer radii of 1.5~cm and 8~cm, respectively, in a 3.5~Tesla field.
 \item
 The Time Projection Chamber (TPC) for charged particles tracking is similar to the one being developed for the ILD detector. It is a 3~m long cylinder with an inner and outer radii of 20~cm and 1.4~m, respectively.
 \item
 The electromagnetic calorimeter (ECAL) is build of 2$\times$2$\times$30~cm$^{3}$ crystals, providing readout of the scintillator and Cerenkov light. It is a barrel with inner (outer) radius of 1.5~m (1.8~m), completely surrounding the TPC.
 \item
 The hadron calorimeter (HCAL), made of tungsten or brass, is 100~cm deep (10 interactions length). It is divided onto modules of 5$\times$5$\times$100~cm$^{3}$. In each module a number of channels is drilled. The latter, distributed uniformly with an interval of 2~mm, contain fibres of two or three types: scintillator fibres, Cerenkov fibres or fibres enriched with lithium (Li) or boron (B) for low energy neutron detection. The fibres are aiming to the interaction point with an approximately accuracy of 1$^{\circ}$. The fibres are readout with photon detectors located at the HCAL outer radius.
 \item
 The dual-solenoid system is responsible for providing magnetic field in the 4-th detector. The inner solenoid of 3.01 - 3.15~m radius generates magnetic field of 5.17~Tesla. The outer one of 5.39 - 5.53~m provides inverse magnetic field of 1.6~Tesla and returns the flux of the first solenoid. Due to the superposition of the both fields the resultant magnetic field inside TPC is 3.55~Tesla. The dual-solenoid field is terminated by a novel ''wall of coils`` placed on the both sides of the detector.
 \item
 The muon system is based on the precise tracking, consisting of high precision drift tubes of 23~mm radii with cluster counting electronics. The 20 layers of 4~m long drift chambers are placed in the gap between solenoids in the magnetic field of 1.5~Tesla. The low material budget (no iron absorber as distinct from the conventional muon detectors) and the dedicated readout electronics placed on the both sides of the muon detector provides unprecedented precision of muon properties measurements:
\begin{itemize}
 \item
 the muon track position with respect to the centre of the single drift chamber is known with an approximately accuracy of 50~$\mu$m.
 \item
 the muon angular orientation with respect to the drift chambers is determined with a precision of about 100~mrad.
 \item
 measurement of the track position along the drift chamber is performed with a precision below 1~mm.
 \item
 measurement of the $dE/dx$ done with an accuracy of about 3\%. 
\end{itemize}
The muon detector is a 12~m long barrel with an inner and outer radii of 3.5 and 5.5~m, respectively.
\end{itemize}

The ILD, SiD and 4-th detector groups submitted Letters of Intent (LOI) by March 2009. The proposed concepts were evaluated on common grounds by the International Detector Advisory Group (IDAG) \cite{ilc:IDAG}. The ILD and SiD designs were validated while the 4-th concept not, however the R\&D on dual readout calorimetry was found to be supported in view of its potential for high energy colliders.

\chapter{The ILC Vertex Detector}
\label{ch:vertex}

The ILC vertex detector (VTX) will be based on thin silicon pixel sensors of very high granularity, i.e. small pixel pitch. The pixel matrices will be arranged in ladders, distributed in several coaxial, cylindrical layers surrounding the primary interaction point (IP). The sensitive areas in the neighbouring ladders overlap to avoid dead zones. Two geometrical designs which provide almost full solid angle coverage are considered: the extended barrel or a short barrel with endcaps (disks).\\
The VTX is a key component for a precise flavour identification which is achieved by reconstruction of secondary decay vertices. Weakly decaying particles, like $\tau$ leptons or hadrons containing $b$ or $c$ quarks, can pass on average a distance of a few hundred micrometers from the IP. Since the jet flavour reconstruction (beauty or charm) or identification of $\tau$ lepton decays improve with decreasing radius of the first VTX layer, the innermost layer has to be located close to the interaction point. This creates a major technical challenge in view of intense background near the beam pipe.\\
The VTX is also used for reconstructing the primary vertex as well as track reconstruction, especially for low momentum particles which do not reach the main tracker. It also plays an important role in physics studies requiring knowledge of the total vertex charge.\\
The accurate measurement of the impact position in the VTX requires, beside excellent single position resolution (below 5~$\mu$m), a low material budget of the sensors and supporting structures, to minimise the multiple Coulomb scattering. Precision of the impact position measurement can be expressed in terms of two parameters $a$ and $b$:
\begin{equation}
\label{eq:vxd:sigma_ip}
	\sigma_{ip} = a \oplus \frac{b}{p\cdot\sin{^{3/2}\theta}},
\end{equation}
which are required to be below 5~$\mu$m and 10~$\mu$m$\cdot$GeV/c, respectively. To illustrate how challenging this requirement is a comparison of the $a$ and $b$ parameters for the ILC VTX and the vertex detectors build so far (LEP, SLC, LHC) or being under development (RHIC-II) are presented in tab.~\ref{tab:vxd:ip_res_accelerators}.
\begin{table}[!h]

\begin{center}
	\begin{tabularx}{0.5\textwidth}{@{\extracolsep{\fill}} |>{\small}c|>{\small}c >{\small}c|} \hline

	Accelerator & a [$\mu$m] & b [$\mu$m$\cdot$GeV/c] \\ \hline \hline
	LEP & 25 & 70 \\
	SLC & 8 & 33 \\
	LHC & 12 & 70 \\
	RHIC-II & 13 & 19 \\
	ILC & $<$ 5 & $<$ 10 \\ \hline

	\end{tabularx}

\caption[Values of the parameters $a$ and $b$ in the formula (\ref{eq:vxd:sigma_ip}) expressing the impact position resolution at the ILC, compared to other experiments]{Values of the parameters $a$ and $b$ in the formula (\ref{eq:vxd:sigma_ip}) expressing the impact position resolution at the ILC, compared to other experiments (from \cite{ilc:ILD_LoI}).}
\label{tab:vxd:ip_res_accelerators}

\end{center}
\end{table}\\
Low material budget of the VTX is also an important issue for calorimetry, since secondary particles produced in the VTX layers deteriorate the calorimeter performances.\\
All of the above requirements drive an ambitious R\&D program for pixel sensors and its mechanics. Due to the beam related background, dominated by low $p_{t}$ e$^{+}$e$^{-}$ pairs, a special emphasis is put on improving radiation tolerance and readout speed of the sensors.\\
The intersection of 2625 bunches during one bunch train will generate approx. 80 - 100 hits per mm$^{2}$ in the innermost VTX layer. In order to handle such a high detector occupancy, several readout strategies are considered:
\begin{itemize}
 \item
 Hits can be accumulated during one bunch train and read out afterwards during the time interval between two consecutive trains. In this approach the beam related electrical interferences during read out are avoided however an overlap of hits and related clusters occurs. Each physics event is accompanied by background hits from 2625 bunches.
 \item
 Multiple and consecutive readout of sensors during a beam train duration (approx. 20 times per train) resulting in a lower number of accumulated hits and a smaller probability of their overlaps. This approach however is sensitive to beam related electrical interferences.
 \item
 In the third method hits referring to consecutive subsets of the bunch crossings (one subset refers to 131 bunch crossings) are stored in 20 cells of the in-pixel memory that are read out in the time interval between two consecutive trains. This approach provides reduction of accumulated hits in one readout sequence and is not sensitive to beam related electrical interferences. However it requires implementation of 20 in-pixel memory cells in a singe pixel which requires more advanced and expensive technology.
\end{itemize}
The pixel detector technologies taken into consideration are: CPCCD (Column Parallel CCD), FPCCD (Fine Pixel CCD), SCCCD (Short Column CCD), MAPS (Monolithic Active Pixel Sensor), FAPS (Flexible Active Pixel Sensor), CAP (Continuous Acquisition Pixel), DEPFET (DEPleted Field Effect Transistor), SOI (Silicon On Insulator) Chronopixel and 3D. They are described in details in the next chapter \ref{ch:silicon}.

\section{Geometry}
\label{ch:vertex:geometry}

\begin{figure}[!h] 
	\begin{center}
	\subfigure[]{
		\includegraphics[width=0.7\textwidth]{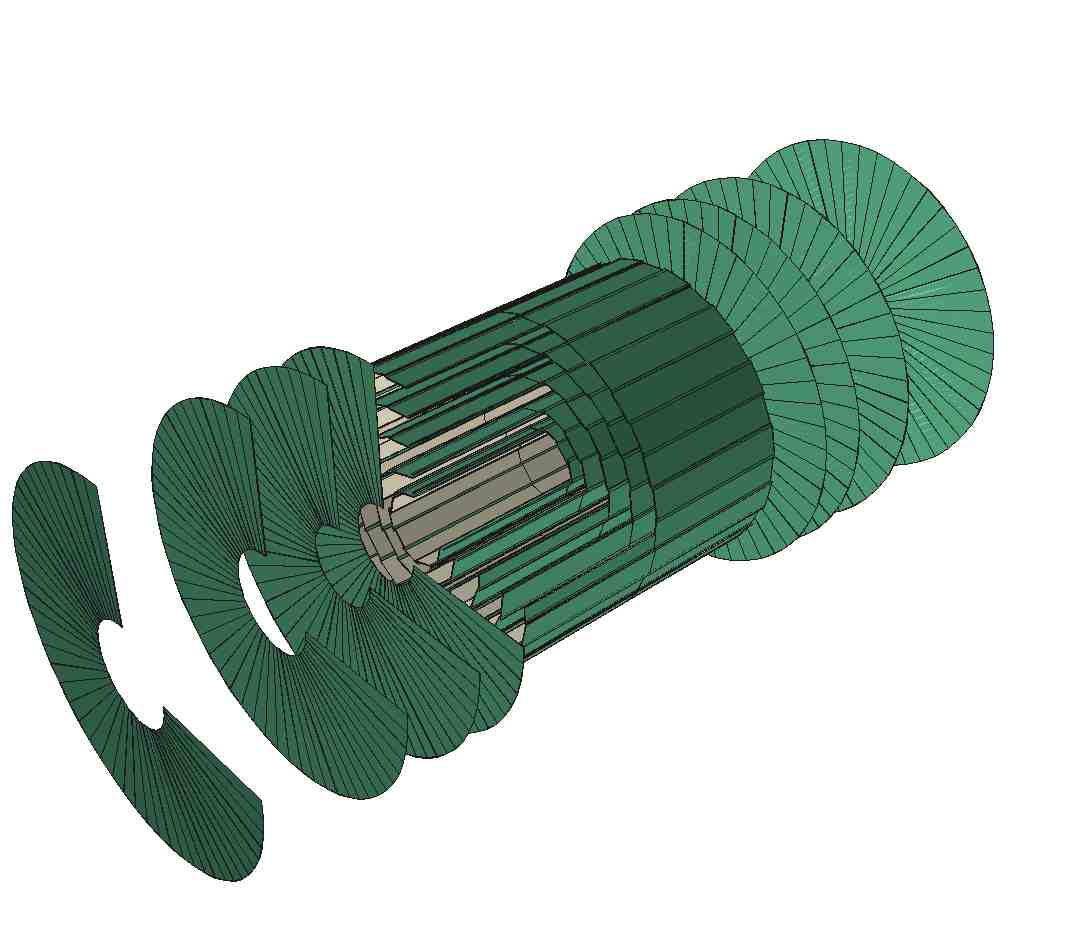}
		\label{fig:vertx:SiD}
	}
	\hspace{0.1cm}
	\subfigure[]{
                \includegraphics[width=0.7\textwidth]{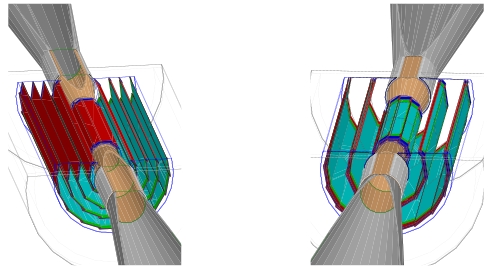}
		\label{fig:vertx:ILD}
	}
	\caption[The ILC vertex detector layouts]{The ILC vertex detector layouts proposed by (a) the SiD group ``short barrel plus forward endcaps (disks)'' \cite{ilc:GDE_Detector} and (b) the ILD group ``long cylinders'' composed of 5 single layers (VTX-SL) or 3 double layers (VTX-DL) \cite{ilc:ILD_LoI}.}
	\label{fig:vertx}
	\end{center}
\end{figure}

Two distinct VTX geometries, providing a wide polar angle ($\theta$) coverage, are considered. The one proposed by the SiD collaboration consists of a short cylindrical barrel of 5 layers completed with 4 discs on both sides \cite{ilc:SiD_LoI} (see fig.~\ref{fig:vertx:SiD}). This design provides very high hermeticity, $|\cos{\theta}| \lesssim 0.984$, however it requires a complicated supporting structures for the outer disks which introduce additional material. The second concept proposed by the ILD group is based on long cylinders without endcaps and two versions are proposed: 5 single layers (called VTX-SL) or 3 double layers (called VTX-DL) \cite{ilc:ILD_LoI} (see fig.~\ref{fig:vertx:ILD}). The VTX-DL allows spatial correlations between hits generated by the same particle in the two sensor layers located on the opposite sides of the same ladder, even if the occupancy is high. This makes the design more robust against (low momentum) beamsstrahlung background. Moreover this solution should improve reconstruction of tracks at low $\theta$ angle in the very forward region. The VTX-DL geometry may be however less efficient in reconstructing long lived $B$ mesons decaying outside of the beam pipe or low momentum tracks (due to high multiple scattering in the first layer). The expected $\theta$ angle coverage for both ILD VTX concepts is $|\cos{\theta}| \lesssim 0.97$. 
\\
Detailed informations on the layers and disks arrangements in the SiD concept and in the ILD concepts (VTX-SL and VTX-DL) are presented in tables~\ref{tab:vertex:sid_VTX} and \ref{tab:vertex:ild_VTX}, respectively.
\begin{table}[!h]

\begin{center}
	\begin{tabularx}{0.6\textwidth}{@{\extracolsep{\fill}} |>{\small}c|>{\small}c|>{\small}c|} \hline

	Barrel region & R [mm] & Length [mm] \\ \hline \hline
	Layer 1 & 14 & 125 \\
	Layer 2 & 21 & 125 \\
	Layer 3 & 34 & 125 \\
	Layer 4 & 47 & 125 \\
	Layer 5 & 60 & 125 \\ \hline

	\end{tabularx}

	\begin{tabularx}{0.6\textwidth}{@{\extracolsep{\fill}} |>{\small}c|>{\small}c|>{\small}c|>{\small}c|} \hline

	Disk & $R_{inner}$ [mm] & $R_{outer}$ [mm] & Z [mm] \\ \hline \hline
	Disk 1 & 15 & 75 & 76\\
	Disk 2 & 16 & 75 & 95\\
	Disk 3 & 18 & 75 & 125\\
	Disk 4 & 21 & 75 & 180\\ \hline

	\end{tabularx}

\caption[The SiD vertex detector parameters]{The SiD vertex detector parameters \cite{ilc:SiD_LoI}.}
\label{tab:vertex:sid_VTX}

\end{center}
\end{table}
\begin{table}[!h]

\begin{center}

	\begin{tabularx}{0.6\textwidth}{@{\extracolsep{\fill}} |>{\small}c|>{\small}c >{\small}c|>{\small}c >{\small}c|} \hline

          & \multicolumn{2}{c|}{radius~[mm]} & \multicolumn{2}{c|}{ladder length~[mm]}\\
	Layer & VTX-SL & VTX-DL & VTX-SL & VTX-DL \\ \hline \hline
	Layer 1 & 15 & 16/18 & 125 & 125\\
	Layer 2 & 26 & 37/39 & 250 & 250\\
	Layer 3 & 37 & 58/60 & 250 & 250\\
	Layer 4 & 48 &  & 250 &\\
	Layer 5 & 60 &  & 250 &\\ \hline

	\end{tabularx}

\caption[The ILD vertex detector parameters]{The ILD vertex detector parameters \cite{ilc:ILD_LoI}.}
\label{tab:vertex:ild_VTX}

\end{center}
\end{table}
The vertex detector surrounds the beryllium cylindrical beam pipe which radius is determined by the magnetic field provided by the solenoid. Since higher field is more effective for suppressing the e$^{+}$e$^{-}$ pair background, the beam pipe radius in case of the SiD (5~Tesla) is 12~mm while in case of the ILD (3.5~Tesla) it is 14~mm. At $Z=\pm6.25$~mm the beam pipe radius starts to increase conically in order to stay safely beyond the envelope of beam related e$^{+}$e$^{-}$ background. Thus the VTX innermost layer is only 125~mm long.

\section{Beamsstrahlung background}
\label{ch:vertex:bemastrahlung}

\begin{figure}[!h] 
	\begin{center}
	\subfigure[Breit-Wheeler]{
		\includegraphics[width=4cm,angle=90]{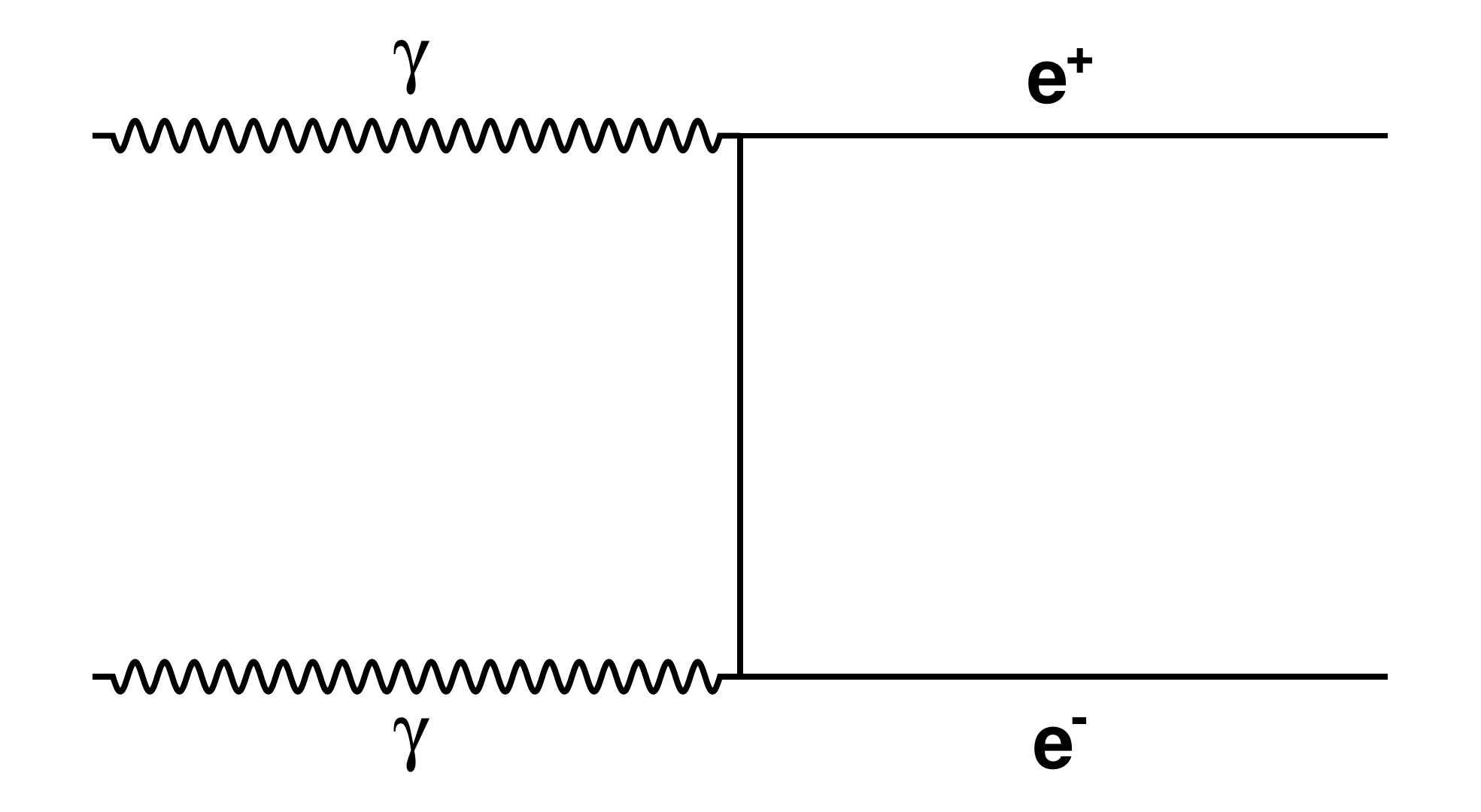}
		\label{fig:vertx:IPC:BW}
	}
	\hspace{0.1cm}
	\subfigure[Bethe-Heitler]{
		\includegraphics[width=4cm,angle=90]{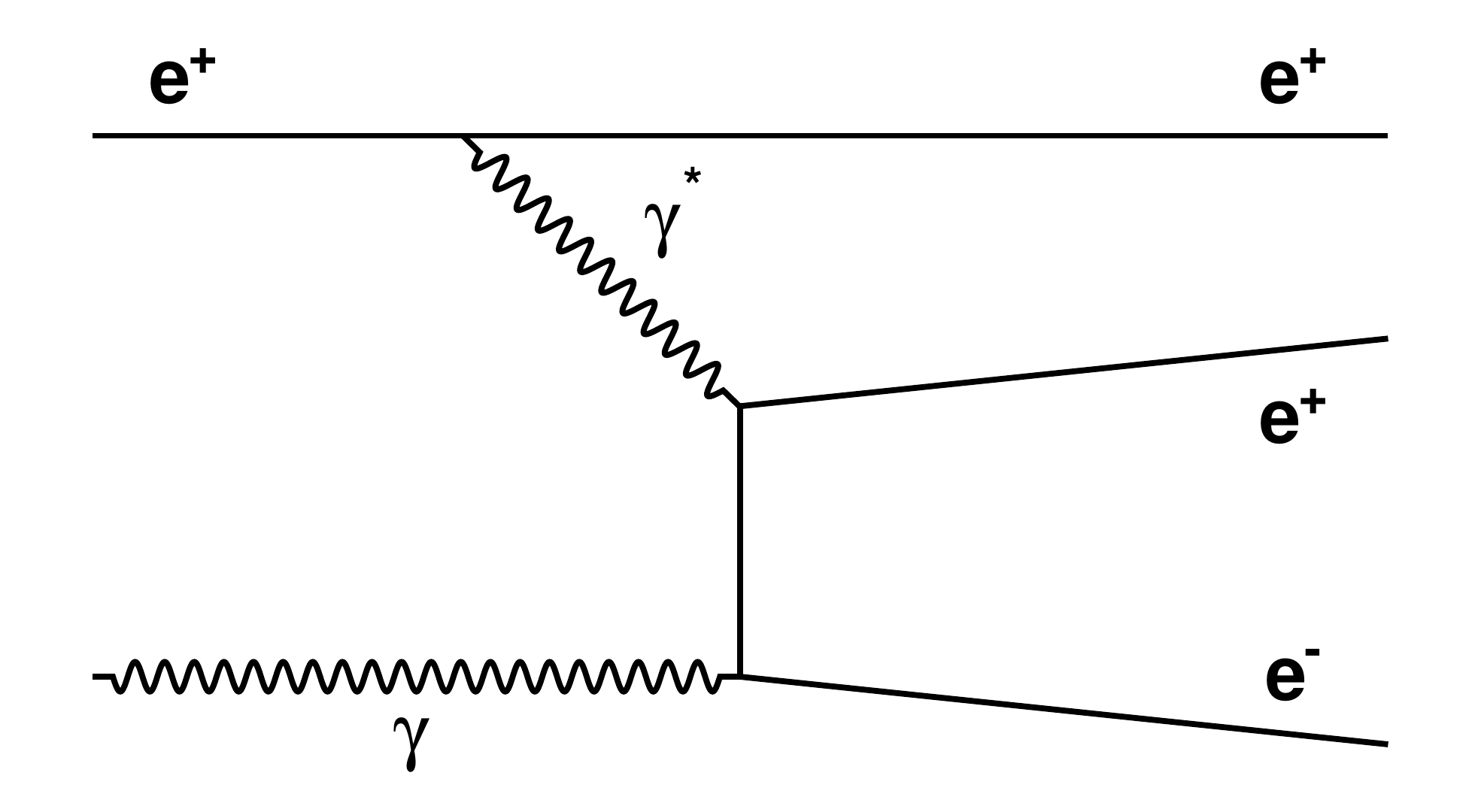}
		\label{fig:vertx:IPC:BH}
	}
	\hspace{0.1cm}
	\subfigure[Landau-Lifshitz]{
		\includegraphics[width=4cm,angle=90]{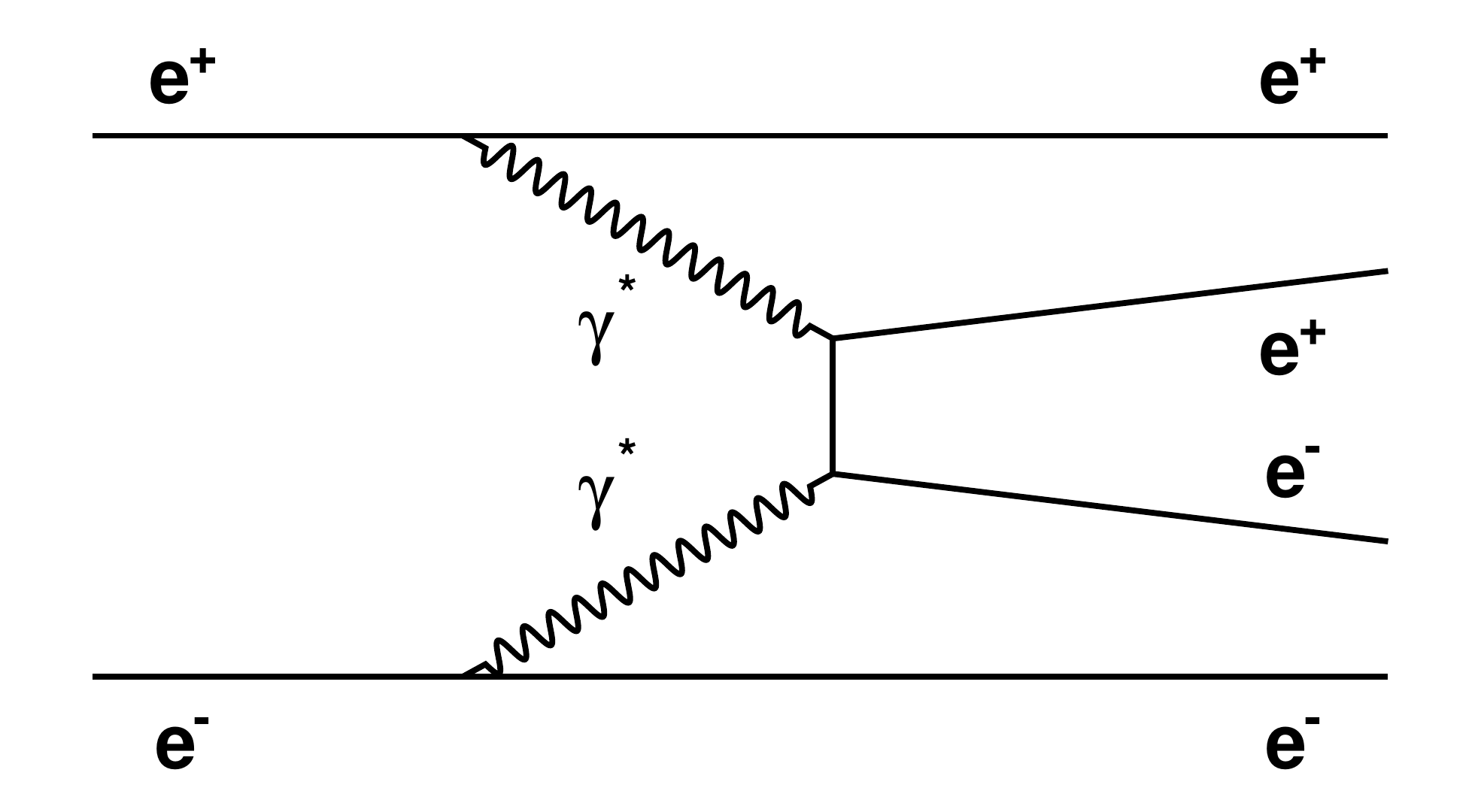}
		\label{fig:vertx:IPC:LL}
	}
	\caption[Feynman diagrams of the incoherent pair creation processes]{Feynman diagrams of the incoherent pair creation processes.}
	\label{fig:vertx:IPC}
	\end{center}
\end{figure}

\begin{figure}[!h] 
	\begin{center}
	\subfigure[]{
		\includegraphics[width=0.5\textwidth]{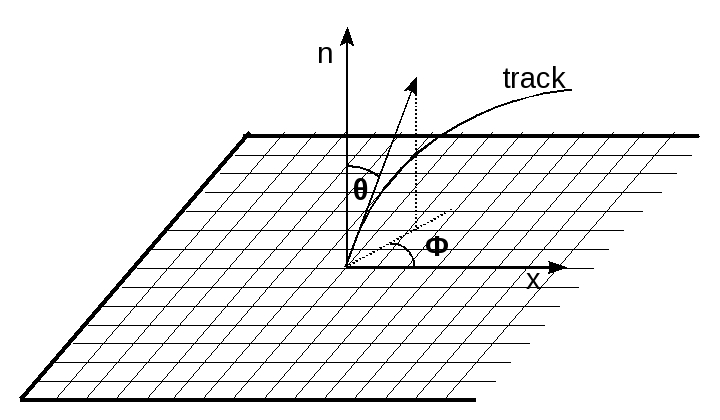}
		\label{fig:vertx:Beamstarhlung:TrackOrient}
	}
	\hspace{0.1cm}
	\subfigure[]{
                \includegraphics[width=0.4\textwidth]{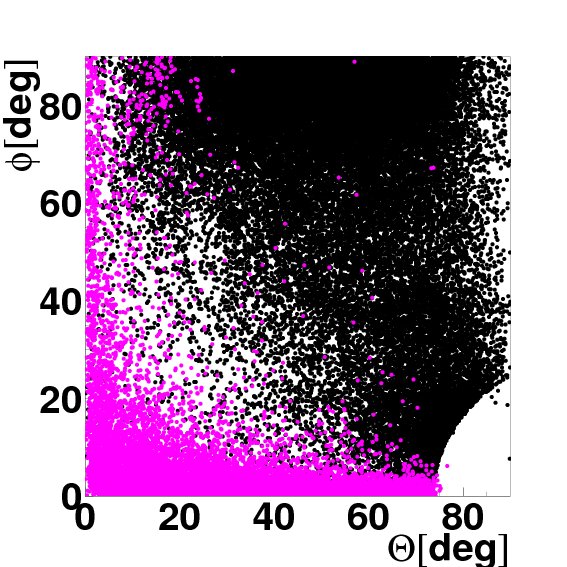}
		\label{fig:vertx:Beamstarhlung:ThetaPhi}
	}
	\caption[Definition of the polar and azimuthal angels, $\theta$ and $\phi$, and simulated distribution on the $\theta$ - $\phi$ plane of tracks from beamsstrahlung and tracks from the beamsstrahlung process in the first layer of the vertex detector]{(a) Definition of the polar and azimuthal angels, $\theta$ and $\phi$, in the pixel matrix coordinate system with the X axis parallel to the matrix edge and aligned with the beam direction; n - normal to the pixel plane. (b) Simulated distribution on the $\theta$ - $\phi$ plane of tracks from beamsstrahlung (black points) and tracks from the beamsstrahlung process, $e^{+}e^{-}\rightarrow ZH$, (purple points), in the first layer of the vertex detector (radius 15~mm). The beamsstrahlung entries correspond to the nominal 131 bunch crossings while the physics entries were scaled up by a factor of $10^{3}$ for better visibility.}
	\label{fig:vertx:Beamstarhlung}
	\end{center}
\end{figure}

Most of the particles generated during bunch crossings at the ILC will be related to the machine induced backgrounds. A single bunch crossing will deliver about $10^{5}$ background particles while the expected rate for hard electroweak interactions at the nominal luminosity of 2$\cdot$10$^{34}$~cm$^{-2}$s$^{-1}$ and cms energy of 500~GeV is below 1~Hz, even for processes that are not in the main focus of physical analyses. A major contribution to the machine induced backgrounds is due to e$^{+}$e$^{-}$ pairs of low transverse momentum $p_{t}$. They are created due to scattering of beamsstrahlung photons, radiated by the beam particles which have been deflected in the electromagnetic field of the other beam. Since the beamsstrahlung photons are strongly focused in the forward direction, most of them exit the detector through the beam pipe and only a fraction of them create e$^{+}$e$^{-}$ pairs. The beamsstrahlung pairs can be created in coherent (CPC) and  incoherent (IPC) processes. In the CPC process, which is negligible in the ILC, the beamsstrahlung photons convert in the collective electric field of the bunch. In the IPC processes, which are dominant in the ILC, the electron positron pairs are created due to scattering of two photons. Three different contributions to the IPC processes can be distinguished: collision of two real beamsstrahlung photons (the Breit-Wheeler process fig.~\ref{fig:vertx:IPC:BW}), the collision of one real and one virtual photon that is emitted by an electron or a positron in the bunch (the Bethe-Heitler process fig.~\ref{fig:vertx:IPC:BH}) and the collision of two virtual photons (the Landau-Lifshitz process fig.~\ref{fig:vertx:IPC:LL}).
\\
The beamsstrahlung e$^{+}$e$^{-}$ pairs are produced at very low polar angles and afterwords they are deflected by the beam related electromagnetic field. This leads to an increase of their transverse momentum without affecting the energy spectrum. Majority of the generated electrons and positrons stay confined to the beam pipe due to the magnetic field. However a fraction of them, with high enough $p_{t}$, can reach the vertex detector. This is the main source of background hits affecting the vertex detector. Additionally the VTX can be hit by beamsstrahlung particles which are backscattered in the forward calorimeters.\\
Simulations have shown that presence of the beamsstrahlung related hits deteriorates the VTX tracking performances and influences the heavy flavour identification \cite{vertex:PLuzniak}. Moreover, beamsstrahlung tracks intersect the VTX pixel sensor planes at different angles than the tracks of the final state particles from e$^{+}$e$^{-}$ hard interactions \cite{vertex:PLuzniak}. Track orientation with respect to the VTX sensor ladder can be parametrised in terms of two angles, $\theta$ and $\phi$, as shown in fig.~\ref{fig:vertx:Beamstarhlung:TrackOrient}: ($i$) the polar angle $\theta$ is the angle between the track and the normal to the matrix plane and ($ii$) the azimuthal angle $\phi$ is the angle between the direction of the track projected to the matrix plane and one of the matrix axes. 
\\
In fig.~\ref{fig:vertx:Beamstarhlung:ThetaPhi} distribution on the $\theta$ - $\phi$ plane of tracks from beamsstrahlung (black points) and tracks from the higgsstrahlung process, $e^{+}e^{-}\rightarrow ZH$, (grey points), in the first layer of the vertex detector (radius 15~mm) are shown \cite{vertex:PLuzniak}. Tracks of the beamsstrahlung particles, featuring low transverse momentum, having higher curvature than tracks of the final state particles from hard e$^{+}$e$^{-}$ interactions. Thus the beamsstrahlung tracks and tracks of final state particles populate different regions on the $\theta$-$\phi$ plane. According to fig.~\ref{fig:vertx:Beamstarhlung:ThetaPhi} it is expected that clusters arising from the beamsstrahlung tracks are elongated in the direction perpendicular to the beam axis while clusters arising from tracks of the final state particles present elongation in the beam direction. Measurement of the cluster orientation with respect to the pixels netting could be used for distinguishing between clusters originating from beamsstrahlung or physics tracks. This method could be exploited online or offline to reject a fraction of beamsstrahlung clusters from the data sample.

\chapter{Silicon detectors}
\label{ch:silicon}

\section{Properties of silicon}
\label{ch:silicon:properties}

Silicon has the energy gap of 1.1~eV which is low enough to produce large numbers of charge carriers by minimum ionising particles (about 80 electron-hole pairs per one micron of track length). Thus it is possible to build thin detectors that produce signals large enough to be measured. On the other hand, the energy gap of 1.1~eV is high enough to avoid large dark currents at room temperature. High mobility of electrons and holes in silicon at room temperatures results in a very fast charge collection of order of ns. Thus silicon detectors can be used in high-rate environments. Moreover, silicon exhibits excellent mechanical rigidity which allows construction of self-supporting structures.\\
Since silicon is the basic material used in the integrated circuit industry (IC) there is a worldwide experience in growing large silicon crystals of excellent purity, $n$-type and $p$-type doping, growing highly insulating layers like $SiO_{2}$ and building readout and on-chip signal processing microcircuits which can be integrated on the same substrate as a detector.

\subsection{Conduction in pure and doped semiconductors}
\label{ch:silicon:properties:conduction}

\begin{figure}[!h] 
	\begin{center}
	\subfigure[]{
		\includegraphics[width=4.6cm]{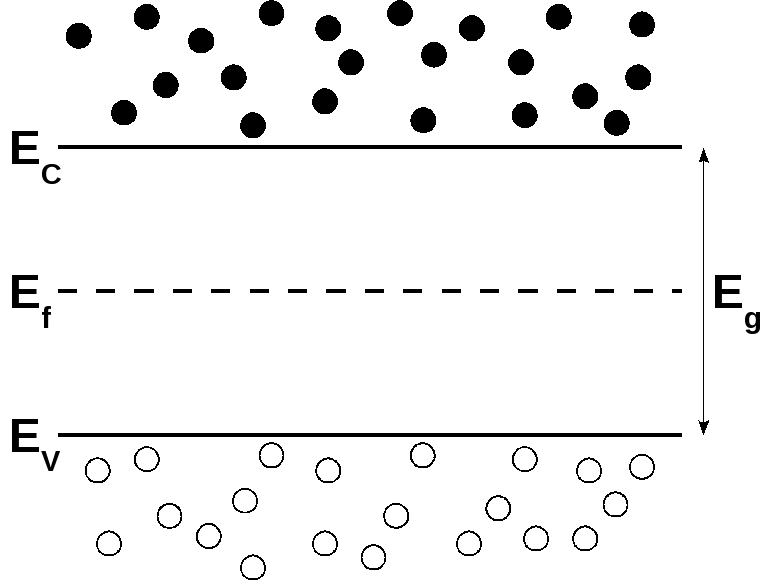}
		\label{fig:silicon:semiconductor:intrisic}
	}
	\hspace{0.1cm}
	\subfigure[]{
		\includegraphics[width=4.6cm]{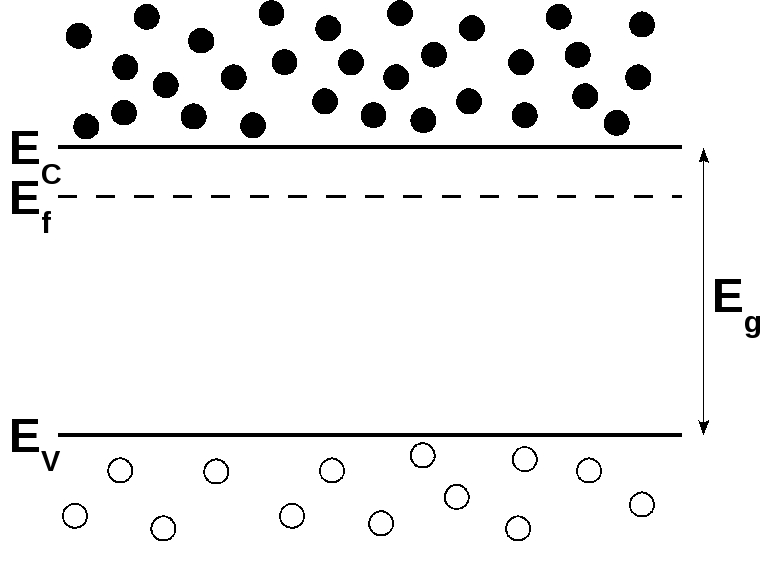}
		\label{fig:silicon:semiconductor:n-type}
	}
	\hspace{0.1cm}
	\subfigure[]{
		\includegraphics[width=4.6cm]{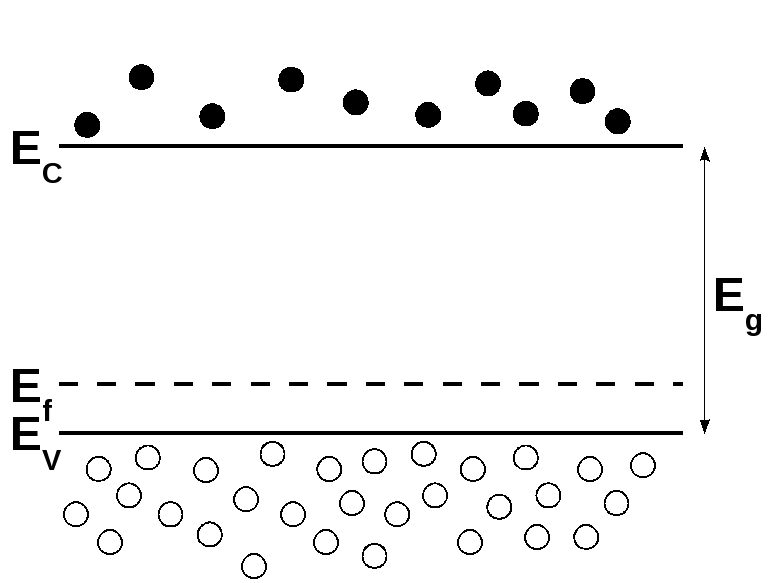}
		\label{fig:silicon:semiconductor:p-type}
	}
	\caption[Structure of energy levels in a semiconductor.]{Structure of energy levels in (a) intrinsic, (b) n-type and (c) p-type semiconductors. $E_{C}$ is the energy at the bottom of the conduction band, $E_{V}$ is the energy at the top of the valence band, $E_{f}$ is the Fermi energy and $E_{g} = E_{C} - E_{V}$ is the energy gap.}
	\label{fig:silicon:semiconductor}
	\end{center}
\end{figure}

Silicon has electrical resistivity in the range between that of a conductor (below $10^{-2}~\Omega\cdot$cm) and an insulator (above $10^{5}~\Omega\cdot$cm). A structure of atomic energy levels in a pure semiconductor (so-called intrinsic) is shown in fig.~\ref{fig:silicon:semiconductor:intrisic}.
\\
Electrons are fermions which subject to the Fermi-Dirac statistics. Probability that state of an energy $E$ is filled with an electron is given by the probability density function:
\begin{equation}
\label{eq:silicon:fermi_dirac}
f(E) = \frac{1}{1+\exp{\left(\frac{E-E_{f}}{k_{B}T}\right)}},
\end{equation}
where $E_{f}$ is the Fermi energy, $k_{B}$ is the Boltzmann constant and $T$ is the absolute temperature. The density $N(E)$ of electron states with energy E in a conduction band is proportional to $(E-E_{C})^{1/2}$. Thus the density of free electrons and holes ($n$ and $p$, respectively) can be calculated by integrating the density of states per unity of energy $N(E)$ times occupation probability $f(E)$, yielding:
\begin{equation}
\label{eq:silicon:n_p_density}
n = N_C\exp{\left(-\frac{E_{C} - E_{F}}{k_{B}T}\right)}~~\textrm{and}~~p = N_V\exp{\left(-\frac{E_{F} - E_{V}}{k_{B}T}\right)},
\end{equation}
where $N_{C}$, $N_{V}$ are the effective densities of states in the conduction and the valence band, respectively, given by:
\begin{equation}
\label{eq:silicon:n_p_density_1}
N_C = 2\left(\frac{2\pi m_{e}^{*}k_{B}T}{h^{2}}\right)^{3/2}~\textrm{and}~~N_V = 2\left(\frac{2\pi m_{h}^{*}k_{B}T}{h^{2}}\right)^{3/2},
\end{equation}
where $m_{e}^{*}$ and $m_{h}^{*}$ are the effective masses of electrons and holes, respectively, and $h$ is the Planck constant.\\
In intrinsic semiconductors thermal agitation excites electrons which leave the valence band and occupy the conduction band, leaving holes in the valence band. In this case $p = n = n_{i}$, where $n_{i}$ is the intrinsic carrier density. Assuming $n=p$ in  (\ref{eq:silicon:n_p_density}) leads to:
\begin{equation}
\label{eq:silicon:fermi_energy}
E_{f} = \frac{E_{C}+E_{V}}{2} + \frac{k_{B}T}{2}\ln{\left(\frac{N_{V}}{N_{C}}\right)}.
\end{equation}
Thus the Fermi level of an intrinsic semiconductor lies very close to the middle of the energy gap. The intrinsic carrier density is given by the formula:
\begin{equation}
\label{eq:silicon:intrisic_density}
n_{i} = \sqrt{pn} = \sqrt{N_{C}N_{V}}\exp{\left(-\frac{E_{g}}{2k_{B}T}\right)} \sim T^{3/2}\exp{\left(-\frac{E_{g}}{2k_{B}T}\right)},
\end{equation}
where $E_{g} = E_{C} - E_{V}$ is the energy band gap.\\
Electrical conductivity of semiconductor materials can be altered by several orders of magnitude by adding small quantities of other substances that are called impurities. A process of replacing atoms in the semiconductor lattice with atoms of other elements is called doping. It leads to creation of additional energy levels within the energy gap. The doped semiconductor is called extrinsic.\\
Pentavalent impurities such as phosphorus are called donors since they donate additional electrons. The four valence electrons of a donor are shared in the covalent bonding with neighbouring silicon atoms, while its fifth electron is loosely bound. At room temperatures those electrons would be free and hence available for conduction. Silicon doped with donors is called $n$-type and its Fermi energy $E_{f}$ is close to the energy of the conduction band $E_{C}$, as shown in fig.~\ref{fig:silicon:semiconductor:n-type}. Alternatively, silicon may be doped with trivalent impurities such as boron. They are called acceptors since they accept electrons from the valence band leaving a hole there. In the case of acceptors three strong covalent bounds are formed with adjacent silicon atoms but the fourth bound is incomplete. This vacancy can be easily filled with an electron from the valence band. This is called a $p$-type silicon and its  Fermi energy $E_{f}$ is close to the energy of the valance band $E_{V}$, as shown in fig.~\ref{fig:silicon:semiconductor:p-type}.

\subsection{The $p$-$n$ junction}
\label{ch:silicon:properties:p-n}

\begin{figure}[!h]
        \begin{center}
                \resizebox{0.7\textwidth}{!}{
                       \includegraphics[]{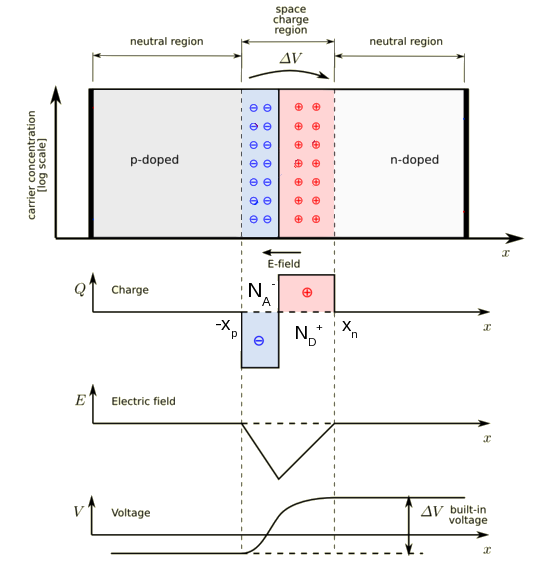}
                }
                \caption[A picture of a $p$-$n$ junction.]{A picture of a $p$-$n$ junction with schematically marked x-dependence of a space charge concentration, electric field and electrostatic potential \cite{silicon:pn_junction}.}
                \label{fig:silicon:pn_junction}
        \end{center}
\end{figure}

Junction between $p$-type and $n$-type semiconductors exhibit interesting electrical properties which are of the great importance for modern electronics as well as ionisation detectors. On contact, electrons diffuse from the $n$-type material into the $p$-type while holes do in the opposite direction. Electrons leave exposed donor ions of $N_{D}^{+}$ concentration over a thickness $x_{n}$ in the $n$-type semiconductor and holes leave exposed acceptor ions of $N_{A}^{-}$ concentration over a thickness $x_{p}$ in the $p$-type semiconductor. Thus at the $p$-$n$ junction a fixed space charge of ionised donors and acceptors is created (the so-called depletion region), as illustrated in fig.~\ref{fig:silicon:pn_junction}. This processes create an electric field that eventually balances the tendency for current to flow by diffusion. The Fermi levels in the materials becomes equal once the static condition is reached.\\
The electrostatic potential $V$ and electric field strength are related by the Poisson equation in one dimension:
\begin{equation}
\label{eq:silicon:poisson}
-\frac{d^{2}V}{dx^{2}} = \frac{dE}{dx} = \frac{\varrho(x)}{\epsilon_{Si}\epsilon_{0}},
\end{equation}
where $\epsilon_{Si}$ and $\epsilon_{0}$ are the dielectric constants of silicon and vacuum, respectively, and $\varrho(x)$ is the charge density function given by:
\begin{equation}
\label{eq:silico:charge_density}
	\varrho(x) = \left\{
	\begin{array}{ll}
	qN_{D}^{+} & \textrm{for $0 < x < x_{n}$},\\ \\
	-qN_{A}^{-} & \textrm{for $-x_{p} < x < 0$}.
	\end{array} \right.
\end{equation}
Thus
\begin{equation}
\label{eq:silico:E_field}
	E(x) = \left\{
	\begin{array}{ll}
	\frac{-qN_{D}^{+}}{\epsilon_{Si}\epsilon_{0}}\left(x_{n} - x\right) & \textrm{for $0 < x < x_{n}$},\\ \\
	\frac{-qN_{A}^{-}}{\epsilon_{Si}\epsilon_{0}}\left(x + x_{p}\right) & \textrm{for $-x_{p} < x < 0$}.
	\end{array} \right.
\end{equation}
and
\begin{equation}
\label{eq:silico:electrostatic_potential}
	V(x) = \left\{
	\begin{array}{ll}
	V_{n} - \frac{-qN_{D}^{+}}{\epsilon_{Si}\epsilon_{0}}\left(x_{n} - x\right)^{2} & \textrm{for $0 < x < x_{n}$},\\ \\
	V_{p} + \frac{-qN_{A}^{-}}{\epsilon_{Si}\epsilon_{0}}\left(x + x_{p}\right)^{2} & \textrm{for $-x_{p} < x < 0$},
	\end{array} \right.
\end{equation}
where $V_{n} = V(x_{n})$ and $V_{p} = V(-x_{p})$ are the integration constants. 
\\
Continuity of the field at $x = 0$ implies:
\begin{equation}
\label{eq:silico:implies1}
N_{A}^{-}x_{p} = N_{D}^{+}x_{n},
\end{equation}
which shows that depth of the depletion region is inversely proportional to the doping concentration on each side of a $p$-$n$ junction.\\
The potential step at the depletion region, the so-called build-in potential barrier $\Delta V = V_{n} - V_{p}$, can be calculated by imposing the continuity of the electrostatic potential at $x = 0$:
\begin{equation}
\label{eq:silico:implies2}
\Delta V = \frac{q}{2\epsilon_{Si}\epsilon_{0}} \left(N_{A}^{-}x_{p}^{2} + N_{D}^{+}x_{n}^{2}\right).
\end{equation} 
Exploiting equations (\ref{eq:silico:implies1}) and (\ref{eq:silico:implies2}) one can find a relation between the depth of the depletion region $w$ and the build-in potential $\Delta V$:
\begin{equation}
\label{eq:silico:implies3}
w(\Delta V) = x_{n} + x_{p} = \sqrt{\frac{2\epsilon_{Si}\epsilon_{0}}{q}\Delta V\left(\frac{1}{N_{A}^{-}} + \frac{1}{N_{D}^{+}}\right)}.
\end{equation}
In tracking detectors the doping concentration is usually much larger on one side of the junction than on the other. Assuming a higher doping concentration of the $p$-type material ($N_{A}^{-} \gg N_{D}^{+}$), the formula (\ref{eq:silico:implies3}) reads:
\begin{equation}
\label{eq:silico:implies4}
w(\Delta V) = \sqrt{\frac{2\epsilon_{Si}\epsilon_{0}}{qN_{D}^{+}}\Delta V}.
\end{equation}
The depth of the depletion region can be increased by applying the bias voltage $V_{b}$ with the same polarity as that of the built-in potential $\Delta V$. The bias voltage needed to deplete the full detector thickness $D$ (the so-called full depletion voltage) is given by:
\begin{equation}
\label{eq:silico:implies5}
V_{dep} = \frac{qN_{D}^{+}D^{2}}{2\epsilon_{Si}\epsilon_{0}} - \Delta V.
\end{equation}
Diodes biased with a voltage of the same polarity as the built-in potential $\Delta V$ are called reverse biased diodes while diodes biased with a voltage of the inverse polarity as the built-in potential $\Delta V$ are called forward biased diodes.

\subsection{Charge generation in silicon}
\label{ch:silicon:properties:charge_generation}

Particles passing through the silicon medium are subject to different process which lead to the loss of their energy. A part of the energy absorbed in the material is used for generation of electron-hole pairs which subsequently can be detected as electrical signals. Free charge carriers are also generated thermally what leads to the so-called leakage current. 

\subsubsection{Interactions of charged particles}
\label{ch:silicon:properties:charge_generation:charged}

\begin{figure}[!h]
        \begin{center}
                \resizebox{0.5\textwidth}{!}{
                        \includegraphics[]{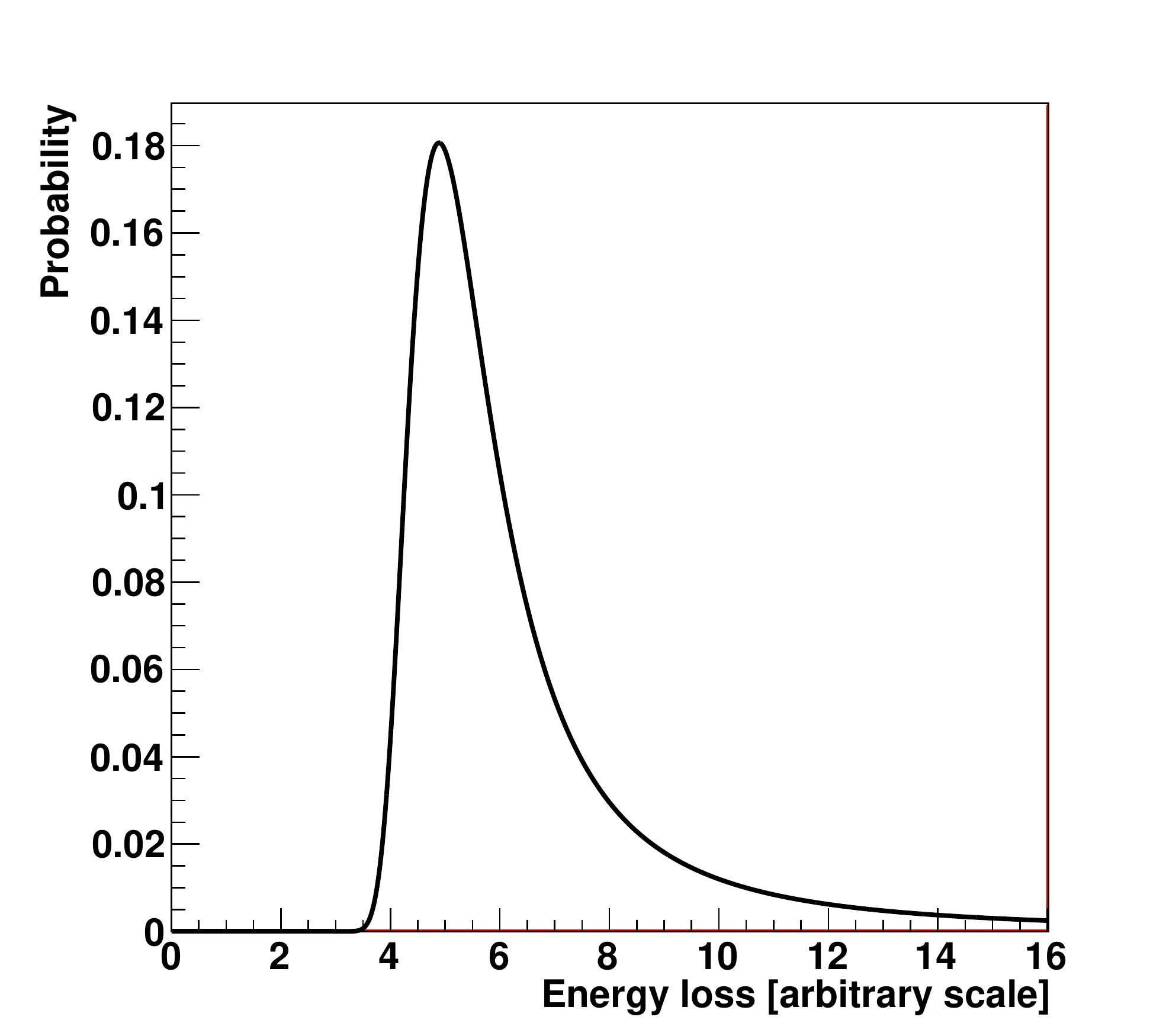}
                }
                \caption[The Landau distribution of the ionisation energy loss in a thin absorber]{The Landau distribution of the ionisation energy loss in a thin absorber.}
                \label{fig:silicon:Landau}
        \end{center}
\end{figure}

Charged particles loose a part of their energy in elastic electromagnetic collisions with shell electrons of the absorbing material. The average energy loose in matter, $-\left<dE/dx\right>$, for charged particles is described by the Bethe-Bloch formula:
\begin{equation}
\label{eq:silico:Bethe-Bloch}
-\left<\frac{dE}{dx}\right>=Kz^{2}\frac{Z}{A}\frac{1}{\beta^{2}}\left(\frac{1}{2}\ln{\frac{2m_{e}c^{2}\beta^{2}\gamma^{2}T_{max}}{I^2}} - \beta^{2} + ... \right),
\end{equation}
where:\\
$\frac{dE}{dx}$~~~~~~is usually expressed in $\frac{eV}{g/cm^{2}}$,\\
$K$~~~~~~~constant equal 0.307075~MeV$\cdot$cm$^{2}$,\\
$z$~~~~~~~~charge of the particle in units of the electron charge,\\
$Z$~~~~~~~atomic number of the absorption medium ($Z = 14$ for silicon),\\
$A$~~~~~~~atomic mass of the absorption medium ($A = 28$ for silicon),\\
$m_{e}c^{2}$~~~rest mass of the electron (0.511~MeV),\\
$\beta$~~~~~~~velocity of the traversing particle in units of the speed of light,\\
$\gamma$~~~~~~~the Lorentz factor $1/\sqrt{1-\beta^{2}}$,\\
$I$~~~~~~~mean excitation energy (137~eV for silicon),\\
$T_{max}$~~~the maximum kinetic energy which can be transferred to an electron by a particle of mass $M$.\\
Dots at the end of~(\ref{eq:silico:Bethe-Bloch}) indicate presence of additional correction terms, omitted here, like the density correction for high particle energies and the atomic shell correction for low energies.\\
If the energy transferred to the atomic electron is large enough it is moved from the valence band to the conduction band and the electron-hole pair is created. The mean energy required for a single electron-hole pair creation in silicon is 3.6~eV. The amount of energy transferred to an electron may be high enough for the electron to causes substantial secondary ionisation. If the energy transferred to the atom electron is not sufficient to cause its ionisation the atom structure gets only excited. Heavy charged particles lose energy in matter primarily by ionisation.\\
Ionisation energy loss is subject to statistical fluctuations and the value given by (\ref{eq:silico:Bethe-Bloch}) is only an average value of the so-called Landau distribution shown in fig.~\ref{fig:silicon:Landau}.
\\
If the particle is not stopped in the active volume of the detector, the energy loss varies around the peak of the distribution with a significant probability of high values (asymmetric distribution). The average value is higher than the most probable value of the distribution. Fluctuations around the maximum become higher in thinner sensors.\\
The asymmetry of the Landau fluctuations stems mainly from $\delta$-electrons, i.e. knocked out electrons which receive enough energy to become ionising particles themselves.\\
For light charged particles, e.g. electrons and positrons, there are two main processes contributing to energy loss -- ionisation and bremsstrahlung. In the bremsstrahlung process the charged particle emits photons due to its acceleration in the electric field of e.g. atomic nucleus. Bremsstrahlung is an important process for electrons since its probability is inversely proportional to the squared mas of the incident particle. Thus it dominates for electron energies of few tens of MeV or higher.\\
A charged particle traversing medium is also subject to multiple Coulomb scattering. As a result, it is deflected after leaving the material.

\subsubsection{Photon interactions}
\label{ch:silicon:properties:charge_generation:photons}

Photons interact with matter mainly via the following three processes, their probability vary with energy: 
\begin{itemize}
\item 
Photoelectric effect in which a photon is absorbed by an atomic electron which moves into the conduction band. It is the dominant process at low photon energies (in silicon below about 100~keV). If the photo-electron gains sufficiently high kinetic energy it may be a source of a secondary ionisation occurring along its trajectory.
\item
Compton scattering of photons on atomic electrons for photon energies much higher that electron binding energies, electrons may be treated as free. The Compton effect leads to ionisation of atoms while the energy of the incident photon partially transferred to an electron.
\item
Conversion of photons into an electron-positron pair (in presence of a third body, usually a nucleus). This process is possible for energies exceeding twice the electron mass for $E_{\gamma} > 1.02~MeV$ and for energies above 10~MeV it becomes the only important process.
\end{itemize}
 
\subsection{Principles of a silicon detector operation}
\label{ch:silicon:properties:principles_of_operation}

The number of generated electron-hole pairs is proportional to the energy transferred by an ionising particle to a medium. Charge carriers move in the semiconductor volume due to an external electric field (drift) and an inhomogeneous distribution of charge carriers (diffusion).\\
If the electric field $\vec{E}$ is present in a semiconductor the average drift velocity of charge carriers is given by:
\begin{equation}
\label{eq:silico:carriers_vel}
\vec{v_{e}} = -\mu_{e}\vec{E}~~\textrm{and}~~\vec{v_{h}} = \mu_{h}\vec{E},
\end{equation}
where $\mu_{e}$ and $\mu_{h}$ are electron and hole mobilities, respectively. The proportionality in (\ref{eq:silico:carriers_vel}) holds only for weak fields such that carrier velocity acquired in the field is lower than its thermal velocity (in the room temperature the thermal velocity is of the order of $10^{7}$~cm/s).\\
If a spatial distribution of charge carriers in silicon is inhomogeneous, electrons and holes diffuse from the region of higher concentration to a region of lower concentration. This results in diffusion currents which are described by equations:
\begin{equation}
\label{eq:silico:diff_currents}
\vec{J_{e}} = qD_{e}\vec{\bigtriangledown}n~~\textrm{and}~~\vec{J_{h}} = -qD_{h}\vec{\bigtriangledown}p,
\end{equation}
where $D_{e}$, $D_{h}$ are the diffusion constants and $\vec{\bigtriangledown}n$, $\vec{\bigtriangledown}p$ are gradients of the carrier concentration. The total current densities for electrons and holes are given by a sum of contributions from drift and diffusion effects:
\begin{equation}
\label{eq:silico:total_currents}
\vec{J_{e}} = q\mu_{e}n\vec{E} + qD_{e}\vec{\bigtriangledown}n~~\textrm{and}~~\vec{J_{h}} =  q\mu_{h}p\vec{E} - qD_{h}\vec{\bigtriangledown}p.
\end{equation}
Movement of the generated charge induce a signal on the detector electrodes. Its height depends on the distance of the charge carrier from the electrode. The current induced in the readout circuity is amplified and integrated by a charge sensitive amplifier resulting in an output voltage which is proportional to the collected charge. 

\subsection{Leakage current}
\label{ch:silicon:properties:leakage}

The leakage or dark current is one of the main sources of detector noise and it is flowing through the $p$-$n$ junction in the absence of ionisation source if a reverse bias is applied. The leakage current is due to the diffusion of free carriers from the undepleted volume into the sensitive space charge region and thermal generation at the generation-recombination centres at the surface of the device and in the depleted volume. The latter usually dominates the $p$-$n$ junction leakage current and is proportional to the developed volume $w$ (\ref{eq:silico:implies4}):
\begin{equation}
\label{eq:silico:leakage}
{J_{vol}} \approx -e\frac{n_{i}}{\tau_{g}}w,
\end{equation}
where $J_{vol}$ is the volume generation current per unit area, $\tau_{g}$ is the carrier generation live time and $n_{i}$ is the intrinsic carrier concentration. The $\tau_{g}$ and $n_{i}$ temperature dependence implies:
\begin{equation}
\label{eq:silico:leakage(T)}
{J_{vol}} \propto T^{2}\exp{\left(-\frac{E_{g}(T)}{2k_{B}T}\right)}.
\end{equation}
Thus the leakage current and related noise can be significantly reduced by cooling the detector.

\subsection{Radiation damages in silicon}
\label{ch:silicon:properties:radiation_damage}

Silicon tracking detectors have to face intense radiation present near the interaction point. It is responsible for causing defects in silicon which deteriorate detector performances. There are two main categories of radiation damages: ($i$) atom displacement in the lattice and ($ii$) ionisation damages. The displacement process affects the properties of the bulk and it is known as the bulk damage. Ionisation damages are responsible for formation of trapped charges and interface defects which are called surface damages.\\
The observed effects of radiation damages in a sensor are: ($i$) increased leakage current translating into increased noise, ($ii$) reduction of the collected charge due to occurrence of trapping centres and ($iii$) decrease of the charge carrier mobility and their lifetime.

\subsubsection{Displacement damages}
\label{ch:silicon:properties:radiation_damage:disp_dam}

A fraction of high energy particles traversing the silicon volume interacts with nuclei, often displacing them from the lattice position. This produces crystal imperfections which may be electrically active and hence change the electric properties of the material. In a crystal there can be point defects for local single atom displacements or cluster defects characterised by large regions of lattice disturbances. Isolated displacements are created mostly by electromagnetic radiation of low energy electrons and X-ray photons that can deliver only small energy to the recoil silicon atom. To remove a silicon atom from its lattice position a minimum recoil energy of about 25~eV is required. Electrons need an energy of at least 260~keV in order to provide such a recoil energy in a collision, while much heavier protons and neutrons require only 190~keV. If the recoil silicon atom gets enough energy through the collision, it can cause further defects. In case this energy exceeds 2~keV the atoms loose most of their energy in a very localised area, creating a cluster of defects. Since the displacements are closely situated most of the interstitials and vacancies recombine and repair the initial defects. The rest of them migrate through the silicon medium and interact with impurities like oxygen, carbon, atoms of dopant or with themselves forming traps which capture and emit charge carriers. The point defects create new energy levels in the energy gap. These energy levels can act as acceptors, donors or charge traps and modify the lifetime of charge carriers.\\
Those of the energy levels which behave as generation-recombination centres result in a decrease of the carrier generation lifetime $\tau_{g}$ and an increase of the volume generation current $J_{vol}$. The carrier generation lifetime $\tau_{g}$ is inversely proportional to radiation fluence $\Phi$:
\begin{equation}
\label{eq:silico:tau_g}
\frac{1}{\tau_{g}} = \frac{1}{\tau_{g,\Phi=0}} + k_{\tau}\Phi,
\end{equation}
where $k_{\tau}$ is the lifetime related damage rate, and thus the volume generation current $J_{vol}$ (\ref{eq:silico:leakage}) increased with the absorbed radiation fluence $\Phi$:
\begin{equation}
\label{eq:silico:j_vol_phi}
J_{vol} \approx J_{vol,\Phi=0} + \alpha\Phi,
\end{equation}
where $\alpha$ is the current related damage rate.\\
Since some of the radiation induced defects act as donor-like or acceptor-like states the effective doping concentration of the material changes with the absorbed radiation fluence. The absorption of a very high radiation dose can even lead to the inversion of the silicon conduction type. Changes in the doping concentration impacts the detector full depletion voltage (\ref{eq:silico:implies5}).\\
The energy levels referring to the charge trapping centres are mostly unoccupied in the depletion region due to the lack of free charge carriers. They can hold or trap parts of the signal charge for a time longer than the charge collection time and so reduce the signal height. The defect trapping propability is defined as the inverse of the trapping time $\tau_{t}$, which represents the mean time an electron or a hole spends in the space-charge region before being trapped. The trapping propability at a given temperate is proportional to the irradiation fluence:
\begin{equation}
\label{eq:silico:tau_t}
\frac{1}{\tau_{t}} = \frac{1}{\tau_{t,\Phi=0}} + \gamma\Phi,
\end{equation}
where $\gamma$ is the effective electron or hole damage constant.

\subsubsection{Ionisation damages}
\label{ch:silicon:properties:radiation_damage:io_dam}

The damages in silicon are also induced by ionisation radiation of charged particles and photons. Ionisation damages concentrate at interfaces between silicon and silicon dioxide, thus they are also called surface damages. They constitute the main concern for the front-end electronics based on the NMOS and PMOS transistors.\\
Most of the electron hole pairs generated in the silicon dioxide recombine immediately and do not cause any negative effects. Rest of the carriers, which did not recombine, drift away from the place of their origin. Since electrons have much higher mobility in the silicon dioxide ($\mu_{e}\approx20~$cm$^{2}/(Vs)$) then holes ($\mu_{h}\approx20\times10^{-4}~$cm$^{2}/(Vs)$), they leave the dioxide volume in a short time following the irradiation while holes get blocked there. Additionally some of the holes moving towards negative electrodes get stuck within several nanometres from the interface between silicon and silicon dioxide where deep hole traps exist. As a result a positive charge builds up in the silicon dioxide region and affects the PMOS and NMOS transistors operation. In case of the NMOS and PMOS transistors, concentration of a positive charge in the silicon dioxide leads to an increase of the absolute value of their threshold voltage which is defined as the minimum voltage between transistor gate and a source to initiate conduction of a drain.\\
A further effect of ionisation radiation is the generation of interface states leading to a surface generation current which contributes to the total dark current of the detector. Some of the surface defects act also as carrier traps resulting in deterioration of the signal charge.

\section{Silicon tracking detectors}
\label{ch:silicon:tracking_detectors}

\subsection{Microstrip detectors}
\label{ch:silicon:tracking_detectors:strip_det}

\begin{figure}[!h] 
	\begin{center}
	\subfigure[]{
		\includegraphics[width=0.47\textwidth]{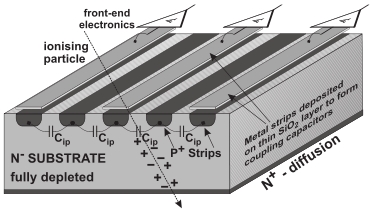}
		\label{fig:silicon:strip_detectors:single}
	}
	\hspace{0.1cm}
	\subfigure[]{
		\includegraphics[width=0.47\textwidth]{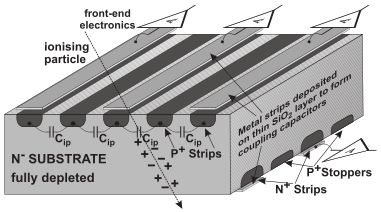}
		\label{fig:silicon:strip_detectors:double}
	}
	\caption[Cross section of the silicon strip detectors]{Schematic view of a (a) single-sided and (b) double-sided, AC coupled silicon strip detectors with interleaved strips (from~\cite{tests:DeptuchTh}).}
	\label{fig:silicon:strip_detectors}
	\end{center}
\end{figure}

In single sided silicon microstrip detectors \cite{silicon:microstrip_detectors} an active volume of $n$-type silicon is covered on one side with a strongly doped $p$-type ($p+$) silicon implants of a strip form fig.~\ref{fig:silicon:strip_detectors:single}. The segmented side is usually covered by a few $\mu$m layer of $SiO_{2}$ or $Si_{3}N_{4}$ for detector protection. Strips are usually from 10~$\mu$m to 50~$\mu$m wide and several centimetres long. Every strip can be connected to its readout channel or intermediate strips can be left floating, being only capacitively coupled to the neighbouring strips. A good ohmic connection between readout electronics and a $p+$ implant is provided with an aluminium strip placed on top of it. The aluminium strips can be connected directly to the $p+$ implants (DC coupled detectors) or capacitively by putting them on a thin oxide or nitride layer generated on the segmented side of the detector (AC coupled detector). The AC coupling prevents leakage current to flow through the electronics however it is more expensive since an additional step is needed in the production process.
\\
The single sided detector, presented above, delivers only one dimensional information on the charged particle track. In order to measure both coordinates at the same time a double sided microstrip detector has to be used. Double sided microstrip detectors can be constructed in to ways. In the first approach two single sided microstrip detectors of different strip orientation can be mounted back to back. In the second approach a strongly doped $n$-type ($n+$) silicon layer, placed on the sensor bottom side, is divided into strips under some angle with respect to the $p+$ strips on the top side fig~\ref{fig:silicon:strip_detectors:double}. Since the $n+$ strips form an ohmic contact to the active volume they have to be separated by $p+$ strips so-called $p$-stoppers. The advantage of using double sided silicon strip detectors against the combination of two planes of single sided silicon strip detectors is a significant reduction of material that particle has to traverse.\\
Unfortunately strip detectors are not applicable for the new generation of vertex detectors which are going to work in a very dense environment, i.e. sensors are traversed by heavy fluxes of particles. In case of $N$ tracks passing a single silicon strip sensor there are $N!$ possible combinations of hits, which introduces important ambiguities. Thus pixel detectors are most often used in modern vertex detectors.

\subsection{Pixel detectors}
\label{ch:silicon:tracking_detectors:pixels}

\subsubsection{Hybrid pixel detectors}
\label{ch:silicon:tracking_detectors:pixels:hybrid}

\begin{figure}[!h]
	\begin{center}
		\includegraphics[width=0.7\textwidth]{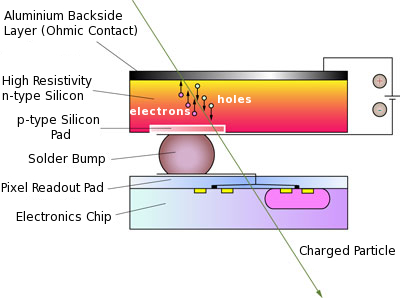}
		\caption[Principle of hybrid pixel detector]{Principle of hybrid pixel detectors in which readout electronics is connected with the silicon sensor by means of bump-bonding technique (from \cite{silicon:hybrid_pixel_CMS}).}
        	\label{fig:silicon:hybrid}
        \end{center}
\end{figure}

Hybrid pixel detectors have been developed for the LHC detectors that require very fast and radiation tolerant devices. Similarly to the microstrip detectors the hybrid pixel detectors use high resistivity substrate. The pixel sensor and the readout chip are developed and produced independently and they are connected together in the final step. This solution enables combination of the radiation hard sensor with a fast readout chip. The mechanical and electrical connections between sensor and readout circuity is established with small balls of solid, indium or gold as shown in fig.~\ref{fig:silicon:hybrid}. This is the so-called bump bonding technique.
\\
Pixel size in a hybrid pixel detector is determined by the size of the readout chip, e.g. pixels used in the CMS and ATLAS experiments have sizes of $150\times150~\mu$m$^{2}$ and $50\times400~\mu$m$^{2}$, respectively. The readout circuits are usually built in the standard CMOS technology which has high integration density and the possibility of combining the analogue and digital circuits on the same chip. Using modern sub-micrometer processes it is possible to integrate complex and fast circuits providing initial data processing and significant reduction of the data transferred to the data acquisition system.\\
The disadvantages of hybrid pixel detectors, beside limited granularity, are the complexity of millions of interconnections introducing extra material in the active area and the relatively high power dissipation reaching a few hundred mW/cm$^{2}$.

\subsubsection{Charge Coupled Devices (CCDs)}
\label{ch:silicon:tracking_detectors:pixels:CCD}

\begin{figure}[!h]
        \begin{center}
                \resizebox{0.55\textwidth}{!}{
                       \includegraphics[]{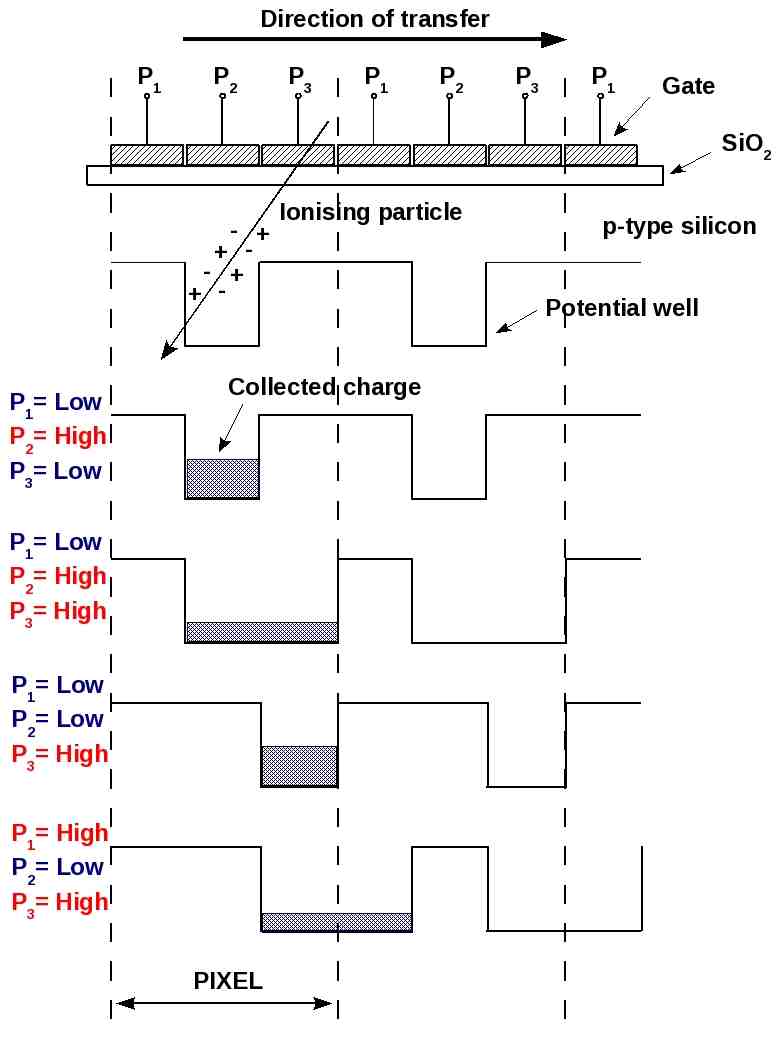}
                }
                \caption[Principles of CCDs operation]{Principles of CCDs operation. The charge generated by an ionising particle is trapped in a potential well below middle MOS gate. Afterwords charge is transferred across the detector by means of a proper sequence of voltages applied to the MOS gates.}
                \label{fig:silicon:ccd_transfer}
        \end{center}
\end{figure}

CCDs have been developed as photon detectors in the visible light band. They have been also successfully used as tracking detectors in high energy physics, e.g. in the vertex detector of the SLD experiment \cite{silicon:SLD}. CCDs used in the SLD had $20\times20~\mu$m$^{2}$ pixels, providing intrinsic space-point resolution better than $4~\mu$m. Except high detector granularity CCDs have small thickness which can be decreased down to its active volume depth of approx.~$20~\mu$m. The low material budget is favourable for reduction of the multiple scattering effects. Moreover, CCDs can be fabricated in a form of large areas allowing an elegant VXD geometry with very few gaps (the second generation of CCDs used in SLD were of size $80\times16$~mm$^{2}$).\\
The active volume of the CCDs is made of $p$-type silicon grown on a strongly doped $p+$ silicon substrate. On the opposite side an active volume is covered with a $SiO_{2}$ insulating layer on which MOS (Metal-Oxide-Semiconductor) structures are located fig.~\ref{fig:silicon:ccd_transfer}. A single pixel contains three MOS gates creating potential wells in which electrons, released by passage of charge particles, are trapped. The device is read out by sequentially changing potential on three neighbouring gates in such a way that all charges are transformed in parallel from one row to the next, down the device fig.~\ref{fig:silicon:ccd:ccd_standard.jpg}. The charges in the last, bottom row of the array are transferred into the adjacent linear register, from which they are shifted, one at time, onto the output node to the charge sensitive amplifier realising charge-to-voltage conversion.\\
One can distinguish two main types of the CCDs: the surface-channel CCD (SCCD) and buried-channel CCD (BCCD). In the SCCDs the signal charge is stored at the interface between silicon and silicon dioxide while in the BCCDs the signal charge is stored in the bulk of the silicon approx.~$1~\mu$m below the surface.
\\
CCDs do not experience any dead time or dead zones and are continuously sensitive to radiation. However they require fairly long time for readout since the charge has to be serially shifted under one gate to the next across the rows and columns of pixels fig.~\ref{fig:silicon:ccd:ccd_standard.jpg}. This was not a problem in case of the SLD experiment however the ILC high rate environment enforces usage of much faster pixel detectors. Another limitation of the CCDs is their sensitivity to radiation damage, which results in the degradation of the charge transfer efficiency due to trapping of the signal charges at the radiation induced defects in the detector bulk. In order to meet the ILC vertex detector requirements a new types of CCDs are under studies: a Column Parallel CCD (CPCCD), a Fine Pixel CCD (FPCCD) and a Short Column CCD (SCCCD).
\begin{itemize}
 \item 
In the CPCCD \cite{silicon:CPCCD} the bottom serial register is omitted and every column is equipped with its own readout chain as shown in fig.~\ref{fig:silicon:ccd:ccd_column_parallel.jpg}. The columns are read out in parallel which results in a significant increase of the readout speed. Moreover the CPCCD housed in the first layer of the ILC vertex detector are going to be clocked with 50~MHz. Assuming pixels of $20~\mu$m pitch this allows for 20 readouts of the first layer during one bunch train of 1~ms duration. To minimise the power dissipation the CPCCD will operate at a very small amplitude of the clock signal.
\begin{figure}[!h] 
	\begin{center}
	\subfigure[]{
		\includegraphics[width=0.45\textwidth]{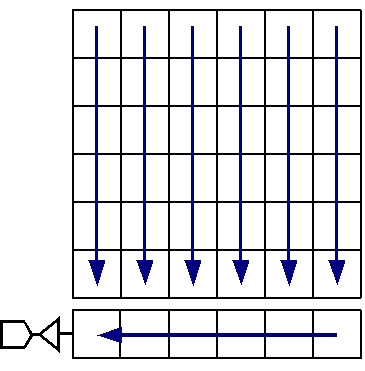}
		\label{fig:silicon:ccd:ccd_standard.jpg}
	}
	\hspace{0.1cm}
	\subfigure[]{
		\includegraphics[width=0.45\textwidth]{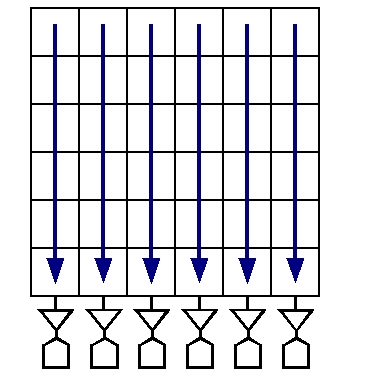}
		\label{fig:silicon:ccd:ccd_column_parallel.jpg}
	}
	\caption[Schematic view of charge transfer in CCDs]{Schematic view of charge transfer in (a) standard CCD and (b) Column Parallel CCD.}
	\label{fig:silicon:ccd}
	\end{center}
\end{figure}
 \item
The FPCCD \cite{silicon:FPCCD} is a fully depleted CCD equipped with fine pixels of $5~\mu$m pitch. The increased number of pixels (with respect to the standard CCD) and reduced charge carrier diffusion result in a lower pixel occupancy in a dense environment of the ILC. Thus hits can be accumulated during the whole bunch train and read out in the 200~ms interval between trains. Therefore a very fast readout is not needed. Moreover since the readout is performed in an absence of the beam the electrical interferences do not influence the signal charge which is transferred through the CCD. The small size of pixels provides a high spatial resolution even with digital readout (approx.~$1.4~\mu$m). The fully depleted fine pixel CCD will also provide an excellent two track separation capability. It is also expected that small pixels of FPCCD will allow for distinguishing between hits originating from the beamsstrahlung background and from high $p_{t}~e^{+}e^{-}$ interactions on the basis of cluster shapes.
\item
The SCCCD \cite{silicon:SCCCD} consists of a CCD sensor bump bonded with a readout chip made in the CMOS technology. Pixels of $15~\mu$m pitch are arranged in columns of 512 pixels. Each column is terminated with a readout node and a bond pad. The charge collected in adjacent columns is transferred through the CCD in the opposite directions, see fig.~\ref{fig:silicon:SCCCD}. The readout layer discriminate pixels with charge above a threshold and record they signal amplitude and the clock time. Since the SCCCD is going to be clocked three times during each bunch crossing (approx. 10~MHz in contrast to 50~MHz of CPCCD) an excellent time resolution is expected. Thus recorded hits can be matched with an adequate bunch crossing resulting in an efficient rejection of background hits.
\begin{figure}[!h]
        \begin{center}
                \resizebox{\textwidth}{!}{
                       \includegraphics[]{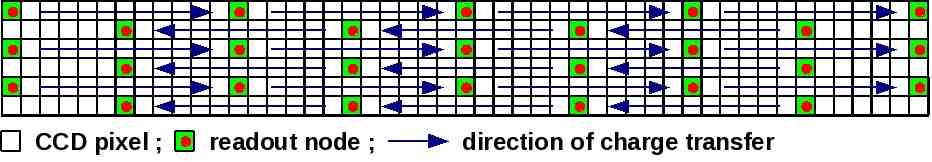}
                }
                \caption[Principles of SCCD operation]{Principles of SCCD operation.}
                \label{fig:silicon:SCCCD}
        \end{center}
\end{figure}\\
Since the SCCCD is a combination of two bump bonded silicon layers it will significantly influence direction of charge particle tracks resulting in deterioration of vertex detector performances. This problem can be overcome with a new 3D integration technology.
\end{itemize}

\subsubsection{Imaging System with In-situ Storage (ISIS)}
\label{ch:silicon:tracking_detectors:pixels:ISIS}

\begin{figure}[!h]
        \begin{center}
                \resizebox{\textwidth}{!}{
                       \includegraphics[]{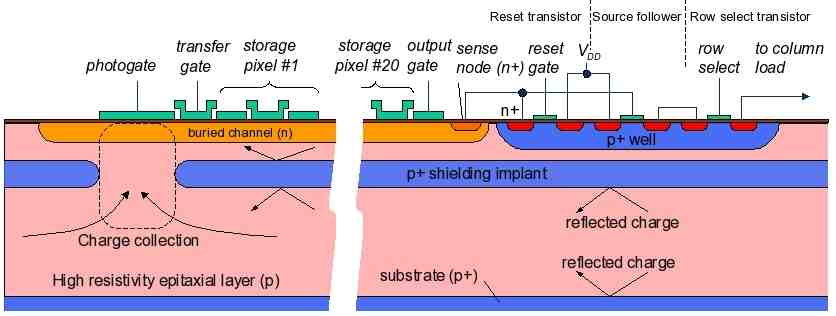}
                }
                \caption[Cross section of the ISIS with a linear CCD storage]{Cross section of the ISIS with a linear CCD storage (from \cite{silicon:ISIS}).}
                \label{fig:silicon:ISIS}
        \end{center}
\end{figure}

The ISIS \cite{silicon:CPCCD} cross section is shown in fig.~\ref{fig:silicon:ISIS}. Its active volume is a high resistivity $p$-type epitaxial layer. Each pixel of the ISIS device is equipped with a short CCD register which is embedded in a $p$-well. The CCD cells are separated from the epitaxial layer with an additional strongly doped $p+$ shielding implant.
\\
The charge liberated by the ionising particle diffuse isotropically in the ISIS active volume and only its fraction is collected by the photodiode. The charge collection efficiency is improved by highly doped $p+$ implant planes which reflect electrons preventing they escape. Signals collected in the consecutive readout cycles, referring to one bunch train, are stored in the CCD register while the charge to voltage conversion is done in the absence of the beam. Thus the ISIS chip exhibits much higher degree of immunity to electromagnetic interferences than designs where the voltage is sampled during a bunch train. Moreover in this approach a readout can be timed with a low frequency clock (20~kHz for reading out pixels during bunch train and 1~MHz for transferring data in the time gap between trains). In addition, since far fewer charge transfers are needed than for traditional CCD the ISIS is much higher radiation tolerant than the standard CCD.\\
A disadvantage of the ISIS is that it needs additional readout integrated circuits made in the CMOS technology which have to be combine with the sensor on the same chip.

\subsubsection{Monolithic Active Pixel Sensor (MAPS)}
\label{ch:silicon:tracking_detectors:pixels:MAPS}

\begin{figure}[!h]
        \begin{center}
                \resizebox{0.65\textwidth}{!}{
                       \includegraphics[]{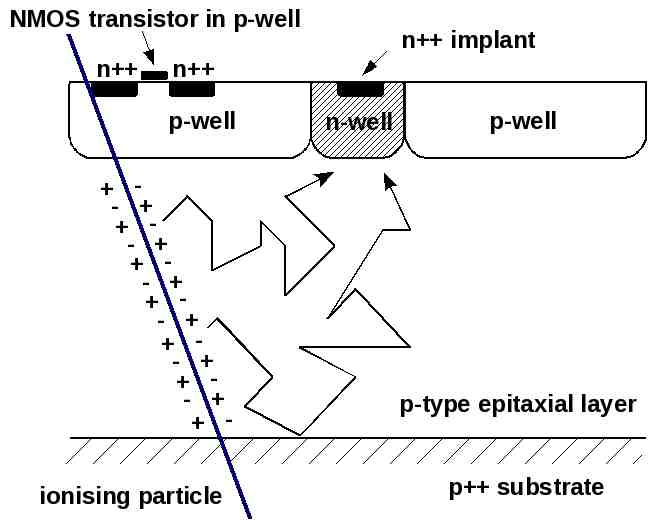}
                }
                \caption[Principles of MAPS operation]{Principles of MAPS operation.}
                \label{fig:silicon:MAPS}
        \end{center}
\end{figure}

The MAPS detectors \cite{tests:MAPS_Turchetta} shown in fig.~\ref{fig:silicon:MAPS} are made using the CMOS (Complementary Metal-Oxide-Semiconductor) commercial technology with an epitaxial layer of a few to 20~$\mu$m thickness.
\\
The epitaxial layer plays a role of the active volume of the detector. It is made of a lightly doped $p$-type silicon grown on a highly doped $p++$ substrate. On the top of the epitaxial layer the complementary $p$-type and $n$-type wells are implanted. A diode established by a $n$-well/$p$-type epitaxial layer junction is responsible for the collection of electrons liberated by the ionising particle. Since most of the epitaxial layer is free of the electric field, the charge carriers reach the collection diodes by the thermal diffusion. Thus a MAPS detector exhibit long collection times of approx. 100~ns. The charge liberated in the highly doped substrate is mostly lost due to a fast recombination of carriers. However some fraction of these charges can diffuse form the substrate to epitaxial layer and contribute to the total collected charge. The 3 orders of magnitude difference between doping levels of lightly doped epitaxial layer and the $p$-well and $p++$ substrate leads to a creation of the potential barrier at the boundaries, which act like mirrors for the excess electrons.\\
Each pixel of the MAPS detector is equipped with its own first signal processing electronics, based on the NMOS transistors. Only NMOS structures are allowed in the active area because of the $n$-well/$p$-epi collection diode that does not permit other $n$-wells. This limits the complexity of the functionalities which can be integrated at the pixel level.\\
It has been already demonstrated that MAPS detectors feature excellent tracking capabilities: ($i$) high detection efficiency exceeding 99\% and ($ii$) excellent single point resolution below $2~\mu$m for a detector equipped with pixels of $20~\mu$m pitch. A MAPS detector can be thinned down to $50~\mu$m, to minimise multiple scattering.\\ 
Due to its readout architecture, MAPS detectors have much faster readout and much lower power dissipation than CCDs. They are also less sensitive to radiation damages than CCDs. Since MAPS detectors are produced with the CMOS technology this are potentially cheap particle detectors.\\
There have been also performed studies on MAPS detectors equipped with pixels containing capacitors for signal storage. The two different approaches are being developed in parallel: the Flexible Active Pixel Sensors (FAPS) and Continuous Acquisition Pixel (CAP). In these approaches detector are read out continuously during bunch train and the acquired signals are stored in the in-pixel memory cells. Afterwords the memory cells are read out in the time interval between consecutive bunch trains avoiding beam induced electromagnetic interferences.

\subsubsection{DEPleted Field Effect Transistor (DEPFET)}
\label{ch:silicon:tracking_detectors:pixels:DEPFET}
\begin{figure}[!h]
        \begin{center}
                \resizebox{0.7\textwidth}{!}{
                       \includegraphics[]{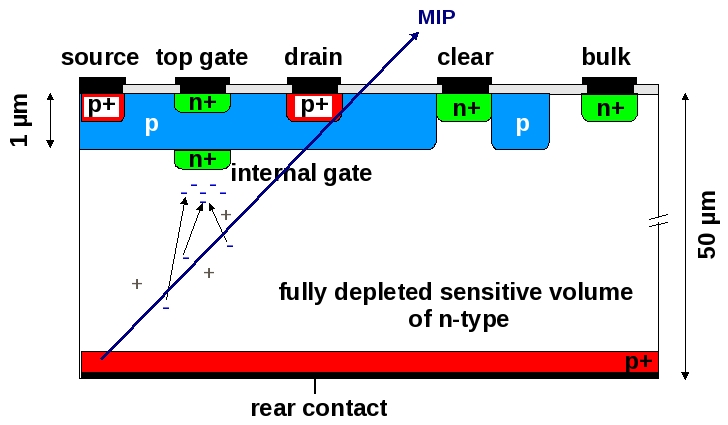}
                }
                \caption[Principles of DEPFET operation]{Principles of DEPFET operation.}
                \label{fig:silicon:DEPFET}
        \end{center}
\end{figure}

The principles of the DEPFET \cite{silicon:DEPFET} operation are shown in fig.~\ref{fig:silicon:DEPFET}. It is a fully depleted pixel detector providing detection and amplification jointly. DEPFET is based on the sidewards depletion in which a sufficiently high negative voltage is applied to a back side $p+$ contact. The charge sensing element exploits MOSFET (Metal-Oxide Semiconductor Field-Effect Transistor) or JFET (Junction Field-Effect Transistor) structures which are implemented on the top of the detector.\\
Sidewards depletion provides a parabolic potential inside the detector volume. The potential minimum for a majority carriers (electrons in the case of $n$-type silicon) is located under transistor channel at a depth of about $1~\mu$m where an additional phosphorus $n+$ implementation is located. The phosphorus implementation constitute the internal gate where liberated electrons are accumulated. The holes drift in the opposite direction of the rear contact. Electrons collected in the internal gate generate a potential that modulates current floating through a transistor. The readout is non-destructive and can be repeated several times. The removal of the signal charge and thermally generated electrons from the internal gate is done by applying a positive voltage to the clear contact.
\\
The DEPFET presents a number of advantages: ($i$) a fully depleted bulk results in a small collection times and high signal amplitude, ($ii$) low internal gate capacitance, of the order of 10~fF, provide a very low noise operation and ($iii$) amplification of the signal charge at the position of its generation prevents charge losses during its transfer. Moreover, DEPFETs consume very little power since pixels are powered only during readout.\\
The major disadvantage of the DEPFET detector is its complicated and expensive fabrication process. 

\subsubsection{Silicon On Insulator (SOI)}
\label{ch:silicon:tracking_detectors:pixels:SOI}

\begin{figure}[!h]
        \begin{center}
                \resizebox{0.7\textwidth}{!}{
                       \includegraphics[]{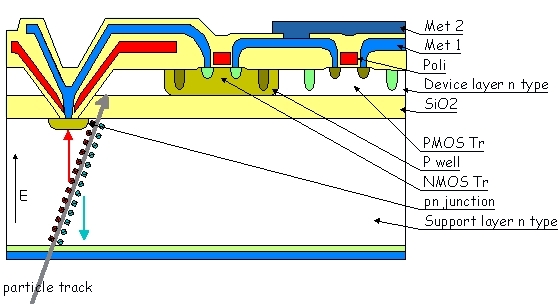}
                }
                \caption[Principles of the SOI operation]{Principles of the SOI operation (from \cite{silicon:SOI-1}).}
                \label{fig:silicon:SOI}
        \end{center}
\end{figure}

The cross section of the SOI \cite{silicon:SOI} monolithic pixel cell is shown in fig.~\ref{fig:silicon:SOI}. The high-resistivity low doped $n$-type material of approx. $300~\mu$m thickness constitutes an active volume of the device. The electronics layer is isolated from the sensor active volume with a $1$-$2~\mu$m thick buried oxide layer (BOX). Thus readout circuits can exploit both types of MOS transistors (PMOS and NMOS) resulting in their much higher functionality. The contact between electronics layer and $p+$ implants in the $n$-type sensitive volume is made through the bulk oxide by vias. The detector bottom is metallised to provide electric contact for depletion voltage. Since the detector operates in a full depletion it has high signals and low charge collection times. At present the technology is not a commercial standard but certainly would become very attractive if industrial processing of bonded sensors and CMOS-wafers were available.

\subsubsection{Chronopixel}
\label{ch:silicon:tracking_detectors:pixels:Chronopixels}

The Chronopixel \cite{silicon:SCCCD} is a monolithic CMOS sensor providing an excellent time resolution. The electronics implemented in each pixel records time occurrence of hits above adjustable threshold. A single pixel houses 4 memory cells of 13 bits, each, providing storage of 4 hit times with a single bunch crossing precision. The hits are accumulated during the whole bunch train (approx. 1~ms) and the sensor is read out during 200~ms between bunch trains. Since only the hit coordinates (x,y,t) are read out from the detector, the single point resolution is determined by the pixel size. In order to achieve a precision of the order of 3~$\mu$m, pixels of a 10~$\mu$m pitch are required. To accommodate 4 memory cells of 13 bits, each, in a pixel of $10\times10~\mu$m$^{2}$ area a small feature size CMOS process of 45~nm has to be used. With the Chronopixel it will be possible to associate uniquely each hit with a specific bunch crossing within a bunch train. This would result in a significant reduction of the effective backgrounds.

\subsubsection{3D-pixel detectors}
\label{ch:silicon:tracking_detectors:pixels:3D}

The 3D integrated pixel detector \cite{silicon:3D} consists of 2 or more layers of thinned active semiconductor devices. They are bonded and interconnected to form a monolithic circuit. These layers are also called tiers and can be fabricated according to different processes. Each layer can be optimised independently in order to improve its performance and to increase the in-pixel readout electronics functionalities. The possible structure of the 3D pixel detector could be as follows: CCD, MAPS, DEPFET or SOI sensor followed with an analogue and digital layers fabricated in the CMOS technology. The signal collected in a sensor is amplified in the analogue tier and afterwords digitised with the ADC units placed in the digital tier. One could also considered adding layer handling data sparsification and data storage.

\chapter{MIMOSA detectors and the experimental setup}
\label{ch:tests:exp}

\section{The MIMOSA family}
\label{ch:tests:exp:MIMOSA}

The MAPS (Monolithic Active Pixel Sensor) \cite{tests:MAPS_Turchetta} devices are under consideration as active components of the ILC vertex detector. The MAPS technology provides a good spatial resolution, high signal-to-noise ratio, low material budget, low costs of fabrication and high radiation tolerance. The MAPS pixel matrices used for studies described in this thesis were the Minimum Ionizing MOS Active Pixel Sensors (MIMOSA) fabricated in the CMOS (Complementary MOS) technology. Numerous MIMOSA models were designed at IReS (Institut de Recherches Subatomiques) in Strasbourg (France), differing in physical properties, sensing elements, readout architectures and manufacturing processes. The main features of those, fabricated up to date, are summarised in table.~\ref{tab:tests:MIMOSA_Family} \cite{tests:IresTable}.\\
The first four prototypes were focused on the technology demonstration and on exploration of different manufacturing processes. The MIMOSA-1 was fabricated with a CMOS AMS\footnote[1]{AMS stands for austriamicrosystems} 0.6 process featuring a ``thick'' epitaxial layer of 14~$\mu$m. The MIETEC 0.35 process used for the production of the MIMOSA-2 provides a ``thin'' epitaxial layer of 4.2~$\mu$m. Moreover the electronics implemented in the MIMOSA-2 was designed according to stand high radiation doses. The MIMOSA-3 was the first attempt at the use of a deep-submicron fabrication technology which allowed the design of a large number of small pitch pixels in a small area. It also featured a very thin epitaxial layer thickness of 2~$\mu$m, resulting in a small signal. The MIMOSA-4 prototype was manufactured with a technology without epitaxial layer, using a low-doping substrate (a low-doping means concentration of $\sim$10$^{14}$~cm$^{-3}$). The MIMOSA-5 was the first large-scale prototype of 1.7$\times$1.7~cm$^{2}$ reticle-size. The next prototypes MIMOSA-6, 7 and 8 were focused on optimisation of the readout architecture and on the integration of on-chip functionalities. The MIMOSA-6, 7 and 8 had a column-parallel readout architecture, with the Correlated Double Sampling (CDS) operation performed on pixel. At the end of each column a discriminator was added providing a binary output. In addition, MIMOSA-7 features a novel charge-sensing element, the photoFET diode. The latter aims at the increase of the pixel response to the charge generated inside of the detector active volume. The MIMOSA-9 was fabricated with an Opto technology which currently seems to be the best choice for the fabrication of CMOS monolithic pixels, since it features a 20~$\mu$m thick epitaxial layer. An important feature of this prototype was the presence of different pixel pitches in different subarrays. The MIMOSA-10 was fabricated as the first chip prototype for the upgrade of the vertex detector in the STAR experiment at RHIC (Brookhaven National Laboratory) \cite{tests:STAR}. The MIMOSA-11 was based on the design of the MIMOSA-9 but with a modified layout of the charge collecting diode with minimum field oxide around the junction, in order to improve radiation tolerance. The goal of the MIMOSA-12 and 13 fabrication was to validate the usefulness of in-pixel memory cells. With such structures it will be possible to store quickly several frames in the in-pixel memory cells during beam interactions and in a second step read them during the dead time. This should minimise the impact of speed for the readout. In addition, further test structures were implemented in the prototypes to explore AMS 0.35~$\mu$m technology. In particular the goal was to increase the knowledge about n-well-Polysilicon capacitors and limits of an AC coupling principle needed, in the pixel, between sensitive element and first amplifier stage. The MIMOSA-14 was the enhanced prototype for the STAR experiment vertex detector upgrade. The MIMOSA-15 and 16 were a translation of the MIMOSA-8 design to the AMS~0.35~$\mu$m Opto technology. The advantage of the AMS~0.35~$\mu$m Opto process over the TSMC~0.25~$\mu$m, used for the MIMOSA-8 fabrication, is the thicker epitaxial layer of the order of 14~$\mu$m, resulting in a higher signal. The MIMOSA-17 was first the prototype of the chip devoted to the charged particle tracking in the EUDET telescope \cite{tests:eutel}. The next detector in the MIMOSA family was MIMOSA-18 chip designed for a precise charged particle tracking. The MIMOSA-18 pixels of 10~$\mu$m pitch are equipped with collecting diodes optimised for low dark current at the room temperature. The MIMOSA-19, which is dedicated to bio-medical applications, is equipped with collecting diodes of shapes optimized for better charge collection. The MIMOSA-20 was the third prototype for the vertex detector upgrade in the STAR experiment. With the MIMOSA-21 a STM~0.25~$\mu$m~biCMOS process was explored. The MIMOSA-22 was an intermediate prototype before the final sensor chip of the EUDET beam telescope for the ILC vertex detector studies. Its architecture is based on the MIMOSA-16 with a fast binary readout. The MIMOSA-23 is the fourth version of the monolithic integrated detector to be used in the STAR experiment at RICH. Its architecture is based on the MIMOSA-22 with a faster readout and larger pixel matrix. With the MIMOSA-24 and 25 the XFAB technology was explored. This process provides partially depleted substrate what results in an increase of a signal collected in the pixels. The MIMOSA-26 is the final chip for the EUDET beam telescope. It combines the sensitive area based on the MIMOSA-22 design and the SUZE-01 chip featuring the zero-suppression micro-circuit and output memories.\\
For the purpose of the presented study the models MIMOSA-5 and MIMOSA-18 were used at electron test beams at DESY (test beam area 22) and Frascati (DAFNE Beam-Test Facility). The DESY and the DAFNE infrastructures provides electrons in two complementary energy ranges, from 1 to 6 GeV and from 25 MeV to 750 MeV, respectively. The MAPS detectors were tested with electrons of a different track inclinations with respect to the detector surface.

\begin{table}[!htbp]

	\begin{tabularx}{1.1\textwidth}{@{\extracolsep{\fill}} |>{\small}c|>{\small}c|>{\small}c|>{\small}c|>{\small}c|>{\small}c|} \hline

	& & Manufacturing & Epilayer & Pixel & Pixel \\ 
	Prototype & Year & process & thickness & pitch & arrangement \\
	& & [$\mu$m] & [$\mu$m] & [$\mu$m] & (arrays/pixels)  \\ \hline \hline

	MIMOSA-1 & 1999 & AMS 0.6 & 14 & 20 & 4/64$\times$64 \\ \hline

	MIMOSA-2 & 2000 & MIETEC 0.35 & 4.2 & 20 & 6/64$\times$64\\ \hline	

	MIMOSA-3 & 2001 & IBM 0.25 & 2 & 8 & 1/128$\times$128 \\ \hline	

	MIMOSA-4 & 2001 & AMS 0.35 & none & 20 & 4/64$\times$64 \\ \hline

	\textbf{MIMOSA-5} & \textbf{2001/03} & \textbf{AMS 0.6} & \textbf{14} & \textbf{17} & \textbf{4/510$\times$512} \\ \hline

	MIMOSA-6 & 2002 & MITEC 0.35 & 4.2 & 28 & 1/30$\times$128 \\ \hline

	MIMOSA-7 & 2003 & AMS 0.35 & none & 25 & 1/64$\times$64 \\ \hline

	MIMOSA-8 & 2003 & TSMC 0.25 & 8 & 25 & 1/32$\times$128 \\ \hline

	MIMOSA-9 & 2004 & AMS 0.35 opto & 20 & 20, & 1/64$\times$64 \\ 
	& & & & 30,40 & 3/32$\times$32 \\ \hline

	MIMOSA-10 & 2004 & TSMC 0.25 & 8 & 30 & 2/64$\times$128 \\ \hline

	MIMOSA-11 & 2005 & AMS 0.35 opto & 20 & 30 & 4/42$\times$42 \\ \hline

	MIMOSA-12 & 2005 & AMS 0.35 hires & none & 35 & 6/14$\times$8 \\ \hline

	MIMOSA-13 & 2005 & AMS 0.35 hires & none & 20 & 1.4k \\ \hline	

	MIMOSA-14 & 2005 & AMS 0.35 opto & 14 & 30 & 2/128$\times$64 \\ \hline	

	MIMOSA-15 & 2005 & AMS 0.35 opto & 14 & 20,30 & 4/42$\times$42 \\ \hline

	MIMOSA-16 & 2006 & AMS 0.35 opto & 14,20 & 25 & 4/32$\times$32 \\ \hline

	MIMOSA-17 & 2006 & AMS 0.35 opto & 14,20 & 30 & 4/256$\times$64 \\ \hline

	\textbf{MIMOSA-18} & \textbf{2006} & \textbf{AMS 0.35 opto} & \textbf{14,20} & \textbf{10} & \textbf{4/512$\times$512} \\ \hline

	MIMOSA-19 & 2006 & AMS 0.35 opto & 14,20 & 12 & 80k \\ \hline

	MIMOSA-20 & 2006/08 & AMS 0.35 opto & 14,20 & 30 & 10/320$\times$164 \\ \hline

	MIMOSA-21 & 2006 & STM 0.25 biCMOS & none & 10 & 16/64$\times$24 \\
	& & & & 20 & 16/32$\times$12 \\ \hline	

	MIMOSA-22 & 2007 & AMS 0.35 opto & 14 & 18.4 & 73k digi./4.6k analog \\ \hline

	MIMOSA-23 & 2008 & AMS 0.35 opto & 14 & 30 & 1/640$\times$640 \\ \hline

	MIMOSA-24 & 2008 & XFAB 0.35 & 14 & 20,30 & 13k \\ \hline

	MIMOSA-25 & 2008 & XFAB 0.6 PIN & 14 & 20,30,40 & 4.8k \\ \hline

	MIMOSA-26 & 2008 & AMS 0.35 opto & 14 & 18.4 & 9/128$\times$576 \\ \hline

	\end{tabularx}

\caption[Design features of the fabricated MIMOSA prototypes]{Design features of the fabricated MIMOSA prototypes \cite{tests:IresTable}.}
\label{tab:tests:MIMOSA_Family}

\end{table}

\subsection{The MIMOSA-5 pixel matrix}
\label{ch:tests:exp:MIMOSA:5}

\begin{figure}[!h]
        \begin{center}
	\subfigure[]{
		\includegraphics[width=0.47\textwidth]{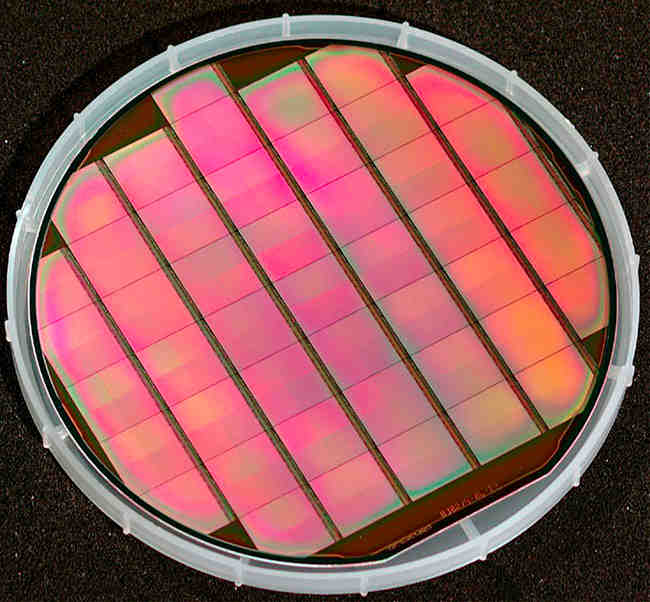}
		\label{fig:tests:MIMOSA-5_detector:wafer}
	}
	\hspace{0.1cm}
	\subfigure[]{
		\includegraphics[width=0.47\textwidth]{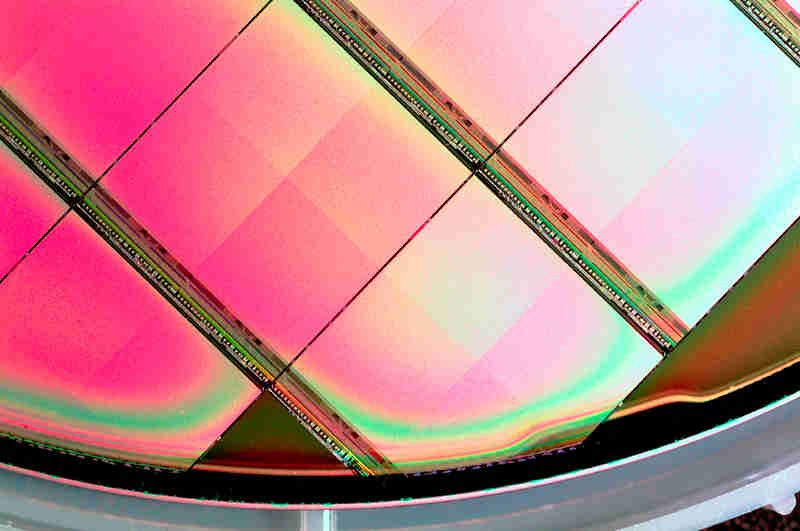}
		\label{fig:tests:MIMOSA-5_detector:matrix}
	}
        \caption[The MIMOSA-5 wafer before cutting]{(a) The MIMOSA-5 wafer before cutting, (b) detail of the MIMOSA-5 wafer (courtesy IReS, Strasbourg).}
        \label{fig:tests:MIMOSA-5_detector}
        \end{center}
\end{figure}

\begin{figure}[!h]
        \begin{center}
                \includegraphics[width=0.5\textwidth]{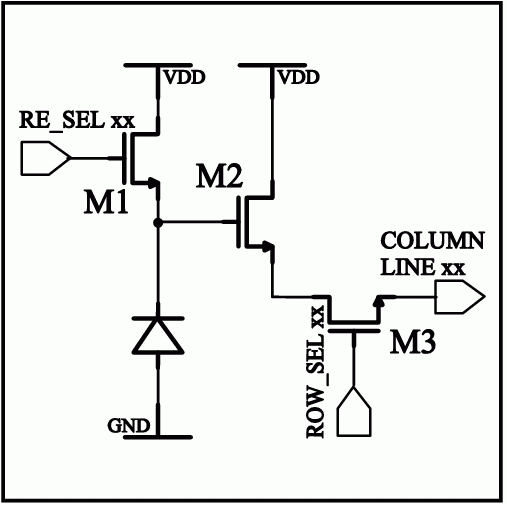}
                \caption[The single pixel readout architecture of the MIMOSA-5 detector]{The single pixel readout architecture of the MIMOSA-5 detector.}
                \label{fig:tests:3Tarchitecture}
        \end{center}
\end{figure}

The MIMOSA-5 \cite{tests:DeptuchTh} is the first large-scale MAPS detector with dimensions of 19.4 $\times$ 17.35~mm$^{2}$. The chip was designed in order to study the issue whether the detector parameters like noise level, charge particle tracking efficiency or spatial resolution deteriorate with the increasing scale of the device. Additionally the large scale of the MIMOSA-5 allows testing the device thinning procedure, difficult in case of small prototypes.\\
The MIMOSA-5 matrix is build in an AMS~0.6~$\mu$m CMOS technology (5~V maximum voltage) which has a 14~$\mu$m thick epitaxial layer. Many modules are fabricated on one common silicon wafer, as shown in fig.~\ref{fig:tests:MIMOSA-5_detector:wafer}. Due to ``clever dicing'', they are aligned in one direction forming a structure of five or seven matrices which are not electrically connected. The dead area between the consecutive chips is approximately 200~$\mu$m. The readout electronics, 2~mm wide, is placed at the bottom of each unit fig.~\ref{fig:tests:MIMOSA-5_detector:matrix}. All bonding pads and readout control logic cells are grouped along one edge. No rules for radiation tolerant layout were applied in the detector design.
\\
A single MIMOSA-5 matrix consists of 4 subarrays of 510$\times$512 active pixels, which gives approx. 1~million pixels in the active area of the device. Pixels in each submatrix have the size of 17$\times$17~$\mu$m$^{2}$. Each pixel is equipped with a collecting diode. The only difference between submatrices are in different sizes of the collecting diodes. Pixels with ``small'' collecting diodes (3.1$\times$3.1~$\mu$m$^{2}$) are placed in two submatrices called T01 and B01 and pixels with ``big'' collecting diodes (4.9$\times$4.9~$\mu$m$^{2}$) are placed in the other two submatrices called T02 and B02. Each chip has four independent analog outputs, i.e. one output per submatrix. The readout electronics is optimised to accept maximum readout clock frequency of 40~MHz.\\
The readout of the MIMOSA-5 matrix is designed in the so called 3-transistor architecture, as shown in fig.~\ref{fig:tests:3Tarchitecture}. A photo-diode is integrated on an individual pixel together with the first signal processing electronic. The M1 transistor resets the the photo-diode to reverse bias, the M2 transistor is part of the source follower and with the M3 transistor a row selection is performed. The current source for the source follower and the column selection switch are located outside the pixel. Such a pixel configuration provides continuous charge integration between two consecutive reset operations. 
\\
The reset operation is needed for removing the charge captured by the collecting diodes in order to avoid its saturation. This introduces reset noise which can be effectively removed by means of the Correlated Double Sampling technique \cite{tests:CDS_Hyneck}. The latter reduces also the influence of the low-frequency (i.e. 1/f) noise and of the noise component deriving from non-uniformities in the array, the so-called Fixed Pattern Noise.\\
The analog readout of the MIMOSA-5 chip requires, apart from a few biasing lines, only two digital signals to operate: ($i$) the \texttt{CLOCK} signal used for addressing pixels and for selecting the columns to restore the reverse bias on the charge-sensing node during the reset phase, ($ii$) the \texttt{RESET} signal applied to to the gate of the M1 transistor which initiates the reset phase. Analog power supplies, bias signals and analog outputs are separated for the individual subarrays of each chip, and the corresponding lines are routed independently. Digital parts used for control and addressing are also independent for each array, but are powered from common digital power supplies and are driven from common control lines \texttt{CLOCK} and \texttt{RESET}.\\
The pixels during readout are addressed sequentially by means of an appropriate row and column selection. The output of each pixel is sent alternatively to 6 horizontal readout lines through a p-MOS source follower which is placed at the bottom of every column. Each readout line is terminated with a voltage amplifier with a gain of 5, whose output is multiplexed to the common output buffer. The columns are selected in groups of 3. When a chosen group of 3 columns is read out, the following group is being ``prepared'' by switching on the bias current passing through the source follower transistors of the corresponding pixels. At the end of every row two additional clock cycles are necessary in order to provide enough time for preparation of the first pixel of the next one. Thus two last pixels in each row are readout twice.

\subsection{The MIMOSA-18 pixel matrix}
\label{ch:tests:exp:MIMOSA:18}

\begin{figure}[!h]
        \begin{center}
                \includegraphics[width=0.5\textwidth]{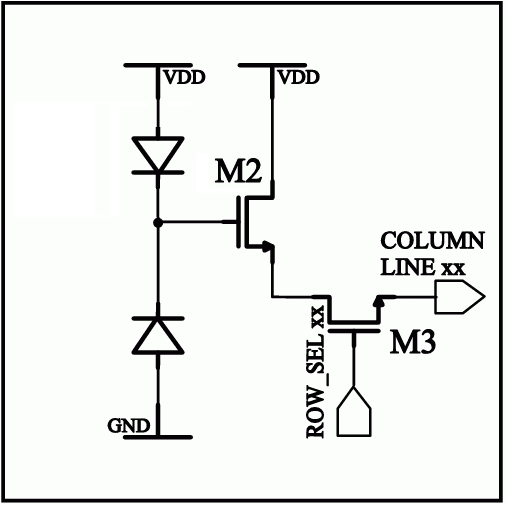}
                \caption[The single pixel readout architecture of the MIMOSA-18 detector]{The single pixel readout architecture of the MIMOSA-18 detector, based on the self-bias diode.}
                \label{fig:tests:self-bias}
        \end{center}
\end{figure}

The MIMOSA-18 chip is optimised for high precision tracking. It was fabricated using the AMS~0.35~$\mu$m OPTO process. The latter is an advanced mixed-signal CMOS process, providing four metal layers, two polysilicon layers, high-resistivity polysilicon and two types of transistor gates (3.3 V and 5 V). The MIMOSA-18 prototypes are available in two versions: ``standard'' with 14~$\mu$m thick epitaxial layer and ``experimental'' with 20~$\mu$m thick epitaxial layer. Presently it is the thickest epitaxial layer available through a commercial CMOS process. The performed measurements refer to the standard version of the MIMOSA-18 detector equipped with the 14~$\mu$m epitaxial layer.
\begin{figure}[!h]
        \begin{center}
          \includegraphics[width=0.5\textwidth]{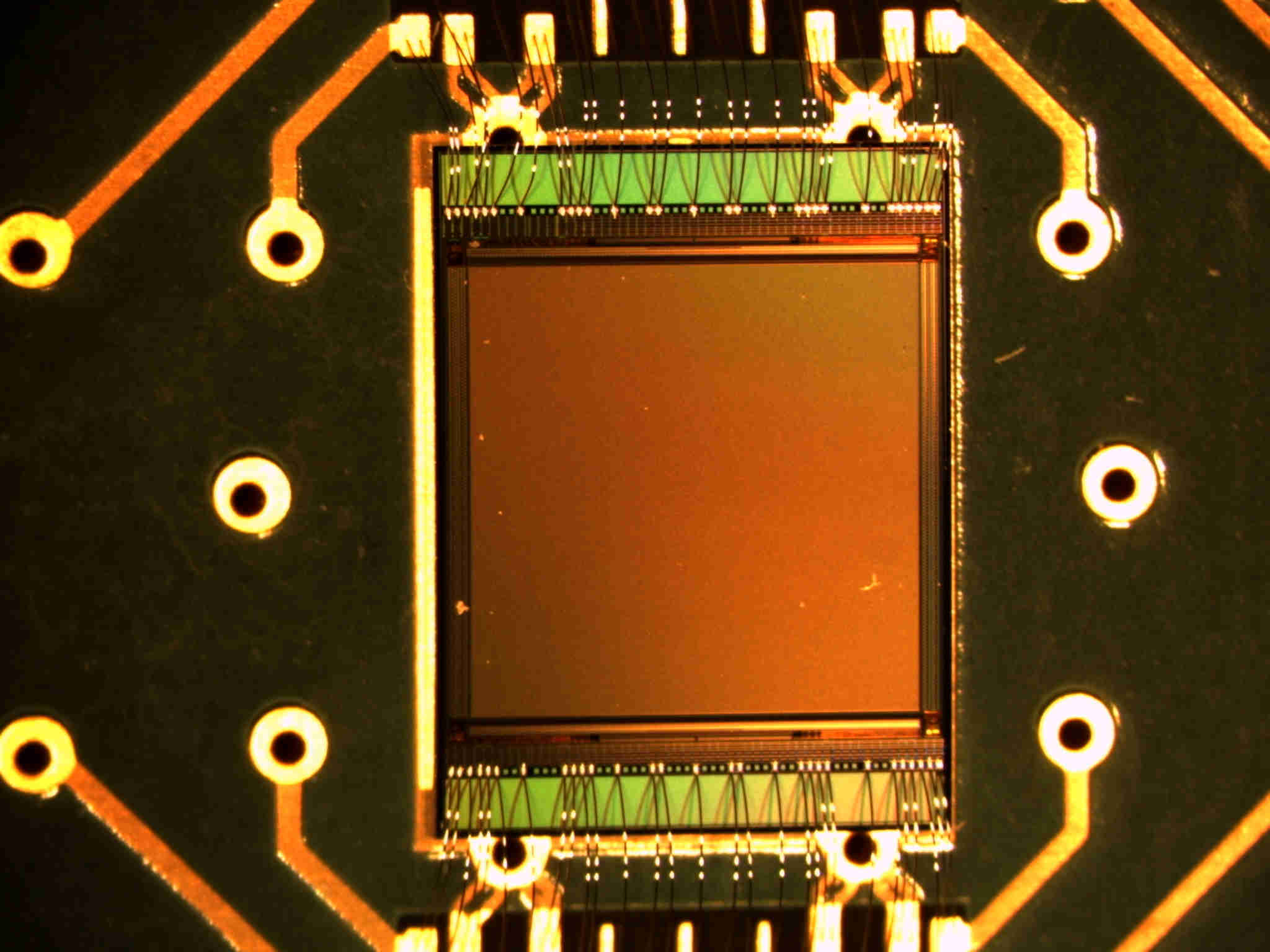}
        \caption[MIMOSA-18 prototype mounted on a Printed Circuit Board]{MIMOSA-18 prototype mounted on a Printed Circuit Board (courtesy IReS, Strasbourg).}
        \label{fig:tests:ReticalM18}
        \end{center}
\end{figure}\\
A MIMOSA-18 matrix fig.~\ref{fig:tests:ReticalM18} consists of 4 subarrays of 256$\times$256 active pixels each. This gives $\sim$26~k pixels distributed over an 5$\times$5~mm$^{2}$ active area of the device. All submatrices are equipped with pixels of 10~$\mu$m pitch with a collecting node of 4.4$\times$3.4~$\mu$m$^{2}$ each. The n-well/p-epi diodes responsible for the charge collection are optimised for a low dark current at the room temperature but they are not designed to be radiation resistant. The pixel readout scheme is based on the self-bias diode architecture, which is shown fig.~\ref{fig:tests:self-bias}. In this design the charge collecting n-well diode is continuously biased by an another diode (forward biased) implemented inside the sensing n-well. Thus the charge collected in the sensing node of a pixel is removed continually and no reset operation is required. The M2 transistor is a part of the source follower and the row selection is performed with the M3 transistor. The current source for the source follower and the column selection switch are located outside the pixel. Such a pixel configuration provides continuous charge integration.\\
Bonding pads and readout control logic cells occupy the two opposite edges of the chip of 1~mm width each.
\\
Every chip has four parallel outputs. The signal information from each pixel is serialised by a circuit (one per subarray), which can operate up to a 25 MHz readout clock frequency. The data readout from the device are submitted to the CDS processing.\\
The analog readout of the MIMOSA-18 chip requires, apart from a few biasing lines, only two digital signals to operate: the \texttt{CLOCK} and \texttt{RESET}. The \texttt{CLOCK} signal is used for addressing pixels while \texttt{RESET} signal is needed for a digital control of the Mimosa-18 readout. The \texttt{RESET} pulse, occurring before each acquisition, resets the shift registers used for the pixel addressing and causes the overwriting of the previous events stored in the SRAM memory until a trigger comes.

\subsection{Readout of the MIMOSA devices}
\label{ch:tests:exp:MIMOSA_readout}

\begin{figure}[!h]
        \begin{center}
                \includegraphics[width=\textwidth]{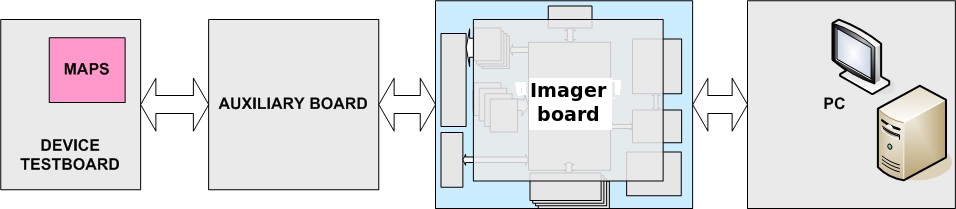}
                \caption[A schematic picture of the MIMOSA readout chain]{A schematic picture of the MIMOSA readout chain (courtesy IReS, Strasbourg).}
                \label{fig:tests:MIMOSA_readout_scheme}
        \end{center}
\end{figure}

A schematic picture of the MIMOSA detectors readout chain is shown in fig.~\ref{fig:tests:MIMOSA_readout_scheme}. The MIMOSA chips are mounted and wire-bonded on a Device Test Board (front-end board) which is used as a support. It contains the first stage of external amplifiers and current source needed for the detector operation. The front-end board is connected to an Auxiliary Board (repeater board). The latter is powered by an external power supply. It also generates reference voltages for the on-chip and external amplifiers. Moreover the repeater board is used for the two-direction transmission of digital control signals between the chip and the Imager Board, and for a transmission of the analog output data for digitisation.
\\
The Imager Boards are housed in the VME crate. A single Imager Board generates the needed digital signals, i.e. the readout \texttt{CLOCK} and \texttt{RESET} signals, which are then transmitted to the front-end board via the repeater board. The digital control of the data acquisition is handled by a Xilinx Field Programmable Gate Array (FPGA) installed on the Imager Board. The Xilinx FPGA logic units are also programmed to implement the on-line CDS data processing. The Imager Board contains also the SRAM memory for storing the two consecutive images (frames) of the full detector which are next used in the CDS calculation.\\ 
In the MIMOSA-5 setup the Imager Board is controlled by the CETIA VME PowerPC CPU running the LynxOS real-time operating system. An ethernet link connects the VME CPU to a remote Linux PC. The data acquisition software (DAQ), provided by the IReS/LEPSI group, runs on the VME CPU and reads data from the SRAM memory. The readout data are sent through the ethernet link to the hard disk of the Linux PC and they are stored there. In case of the MIMOSA-18 setup the VME crate provides only a power supply for the Imager Board. The software based control and data collection, developed by the IReS/LEPSI, was running under Window PC. The latter is connected to the Imager Board through the USB 2.0 link. The data read from the SRAM memory are transmitted via the USB 2.0 link from the Imager Board to the disk of the Windows PC. Both of the above Imager Boards contains 4 independent Flash ADC Units with 12 bit resolution which are used for digitisation of the analog output data from the MAPS devices.\\
There are two major methods of reading out the MIMOSA detectors: the free running mode and the trigger mode. The first one is used in pedestal runs and during irradiation of the device with a radioactive source when no trigger information is available. The second method is used during beam tests when a trigger is provided to the data acquisition system. More details on the free-running mode and the trigger mode are given in section~\ref{ch:tests:exp:CDS}.

\section{Correlated Double Sampling (CDS)}
\label{ch:tests:exp:CDS}

\begin{figure}[!h] 
	\begin{center}
	\subfigure[]{
		\includegraphics[width=0.28\textwidth,angle=90]{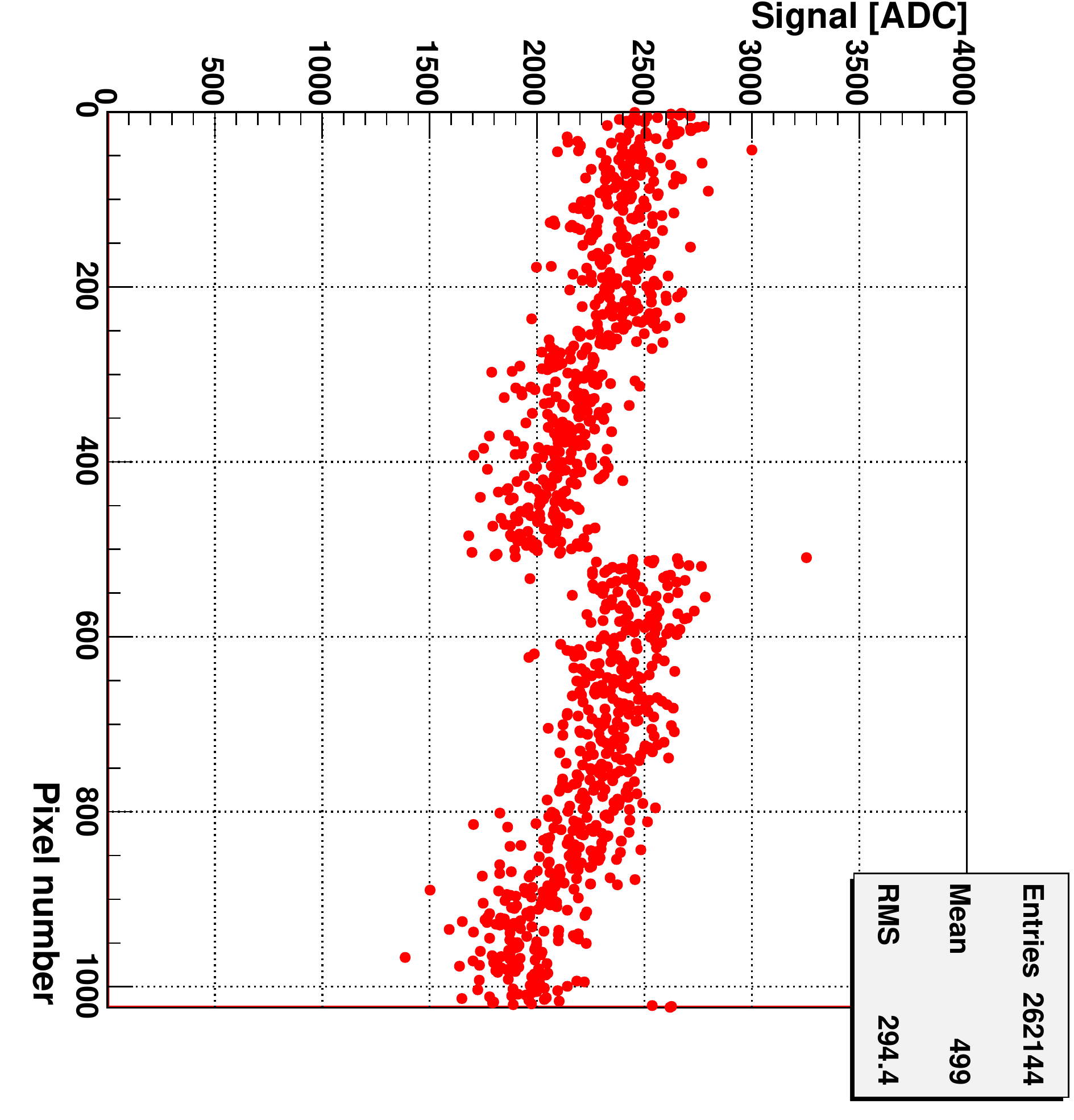}
		\label{fig:tests:CDS:F0}
	}
	\hspace{0.1cm}
	\subfigure[]{
		\includegraphics[width=0.28\textwidth,angle=90]{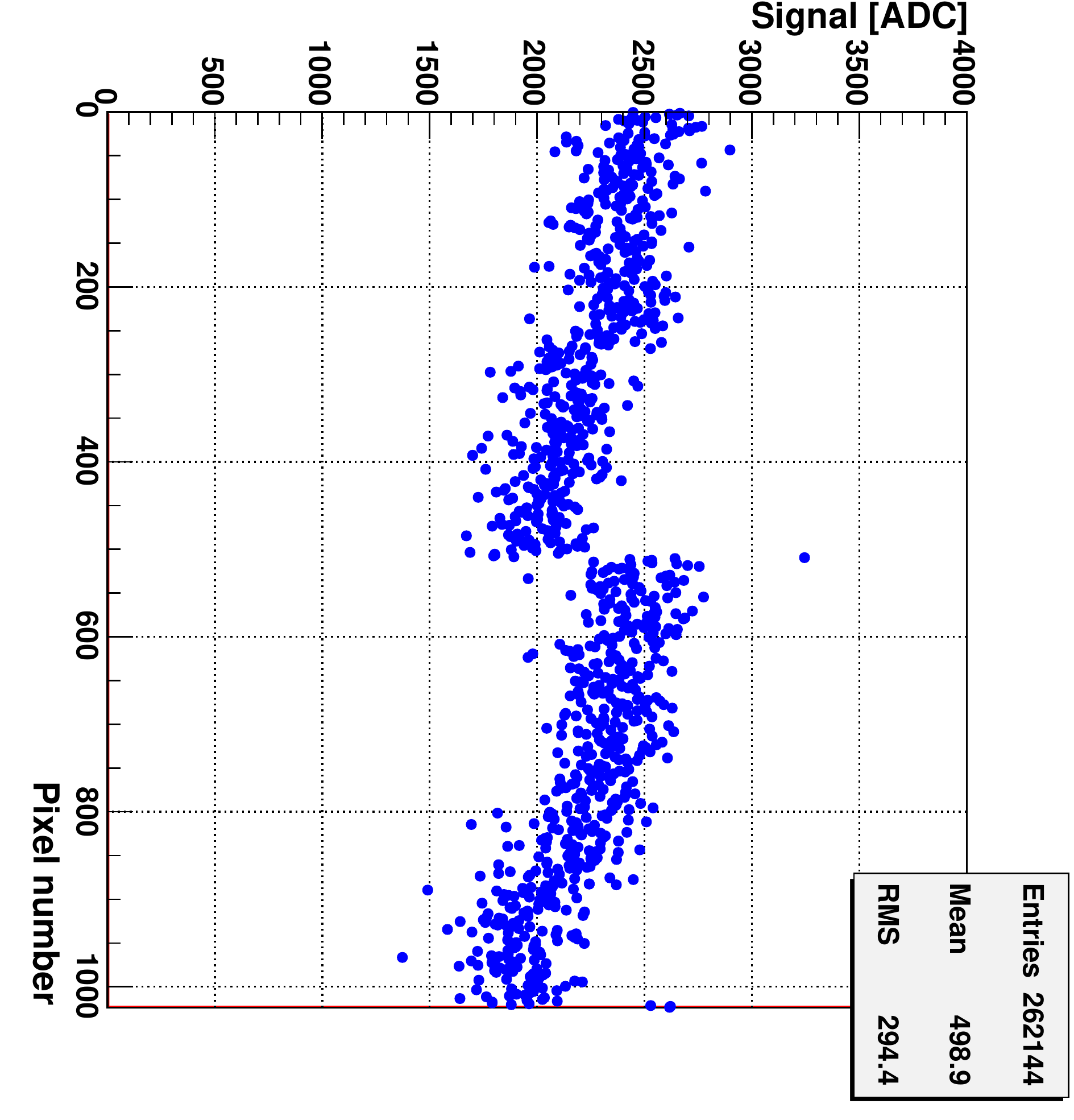}
		\label{fig:tests:CDS:F1}
	}
	\hspace{0.1cm}
	\subfigure[]{
		\includegraphics[width=0.28\textwidth,angle=90]{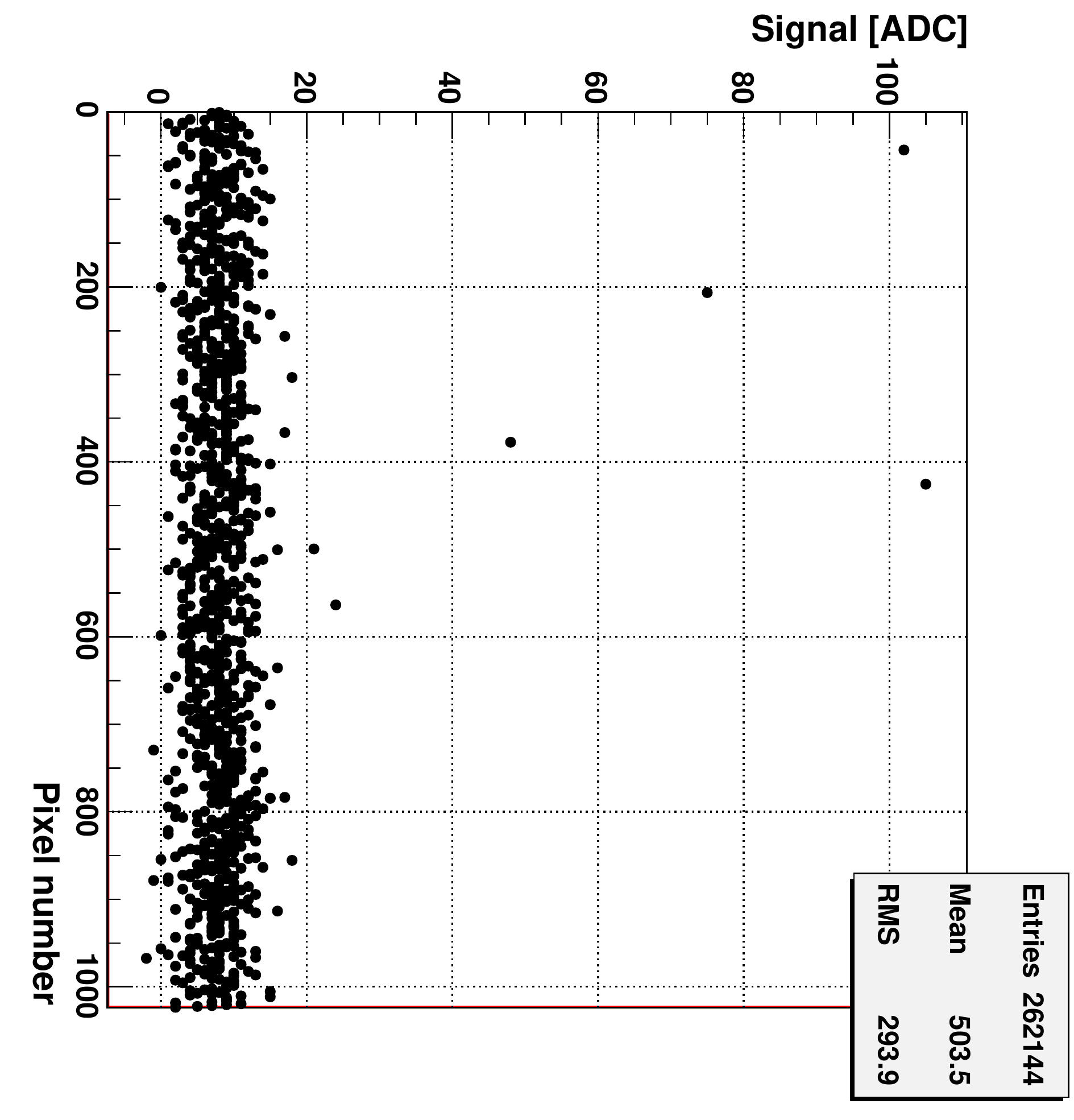}
		\label{fig:tests:CDS:CDS}
	}
	\caption[Example of raw signals acquired from the MIMOSA-5 detector, before and after CDS processing]{Example of raw signals acquired from the MIMOSA-5 detector: (a) first frame read out from the detector, (b) second frame read out from the detector, (c) result of the two frames subtraction (CDS technique). The plots refer to a subset of pixels of one MIMOSA-5 matrix (two consecutive rows of the MIMOSA-5 prototype).}
	\label{fig:tests:CDS}
	\end{center}
\end{figure}

In the free running mode all pixels are read out consecutively and the procedure is continuously repeated until a required number of frames is collected. It takes approximately 26~ms to read one frame in the MIMOSA-5, which consists of 261120 pixels (i.e. 100~ns per pixel). Physical and noise signals appear on pixels during a readout cycle. Information from two consecutive frames is used to extract physical signals. In short, this is done by subtracting signals from these two frames which leaves the physical signal and cancels noise signals. In the trigger mode the procedure is similar. The array is read out continuously, as above. Assume that the trigger signal appears when the $n$-th pixel has been read out. Then the new readout cycle is initiated and starts with the $(n+1)$-th pixel, passes the last, goes to the first, and ends with the $n$-th pixel (one frame). The previous frame is extracted from the memory and the comparison is done as above. This type of signal processing is called Correlated Double Sampling.\\
An example of the two consecutive frames sampled in a subset of pixels contained in two firs rows of the the MIMOSA-5 detector and the result of the CDS processing is shown in fig.~\ref{fig:tests:CDS}. The signal remaining after CDS fig.~\ref{fig:tests:CDS:CDS}, of non-zero mean value, is generated by the leakage current and charge particles traversing detector. In case of the MIMOSA-18 detector, which is equipped with self-bias diodes, the average signal after CDS is close to zero. The pixels substantially diverging from the mean value refer to pixels exhibiting increased leakage current or pixels affected by the ionising particle impinging the detector. A collection of CDS processed signals cumulated during one readout cycle of the whole detector is called an event.

\section{Experimental setup}
\label{ch:tests:exp:setup}

The Device Test Board with the MAPS detector was mounted on an adjustable mechanical support, enabling rotations around the $X$ and $Y$ axes, perpendicular to the beam direction. The matrix was oriented  manually to the desired angles before each data taking run using the angular scale with an accuracy of approximately $\pm 1^\circ$. A temperature stabilisation and a cooling of the detector under test (DUT) was provided by the cooling liquid flowing through the channels drilled in the walls of the aluminium support. The temperature of the cooling liquid was controlled by a thermostat mounted in the cooling unit. The latter was connected to the support with a system of pipes. The DUT temperature was measured with a sensor placed next to it. The support together with the DUT was contained in a polystyrene box for thermal insulation. The latter provides also shielding against light. The MIMOSA-5 chip required cooling to a temperature close or lower than 0$^{\circ}$C. Thus nitrogen was flown through the cooling box in order to maintain the environment dry and avoid steam condensation on the detector surface and electronics. The MIMOSA-18 required only thermal stabilisation (no cooling) since it works at room temperatures.\\
The test experimental setup at DESY contains, apart from the MIMOSA chip, its mechanics and dedicated electronics as well as the position telescope described in section \ref{ch:tests:exp:setup:telescope}. The latter provides reconstruction of the beam particle tracks passing the DUT. The MIMOSA chips were mounted between the first two planes of the telescope. The trigger, provided from a system of small plastic scintillators, placed on both sides of the telescope, was shared between the MIMOSA and the telescope readout systems. First, the trigger signal was send to the MIMOSA Imager Board and afterwards transmitted to the telescope sequencer driving the telescope readout. The measurements at the DAFNE beam test facility were done without an external trigger and tracking system. Both of the MIMOSA chips were clocked with 10~MHz during all measurements.

\subsection{The DESY beam test facility}
\label{ch:tests:exp:setup:DESY}

\begin{figure}[!h]
	\begin{center}
		\includegraphics[width=0.8\textwidth]{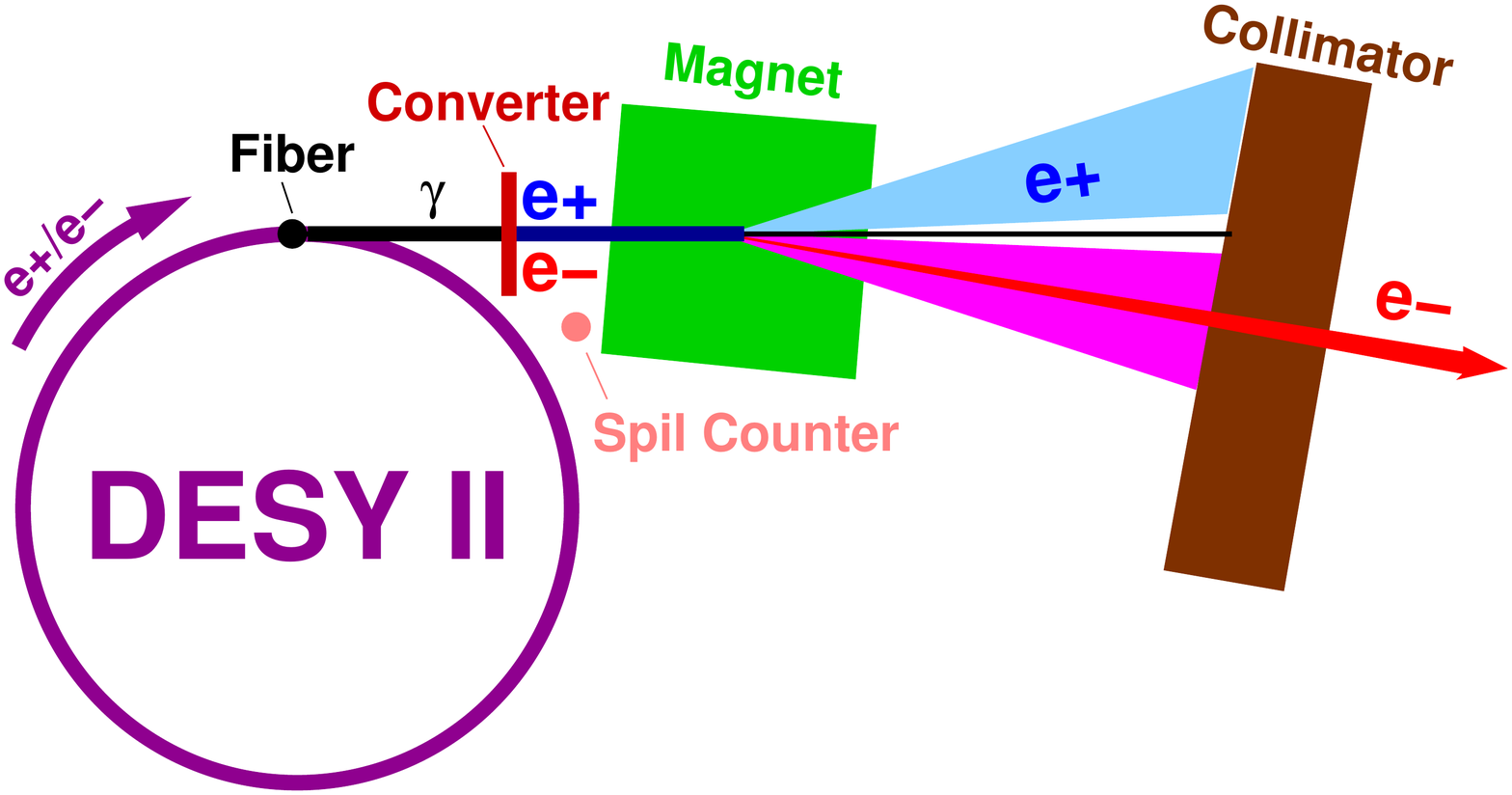}
		\caption[The beam delivery system at DESY]{The beam delivery system at DESY (courtesy DESY, Hamburg).}
		\label{fig:tests:testbeamlayout_DESY}
	\end{center}
\end{figure}

The DESY-II test beam infrastructure provides electrons or positrons of energies from 1~to~6~GeV with a rate from $\sim$~1~kHz~to~$\sim$~1~Hz, respectively. The beam delivered into the experimental hall consists of  secondary electrons or positrons produced in two conversions of the primary electron/positron beam circulating in the DESY-II ring. A schematic picture of the beam delivery system is shown in~fig.~\ref{fig:tests:testbeamlayout_DESY}.
First, a bremsstrahlung beam is generated by a primary target composed of 7~$\mu$m carbon fibre put in the electron/positron beam of the DESY-II accelerator. Afterwards bremsstrahlung photons are converted into electron-positron pairs (e$^{+}$e$^{-}$) on a secondary target. In case of the secondary target one can choose among aluminium or copper plates of various thickness. The beam energy selection is done by the bending magnet placed in front of the collimator slit. The adjustable collimator slit allows to influence the lateral size of the beam and its intensity.

\subsection{The telescope}
\label{ch:tests:exp:setup:telescope}

\begin{figure}[!h]
        \begin{center}
                \includegraphics[width=\textwidth]{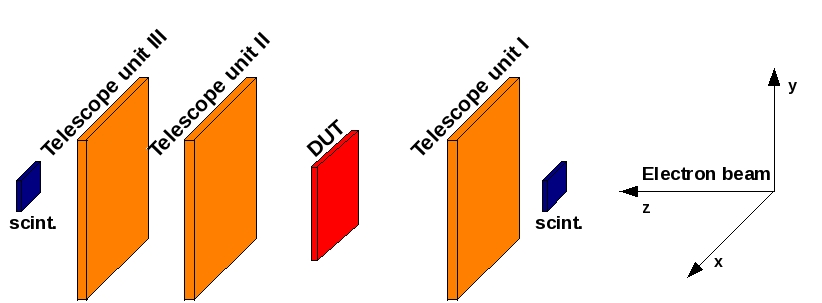}
                \caption[Layout of the beam telescope and the reference coordinate system]{Layout of the beam telescope and the reference coordinate system.}
                \label{fig:tests:ref_sys}
        \end{center}
\end{figure}

The DESY-II test beam area is equipped with a reference telescope which provides tracking for beam particles. The telescope consists of three silicon microstrip units mounted on an optical bench along the beam axis (Z direction), as shown in fig.~\ref{fig:tests:ref_sys}. Each telescope unit is built of two planes of perpendicularly oriented microstrips and gives information on the X and Y coordinates of the track position (below called X-plane and Y-plane). The trigger is provided by the coincidence of signals from three small plastic scintillators placed on both sides of the beam telescope (two in front and one behind). The strip telescope readout is controlled by the data acquisition system developed by the author. The detector under test, DUT, is mounted between the first and the second telescope unit.

\subsubsection{The microstrip unit}
\label{ch:tests:exp:setup:MVDTele}

A single telescope microstrip unit consists of two high resolution, single-sided, AC-coupled silicon microstrip detectors, mounted close to each other ($\sim$~2~mm) with horizontal and vertical strip orientations, respectively. Both microstrip detectors, approx. 300~$\mu$m thick, are encapsulated in an electrically shielded metal box with thin aluminium windows and have the sensitive area of 32$\times$32~mm$^{2}$ each. A single strip detector is equipped with 1280~strips of 25~$\mu$m pitch, however only every second strip has readout circuit, what gives 640~readout channels (with the effective pitch of 50~$\mu$m). The readout is handled by five VA chips \cite{tests:vikingVA} of 128 channels each. Every channel has a charge sensitive pre-amplifier followed by a CD-RC shaper, with a 2~$\mu$s peaking time, and a sample-and-hold circuitry. The data for all channels are serially sent to the output buffer through an on-chip multiplexer. The analog signals are digitised by CAEN~V550 flash ADCs~\cite{tests:v550} (with 10 bits resolution and a maximum conversion rate of 5~MHz) which are controlled by the CAEN~V551 sequencer \cite{tests:v551}. 
Microstrip detectors are operating in a fully depleted mode which is achieved at the bias voltage~of~$\sim$50~V. 

\subsection{Telescope data processing}
\label{ch:tests:exp:setup:MVDAnalysis}

\subsubsection{Cluster selection}
\label{ch:tests:exp:setup:MVDAnalysis:clust}

\begin{figure}[!h] 
	\begin{center}
	\subfigure[]{
		\includegraphics[width=0.45\textwidth,angle=90]{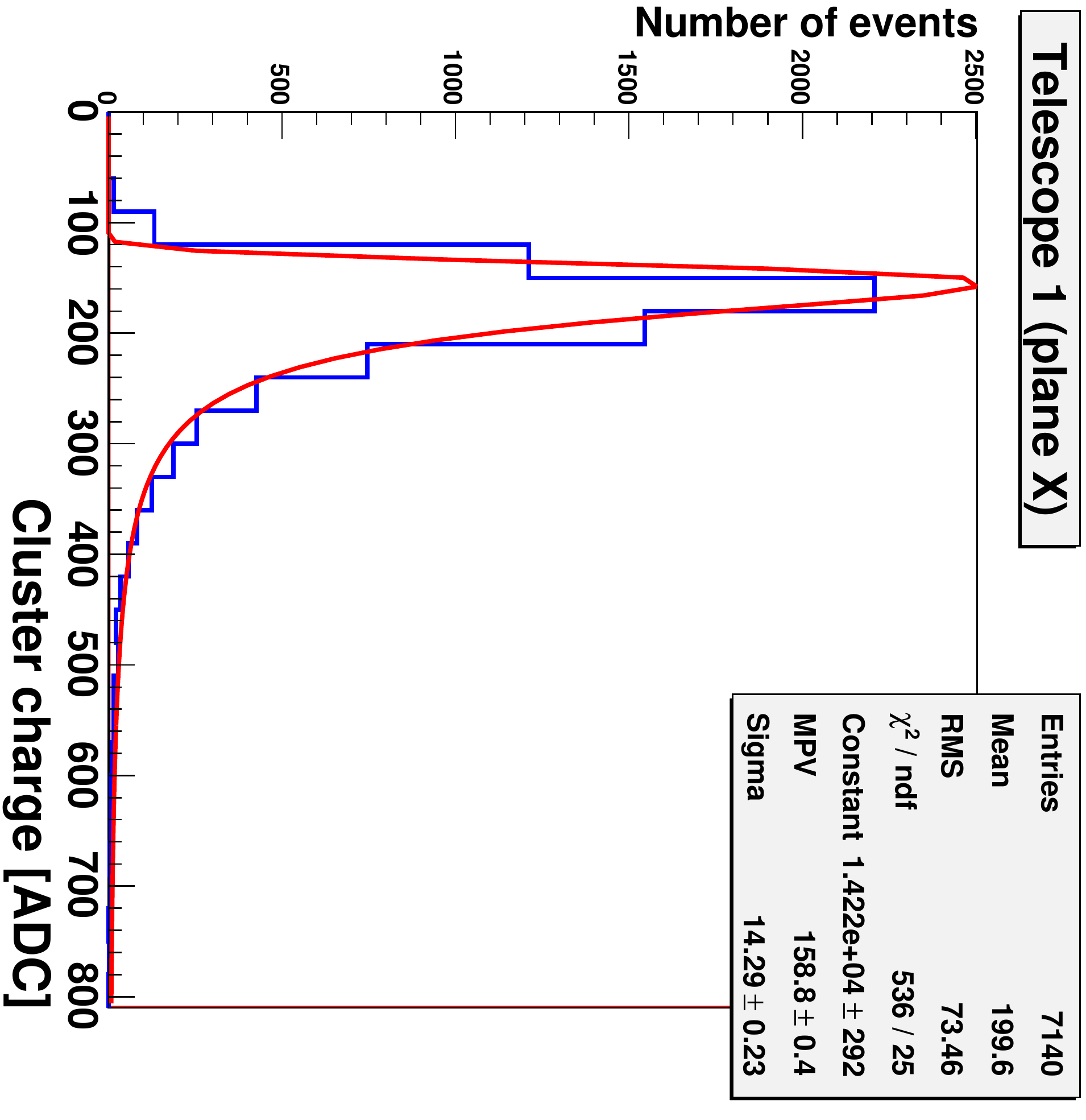}
		\label{fig:tests:ClusterCharge_2012:X}
	}
	\hspace{0.1cm}
	\subfigure[]{
		\includegraphics[width=0.45\textwidth,angle=90]{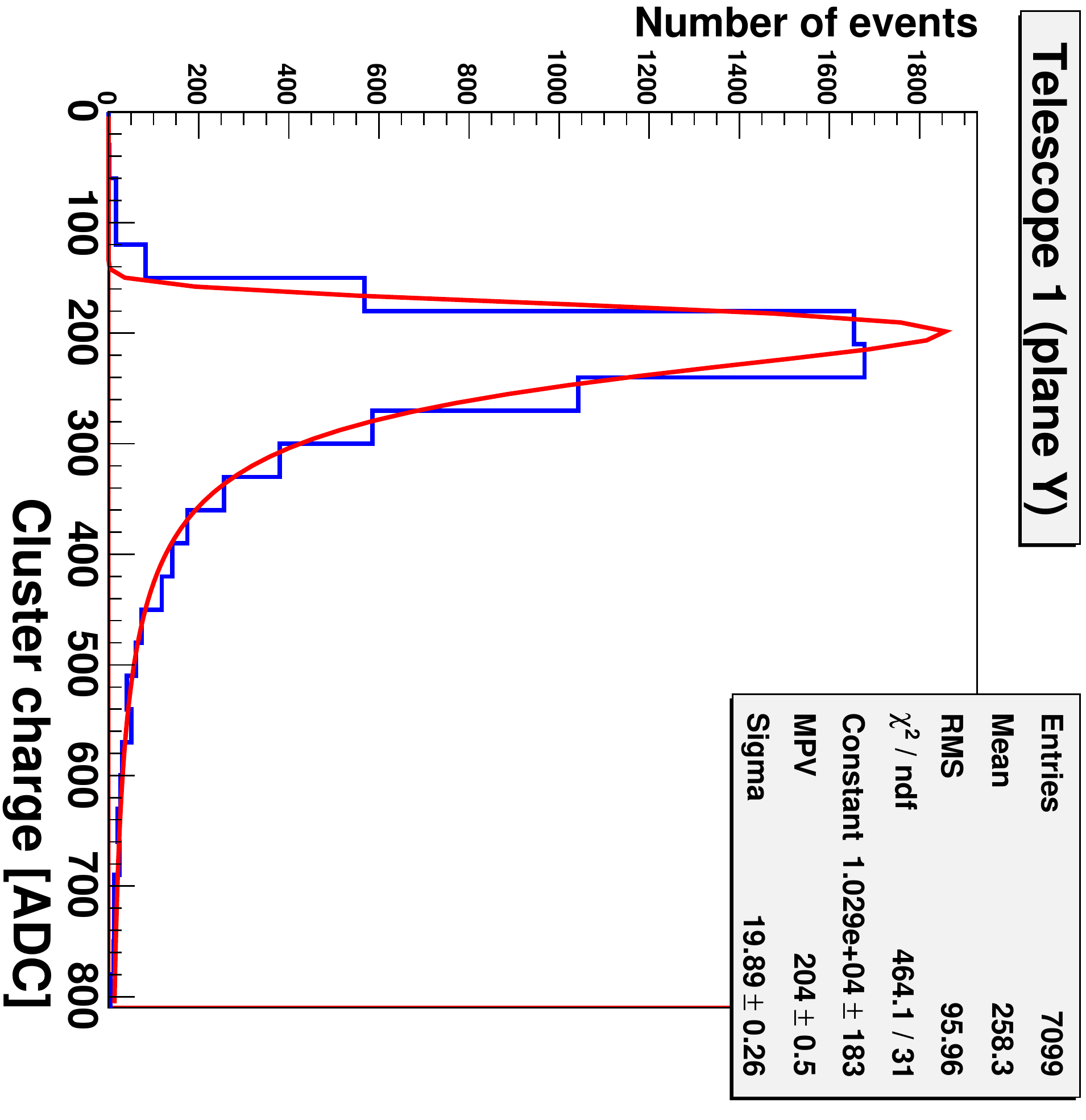}
		\label{fig:tests:ClusterCharge_2012:Y}
	}
	\caption[Cluster charge distributions for the first telescope plane and the corresponding Landau fits]{Cluster charge distributions for the first telescope plane and the corresponding Landau fits, (a) X-plane, (b) Y-plane.}
	\label{fig:tests:ClusterCharge_2012}
	\end{center}
\end{figure}

\begin{figure}[!h] 
	\begin{center}
	\subfigure[]{
		\includegraphics[width=0.45\textwidth,angle=90]{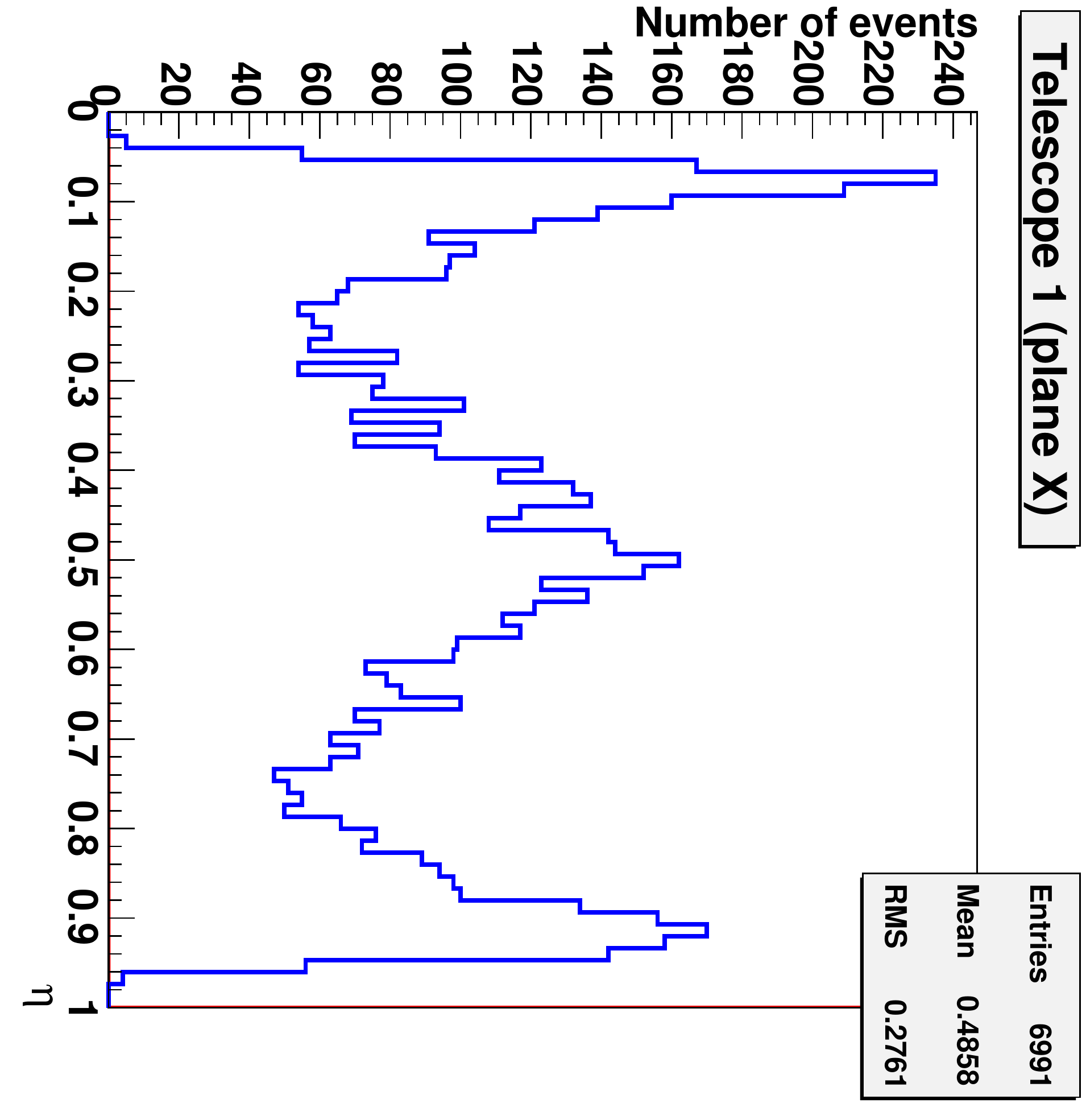}
		\label{fig:tests:Eta:Eta}
	}
	\hspace{0.1cm}
	\subfigure[]{
		\includegraphics[width=0.45\textwidth,angle=90]{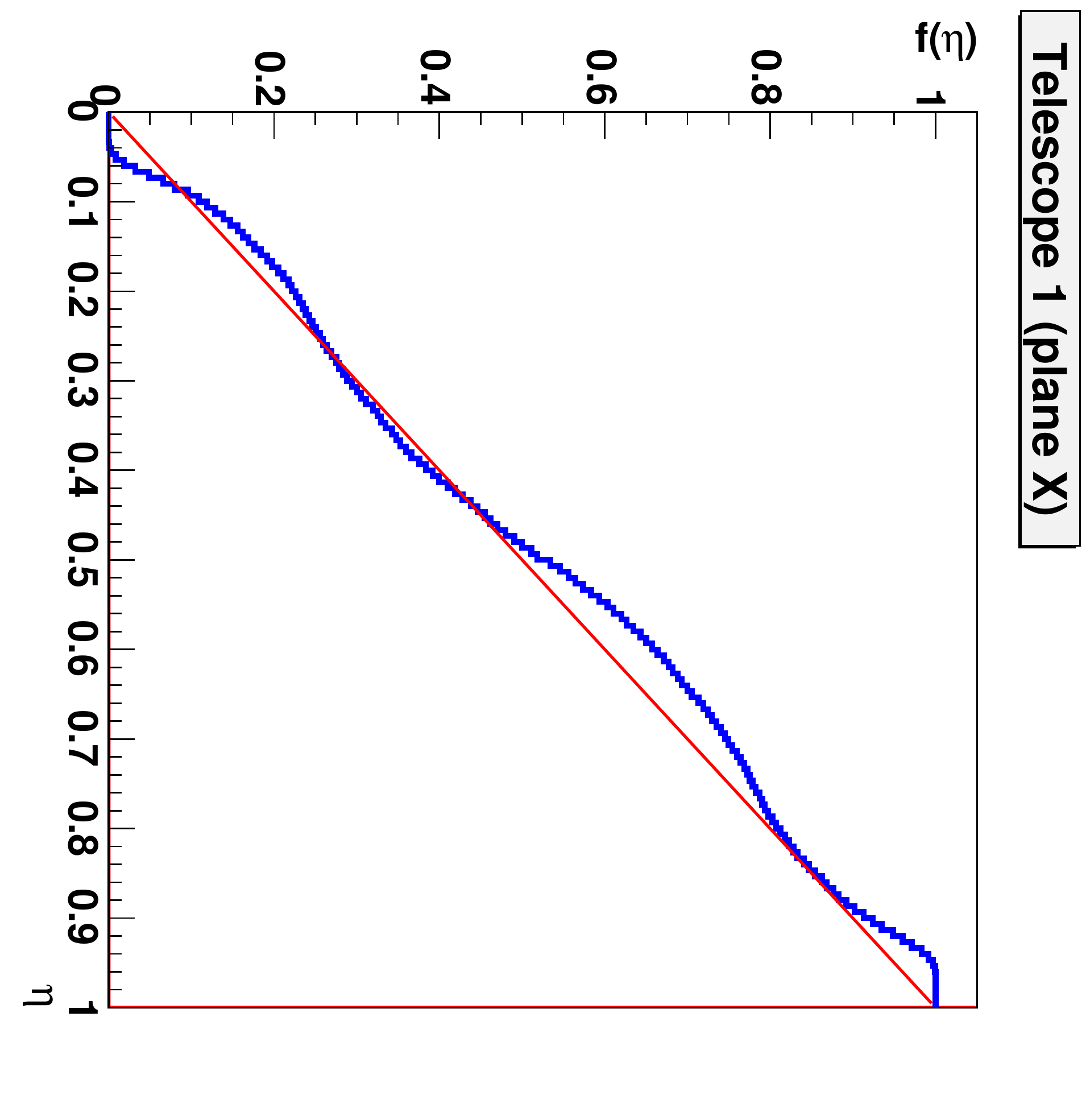}
		\label{fig:tests:Eta:Fun}
	}
	\caption[The $\eta$-algorithm correction]{The $\eta$-algorithm correction: (a) distribution of the $\eta$ variable, (b) the cumulative distribution function of the $\eta$ variable.}
	\label{fig:tests:Eta}
	\end{center}
\end{figure}

\begin{figure}[!h]
        \begin{center}
                \resizebox{0.7\textwidth}{!}{
                        \includegraphics[]{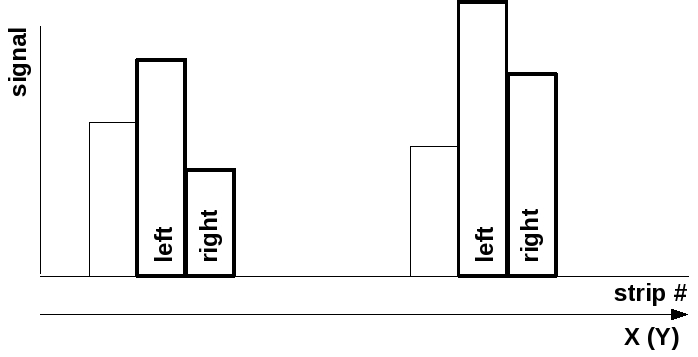}
                }
                \caption[Schematic picture of a cluster from the telescope]{Schematic picture of a cluster from the telescope.}
                \label{fig:tests:tele_cluster}
        \end{center}
\end{figure}

\begin{figure}[!h] 
	\begin{center}
	\subfigure[]{
		\includegraphics[width=0.45\textwidth,angle=90]{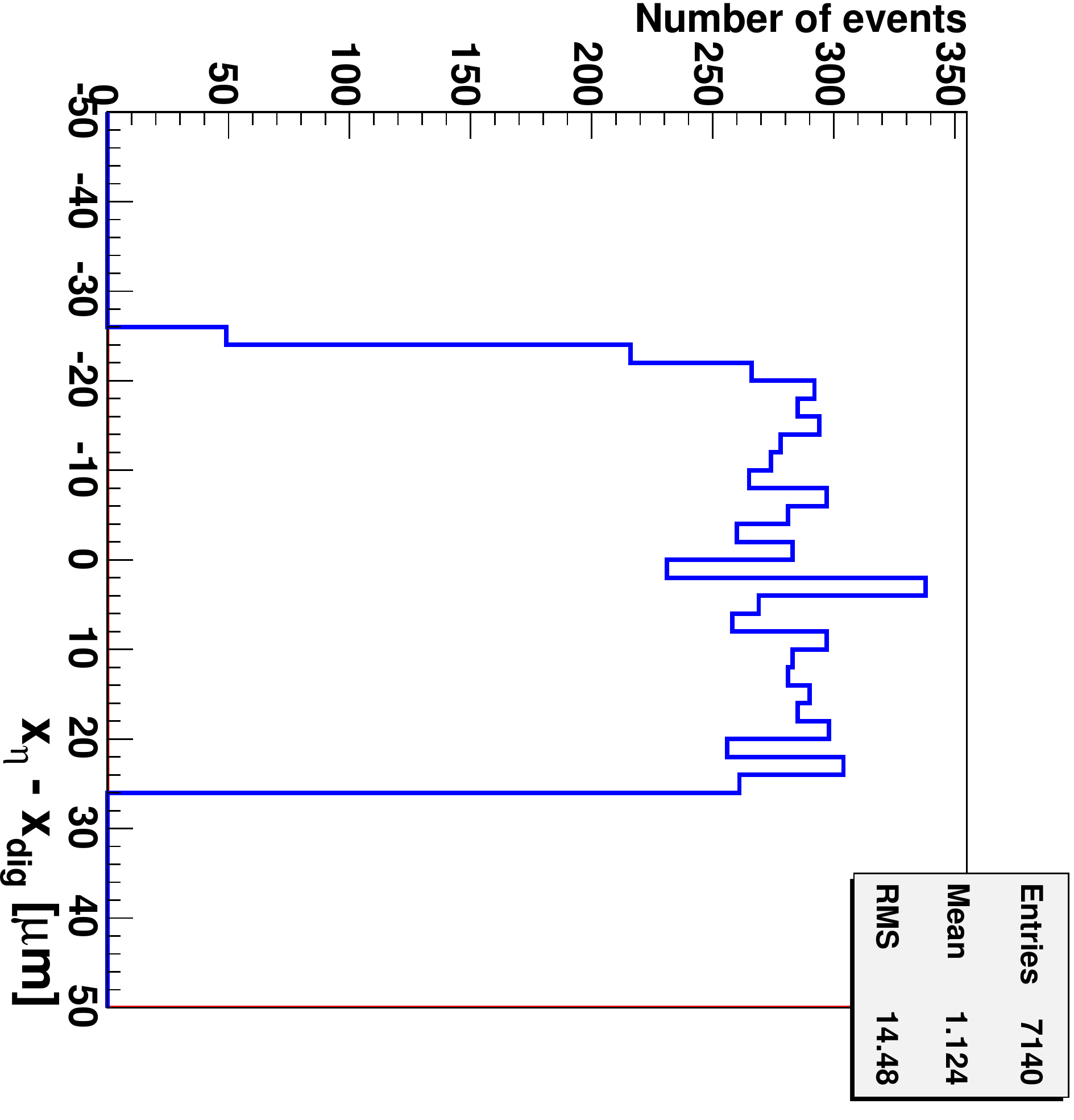}
		\label{fig:tests:Eta_Position:X}
	}
	\hspace{0.1cm}
	\subfigure[]{
		\includegraphics[width=0.45\textwidth,angle=90]{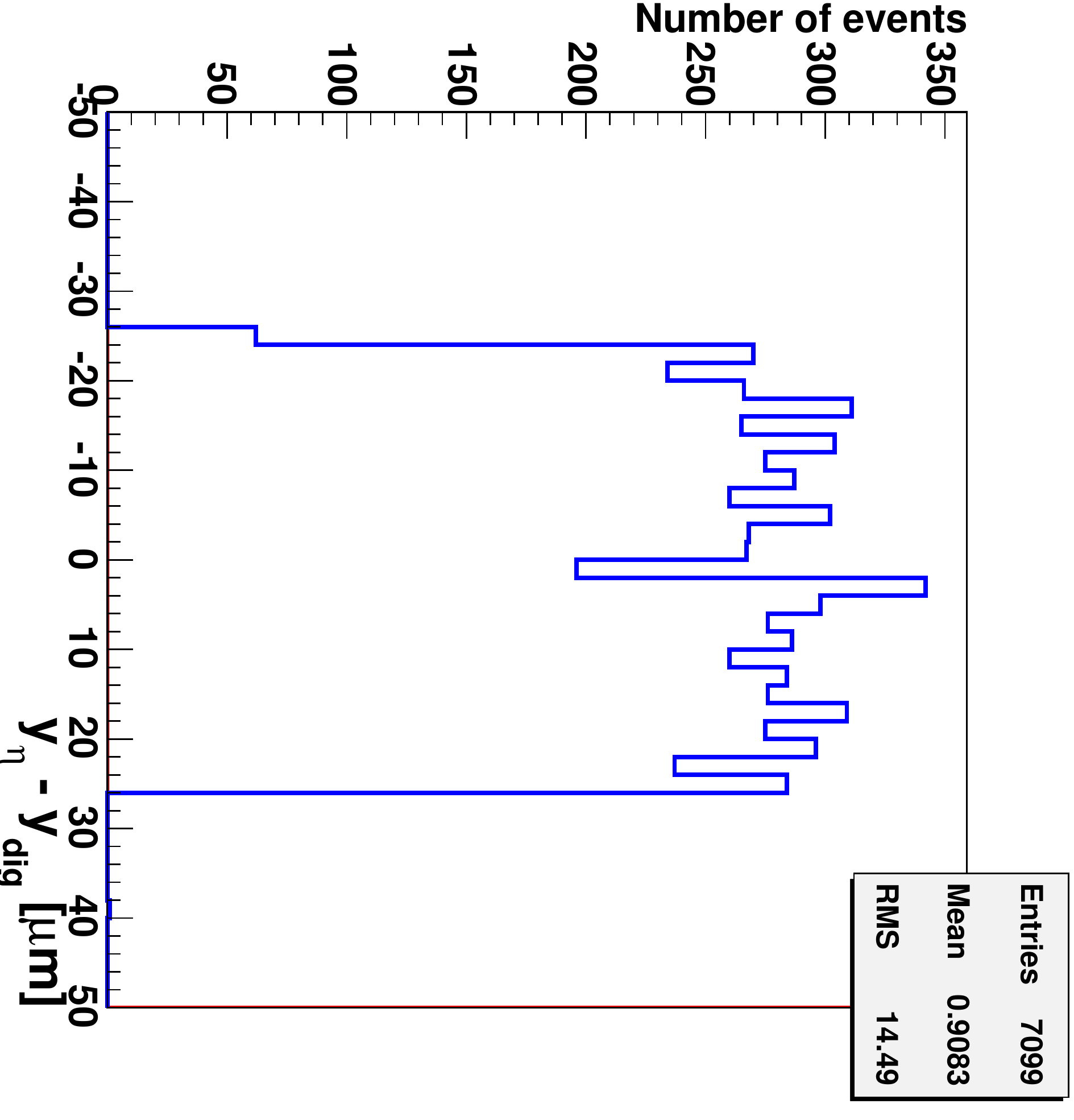}
		\label{fig:tests:Eta_Position:Y}
	}
	\caption[Distribution of the cluster position reconstructed with the $\eta$-algorithm with respect to its seed in the first telescope unit]{Distribution of the cluster position reconstructed with the $\eta$-algorithm with respect to its seed in the first telescope unit, (a) for X-plane, (b) for Y-plane.}
	\label{fig:tests:Eta_Position}
	\end{center}
\end{figure}

The CAEN~V550 module, handling the ADC conversion of the analog signals from the microstrips, provides zero suppression and pedestal subtraction. Accordingly only pulses with amplitudes exceeding thresholds are stored on the disk and submitted for further analysis. The threshold for a strip is defined as a mean value of its signal distribution, recorded in the absence of the beam, enlarged by one standard deviation. In order to evaluate the thresholds, a calibration run of 1000~events was performed before each data taking.\\
During the passage of a charged particle through a silicon detector a number of electron-hole pairs is generated ($e$-$h$). The electric field, present in the active volume of the detector, separates the electrons from the holes and guides them to the collecting strips. The recorded signal is shared among strips close to the impact point. The latter form a cluster. In the following analysis only clusters consisting of tree strips were considered. In order to reconstruct a cluster, a \textit{seed strip} was searched first. It is defined as a strip with the signal exceeding a threshold which is 50, 150 and 100~ADC~counts for telescope modules 1, 2 and 3, respectively. Moreover the seed strip was expected to contain the highest signal among its two adjacent strips. Afterwards a cluster was formed of the seed and its two neighbours. The signal distributions of the reconstructed clusters for first telescope plane are shown in~fig. \ref{fig:tests:ClusterCharge_2012}. The histograms are fitted with the Landau function.
\\
The impact positions of tracks as measured in the telescope is determined according to the $\eta$-algorithm~\cite{tests:eta_1,tests:eta_2}. It is based on the observation that the signal from strips depends on the track position w.r.t. the centre of a strips. The $\eta$ variable is defined as follows:
\begin{equation}	
\label{eq:tests:eta}
	\eta = \frac{S_{right}}{S_{right} + S_{left}},
\end{equation}
where $S_{left}$ and $S_{right}$ are signals of two strips (out of three) with the highest signal in the cluster. The distribution of the $\eta$ variable for the X-plane of the first telescope plane is presented in~fig.~\ref{fig:tests:Eta:Eta}. Due to the fact that only every second strip is read out, the distribution displays the three peak structure despite a uniform beam distribution. In order to correct for this non-uniformity, the cumulated distribution function of the $\eta$ variable is constructed and shown in fig.~\ref{fig:tests:Eta:Fun}: 
\begin{equation}	
\label{eq:tests:eta_fun}
	f(\eta) = \frac{1}{N_{0}}\int^{\eta}_{0}\frac{dN}{d\eta'}d\eta',
\end{equation}
where the $N_{0}$  is the total number of entries in the analysed data sample ($N_{0}~=~\int^{1}_{0}\frac{dN}{d\eta'}d\eta'$).
The corrected impact position of a track may be expressed as follows:
\begin{equation}
\label{eq:tests:x_eta}
	x_{\eta} = x_{left} + P_{x} \cdot f(\eta_{x}),~y_{\eta} = y_{left} + P_{y} \cdot f(\eta_{y}),
\end{equation}
where $P_{x}$ ($P_{y}$) is the value of the effective pitch (50~$\mu$m) and $x_{left}$ ($y_{left}$) denotes the position of the left strip (see fig.~\ref{fig:tests:tele_cluster}).
\\
A distribution of the hit positions reconstructed according to the $\eta$-algorithm with respect to the position of the corresponding cluster seed is shown in fig.~\ref{fig:tests:Eta_Position}. 

\subsubsection{Beam telescope alignment}
\label{ch:tests:exp:setup:MVDAnalysis:alig}

The goal of the alignment is to determine the absolute position of the telescope components in the common reference system.\\
There are six parameters describing the position and the orientation of the telescope units 2 and 3 with respect to the reference system (i.e. unit 1). These are: three offsets $v_{x}$, $v_{y}$ and $v_{z}$ along respective axes and three rotations $\phi_{x}$, $\phi_{y}$ and $\phi_{z}$ around the three axes. The transformation of the hit positions measured in the local coordinate system of the telescope unit 2 or 3 ($x_{loc}$, $y_{loc}$, $z_{loc}$) to the reference system of the telescope unit 1 ($x_{ref}$, $y_{ref}$, $z_{ref}$) is given by:
\begin{equation}
\label{eq:tests:align_formula1}
	\left( \begin{array}{c}
	x_{ref} \\
	y_{ref} \\
	z_{ref}
	\end{array} \right) = 
	\hat{R}(\phi_{x})\hat{R}(\phi_{y})\hat{R}(\phi_{z}) \cdot	\left( \begin{array}{c}
	x_{loc} \\
	y_{loc} \\
	z_{loc}
	\end{array} \right) + 
	\left( \begin{array}{c}
	v_{x} \\
	v_{y} \\
	v_{z}
	\end{array} \right),
\end{equation}
where $\hat{R}(\phi_{x})$, $\hat{R}(\phi_{y})$ and $\hat{R}(\phi_{z})$ are the rotation matrices around the X, Y and Z axes, respectively.\\ 
The offset $v_{z}$ is obtained from the direct measurement of the distance in the Z direction between a given telescope module and the reference one (using a ruler). In order to determine the other five parameters, the method based on the $\chi^{2}$ minimisation is used. Therefore for each telescope plane the $\chi^{2}$ is calculated:
\begin{equation}
\label{eq:tests:align_chi2}
	\chi^{2} = \sum_{i}\frac{\left(x_{i,ref} - x_{i,pred}\right)^{2} + 
	\left(y_{i,ref} - y_{i,pred}\right)^{2}}{\sigma^{2}}.
\end{equation}
The predicted coordinates ($x_{pred}$,$y_{pred}$) are obtained by a straight line extrapolation, parallel to the beam direction, of the hit position reconstructed in the first (reference) telescope plane. The $\sigma$ is the error of the hit position in the studied telescope module and is assumed 9~$\mu$m \cite{tests:MiliTh}. The sum in the equation (\ref{eq:tests:align_chi2}) runs over all events in the sample. The minimisation is done with respect to the following parameters: $v_{x}$, $v_{y}$, $\phi_{x}$, $\phi_{y}$ and $\phi_{z}$.

\subsubsection{Tracking of the beam electrons}
\label{ch:tests:exp:setup:MVDAnalysis:tracking}

\begin{figure}[!h]
        \begin{center}
                \includegraphics[width=0.7\textwidth]{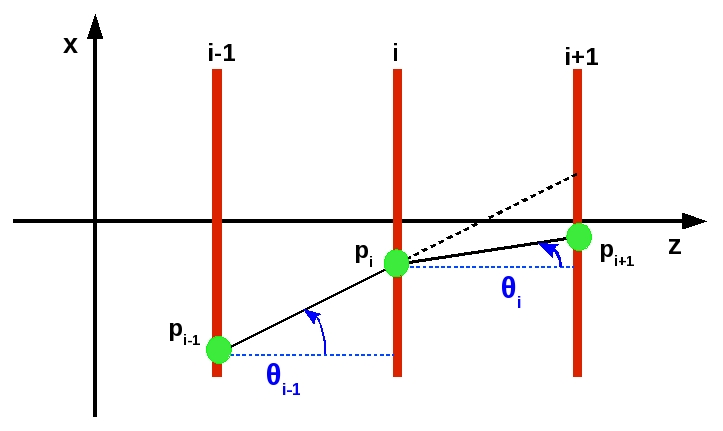}
                \caption[Definition of the scattering angle in the $i$-th telescope plane]{Definition of the scattering angle in the $i$-th telescope plane.}
                \label{fig:tests:scattering_angle}
        \end{center}
\end{figure}

Assuming that there is no correlation between the horizontal and the vertical position measurements, the track reconstruction in the telescope can be separated into two independent procedures: fitting in the horizontal and vertical directions.\\
The tracks measurements were performed in an absence of the magnetic field. In the first approximation the track can be approximated by a straight line fit to the cluster positions, reconstructed in the consecutive telescope planes. It is however necessary to account for multiple Coulomb scattering of electrons in silicon. The distribution of the scattering angle (after passing the detector) is assumed to be Gaussian with an expected width, $\Delta\Theta$, expressed by the formula \cite{tests:theta_ms1,tests:theta_ms2,tests:theta_ms3}:
\begin{equation}
\label{eq:tests:theta_ms}
	\Delta\Theta = \frac{13.6 MeV}{\beta c p}Z\sqrt{\frac{dx}{X_{0}}} \left[1 + 0.038 \ln{\left(\frac{dx}{X_{0}}\right)}\right],
\end{equation}
where $p$, $\beta c$ and $Z$ are the momentum, velocity and charge of the incident particle. The $dx/X_{0}$ is thickness of the scattering medium in radiation lengths ($X_{0}=$9.36~cm for silicon). The value of $\Delta\Theta$ for 6~GeV electrons in the telescope module of 600~$\mu$m thickness amounts to 0.15~mrad.\\
In order to improve precision of the track reconstruction, multiple scattering of charged particles in the telescope planes and in the DUT has to be taken into account. A tracking method  proposed by the EUDET collaboration fulfills this requirement \cite{tests:afz_fit}. Determination of the particle positions ($p_{i}$, $i=$1...4) in three telescope planes and in the DUT is obtained from the $\chi^{2}$ minimisation. The contribution of the $i$-th plane to the $\chi^{2}$ can be written as:
\begin{equation}
\label{eq:tests:afz_fit}
	\Delta\chi^{2}_{i} = \alpha_{i}\cdot(x_{i} - p_{i})^{2} + \beta_{i}\cdot(\theta_{i} - \theta_{i-1})^{2},
\end{equation}
where $x_{i}$ is the position measured in one of the three telescope planes ($i \neq i_{DUT}$) and $\theta_{i}$ denotes the angle between direction perpendicular to the telescope planes and the particle track direction between planes $i$ and $i+1$, as shown in fig.~\ref{fig:tests:scattering_angle}. The $\theta_{i}$ is calculated according to the formula: 
\begin{equation}
\label{eq:tests:afz_theta_i}
	\theta_{i} = \frac{p_{i+1} - p_{i}}{z_{i+1} - z_{i}}.
\end{equation}
The coefficients $\alpha_{i}$ and $\beta_{i}$ are expressed as follows:
\begin{equation}
\label{eq:tests:alpha_i}
	\alpha_{i} = \left\{
	\begin{array}{ll}
	1/\sigma_{i}^{2} & \textrm{for $i \neq i_{DUT}$},\\
	0 & \textrm{for $i = i_{DUT}$},
	\end{array} \right.
\end{equation}
where $\sigma_{i}$ is the resolution of the $i$-th telescope plane \cite{tests:MiliTh} and
\begin{equation}
\label{eq:tests:beta_i}
	\beta_{i} = \left\{
	\begin{array}{ll}
	\frac{1}{\Delta\theta_{i}^{2}} & \textrm{for the internal planes and the DUT},\\
	0 & \textrm{for two outer planes}.
	\end{array} \right.
\end{equation}
\\
The first term in (\ref{eq:tests:afz_fit}) is related to the uncertainty of the position measurement and the second one to the expected distribution of the scattering angle in the $i$-th plane. The $\chi^{2} = \sum_{i=1}^{4}{\Delta\chi^{2}_{i}}$ is minimised with respect to the positions $p_{i}$, $i=1,2,3,4$. The fitted particle position in the $i$-th plane of the system is a linear combination of measured positions $x_{j}$ in all active layers, given by the formula:
\begin{equation}
\label{eq:tests:afz_position}
	p_{i} = \sum_j{\mathcal{S}_{ij}\alpha_{j}x_{j}},
\end{equation}
where $\hat{\mathcal{S}}$ is the inverse matrix to the matrix $\hat{\mathcal{A}}$, defined as follows:
\begin{equation}
\label{eq:tests:afz_matrix}
	\mathcal{A}_{ij} = \frac{1}{2}\frac{\partial^{2}\chi^{2}}{\partial p_{i} \partial p_{j}}.
\end{equation}
The uncertainties of the track position at the $i$-th plane is expressed in terms of the diagonal elements of the matrix $\hat{\mathcal{S}}$:
\begin{equation}
\label{eq:tests:afz_error}
	\tilde{\sigma_{i}} = \sqrt{\mathcal{S}_{ii}}.
\end{equation}
More details of the presented approach can be found elsewhere \cite{tests:afz_fit}.

\subsection{The DAFNE beam test facility}
\label{ch:tests:exp:setup:DAFNE}
The DAFNE Beam Test Facility (BTF) is a part of the DAFNE $\phi$-factory complex \cite{tests:dafne_BTF}. It provides electrons and positrons in the energy range from 25~MeV up to 750~MeV. The BTF was designed to provide intensities ranging from a single electron up to 10$^{10}$ electrons per pulse. The pulse duration is 10~ns and the maximum repetition rate is 50~Hz. The acceleration of the electrons (positrons) is performed in the high current LINAC. The bending magnet assembled on the transfer line after the LINAC drives the beam to the BTF area and together with the slit system provides energy selection. In front of the bending magnet a tungsten target is placed in order to attenuate the LINAC beam for the BTF purposes. Three different radiation lengths of the target can be selected, 1.7, 2.0, 2.3~$X_{0}$. An additional bending magnet is placed at the end of the BTF transfer line with the purpose to split the beam into two separate test stations.

\chapter{Data analysis and results}
\label{ch:tests:exp_results}

\section{Pedestal and noise evaluation}
\label{ch:tests:exp_results:ped_noi}

\begin{figure}[!h] 
	\begin{center}
	\subfigure[MIMOSA-5]{
		\includegraphics[width=0.45\textwidth,angle=90]{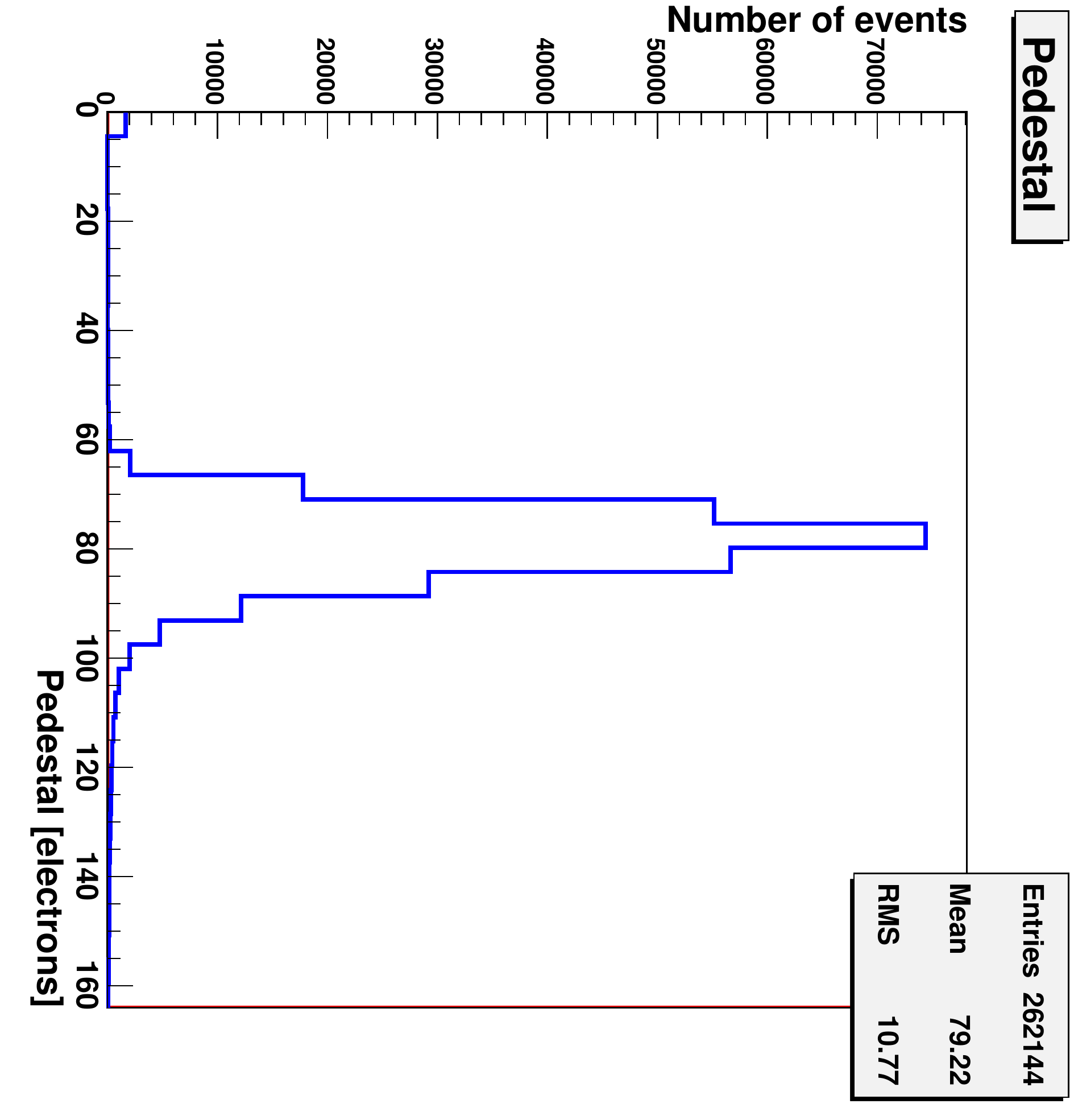}
		\label{fig:tests:Pedestal:M5}
	}
	\hspace{0.1cm}
	\subfigure[MIMOSA-5]{
		\includegraphics[width=0.45\textwidth,angle=90]{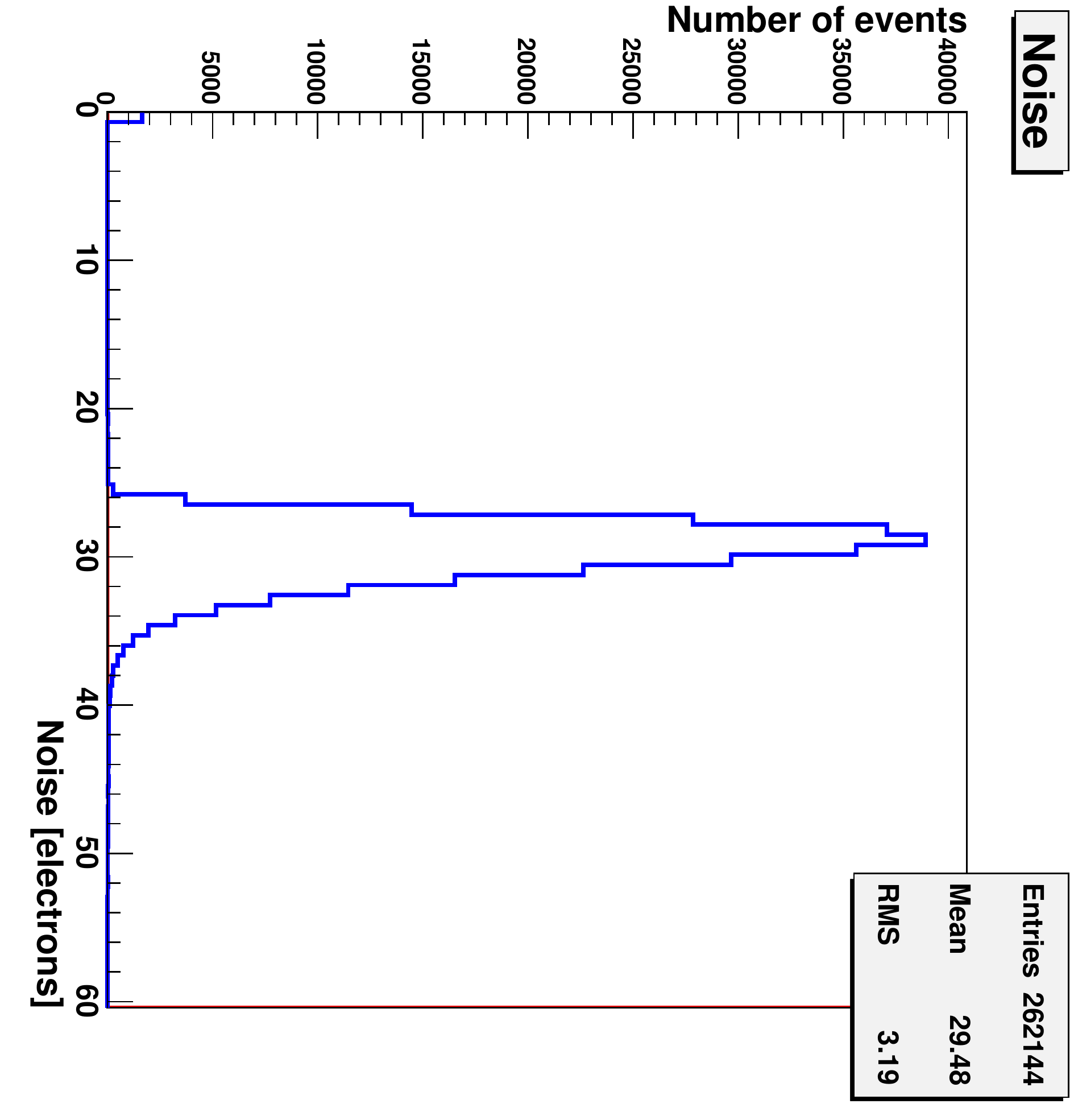}
		\label{fig:tests:Noise:M5}
	}
	\subfigure[MIMOSA-18]{
		\includegraphics[width=0.45\textwidth,angle=90]{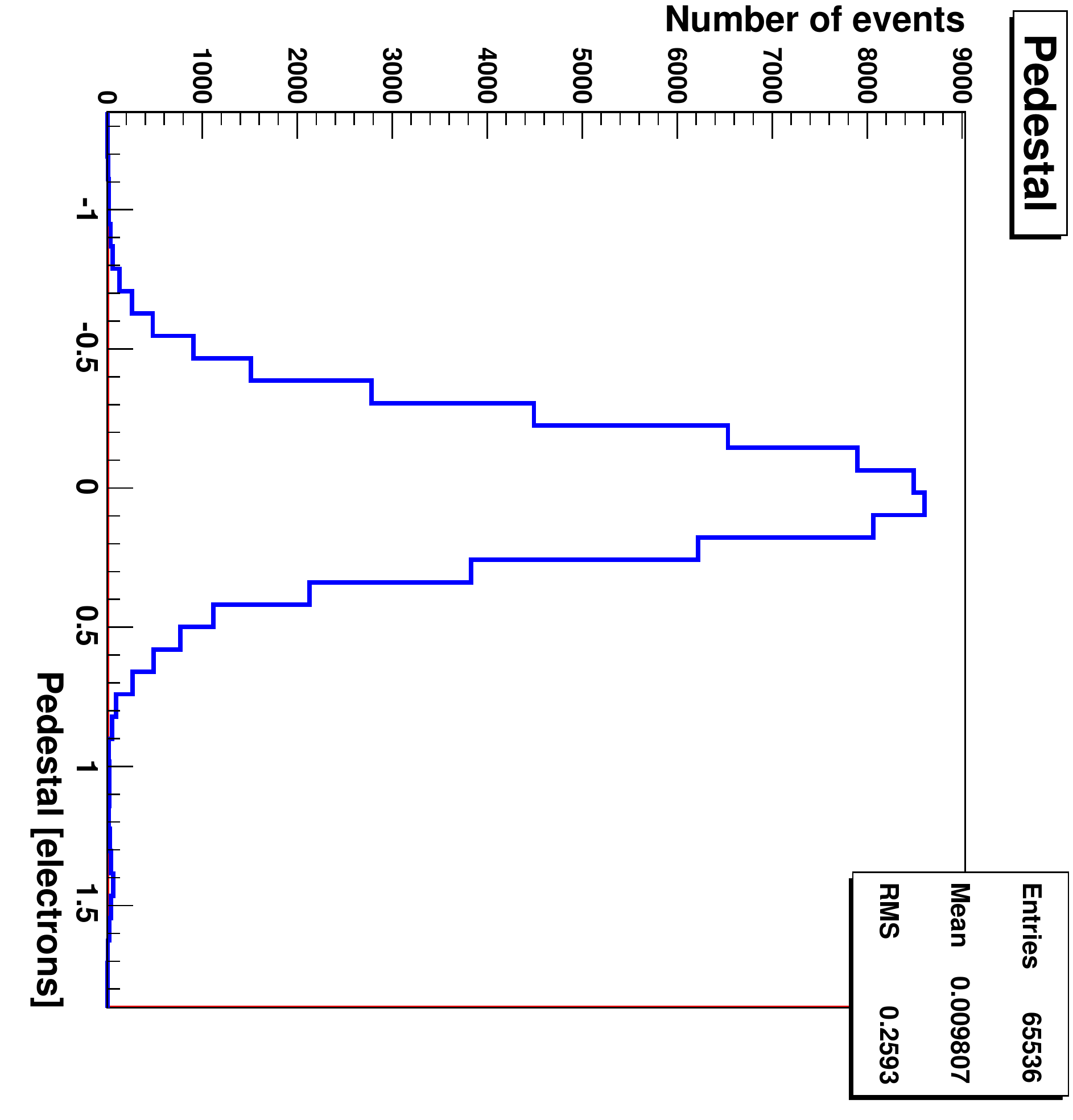}
		\label{fig:tests:Pedestal:M18}
	}
	\hspace{0.1cm}
	\subfigure[MIMOSA-18]{
		\includegraphics[width=0.45\textwidth,angle=90]{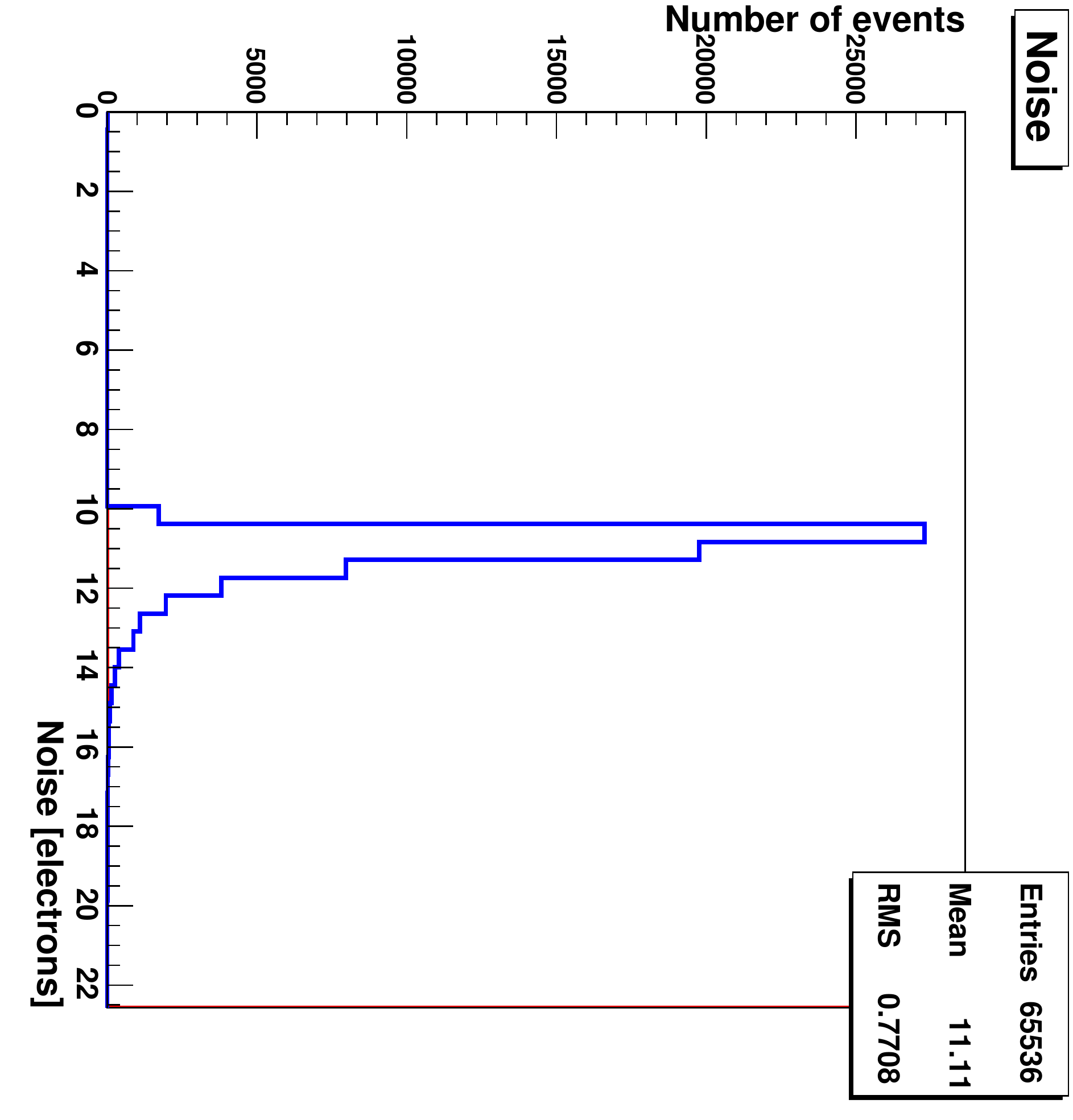}
		\label{fig:tests:Noise:M18}
	}
	\caption[Distributions of pedestals and the noises measured for the MIMOSA-5 and MIMOSA-18 matrices.]{Distributions of pedestals (a), (c) and noises (b), (d) measured for the MIMOSA-5 and MIMOSA-18 matrices, respectively.}
	\label{fig:tests:Ped_Noi}
	\end{center}
\end{figure}

\begin{figure}[!h] 
	\begin{center}
	\subfigure[]{
		\includegraphics[width=0.45\textwidth,angle=90]{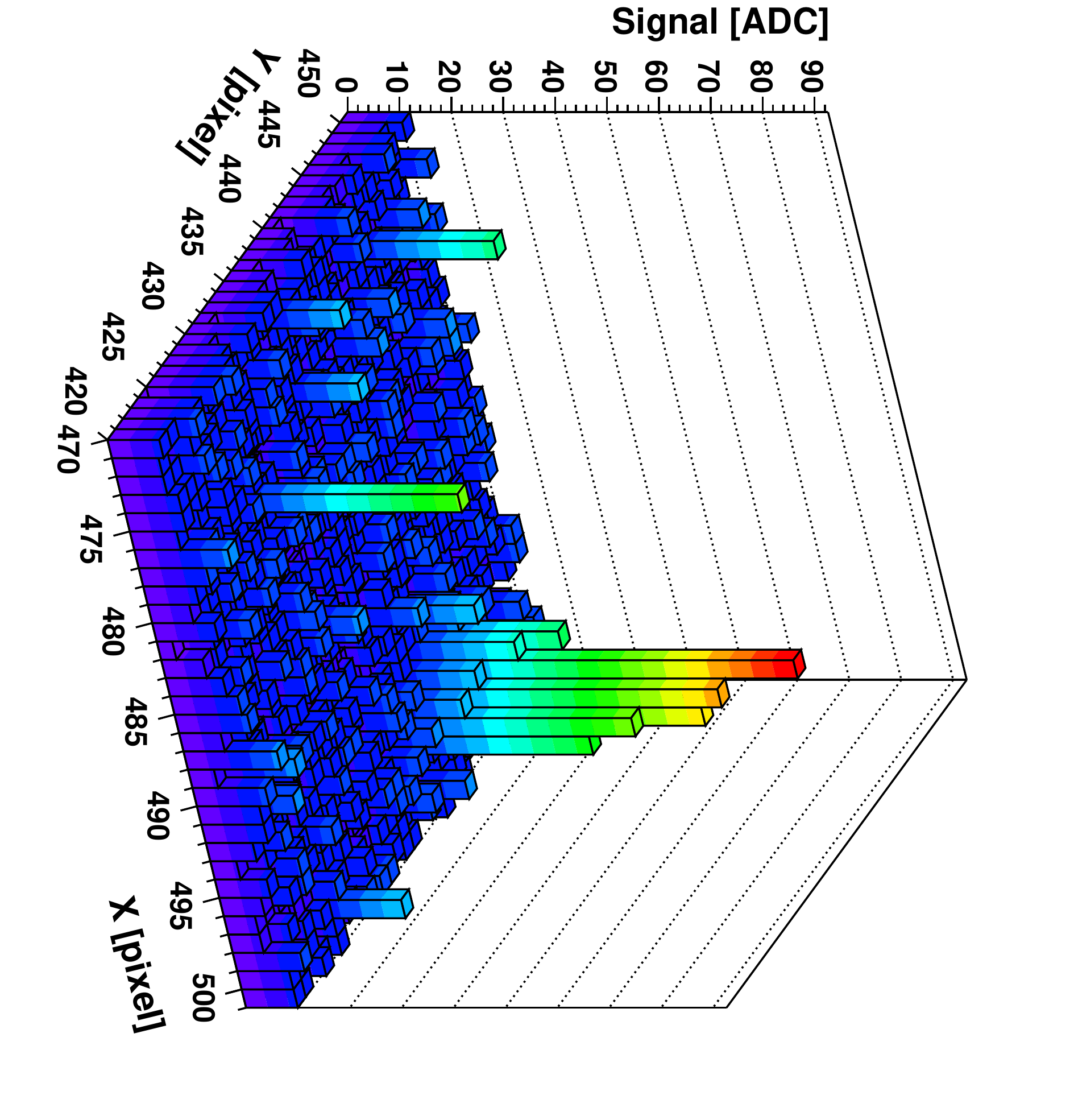}
		\label{fig:tests:MIMOSA-5_Hits_Raw}
	}
	\hspace{0.1cm}
	\subfigure[]{
		\includegraphics[width=0.45\textwidth,angle=90]{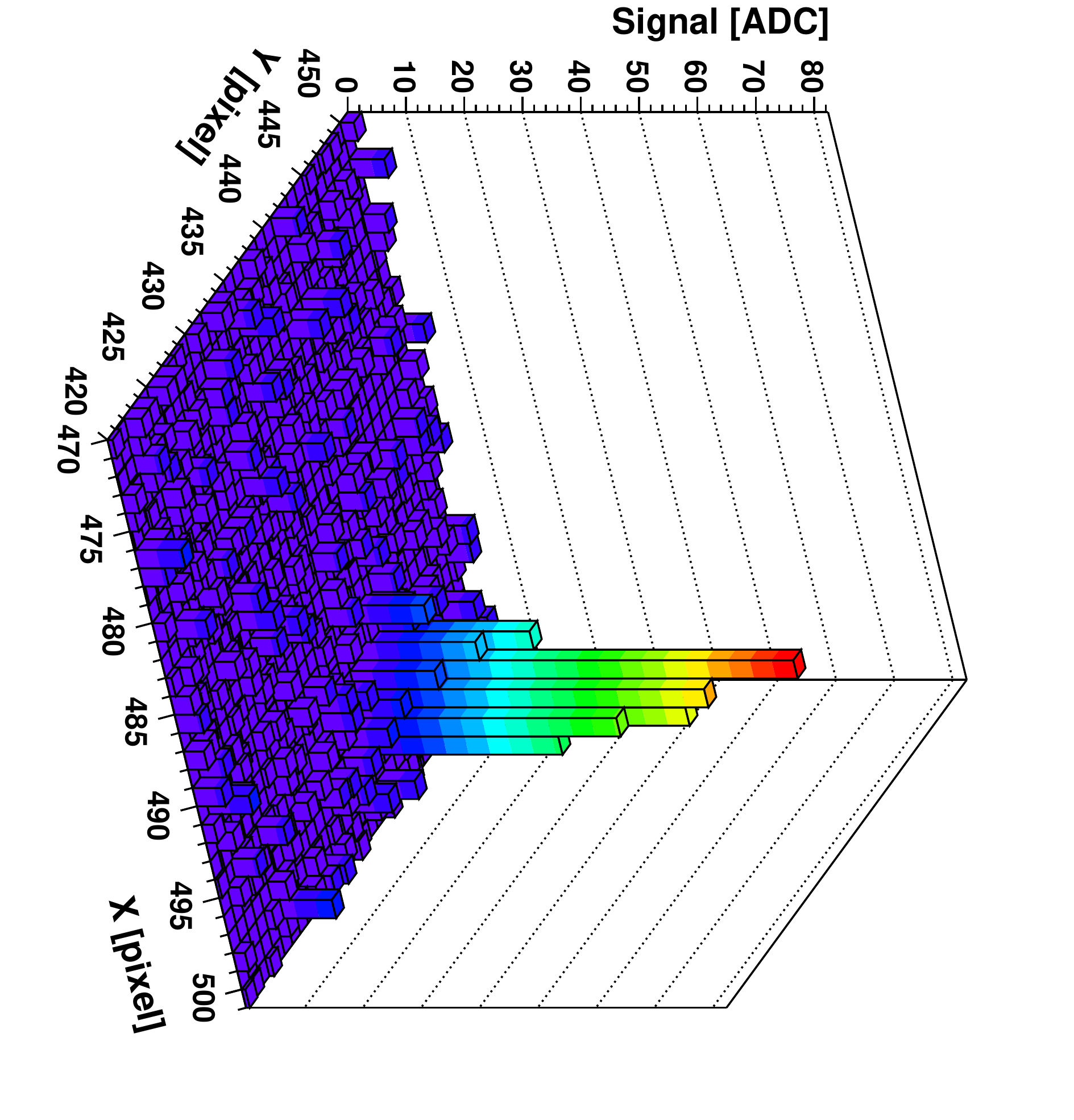}
		\label{fig:tests:MIMOSA-5_Hits_Raw_PedSub}
	}
	\caption[The signal distribution in a random 30$\times$30 pixel subset of the MIMOSA-5 before and after the pedestal subtraction]{The signal distribution in a random 30$\times$30 pixel subset of the MIMOSA-5; (a) raw data before pedestal subtraction, (b) physical signal after pedestal subtraction.}
	\label{fig:tests:MIMOSA-5_Hits}
	\end{center}
\end{figure}

The raw signal in the $k$-th pixel in the event $i$, $r_{k}^{i}$, includes the physical signal $s_{k}^{i}$ and the pedestal $p_{k}$:
\begin{equation}
\label{eq:exper_results:signal}
	r_{k}^{i} = s_{k}^{i} + p_{k}.
\end{equation}
The physical signal $s_{k}^{i}$ originates from charges carriers created by an ionising particle and the pedestal $p_{k}$ is due to the leakage current. The pedestal and noise are determined according to the following procedure:
\begin{enumerate}
\item
A rough estimation of the $k$-th pixel pedestal is obtained as a mean value of raw signals collected in the first N = 1000 events in a run:

\begin{equation}
\label{eq:exper_results:pedestal1}
	p_{k}^{(1)} = \frac{1}{N}\sum_{i = 1}^{N} r_{k}^{i}.
\end{equation}
The justification of this approximation comes from the fact that physical signals $s_{k}^{i}$ in consecutive frames are rare (order of $1/100$) so 99\% of the quantities $r_{k}^{i}$ contain only the pedestal contribution.\\ 
The corresponding noise, $n_{k}^{(1)}$, is assumed to be the standard deviation of the pedestal:
\begin{equation}	
\label{eq:exper_results:noise1}
	n_{k}^{(1)} = \sqrt{\frac{N}{N-1}\left(\frac{1}{N}\sum_{i=1}^{N} \left(r_{k}^{i}\right)^2 - \left(p_{k}^{(1)}\right)^{2}\right)}.
\end{equation}
\item
Next, the contribution to the sums~(\ref{eq:exper_results:pedestal1}) and (\ref{eq:exper_results:noise1}) from physical signals, $s_{k}^{i}$, is removed. This is done by comparing the total signal $r_{k}^{i}$ and the first approximation of the pedestal, $p_{k}^{(1)}$, calculated as above. The distribution of differences $\delta_{k}^{i} = r_{k}^{i} - p_{k}^{(1)}$ for all MIMOSA-5 pixels is shown in~fig.~\ref{fig:tests:PedestalEst_SigSupp}. The measured signals are expressed in the ADC counts which can be identified with the equivalent charge integrated in the collecting diodes.
\begin{figure}[!h]
	\begin{center}
		\resizebox{0.5\textwidth}{!}{
			\includegraphics[angle=90]{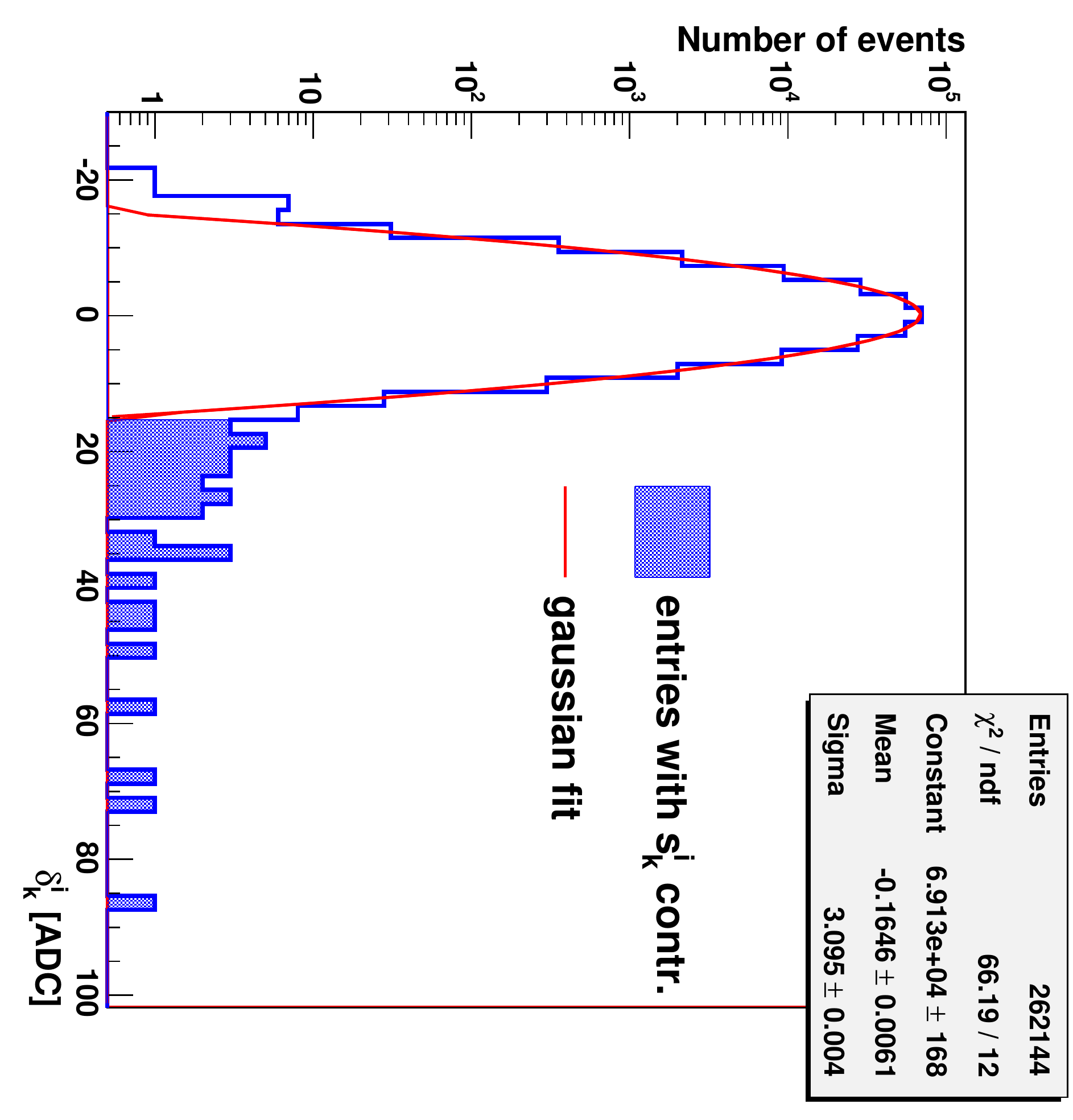}
		}
		\caption[Distribution of the differences $\delta_{k}^{i} = r_{k}^{i} - p_{k}^{(1)}$ for all MIMOSA-5 pixels]{Distribution of the differences $\delta_{k}^{i} = r_{k}^{i} - p_{k}^{(1)}$ for all MIMOSA-5 pixels.}
		\label{fig:tests:PedestalEst_SigSupp}
	\end{center}
\end{figure}
Majority of entries are centred around $\delta_{k}^{i} \approx 0$ and these correspond to pixels containing only the noise. Those pixels which contain physical signals are distributed at larger values of $\delta_{k}^{i}$ (the tail of the histogram). The peak in fig.~\ref{fig:tests:PedestalEst_SigSupp} is fitted with the Gaussian function. The entries for which the $\delta_{k}^{i}$ exceeds the mean value of the Gaussian  by more than five standard deviations (5$\sigma$) are assumed to contain physical signals $s_{k}^{i}$ and are excluded. These entries are marked with blue filling in fig.~\ref{fig:tests:PedestalEst_SigSupp}.
\item
Using all events in a run (approx. 6~k events) the pedestals and noises are recalculated for the second time for each pixel according to (\ref{eq:exper_results:pedestal1}) and (\ref{eq:exper_results:noise1}) but with exclusion of the entries containing physical signals, as described above. The result of the second iteration of the pedestal and noise estimation for MIMOSA-5 and MIMOSA-18 are presented in fig.~\ref{fig:tests:Ped_Noi}. The values are expressed in units of electron charges in order to compare the pedestal and noise for both tested MAPS matrices. The calibration procedure providing conversion of the ADC counts to the units of electron charge is described in section \ref{ch:tests:exp_results:X-rays}.
\end{enumerate}
The distribution of the MIMOSA-5 pedestals differs from that of the MIMOSA-18. These distributions cannot be directly compared because of different readout architectures used in these two matrices. The pedestals of the MIMOSA-18 are close to 0 due to using the self-biasing diodes, while for the MIMOSA-5 they are around 80~electrons. Comparing the average noise in the MIMOSA-18 and MIMOSA-5 shows it is approx. 3 times lower in the former -- 11~electrons and 30~electrons, respectively. The noise of the MIMOSA-18 detector was measured at the room temperature of 15$^{\circ}$C while that of the MIMOSA-5 at -8$^{\circ}$C.\\
Among all pixels in the detector a group may be found with a very high pedestals with respect to the average value in the whole matrix. These are malfunctioning pixels, so-called \textit{hot pixels}, and they are excluded from the analysis. In order to identify the hot pixels, the detector surface is segmented in to regions containing 16$\times$16 pixels. For each region a mean pedestal and its standard deviation is calculated. Afterwards pedestals of all pixels contained in the considered region are compared with the average value for this region. If the pedestal of a given pixel exceeds the mean value by more than 5 standard deviations, such a pixel is assumed to be a hot pixel. In case of the MIMOSA-5 chip hot pixels constitute 0.6\% of all pixels while for the tested MIMOSA-18 only 1 hot pixel was found.\\
In order to estimate the physical signal recorded in the $k$-th pixel, the pedestal has to be subtracted from the raw signal:
\begin{equation}
\label{eq:exper_results:physical_signal}
	s_{k}^{i} = r_{k}^{i} - p_{k}.
\end{equation}
As an example the signal distribution in a random 30$\times$30 pixel subset of the MIMOSA-5 before and after the pedestal subtraction is shown in fig.~\ref{fig:tests:MIMOSA-5_Hits}.
\\
A physical cluster is visible on both figures. The pedestal subtraction removes the isolated sparks in fig.~\ref{fig:tests:MIMOSA-5_Hits_Raw} which correspond to high pedestals in certain pixels. The background in fig.~\ref{fig:tests:MIMOSA-5_Hits_Raw_PedSub} correspond to fluctuating pedestal value, i.e. noise.

\section{Cluster reconstruction}
\label{ch:tests:exp_results:cluster}

Charge carriers generated by an ionising particle in the active volume of a MAPS detector spread among adjacent pixels, forming a cluster. The latter is associated with a hit due to the incoming particle. Clusters of pixels were reconstructed as follows.  First a seed pixel (i.e. with locally the highest charge) was searched. The signal to noise ratio, $S/N$, for the seed was required to be greater than the value $t_{s}$:
\begin{equation}
\label{eq:exper_results:seed_s2n}
	\frac{S}{N} > t_{s}
\end{equation}
and the $t_{s}$ value chosen for the MIMOSA-5 was 4, while for the MIMOSA-18 it was 8 (see section~\ref{ch:tests:exp_results:tracking_performance:effi}). The difference in the $t_{s}$ values for MIMOSA-5 and MIMOSA-18 reflects different noise levels in these two matrices. Additionally it was required that the seed has the highest signal to noise ratio among 8 adjacent pixels.\\
In order to remove fake clusters (formed around malfunctioning pixels) an additional cut was applied. For this purpose the following variables were defined: $(i)$ $\mathcal{S}_{8} = \sum_{i = 1}^{8}q_{i}$ - the total charge collected in 8 pixels adjacent to the seed and $(ii)$ $\mathcal{N}_{8}=\sqrt{\sum_{i=1}^{8}n_{i}^{2}}$ - average noise of these pixels, where $q_{i}$ and $n_{i}$ are the charge and the noise of the $i$-th pixel, respectively. If the condition
\begin{equation}
\label{eq:exper_results:neigh_s2n}
	\frac{\mathcal{S}_{8}}{\mathcal{N}_{8}} > t_{n}
\end{equation}
was fulfilled, the groups of $9\times 9$ and $15\times15$ pixels around the seed in the MIMOSA-5 and MIMOSA-18, respectively, were taken under further consideration. The values of the $t_{n}$ for MIMOSA-5 and MIMOSA-18 were 0.5 and 4, respectively.\\
The cluster position was calculated according to the charge-weighted centre of gravity (CoG) algorithm applied to all pixels contained in the 3$\times$3 pixel cluster: 
\begin{equation}	
\label{eq:tests:cog}
	x_{cog} = \frac{\sum_{i}{q_{i} \cdot x_{i}}}{\sum_{i}{q_{i}}},~y_{cog} = \frac{\sum_{i}{q_{i} \cdot y_{i}}}{\sum_{i}{q_{i}}}.
\end{equation}\\
If the reconstructed cluster position coincided with the track position at the matrix surface predicted by the telescope, the cluster was stored on the disc and submitted to further detailed studies.

\section{ADC calibration}
\label{ch:tests:exp_results:X-rays} 

\begin{figure}[!h]
        \begin{center}
                \includegraphics[width=0.5\textwidth]{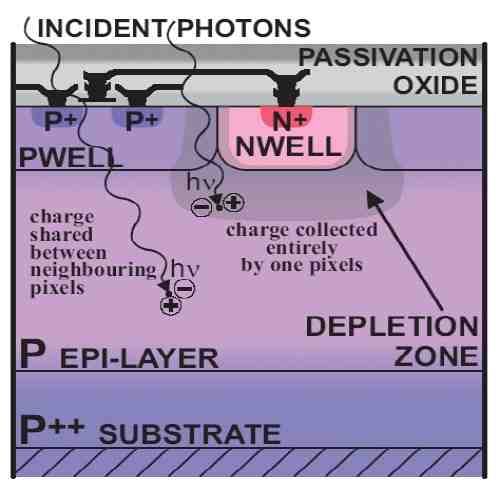}
                \caption[Interaction of the X-ray photons inside the active volume of a MAPS detector]{Interaction of the X-ray photons inside the active volume of a MAPS detector \cite{tests:DeptuchTh}. The $^{55}$Fe source of X-ray photons (not marked in the figure) is placed above the pixel layer.}
                \label{fig:tests:foto_effect_Fe}
        \end{center}
\end{figure}

\begin{figure}[!h] 
	\begin{center}
	\subfigure[]{
		\includegraphics[width=0.45\textwidth,angle=90]{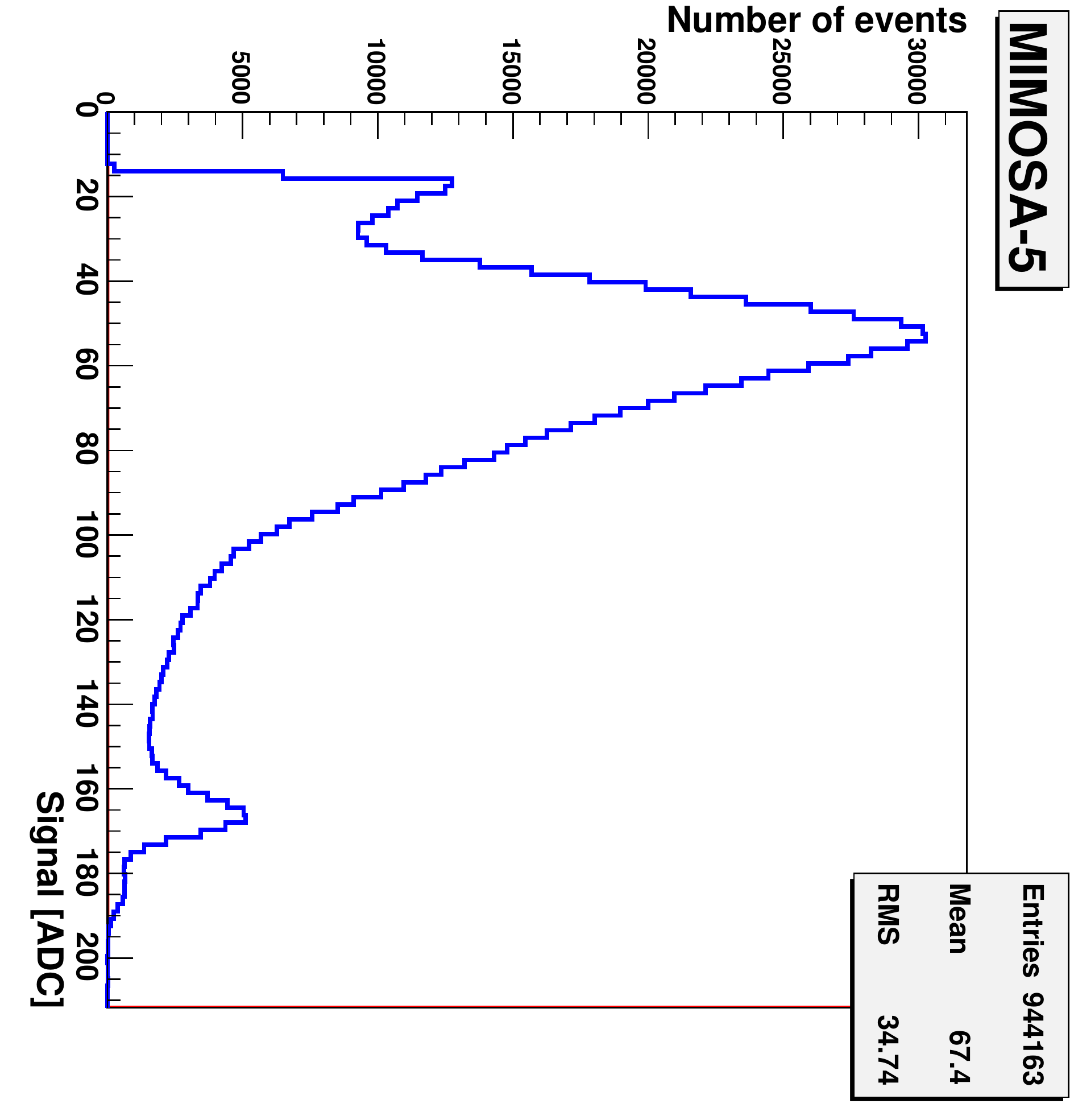}
		\label{fig:tests:MIMOSA-5_Fe55_Seed0}
	}
	\hspace{0.1cm}
	\subfigure[]{
		\includegraphics[width=0.45\textwidth,angle=90]{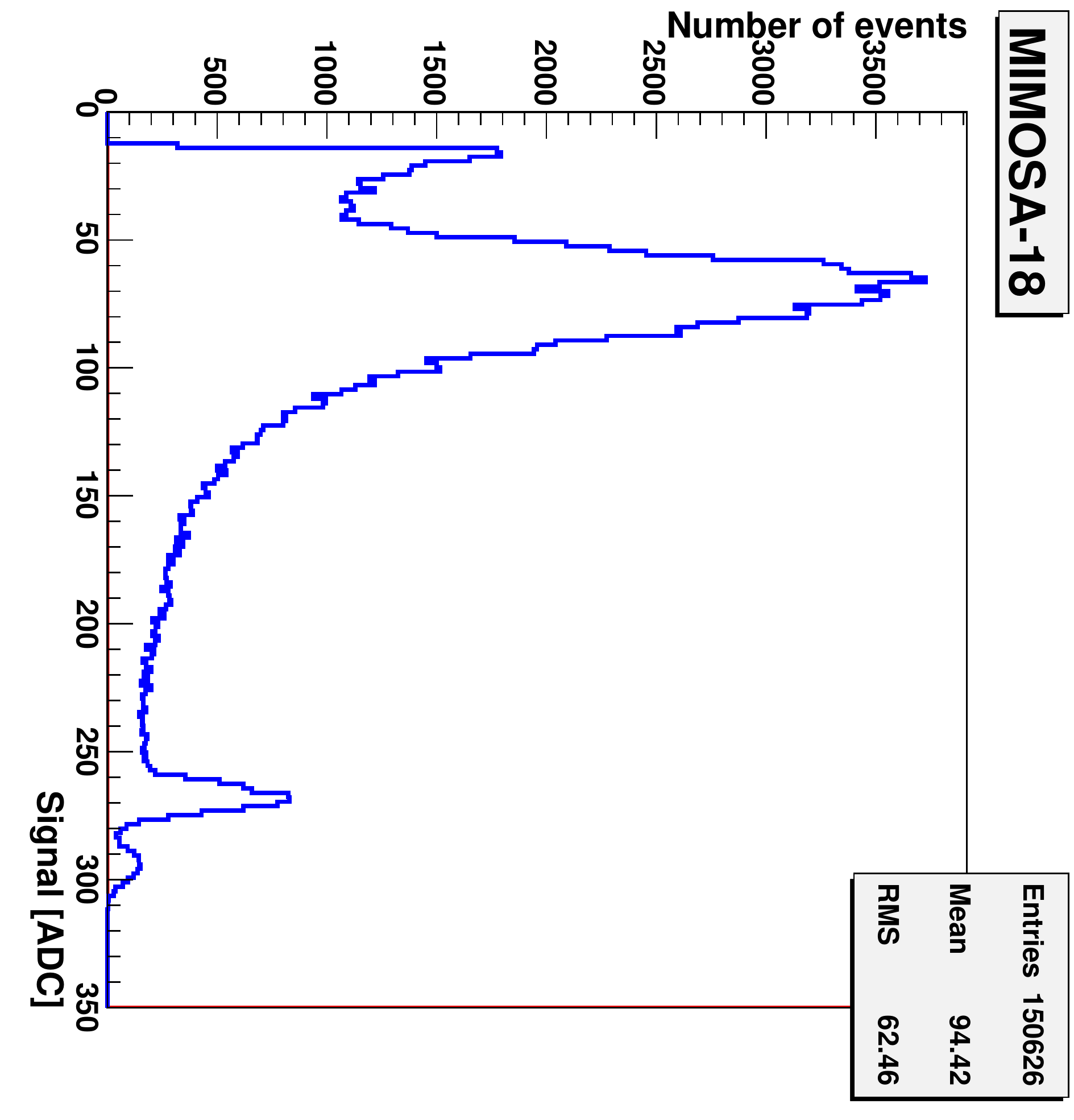}
		\label{fig:tests:MIMOSA-18_Fe55_Seed0}
	}
	\caption[The $^{55}$Fe spectrum measured in the seed pixel of reconstructed clusters]{The $^{55}$Fe spectrum measured in the seed pixel of reconstructed clusters for: (a) MIMOSA-5 and (b) MIMOSA-18.}
	\label{fig:tests:Fe55_Seed0}
	\end{center}
\end{figure}

\begin{figure}[!h] 
	\begin{center}
	\subfigure[]{
		\includegraphics[width=0.45\textwidth,angle=90]{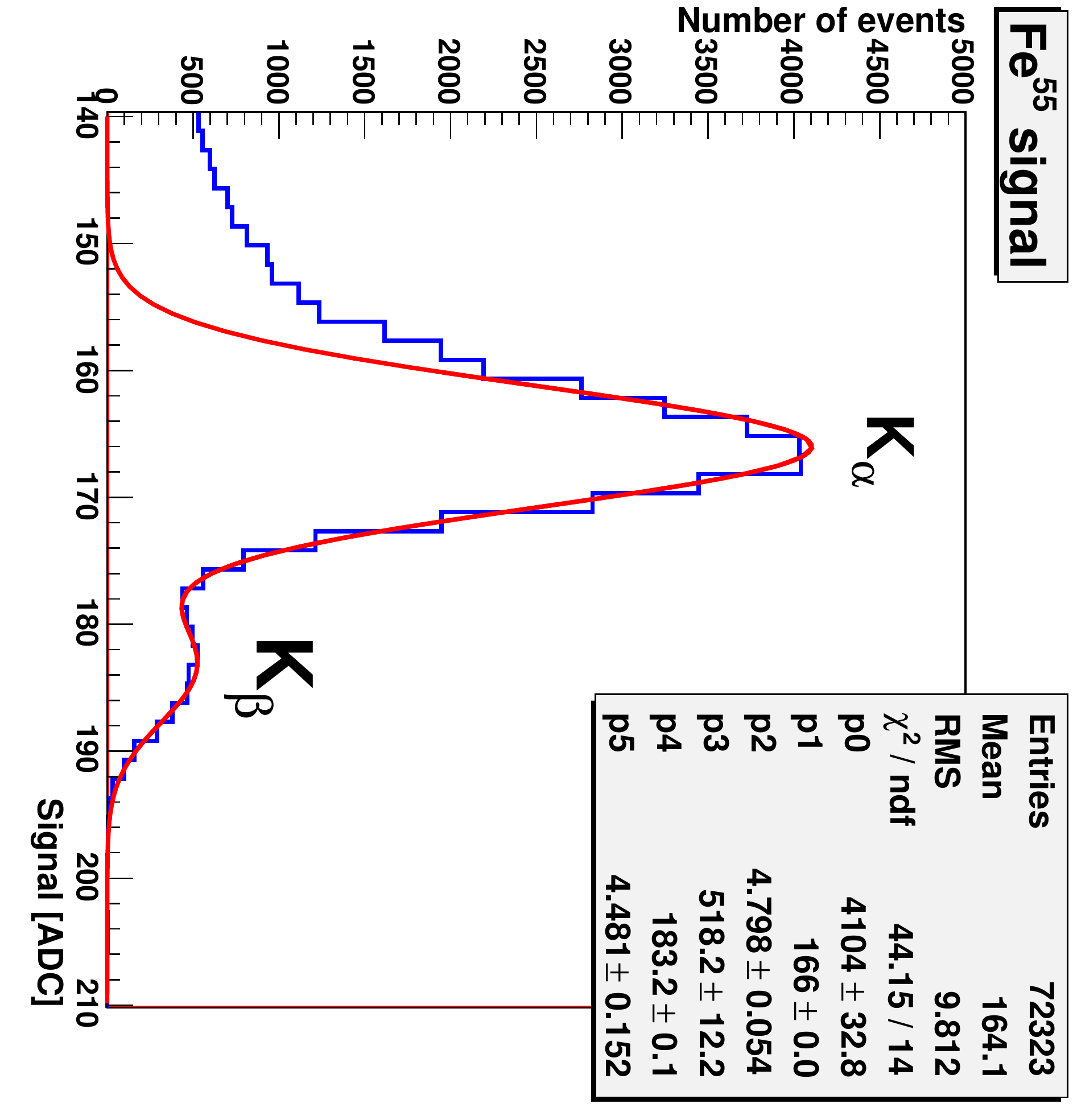}
		\label{fig:tests:MIMOSA-5_Fe55_Seed1}
	}
	\hspace{0.1cm}
	\subfigure[]{
		\includegraphics[width=0.45\textwidth,angle=90]{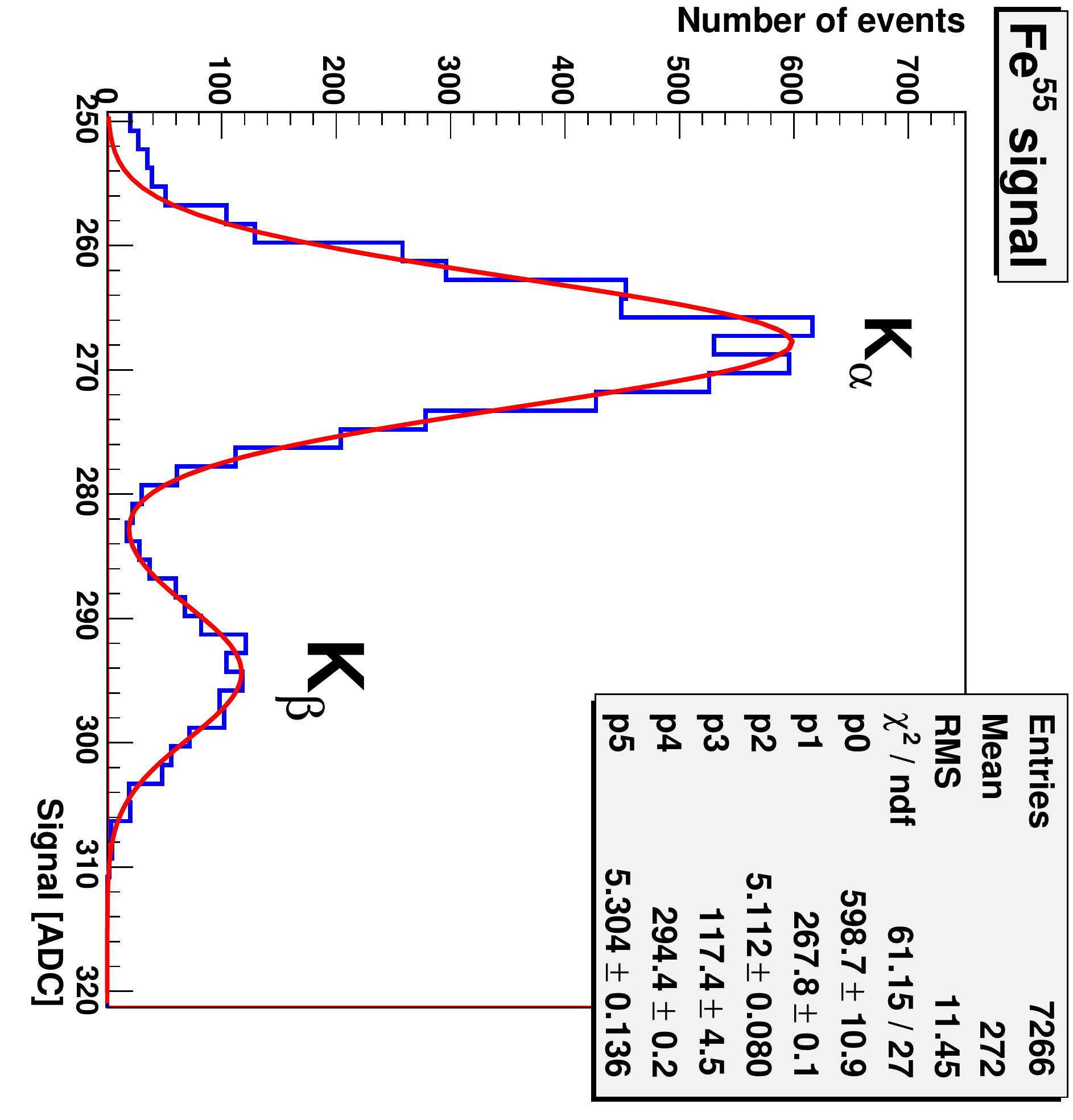}
		\label{fig:tests:MIMOSA-18_Fe55_Seed1}
	}
	\caption[The $^{55}$Fe spectrum measured in single pixel clusters]{The $^{55}$Fe spectrum measured in single pixel clusters in the region corresponding to the 5.9~keV and 6.49~keV emission lines for: (a) MIMOSA-5 and (b) MIMOSA-18. The fitted function is a sum of two Gaussian functions~(\ref{eq:exper:sum_gaussian}).}
	\label{fig:tests:Fe55_Seed1}
	\end{center}
\end{figure}

Signals read out from pixels of a MAPS detector are expressed in ADC counts. In order to convert them to charge, one has to know the proportionality coefficient, $G$, defined as:
\begin{equation}
\label{eq:exper:conv_gain}
G = \frac{Q}{S_{ADC}},
\end{equation} 
where $S_{ADC}$ is the signal in a pixel expressed in ADC counts and $Q$ is the charge collected during the integration time, often expressed as a number of unit charges, $N_{e}$.\\
In the presented work determination of the coefficient $G$, i.e. the calibration, was done using a $^{55}$Fe radioactive source which emits photons mainly in two monochromatic lines $K_{\alpha}$ and $K_{\beta}$ of 5.9~keV and 6.49~keV with the emission probability of 24.4~\% and 2.86~\% \cite{tests:PDG}, respectively. The $^{55}$Fe source was placed above the pixel layer of the detector as shown in fig.~\ref{fig:tests:foto_effect_Fe}. The absorption lengths for the $K_{\alpha}$ and $K_{\beta}$ photons are approx. 27~$\mu$m and 35~$\mu$m, respectively, and signals due to photoelectric effect are generated across the whole thickness of the epitaxial layer.
\\
The mean energy of the electron-hole pair creation in silicon is E$_{pair}=~$(3.66~$\pm$~0.03)~eV \cite{tests:PDG} and photons of 5.9~keV and 6.49~keV generate on average 1612~$\pm$~13 and 1773~$\pm$~14 electrons, respectively. The charge carriers generated in the field-free part of the epitaxial layer diffuse isotropically inside the active detector volume and only a fraction of them reaches the pixel layer. The collected charge is shared by the neighbouring pixels and a cluster is formed. The situation looks different when a photon interacts in the depletion zone of a p-n junction (see fig.~\ref{fig:tests:foto_effect_Fe}). The high electric field present in this region separates rapidly electrons and holes before they recombine and transports electrons into the collecting diodes. In such events the charge collection efficiency is $\approx$100\% and the whole signal is accumulated in a single pixel.\\
Figure~\ref{fig:tests:Fe55_Seed0} shows distributions of the $^{55}$Fe signal in the seed pixel of clusters reconstructed according to the procedure described in section~\ref{ch:tests:exp_results:cluster} using the seed cut $t_{s} = 4$ for MIMOSA-5 and $t_{s} = 8$ for MIMOSA-18. The calibrations of the MIMOSA-5 and the MIMOSA-18 were performed at -13.2$^{\circ}$C and 15.4$^{\circ}$C, respectively.
\\
The measured $^{55}$Fe spectra consist of four peaks. The leftmost peaks originate from noise clusters. The highest peaks with maximum around 50 and 70 counts in the MIMOSA-5 and MIMOSA-18, respectively, correspond to events in which only a fraction of charge was collected. At the high end of the spectrum there are two small peaks originating from the 5.9~keV and 6.49~keV emission lines for which a fully efficient charge collection took place. The calibration is achieved by relating the charges in these peaks and ADC counts according~to~(\ref{eq:exper:conv_gain}).\\
Two rightmost peaks in fig.~\ref{fig:tests:Fe55_Seed0} are once again shown in magnification in fig.~\ref{fig:tests:Fe55_Seed1} after imposing additional cuts. The contribution to the histograms in fig.~\ref{fig:tests:MIMOSA-5_Fe55_Seed1} and fig.~\ref{fig:tests:MIMOSA-18_Fe55_Seed1} is restricted to clusters consisting only of a single, isolated seed pixel. In order to find this type of clusters a cut on the seed neighbours was applied. The ratio between a total signal collected in the 8 pixels adjacent to the seed, $\mathcal{S}_{8} = \sum_{i = 1}^{8}s_{i}$, and they resultant noise $\mathcal{N}_{8}=\sqrt{\sum_{i=1}^{8}n_{i}^{2}}$ was required to be below 2 ($\mathcal{S}_{8}/\mathcal{N}_{8} < 2$), where $s_{i}$ and $n_{i}$ are the signal and the noise of the $i$-th pixel, respectively.
\\
The $K_{\alpha}$ and $K_{\beta}$ peak positions determined by fitting a sum of two Gaussian functions:
\begin{equation}
\label{eq:exper:sum_gaussian}
p0 \cdot \exp{\left(-\frac{(Signal-p1)^{2}}{2\cdot(p2^{2})}\right)} + p3 \cdot \exp{\left(-\frac{(Signal-p4)^{2}}{2\cdot(p5^{2})}\right)},
\end{equation}
where $p0, p1, p2, p3, p4$ and $p5$ are the fitted parameters. The resultant fit parameters are shown in the inserts in those figures. The peak positions, $p1$ and $p4$, were afterwords related with an average number of electrons generated by the 5.9~keV and 6.49~keV photons, respectively, and the ADC-to-charge conversion gain was evaluated. The input numbers and the conversion gains are summarised in table~\ref{tab:tests:exp_ADCGain}.
\begin{table}[!htbp]

	\begin{tabularx}{\textwidth}{@{\extracolsep{\fill}} |>{\small}c|>{\small}c|>{\small}c|>{\small}c|>{\small}c|>{\small}c|} \hline

	 &  Emission line & Conversion & Emission line & Conversion & Mean \\
	Prototype & $K_{\alpha}$ (1612 e$^{-}$) & gain & $K_{\beta}$ (1773 e$^{-}$) & gain & conv. gain\\
	 & p1 [ADC] & $G_{\alpha}$ [e/ADC] & p4 [ADC] & $G_{\beta}$ [e/ADC] & $G$ [e/ADC] \\ \hline \hline

	MIMOSA-5 & 166 & 9.7$\pm$0.05 & 183.4 & 9.7$\pm$0.07 & 9.7$\pm$0.04\\ \hline
	MIMOSA-18 & 267.8 & 6.02$\pm$0.05 & 294.4 & 6.02$\pm$0.06 & 6.02$\pm$0.04\\ \hline

	\end{tabularx}

\caption[Summary of calibration measurements performed for MIMOSA-5 and MIMOSA-18 matrices]{Summary of calibration measurements performed for MIMOSA-5 and MIMOSA-18 matrices. The parameters $p1$ and $p4$ are known from fitting function (\ref{eq:exper:sum_gaussian}) to histograms in fig.~\ref{fig:tests:MIMOSA-5_Fe55_Seed1} and fig.~\ref{fig:tests:MIMOSA-18_Fe55_Seed1}.}
\label{tab:tests:exp_ADCGain}

\end{table}\\
The ADC-to-charge conversion gains $G_{\alpha}$ and $G_{\beta}$ calculated from position of $K_{\alpha}$ and $K_{\beta}$ emission lines, respectively, agree within errors. The resulting value of the ADC-to-charge conversion gain $G$, used for converting ADC units to electrons, was obtained as a weighted mean of the $G_{\alpha}$ and $G_{\beta}$ (last column of table~\ref{tab:tests:exp_ADCGain}).

\subsection{Charge collection efficiency and cluster formation}
\label{ch:tests:exp_results:X-rays:charge_spreading}

\begin{figure}[!h] 
	\begin{center}
	\subfigure[MIMOSA-5]{
		\includegraphics[width=0.45\textwidth,angle=90]{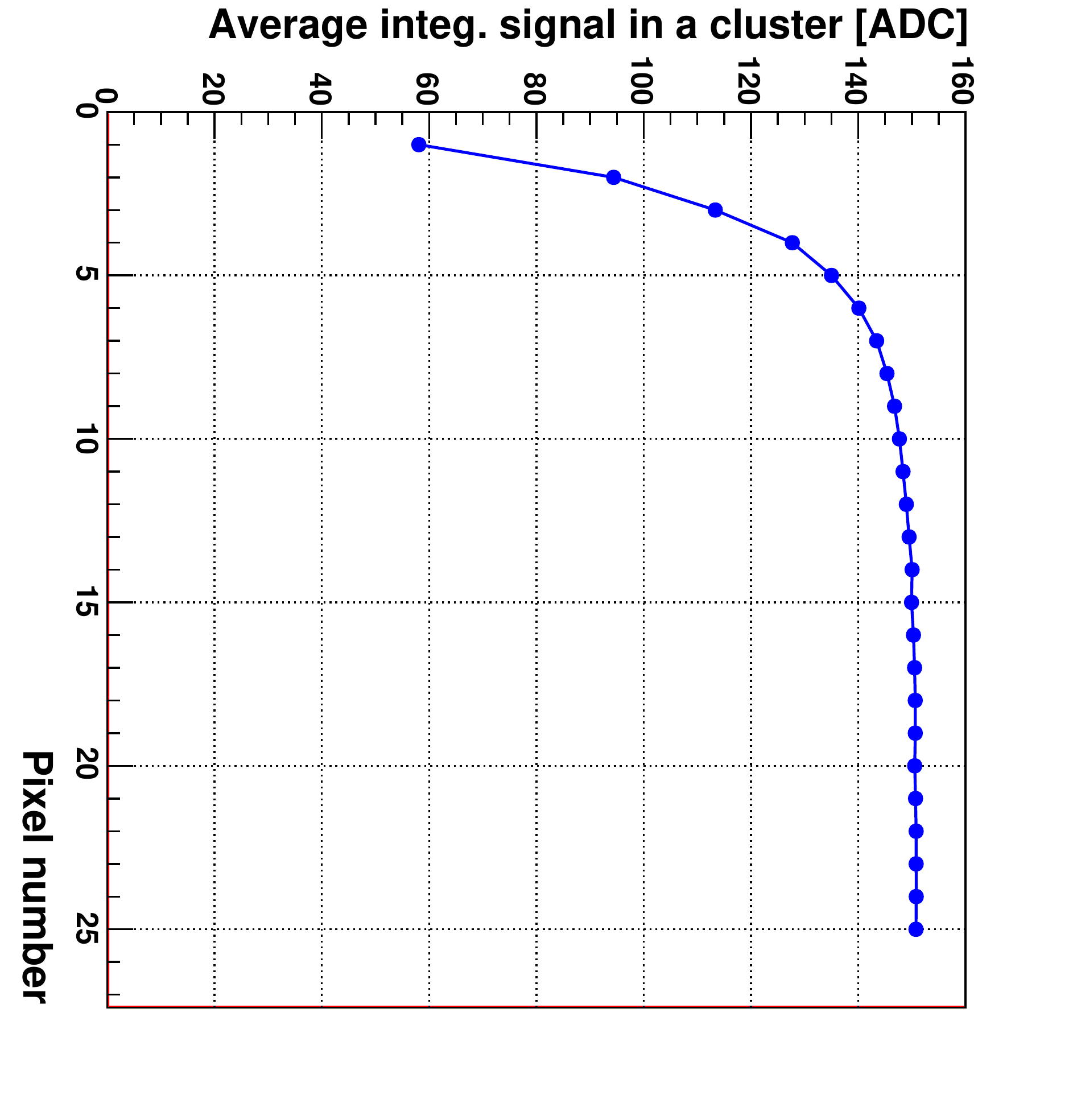}
		\label{fig:tests:MIMOSA-5_AverageFe55}
	}
	\hspace{0.1cm}
	\subfigure[MIMOSA-18]{
		\includegraphics[width=0.45\textwidth,angle=90]{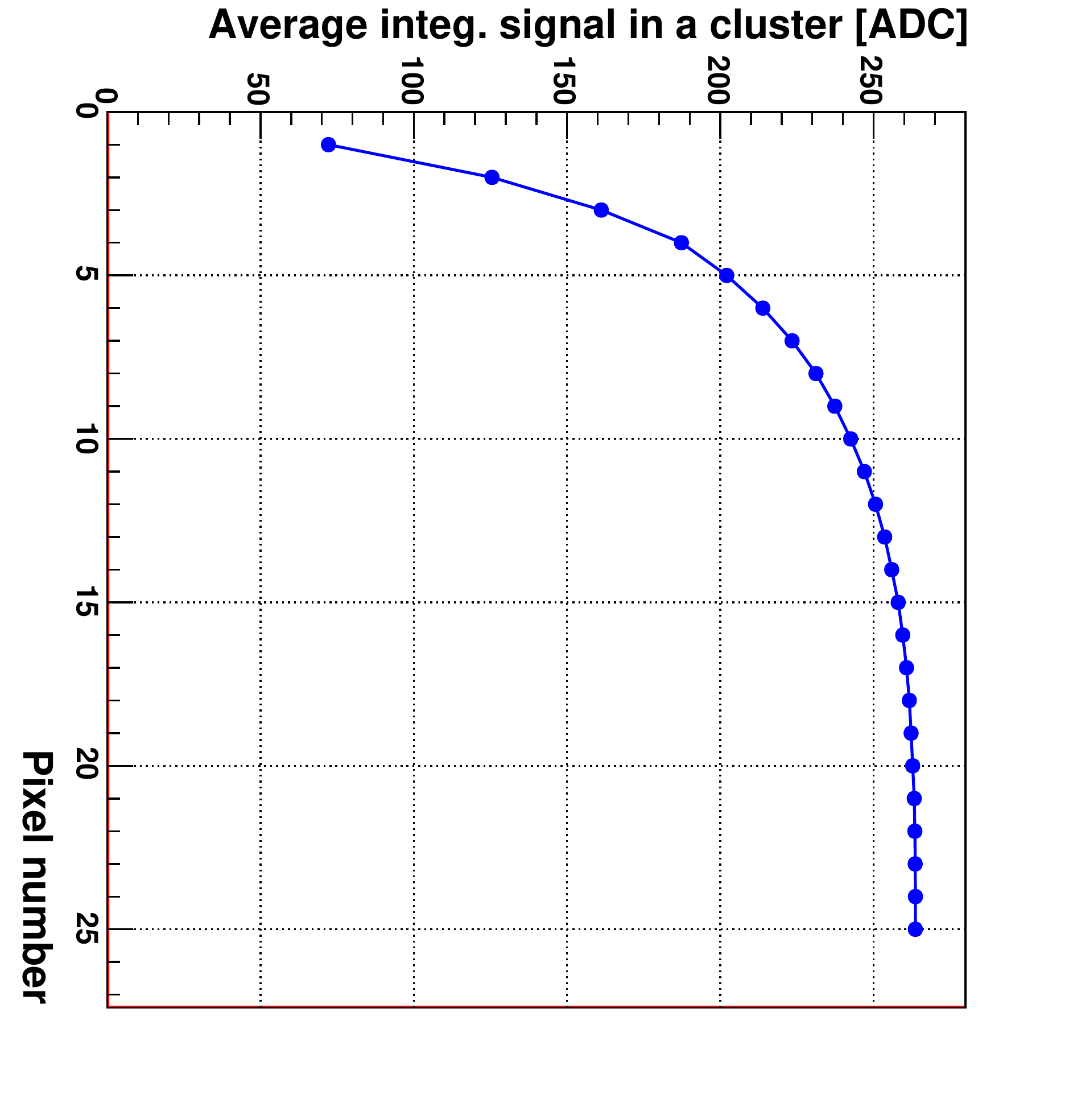}
		\label{fig:tests:MIMOSA-18_AverageFe55}
	}
	\caption[The average integrated cluster signal as a function of a pixel number]{The average integrated cluster signal as a function of a pixel number $n$ (see text) for (a) MIMOSA-5 and in the (b) MIMOSA-18 prototypes. Numbering of pixels corresponds to the order obtained by sorting pixels with respect to the descending signal-to-noise ratio.}
	\label{fig:tests:AverageFe55}
	\end{center}
\end{figure}

\begin{figure}[!h] 
	\begin{center}
	\subfigure[MIMOSA-5]{
		\includegraphics[width=0.45\textwidth,angle=90]{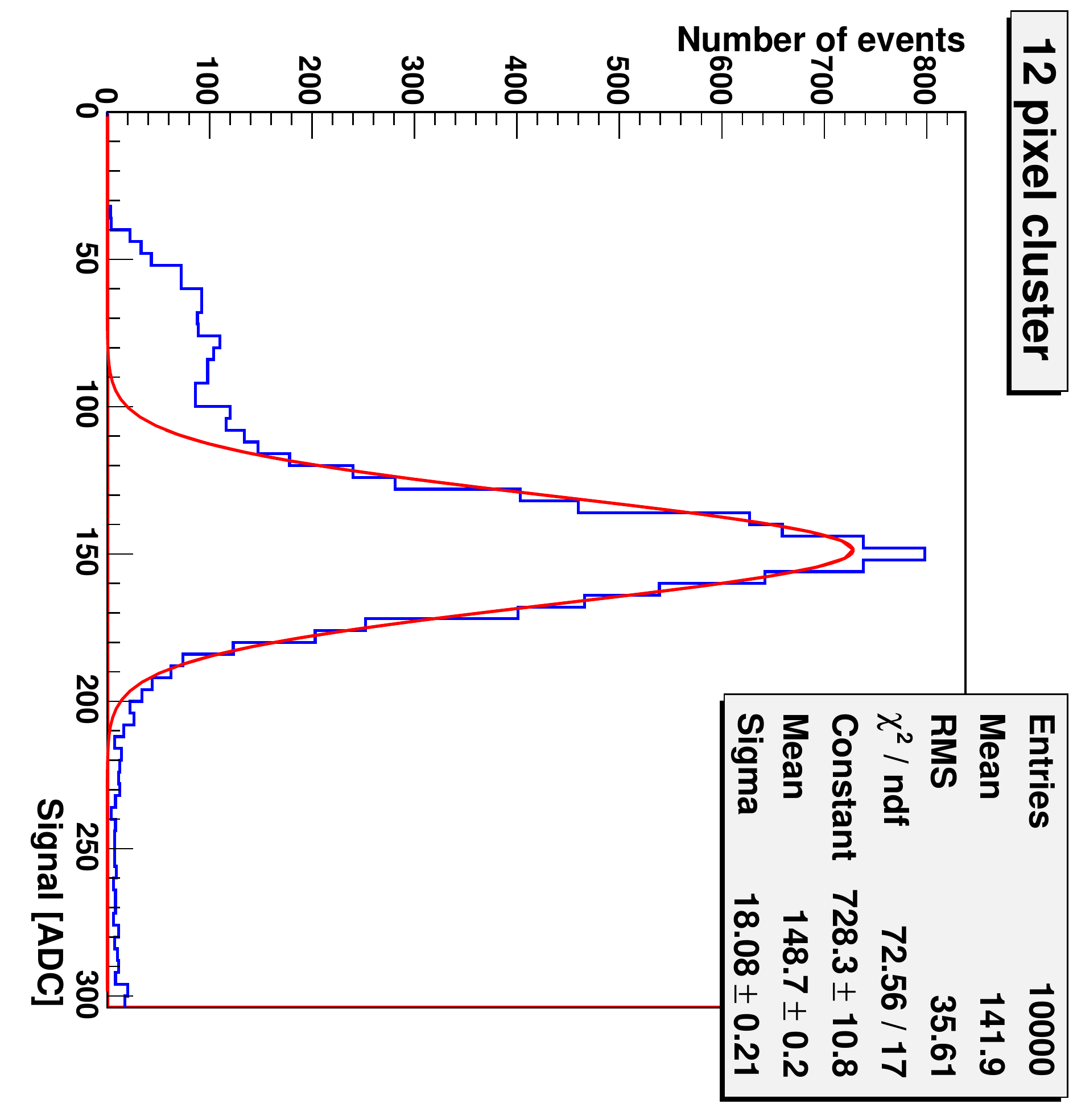}
		\label{fig:tests:MIMOSA-5_Fe55Cluster}
	}
	\hspace{0.1cm}
	\subfigure[MIMOSA-18]{
		\includegraphics[width=0.45\textwidth,angle=90]{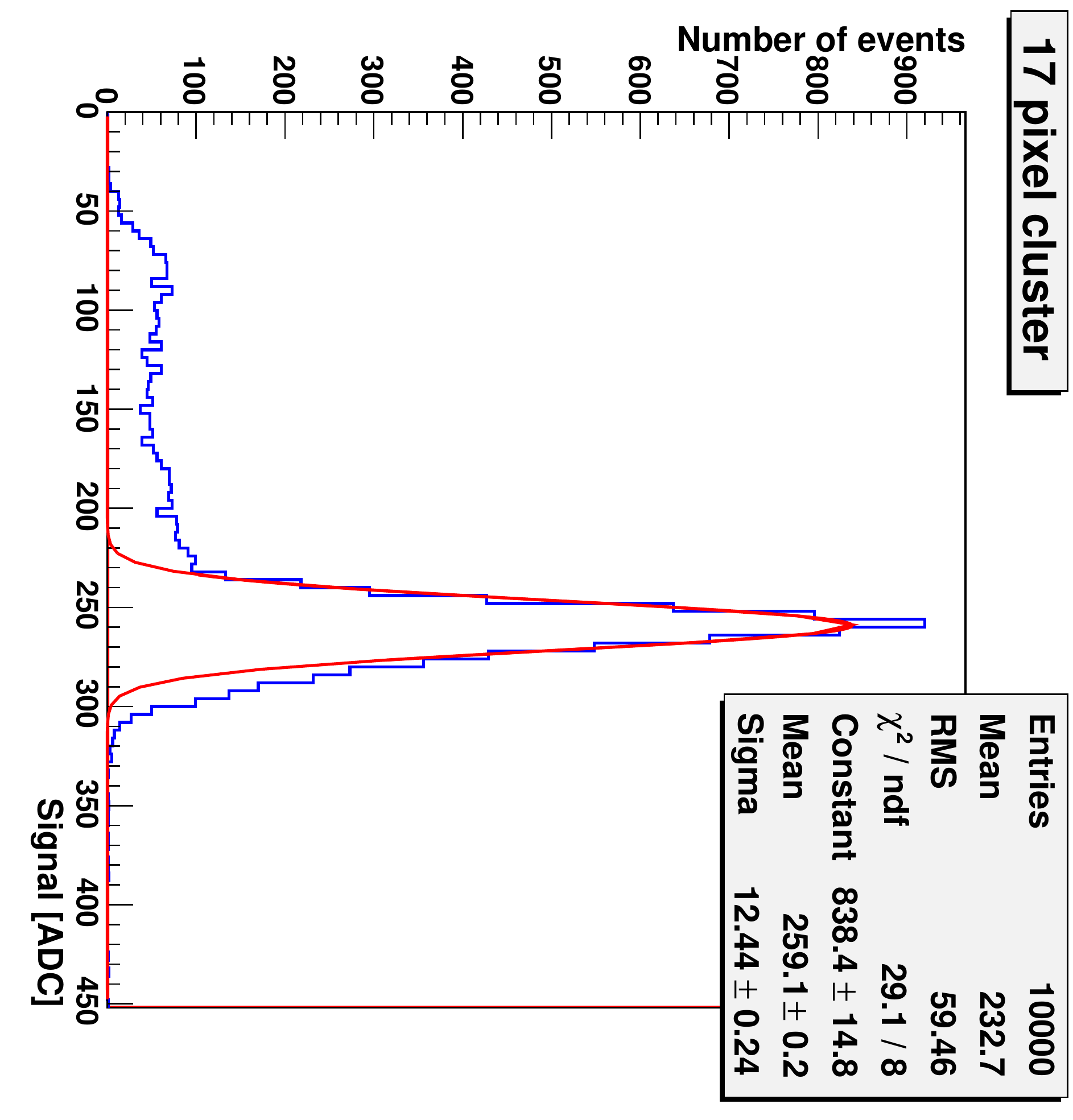}
		\label{fig:tests:MIMOSA-18_Fe55Cluster}
	}
	\caption[Cluster signal (charge) distribution obtained using $^{55}$Fe photons]{Cluster signal (charge) distribution obtained using $^{55}$Fe photons in (a) MIMOSA-5 (clusters composed of the first 12 pixels) and (a) MIMOSA-18 (clusters composed of the first 17 pixels). The pixels were ordered according to descending S/N ratio (see text).}
	\label{fig:tests:Fe55Cluster}
	\end{center}
\end{figure}

Charge carriers generated in the field free epitaxial layer diffuse isotropically inside the active volume of the detector. The collected charge is shared by neighbouring pixels and a cluster is formed. Some of the charge carriers propagating through the epitaxial layer recombine before reaching the collecting diodes which results in the loss of signal. Studies of the charge spread into neighbouring pixels and evaluation of the charge collection efficiency are described below.\\
For the purpose of these investigations clusters of maximum size 5$\times$5 pixels were selected according to the procedure described in section~\ref{ch:tests:exp_results:cluster} with $t_{s} = 4$, $t_{n} = 2$ for MIMOSA-5 and $t_{s} = 8$, $t_{n} = 2$ for MIMOSA-18. The 25~pixels in a given cluster, $c_{i}$, $ i = 1,...,25$ were ordered according to signal-to-noise ratio: $c_{1}, c_{2},..., c_{25}$. The integrated charge (averaged over 10k clusters) in the first $n$ pixels was plotted as a function of $n$ in fig~\ref{fig:tests:MIMOSA-5_AverageFe55} and \ref{fig:tests:MIMOSA-18_AverageFe55} for MIMOSA-5 and MIMOSA-18, respectively. As can be seen from these figures, 99\% of the total charge is collected, respectively, in the first 12 and 17 pixels in the above sequence, after which saturation is observed. The seed pixel in the MIMOSA-5 accounts on average for $\sim$40\% while in the MIMOSA-18 for $\sim$27\% of the total collected charge.
\\
In fig.~\ref{fig:tests:Fe55Cluster} one can see distributions of charges collected in the first 12 and 17 pixels in clusters reconstructed for the MIMOSA-5 and MIMOSA-18, respectively (i.e. containing 99\% of the total charge). These distributions are dominated by the peak corresponding to the 5.9~keV photon emission line, which is fitted with a Gaussian function in order to determine its position. The signals at lower values correspond to noise clusters which also passed the selections. Since a fraction of charge carriers originating from the epitaxial layer recombine before reaching the collecting diodes, the 5.9~keV peak positions in fig.~\ref{fig:tests:MIMOSA-5_Fe55Cluster} and fig.~\ref{fig:tests:MIMOSA-18_Fe55Cluster} are slightly shifted towards lower values with respect to the positions measured for isolated seed pixels, which were shown in fig.~\ref{fig:tests:MIMOSA-5_Fe55_Seed1} and fig.~\ref{fig:tests:MIMOSA-18_Fe55_Seed1}, respectively, in case of which the charge collection efficiency was 100\%. Using the magnitude of this shift one can evaluate the charge collection efficiency for the charge carriers originating from the epitaxial layer as following. The ratio between the positions of the 5.9~keV peaks measured for photons interacting in the epitaxial layer, fig.~\ref{fig:tests:MIMOSA-5_Fe55Cluster} and fig.~\ref{fig:tests:MIMOSA-18_Fe55Cluster}, and those interacting in the depletion zone, fig.~\ref{fig:tests:MIMOSA-5_Fe55_Seed1} and fig.~\ref{fig:tests:MIMOSA-18_Fe55_Seed1}, gives the charge collection efficiency for carriers originating from the epitaxial layer. The results for MIMOSA-5 and MIMOSA-18 are 90\% and 97\%, respectively.

\section{Alignment of the detector in beam tests}
\label{ch:tests:exp_results:alignment_DUT} 

A precise alignment of pixel matrix w.r.t the beam telescope is needed for studies of the spatial resolution of a pixel matrix and its efficiency for track detection. Positions of hits reconstructed in the MAPS detector are expressed in the local coordinate system ($x_{loc}$,$y_{loc}$), while for the comparison with the telescope tracks it is required to transform them to the common reference system ($x_{ref}$,$y_{ref}$), here defined by the first telescope plane. The two sets of coordinates are related as follows:

\begin{equation}
\label{eq:exper:alignDUT}
	\left( \begin{array}{c}
	x_{ref} \\
	y_{ref} 
	\end{array} \right) = 	
	\left( \begin{array}{cc}
	a & b \\
	c & d
	\end{array} \right)
	\cdot	\left( \begin{array}{c}
	x_{loc} \\
	y_{loc}
	\end{array} \right) + 
	\left( \begin{array}{c}
	v_{x} \\
	v_{y}
	\end{array} \right),
\end{equation}
where:
\begin{equation}
\label{eq:exper:alignDUT_a}
a = \cos{\phi_{y}} \cdot \cos{\phi_{z}},
\end{equation}
\begin{equation}
\label{eq:exper:alignDUT_b}
b = \cos{\phi_{y}} \cdot \sin{\phi_{z}},
\end{equation}
\begin{equation}
\label{eq:exper:alignDUT_c}
c = \sin{\phi_{x}} \cdot \sin{\phi_{y}} \cdot \cos{\phi_{z}} + \cos{\phi_{x}} \cdot \sin{\phi_{z}},
\end{equation}
\begin{equation}
\label{eq:exper:alignDUT_d}
d = \cos{\phi_{x}} \cdot \cos{\phi_{z}} - \sin{\phi_{x}} \cdot \sin{\phi_{y}} \cdot \sin{\phi_{z}}.
\end{equation}\\
The six parameters: three offsets $v_{x}$, $v_{y}$, $v_{z}$ and three rotation angles $\phi_{x}$, $\phi_{y}$, $\phi_{z}$ describe the DUT position and angular orientation with respect to the reference system of the telescope. These parameters are determined in the alignment procedure which is based on the $\chi^{2}$ minimisation similar to that described in section \ref{ch:tests:exp:setup:MVDAnalysis:alig}. The $\chi^{2}$ is defined as follows:
\begin{equation}
\label{eq:tests:align_chi2_DUT}
	\chi^{2} = \sum_{i}\frac{\left(x_{i,ref} - x_{i,pred}\right)^{2} + 
	\left(y_{i,ref} - y_{i,pred}\right)^{2}}{\sigma^{2}},
\end{equation}
where $x_{pred}$ and $y_{pred}$ stand for predicted hit positions of the telescope tracks at the DUT plane. The tracks are represented by straight lines, fitted to the hits in the planes of the telescope. The  $x_{pred}$ and $y_{pred}$ are calculated according to the formula:
\begin{equation}
\label{eq:exper:alignDUT_lineFit}
x_{pred} = a_{x} \cdot v_{z} + b_{x},~~~y_{pred} = a_{y} \cdot v_{z} + b_{y}, 
\end{equation}
where $a_{x}$ and $a_{y}$ are the slopes and the $b_{x}$ and $b_{x}$ are the offsets obtained from the fit to the telescope hits. The $\chi^{2}$ minimisation is performed with the C++ version of the \texttt{MINUIT} package \cite{tests:MINUIT}. The result obtained for the parameter uncertainties, using the \texttt{MIGRAD} procedure, are as follows: $\sigma_{v_{x}}\approx$~0.1~$\mu$m, $\sigma_{v_{y}}\approx$~0.1~$\mu$m, $\sigma_{v_{z}}\approx$~15.0~$\mu$m, $\sigma_{\phi_{x}}\approx$~0.01$^{\circ}$, $\sigma_{\phi_{y}}\approx$~0.01$^{\circ}$ and $\sigma_{\phi_{z}}\approx$~0.001$^{\circ}$.

\section{Tracking}
\label{ch:tests:exp_results:tracking_performance}

\begin{figure}[!h] 
	\begin{center}
	\subfigure[]{
		\includegraphics[width=0.45\textwidth,angle=90]{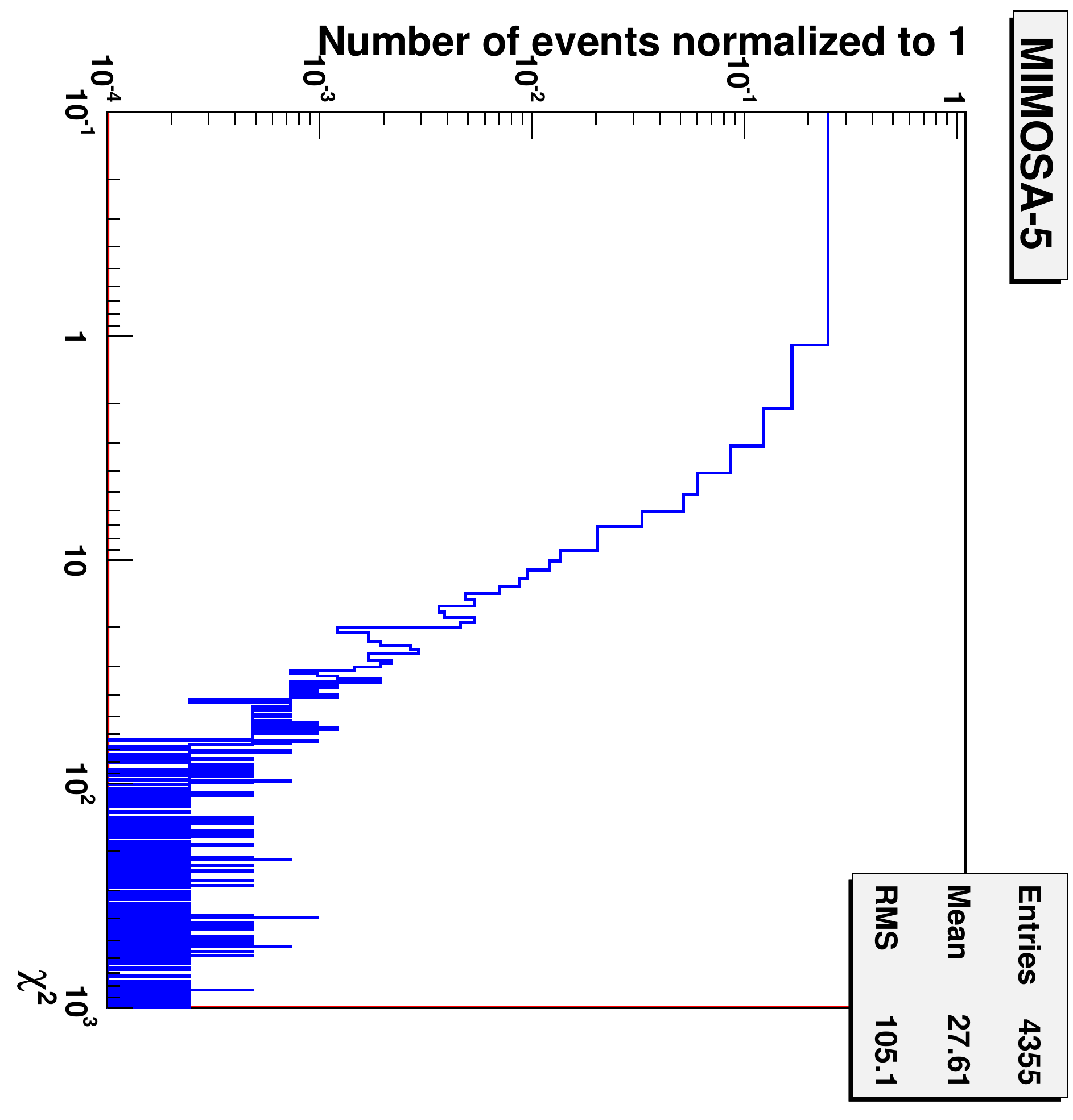}
		\label{fig:tests:TrackChi:M5}
	}
	\hspace{0.1cm}
	\subfigure[]{
		\includegraphics[width=0.45\textwidth,angle=90]{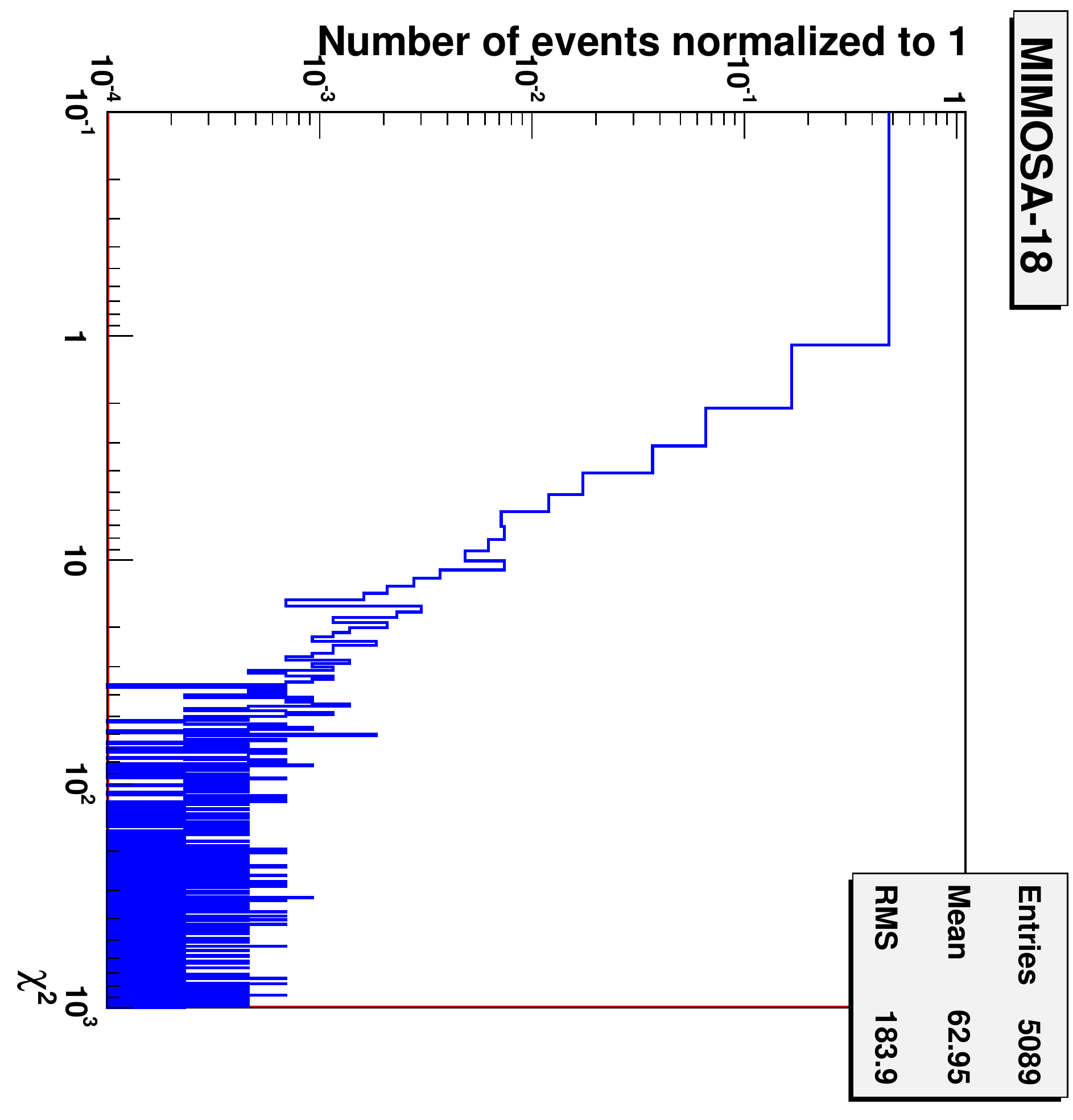}
		\label{fig:tests:TrackChi:M18}
	}
	\caption[The $\chi^{2}$ distribution for tracks passing through the active are of the MIMOSA sensors]{The $\chi^{2}$ distribution for tracks passing through the active are of the (a) MIMOSA-5 and (b) MIMOSA-18.}
	\label{fig:tests:TrackChi2}
	\end{center}
\end{figure}

In order to study tracking capabilities, the MIMOSA-5 and MIMOSA-18 prototypes were exposed to electron beams of 6~GeV and 5~GeV, respectively. Tests of the MIMOSA-18 detector were performed after conversion of the DESY-II accelerator to the low energy mode when only a 5~GeV electron beam was provided.\\ 
At the beam energies available at DESY, multiple Coulomb scattering in the material of the telescope planes and in the DUT itself has a significant influence on electron tracks. Therefore a tracking method described in section~\ref{ch:tests:exp:setup:MVDAnalysis:tracking} was used in which multiple scattering is included in the track finding algorithm. The precision of determining the intersection of the reconstructed track with the DUT plane calculated according to the (\ref{eq:tests:afz_error}) depends on the beam energy and on the amount of material traversed by the track. In the case of the MIMOSA-5 of 150~$\mu$m thickness, the uncertainty of the hit position measurement in both directions was approx. $7.0~\mu$m while for the MIMOSA-18 of 700~$\mu$m thickness it was approx. $8.5~\mu$m.\\
From several thousand of tracks, reconstructed in the telescope, only those were selected which passed through the active area of the DUT. The quality of the fitted truck is defined with a $\chi^{2}$ variable which is a sum of $\Delta\chi^{2}_{i}$ contributions calculated for each plane $i$ (3 telescope plane and 1 DUT plane) according to (\ref{eq:tests:afz_fit}). The $\chi^{2}$ distributions of the accepted tracks for the MIMOSA-5 and MIMOSA-18 are shown in fig.~\ref{fig:tests:TrackChi2}. The number of degrees of freedom for the fitted tracks was 2.
\\
Tracks with the $\chi^{2}$ greater then 8 were removed from the further analysis in order to exclude events in which the position measurements in the telescope planes were affected by strong multiple scattering. This results in 26\% and 32\% reduction of event samples for the MIMOSA-5 and MIMOSA-18, respectively. 

\subsection{Track detection efficiency}
\label{ch:tests:exp_results:tracking_performance:effi}

\begin{figure}[!h] 
	\begin{center}
	\subfigure[]{
		\includegraphics[width=0.45\textwidth,angle=0]{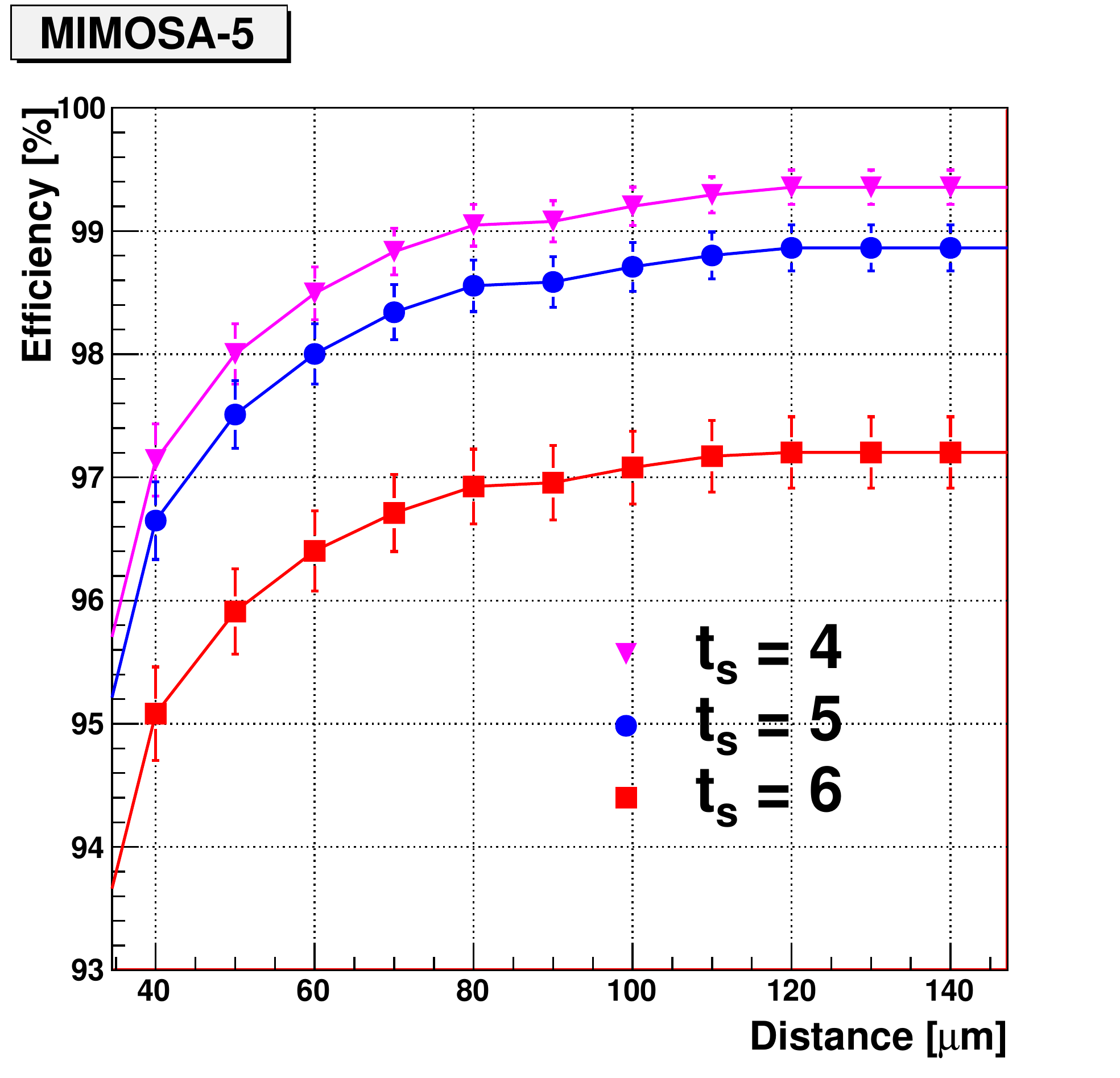}
		\label{fig:tests:Efficiency_Seed:M5}
	}
	\hspace{0.1cm}
	\subfigure[]{
		\includegraphics[width=0.45\textwidth,angle=0]{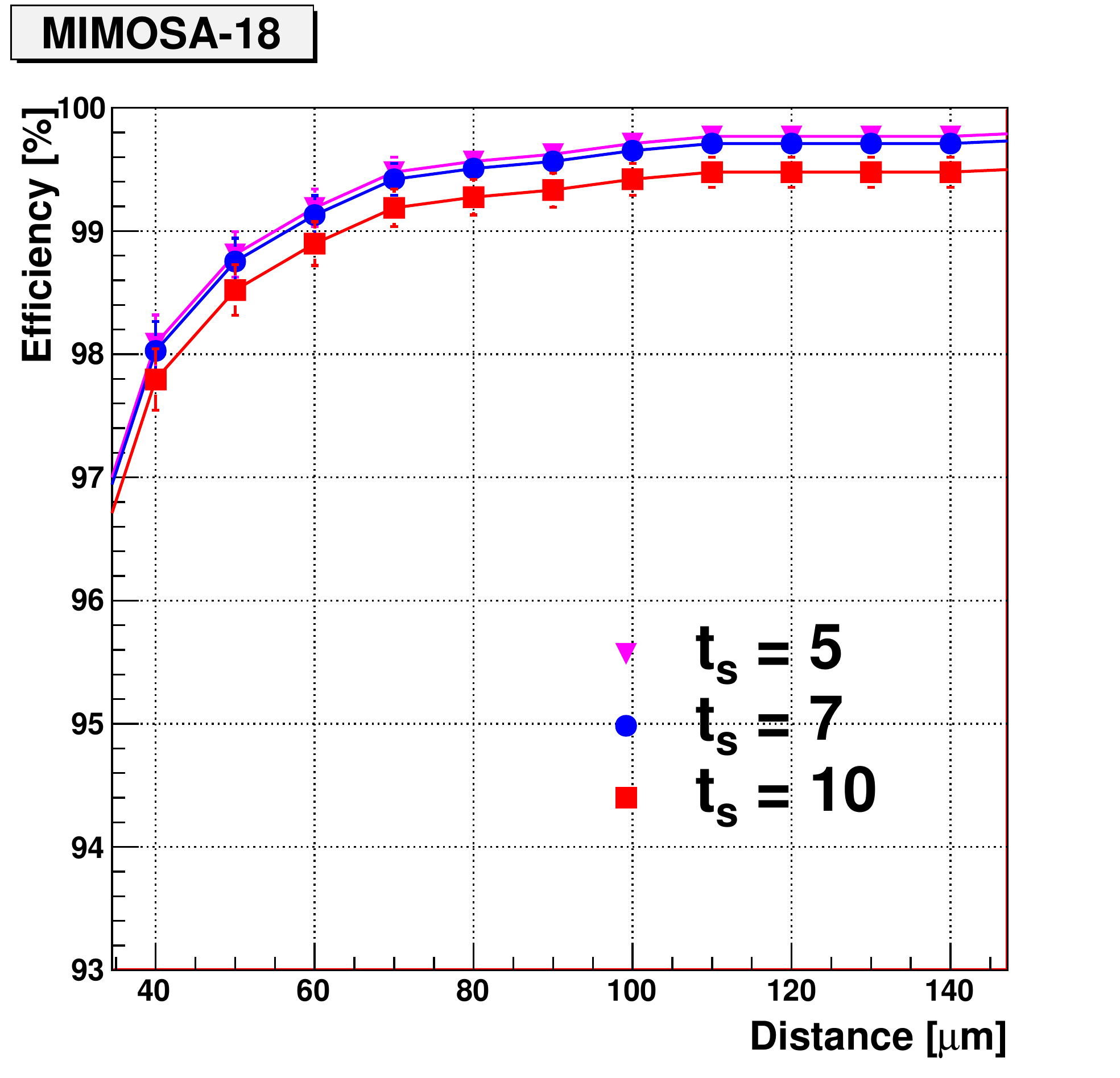}
		\label{fig:tests:Efficiency_Seed:M18}
	}
	\caption[Detection efficiency as a function of cuts]{Detection efficiency as a function of cuts on the signal-to-noise ratio of the seed ($t_{s}$) and on the maximum allowed track-to-hit distance for the (a) MIMOSA-5 and (b) MIMOSA-18 prototypes. The signal-to-noise cut for the seed neighbours was $t_{n}=0.5$.}
	\label{fig:tests:Efficiency_Seed}
	\end{center}
\end{figure}

\begin{figure}[!h] 
	\begin{center}
	\subfigure[]{
		\includegraphics[width=0.45\textwidth,angle=0]{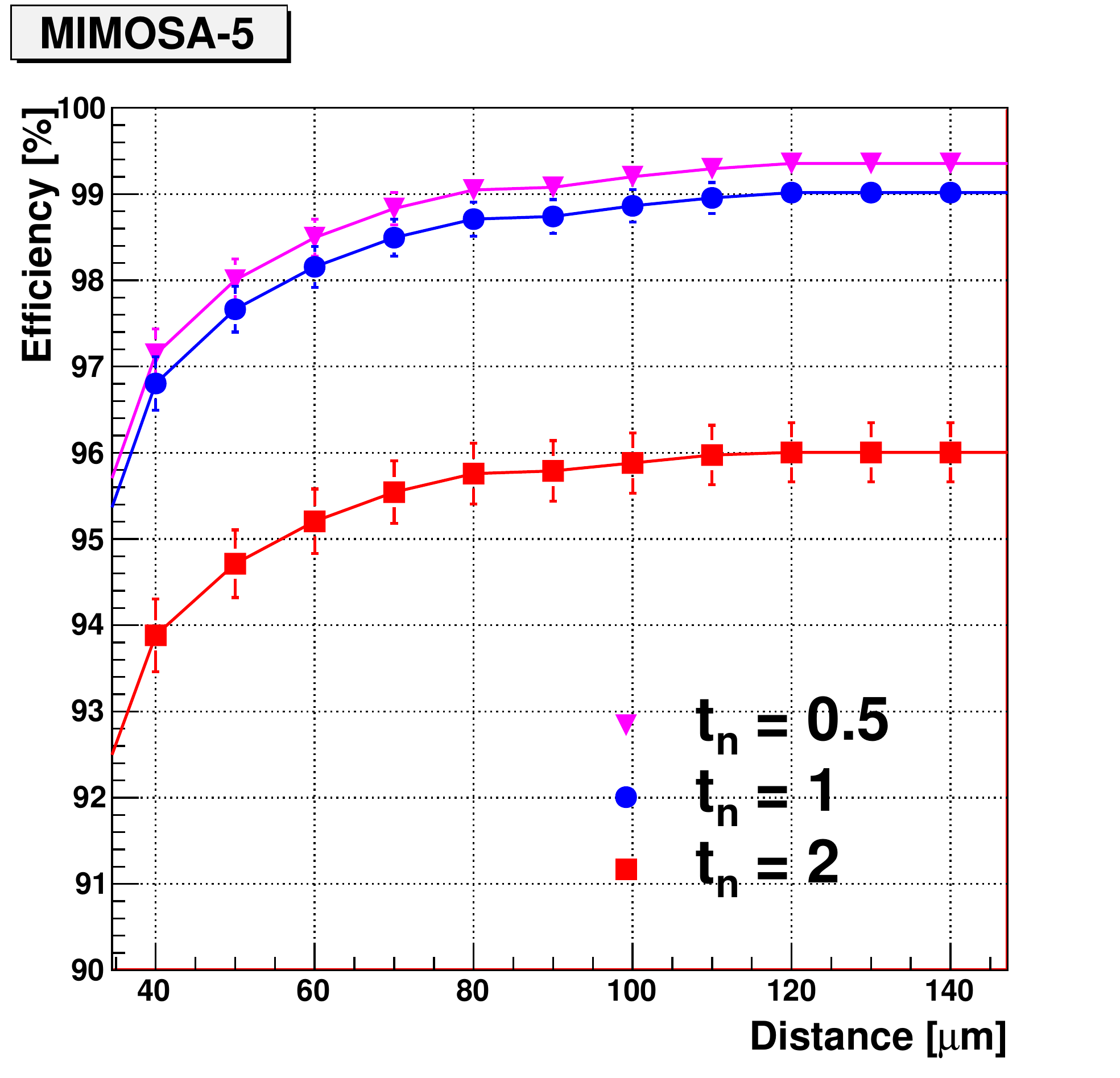}
		\label{fig:tests:Efficiency_Neigh:M5}
	}
	\hspace{0.1cm}
	\subfigure[]{
		\includegraphics[width=0.45\textwidth,angle=0]{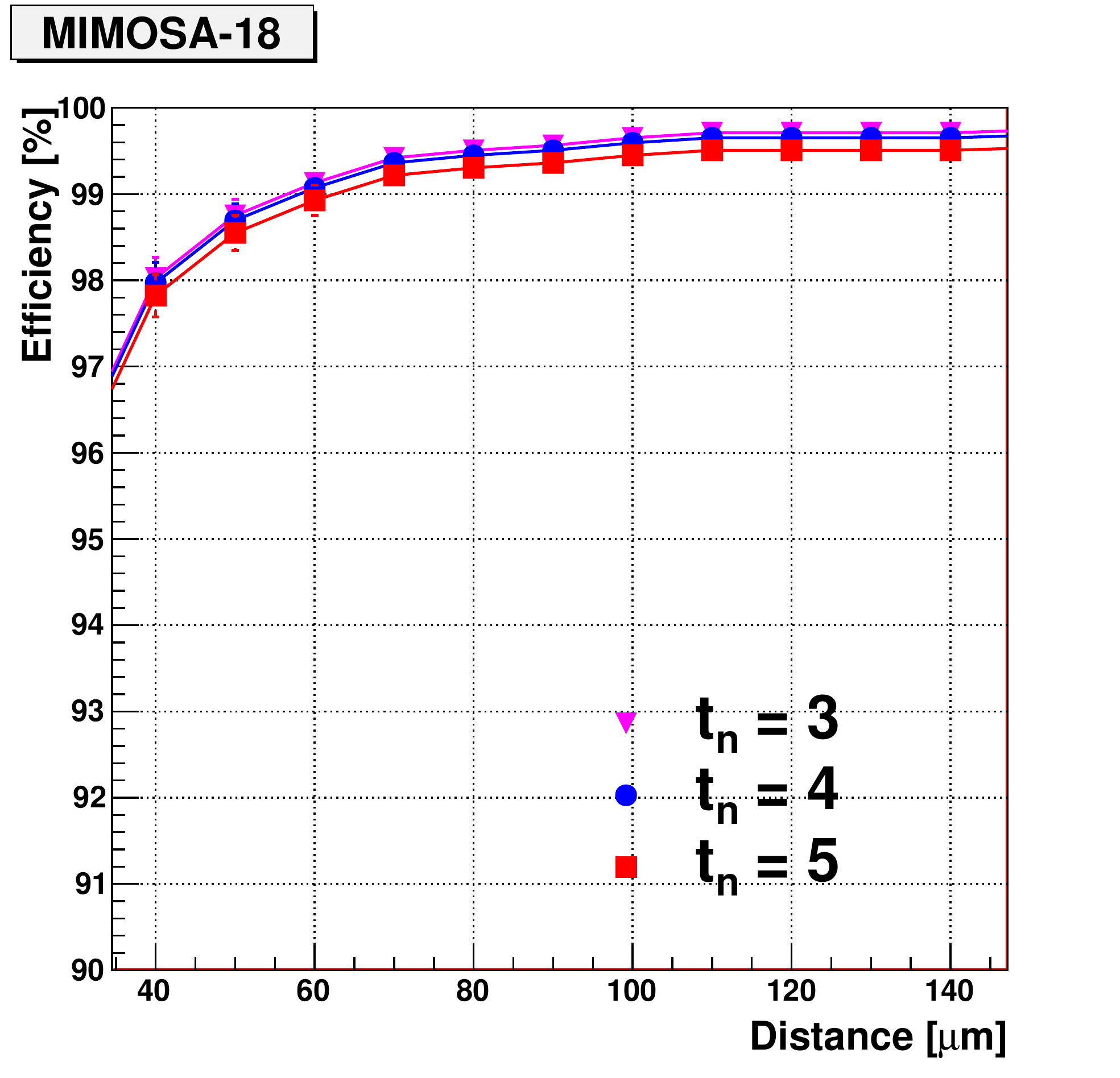}
		\label{fig:tests:Efficiency_Neigh:M18}
	}
	\caption[Detection efficiency as a function of cuts]{Detection efficiency as a function of cuts on the signal-to-noise ratio of the seed neighbours ($t_{n}$) and on the maximum allowed track-to-hit distance for (a) MIMOSA-5 and (b) MIMOSA-18 prototypes. The signal-to-noise cut for the seed was $t_{s}=4$ and $7$ for the MIMOSA-5 and MIMOSA-18, respectively.}
	\label{fig:tests:Efficiency_Neigh}
	\end{center}
\end{figure}

The track detection efficiency, $\varepsilon$, is defined as a ratio of the number of reconstructed clusters, correlated with the corresponding telescope tracks within the geometrical acceptance of a DUT, and the number of all accepted telescope tracks, N. The uncertainty of the efficiency is determined from the variation assuming the binomial distribution:
\begin{equation}
\label{eq:exper:error_effi}
\sigma_{e} = \sqrt{\frac{\varepsilon(1-\varepsilon)}{N-1}}.
\end{equation} 
The detection efficiency was found to depend on the cuts applied during the cluster reconstruction, $t_{s}$ and $t_{n}$ (see section \ref{ch:tests:exp_results:cluster}) and on the maximum allowed track-to-hit distance, which is a distance between predicted and reconstructed impact position of the telescope track in the DUT. The dependence of the detection efficiencies on the $t_{s}$ cut and on the maximum track-to-hit distance for clusters reconstructed with the $t_{n} = 0.5$ cut is shown in fig.~\ref{fig:tests:Efficiency_Seed}.
\\
Detection efficiency increases with the value of the maximal track-to-hit distance and saturates at approx. 100~$\mu$m for both pixel matrices. Such a wide interval in which the clusters are found is due to the effect of multiple scattering which results in a dispersion of predicted cluster positions from the real hit position.\\
The detection efficiency for MIMOSA-5 is much more sensitive to the signal-to-noise cut applied to the seed ($t_{s}$) than that of the MIMOSA-18. This is due to approximately 2.7 times higher noise of the MIMOSA-5 than the noise of the MIMOSA-18. Depending on the $t_{s}$ cut of 4, 5 or 6, the MIMOSA-5 detection efficiency reaches 99.3\%~$\pm$~0.1\%, 98.8\%~$\pm$~0.2\% or 97.2\%~$\pm$~0.3\%, respectively. Similar studies have shown that the MIMOSA-18 detection efficiencies of 99.8\%~$\pm$~0.1\%, 99.8\%~$\pm$~0.1\% and 99.5\%~$\pm$~0.1\% are measured with the $t_{s}$ cuts equal 5, 7 and 10, respectively. It is possible to improve the detection efficiencies by applying less restrictive cuts to the seed pixels. This however results in increasing probabilities of including background or noise fluctuations in the data sample. On the grounds of the above studies, the $t_{s}$ cuts equal 4 and 7 for the MIMOSA-5 and MIMOSA-18, respectively, were chosen as optimal.\\ 
Using the fixed $t_{s}$ cuts, other studies were performed: detection efficiency dependence on the signal-to-noise ratio cut on the seed neighbours ($t_{n}$). Detection efficiency as a function of a $t_{n}$ and a maximal track-to-hit distance are shown in fig.~\ref{fig:tests:Efficiency_Neigh}. The MIMOSA-5 detection efficiency decreases with the increasing $t_{n}$ cut in contrast to the MIMOSA-18 for which the detection efficiency depends rather weakly on the $t_{n}$ in the a wide range of the latter. In order to provide detection efficiency exceeding 99\% the adopted values of the $t_{n}$ cuts for the MIMOSA-5 and MIMOSA-18 are 0.5 and 4, respectively.

\subsection{Remarks on spatial resolution}
\label{ch:tests:exp:tracking_performance:resolution}

\begin{figure}[!h] 
	\begin{center}
	\subfigure[]{
		\includegraphics[width=0.45\textwidth,angle=90]{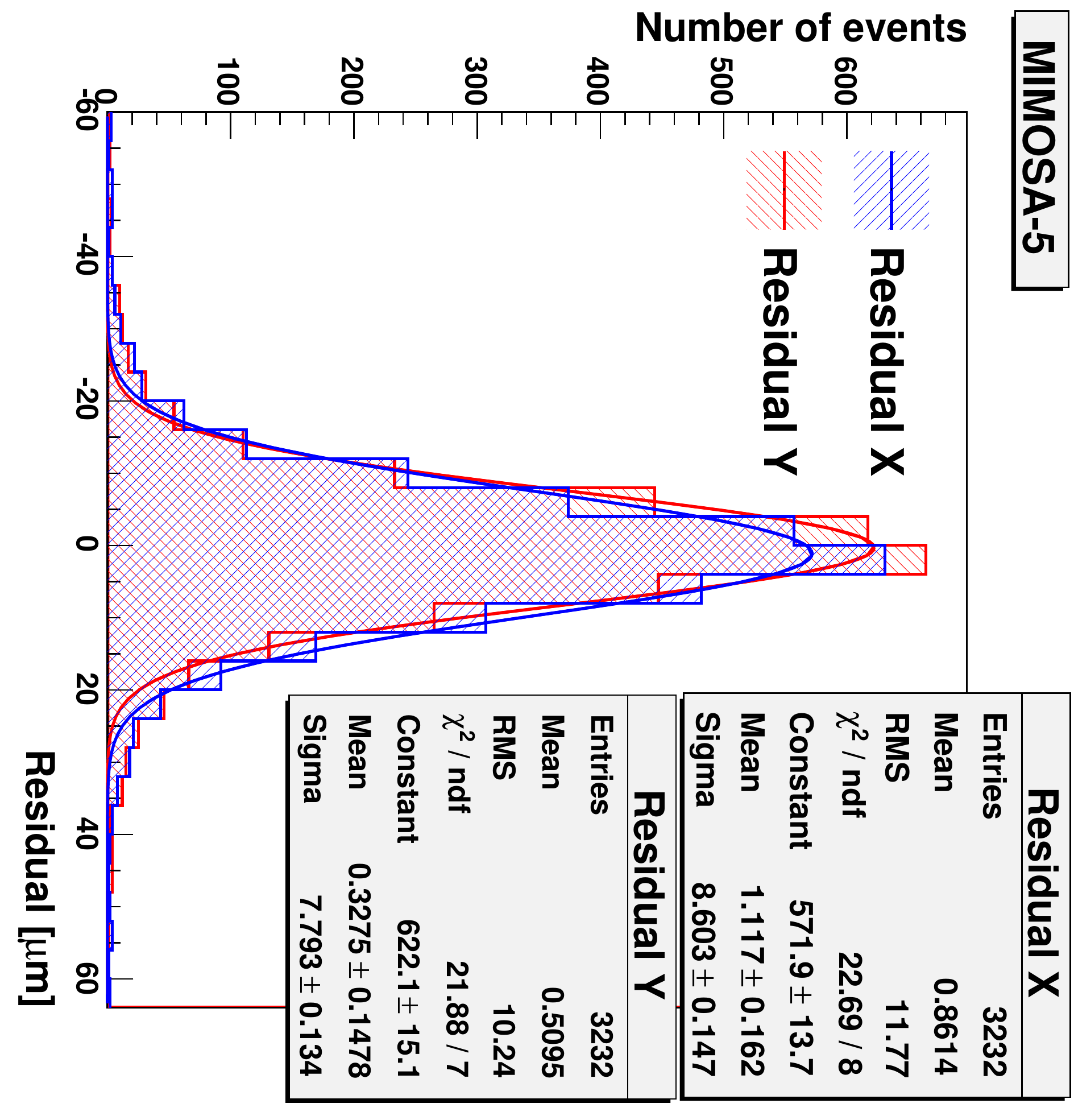}
		\label{fig:tests:Residuals:M5}
	}
	\hspace{0.1cm}
	\subfigure[]{
		\includegraphics[width=0.45\textwidth,angle=90]{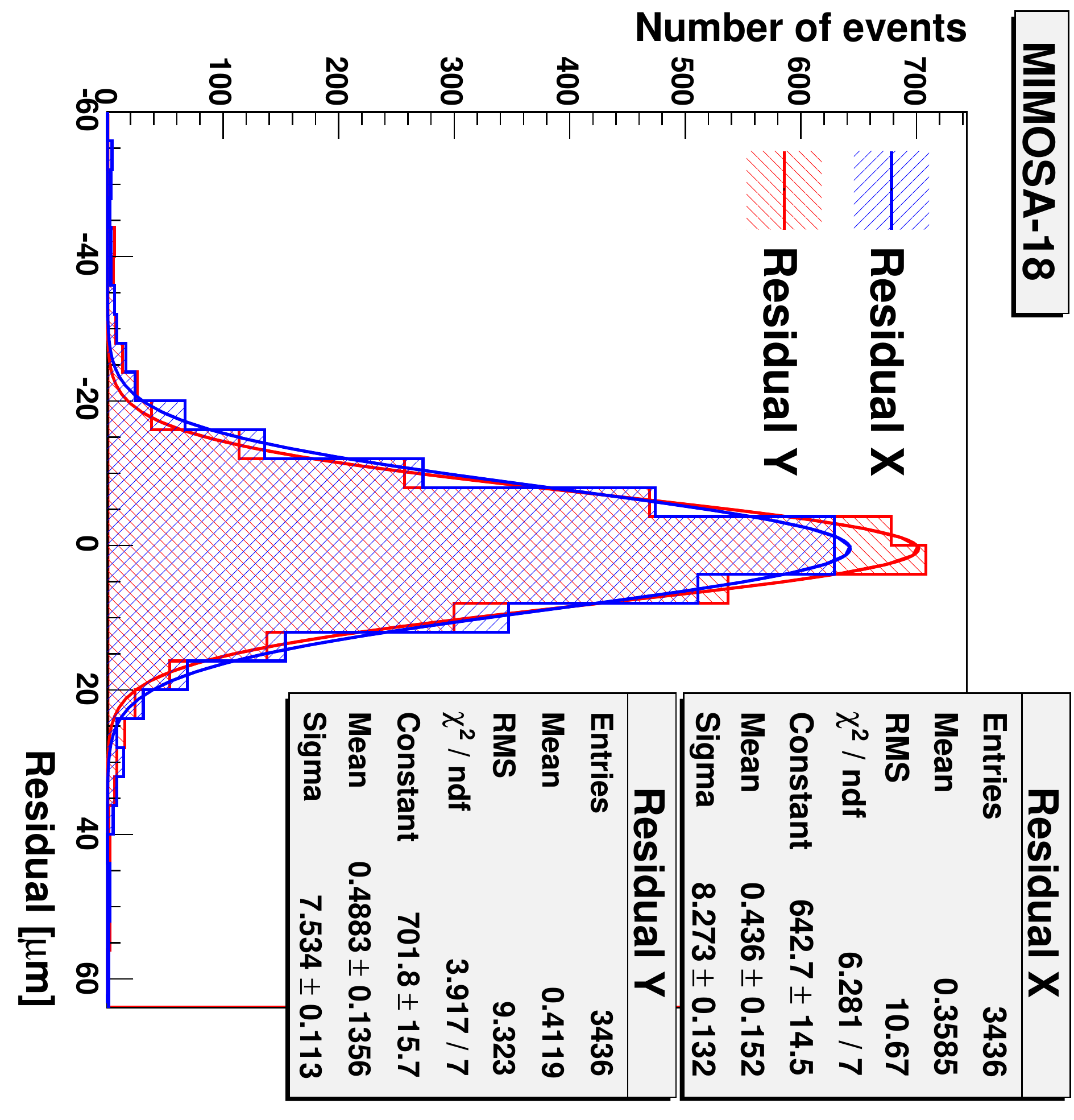}
		\label{fig:tests:Residuals:M18}
	}
	\caption[Distribution of residuals]{Distribution of residuals, fitted with the Gaussian functions for (a) MIMOSA-5 and (b) MIMOSA-18.}
	\label{fig:tests:Residuals}
	\end{center}
\end{figure}

The spatial resolution of the studied pixel matrices was determined from the distribution of residuals. They are defined as differences between the cluster positions reconstructed on the surface of the pixel matrix and the positions of the corresponding tracks interpolated from the telescope to the matrix plane. The distribution of residuals is characterised by a width, $\sigma_{res}$, which is a convolution of the spatial resolutions of the pixel matrix ($\sigma_{MIMOSA}$) and the telescope ($\sigma_{TELE}$):
\begin{equation}
\label{eq:exper:residuals}
\sigma_{res} = \sqrt{\sigma_{MIMOSA}^{2} + \sigma_{TELE}^{2}}.
\end{equation}
In order to determine $\sigma_{MIMOSA}$ one has to use a telescope with much better resolution, i.e. $\sigma_{TELE} \ll \sigma_{MIMOSA}$.\\
The total width of a given measured distribution, determined by fitting the Gaussian, is assumed to be equal to $\sigma_{res}$. In this analysis two values of $\sigma_{res}$ were determined for each matrix, $\sigma_{res,x}$ and $\sigma_{res,y}$, corresponding to distributions of residuals in the $x$ and $y$ projection. The distributions of the MIMOSA-5 and MIMOSA-18 residuals are shown in fig.~\ref{fig:tests:Residuals}. The results presented here were obtained with cluster positions reconstructed using the centre of gravity algorithm applied to the 3$\times$3 pixel clusters.
\\
It is known from earlier measurements that the telescope resolution amounts to 9~$\mu$m for a single plane \cite{tests:MiliTh}. In the present studies 3 telescope planes were used: 1 in front of the DUT and 2 behind it. In this configuration the resolution of the telescope amount to 7~-~8.5~$\mu$m. In view of this, it was impossible to determine precise values of resolution of the pixel matrices. However if $\sigma_{res}\approx\sigma_{TELE}$ then $\sigma_{MIMOSA}\ll\sigma_{TELE}$, i.e. of the order of single microns. The measured widths of the MIMOSA-5 and MIMOSA-18 residual distributions are shown in table~\ref{tab:tests:residuals_summary}.
\begin{table}[!htbp]
\begin{center} 
	\begin{tabularx}{0.5\textwidth}{@{\extracolsep{\fill}} |>{\small}c|>{\small}c|>{\small}c|} \hline

	Prototype & $\sigma_{res,x}$~[$\mu$m] & $\sigma_{res,y}$~[$\mu$m] \\ \hline \hline
	MIMOSA-5 & 8.6$\pm$0.1 & 7.8$\pm$0.1 \\ \hline
	MIMOSA-18 & 8.3$\pm$0.1 & 7.5$\pm$0.1 \\ \hline
	\end{tabularx}
\end{center}
\caption[Widths of the residual distributions in $x$ and $y$ projection]{Widths of the residual distributions in $x$ and $y$ projection. The  $\sigma_{res,x}$ and $\sigma_{res,y}$ are determined by fitting the Gaussian function to the histograms in fig.~\ref{fig:tests:Residuals}.}
\label{tab:tests:residuals_summary}
\end{table}\\
The precise measurements, performed by the Strasbourg group at CERN with 100~GeV pions, for which multiple scattering effects are negligible, gave the resolution $\sigma=1.7\pm0.1~\mu$m for the MIMOSA-5 \cite{tests:Deptuch_MIMOSA-5_resolution} and $\sigma=0.95\pm0.1~\mu$m for the MIMOSA-18 \cite{tests:Deptuch_MIMOSA-18_resolution}, which is consistent with the above result and cannot be measured with the help of the telescope used in the present setup.

\section{Charge collection characteristics}
\label{ch:tests:exp_results:charge_collection}

The response of the MIMOSA matrices to charged particles traversing its active volume was studied at DESY with the electron beam of 5~GeV. The results presented here refer to 5$\times$5 pixel clusters reconstructed in the MIMOSA-5 and MIMOSA-18, according to the reconstruction procedure described in section~\ref{ch:tests:exp_results:cluster}, with $t_{s}=4$, $t_{n}=0.5$ and $t_{s}=7$, $t_{n}=4$, respectively. Additionally, for every selected cluster a distance between position of its centre of gravity and position of the corresponding telescope track intersection with the pixel matrix was required to be smaller than 100~$\mu$m. The MIMOSA-5 and MIMOSA-18 were operated at temperatures of -7.7$^{\circ}$C and 14.4$^{\circ}$C, respectively.

\subsection{Signal-to-noise ratio (S/N)}

\begin{figure}[!h] 
	\begin{center}
	\subfigure[]{
		\includegraphics[width=0.45\textwidth,angle=90]{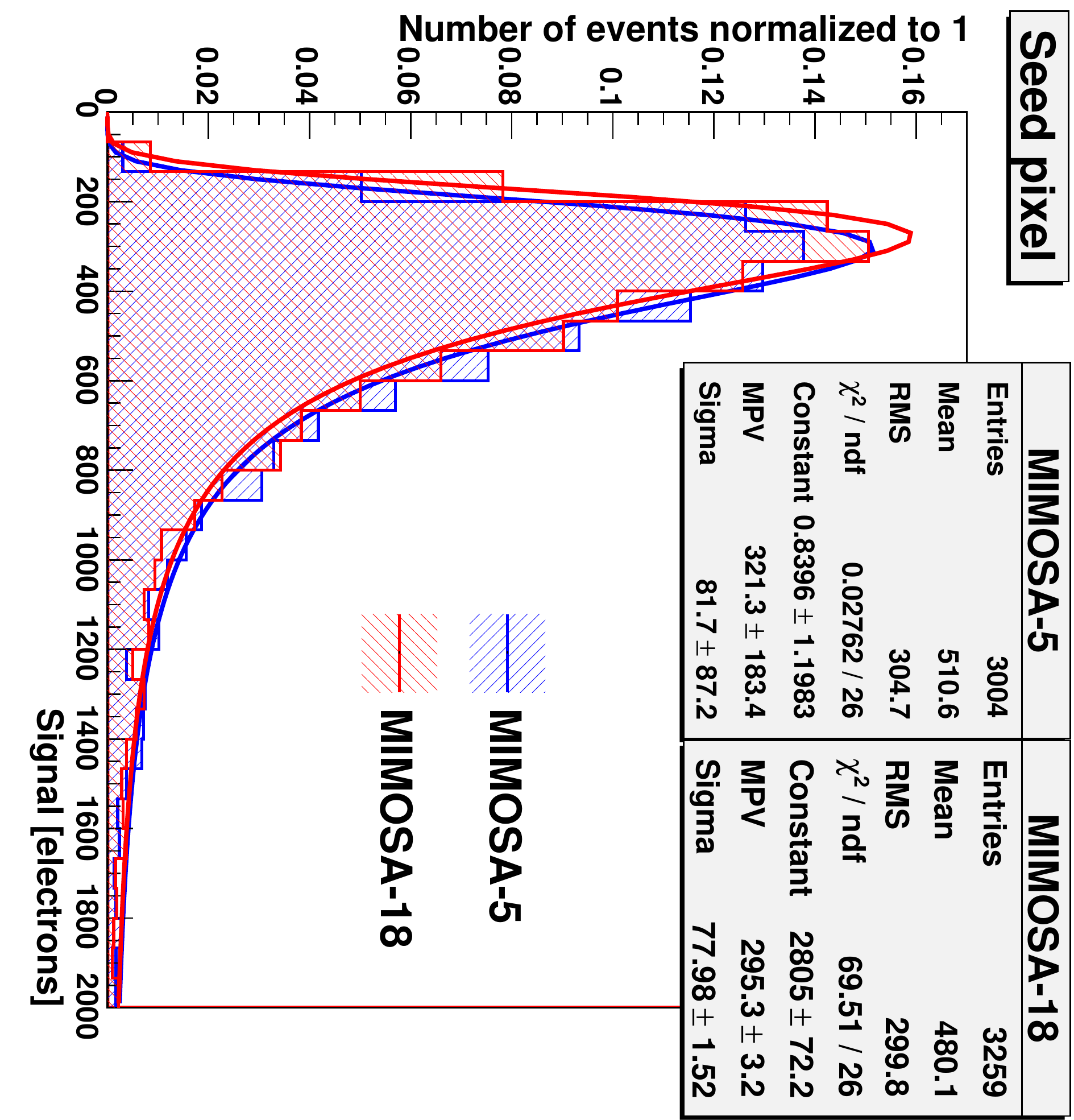}
		\label{fig:tests:Seed:Signal}
	}
	\hspace{0.1cm}
	\subfigure[]{
		\includegraphics[width=0.45\textwidth,angle=90]{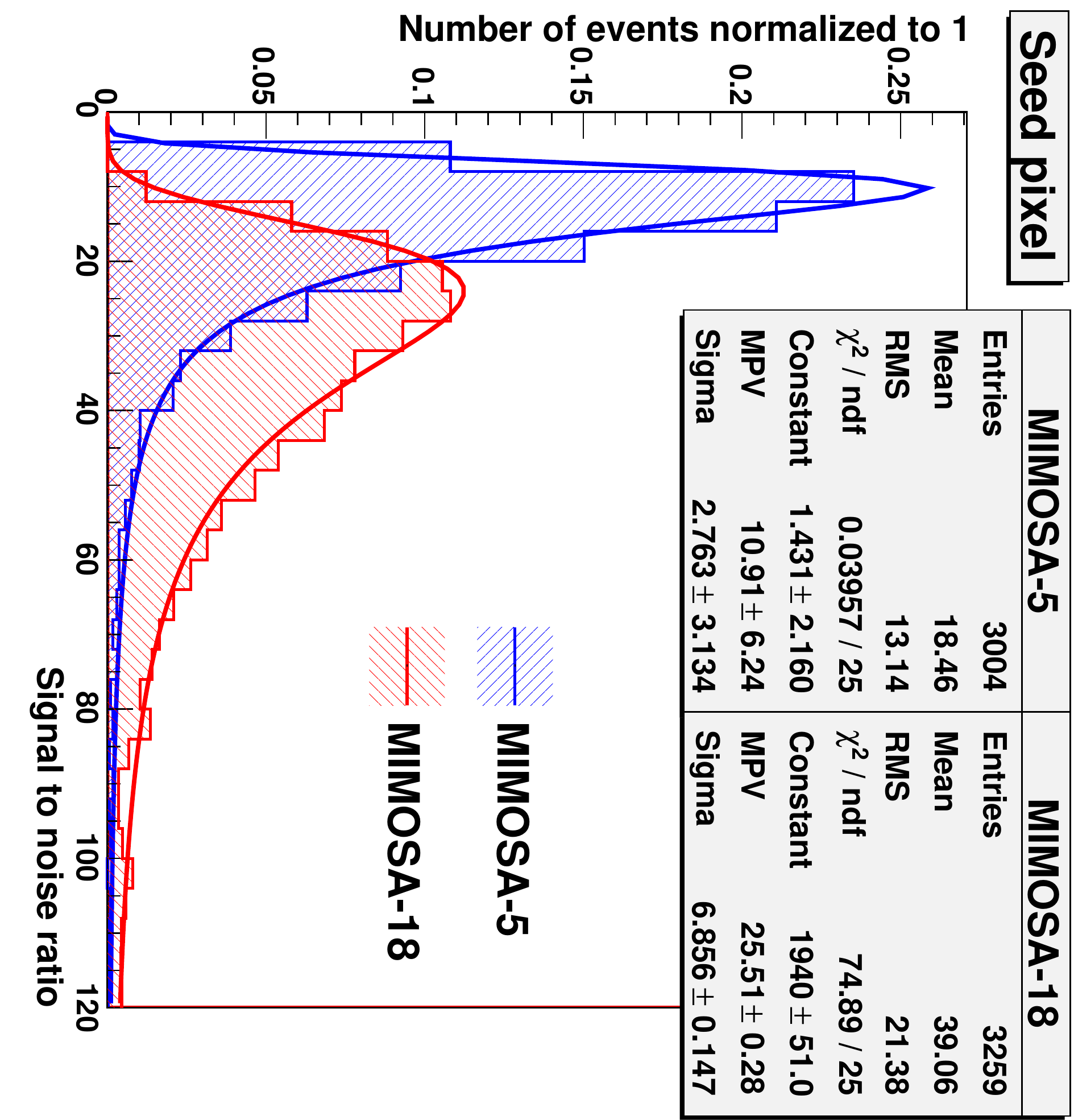}
		\label{fig:tests:Seed:S2N}
	}
	\caption[Distribution of the seed pixel signal]{(a) Distribution of the seed pixel signal for MIMOSA-5 and MIMOSA-18, fitted with the Landau function. (b) The signal-to-noise ratio of the seed pixels in the MIMOSA-5 and MIMOSA-18 prototypes fitted with the Landau function. }
	\label{fig:tests:Seed}
	\end{center}
\end{figure}

Charge distributions for the seed pixels in the MIMOSA-5 and MIMOSA-18, shown in fig.~\ref{fig:tests:Seed:Signal}, have a Landau-like shape with the most probable values (MPV) of 321 and 295 electrons, respectively (in the case of the Landau distribution it is better to use the MPV then the mean value since the latter depends substantially on the long tails of the distribution).
\label{ch:tests:exp:charge_collection:S2N}
\\
The procedure of evaluating noise in each pixel was described in section~\ref{ch:tests:exp_results:ped_noi}. The S/N ratio may be calculated for a single pixel when it becomes a seed due to particle passage. This ratio is evaluated by dividing the charge in that seed pixel by its noise. If one considers the response of a particular pixel to different events of particle passage then the charge of these pixel, being the seed, would have the Landau distribution and thus the S/N ratio would follow the similar Landau trend. If one considers a set of pixels then the single pixel Landau distribution is expected to be smeared due to distribution of noise in those pixels which enter the S/N denominator. The distribution of the S/N ratio for approx. 3k~seed pixels is shown in fig.~\ref{fig:tests:Seed:S2N} for MIMOSA-5 and MIMOSA-18. The MPV for the S/N ratio in the MIMOSA-18 is much larger than in the case of the MIMOSA-5, 25.5 and 10.9, respectively. The signal to noise ratio reflects the detector sensitivity to charge particles which is much better for the MIMOSA-18 than for the MIMOSA-5.

\subsection{Cluster charge}
\label{ch:tests:exp_results:charge_collection:cluster}

\begin{figure}[!h] 
	\begin{center}
	\subfigure[]{
		\includegraphics[width=0.45\textwidth,angle=0]{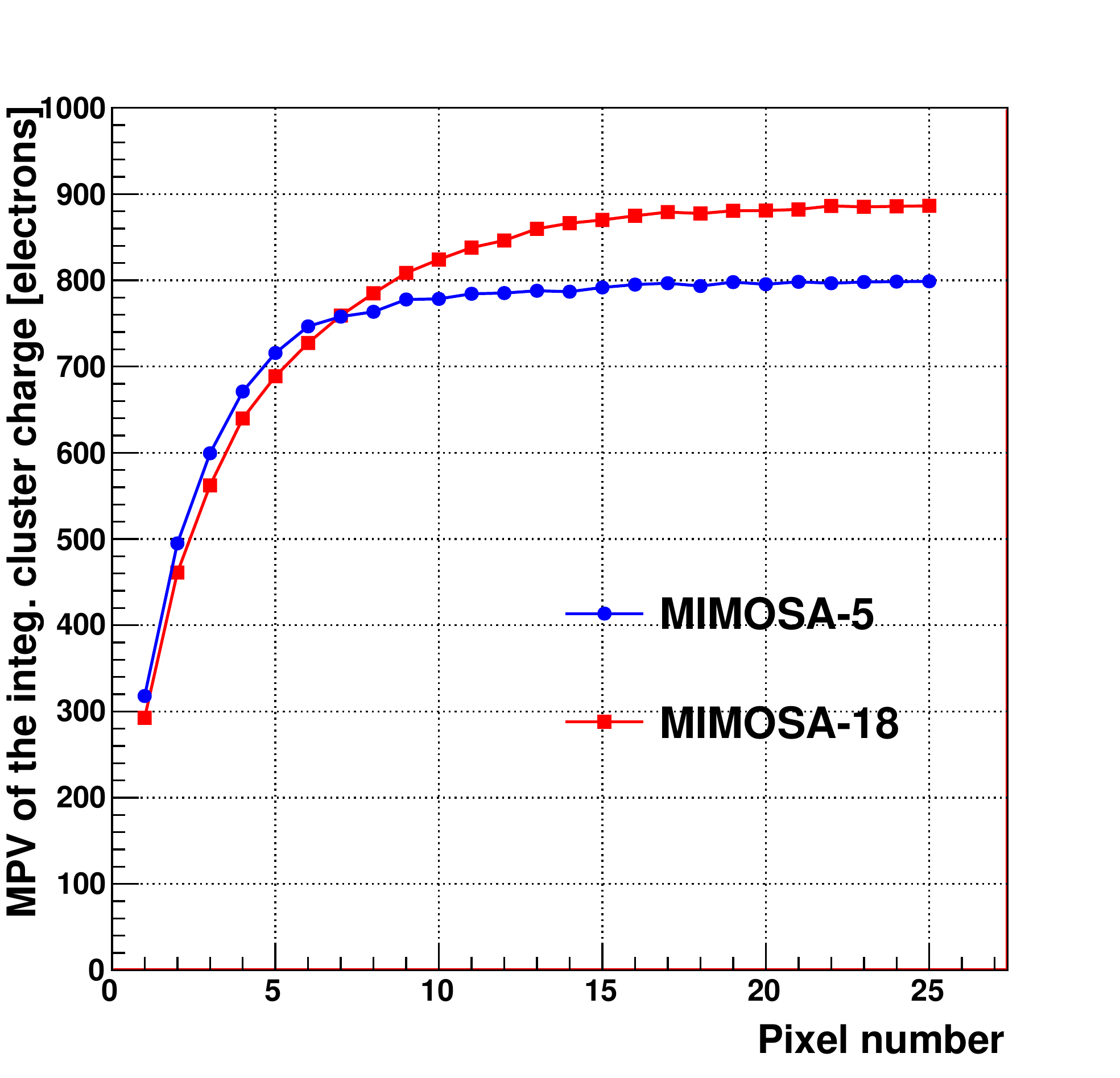}
		\label{fig:tests:Cluster:Multiplicity}
	}
	\hspace{0.1cm}
	\subfigure[]{
		\includegraphics[width=0.45\textwidth,angle=0]{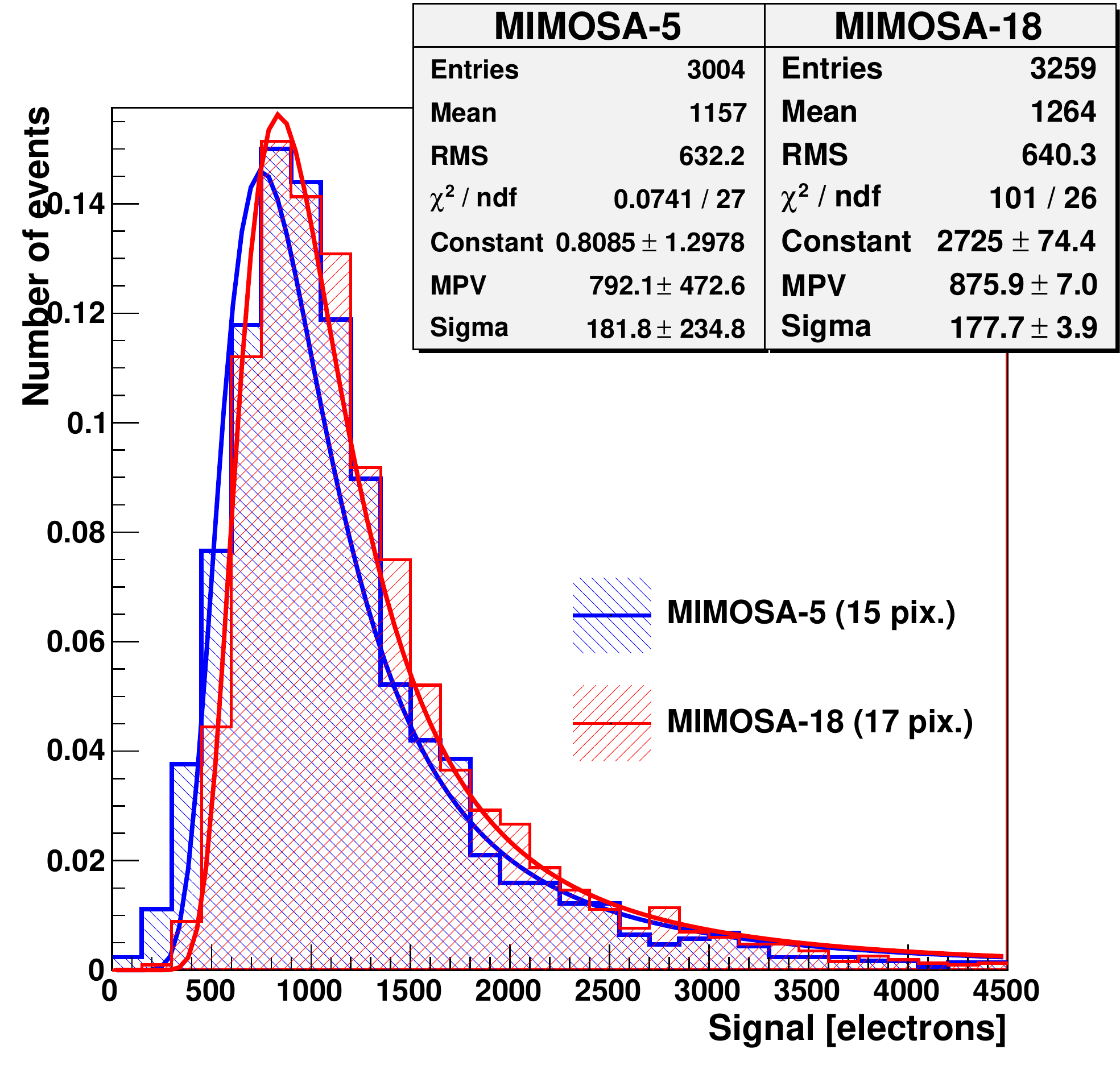}
		\label{fig:tests:Clsuetr:Charge}
	}
	\caption[The MPV of the integrated cluster charge as a function of the pixel number $n$ for MIMOSA-5 and MIMOSA-18]{(a) The MPV of the integrated cluster charge as a function of the pixel number $n$ for MIMOSA-5 and MIMOSA-18 (see text). (b) Signal distribution for the clusters composed of 15 and 17 pixels for the MIMOSA-5 and MIMOSA-18 prototypes.}
	\label{fig:tests:Clsuetr}
	\end{center}
\end{figure}

In order to determine the cluster charge, the 25 pixels in each cluster were sorted in a descending order with respect to the signal-to-noise ratio as described in section~\ref{ch:tests:exp_results:X-rays:charge_spreading}. Next the integrated charge $Q_{n}$ for the first $n$ pixels in each cluster was calculated: $Q_{n} = \sum_{i=1}^{n}q_{i}$, where $q_{i}$ is a charge of the $i$-th pixel. Each quantity $Q_{n}$ follows the Landau distribution. The MPV was determined for each one by fitting the Landau function and the results are shown as a function of $n$ in fig.~\ref{fig:tests:Cluster:Multiplicity} for MIMOSA-5 and MIMOSA-18.
\\
The amount of collected charge increases with $n$ until it flattens out or, equally, reaches a saturation level. The saturation values correspond to approx. 800 and 890 electrons for the MIMOSA-5 and MIMOSA-18, respectively. The seed pixel in the MIMOSA-5 accounts on average for $\sim$40\% while in the MIMOSA-18 for $\sim$33\% of the total collected charge. The 99\% of the total cluster charge in the case of the MIMOSA-5 (pixels of 17$\times$17~$\mu$m$^{2}$) is contained in 15 out of 25 pixels while for the MIMOSA-18 (pixel 10$\times$10~$\mu$m$^{2}$) in 17 out of 25 pixels. These studies shown that almost the entire cluster charge is contained within approx. 4$\times$4 pixel cluster (for perpendicular tracks).\\
Distributions of charges collected in the first 15 and 17 pixels of clusters in the MIMOSA-5 and MIMOSA-18, respectively, are shown in fig.~\ref{fig:tests:Clsuetr:Charge}. These distributions follow the Landau shape, as is demonstrated by the fitted curves. The shapes of the Landau distributions are similar for both matrices as one would expect since the amount of generated charge depends on the thickness of the epitaxial layer, which is the same in both cases (14~$\mu$m). 

\section{Study of cluster shapes}
\label{ch:tests:exp_results:cluster_shape}

\begin{figure}[!h]
        \begin{center}
                \includegraphics[width=0.6\textwidth]{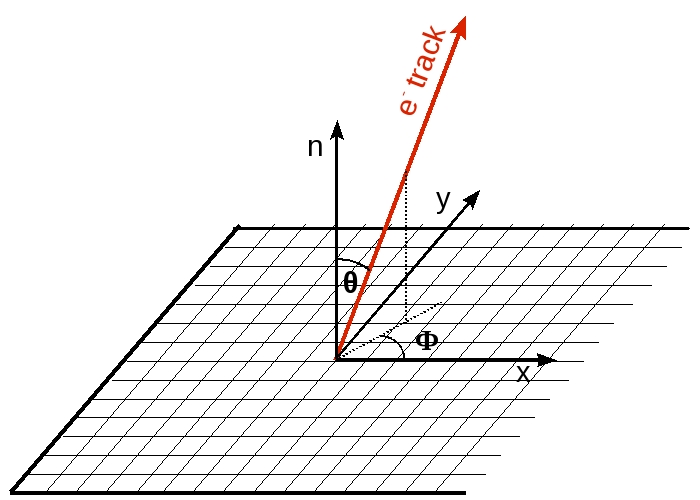}
                \caption[Definition of the polar and azimuthal angles, $\theta$ and $\phi$, in the pixel matrix coordinate system]{Definition of the polar and azimuthal angles, $\theta$ and $\phi$, in the pixel matrix coordinate system with the $X$ axis parallel to the edge; n - normal to the pixel plane.}
                \label{fig:tests:exp:TrackOrientation}
        \end{center}
\end{figure}

\begin{figure}[!h] 
	\begin{center}
	\subfigure[]{
		\includegraphics[width=0.45\textwidth,angle=90]{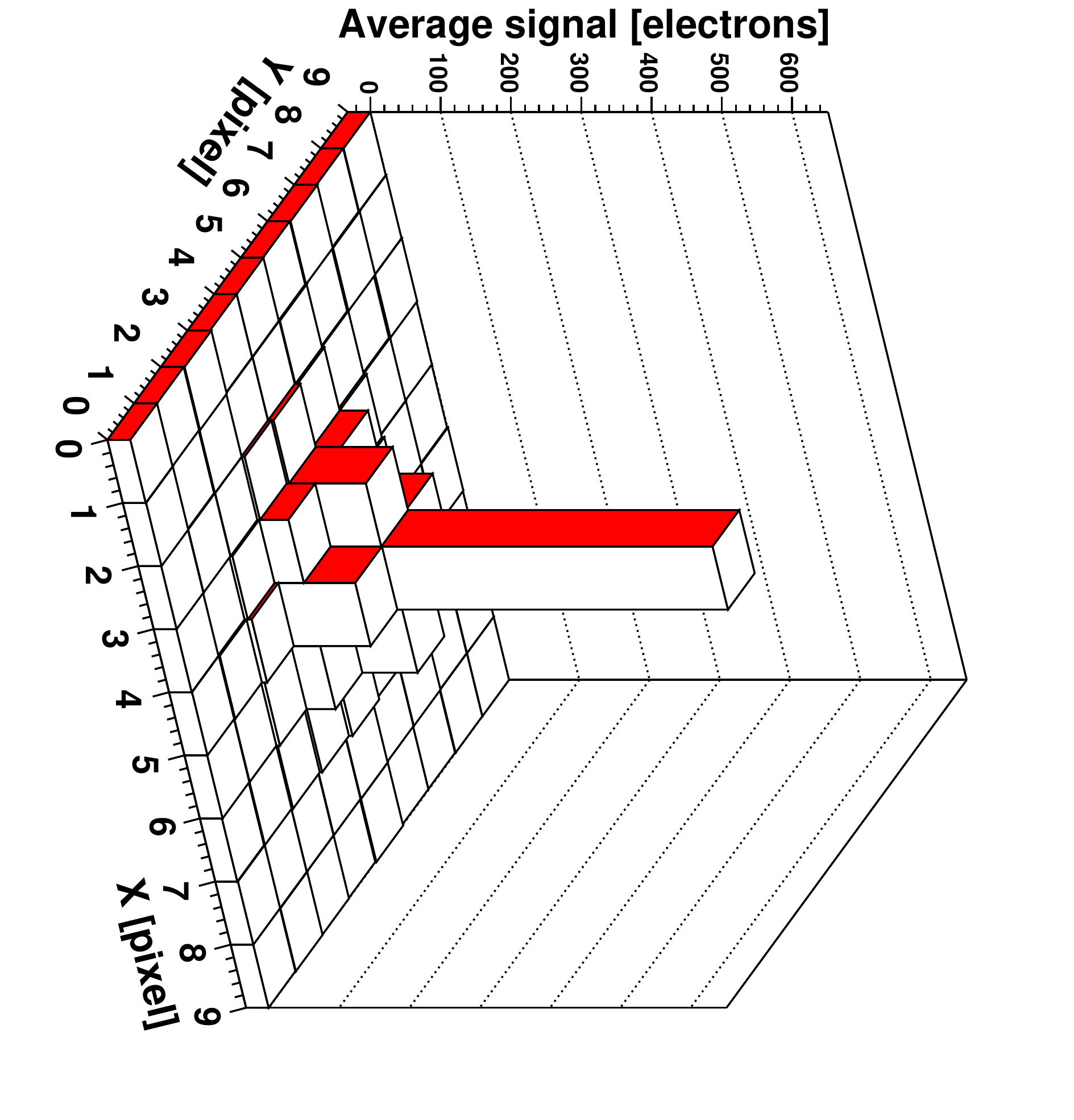}
		\label{fig:tests:MeanCl2012}
	}
	\hspace{0.1cm}
	\subfigure[]{
		\includegraphics[width=0.45\textwidth,angle=90]{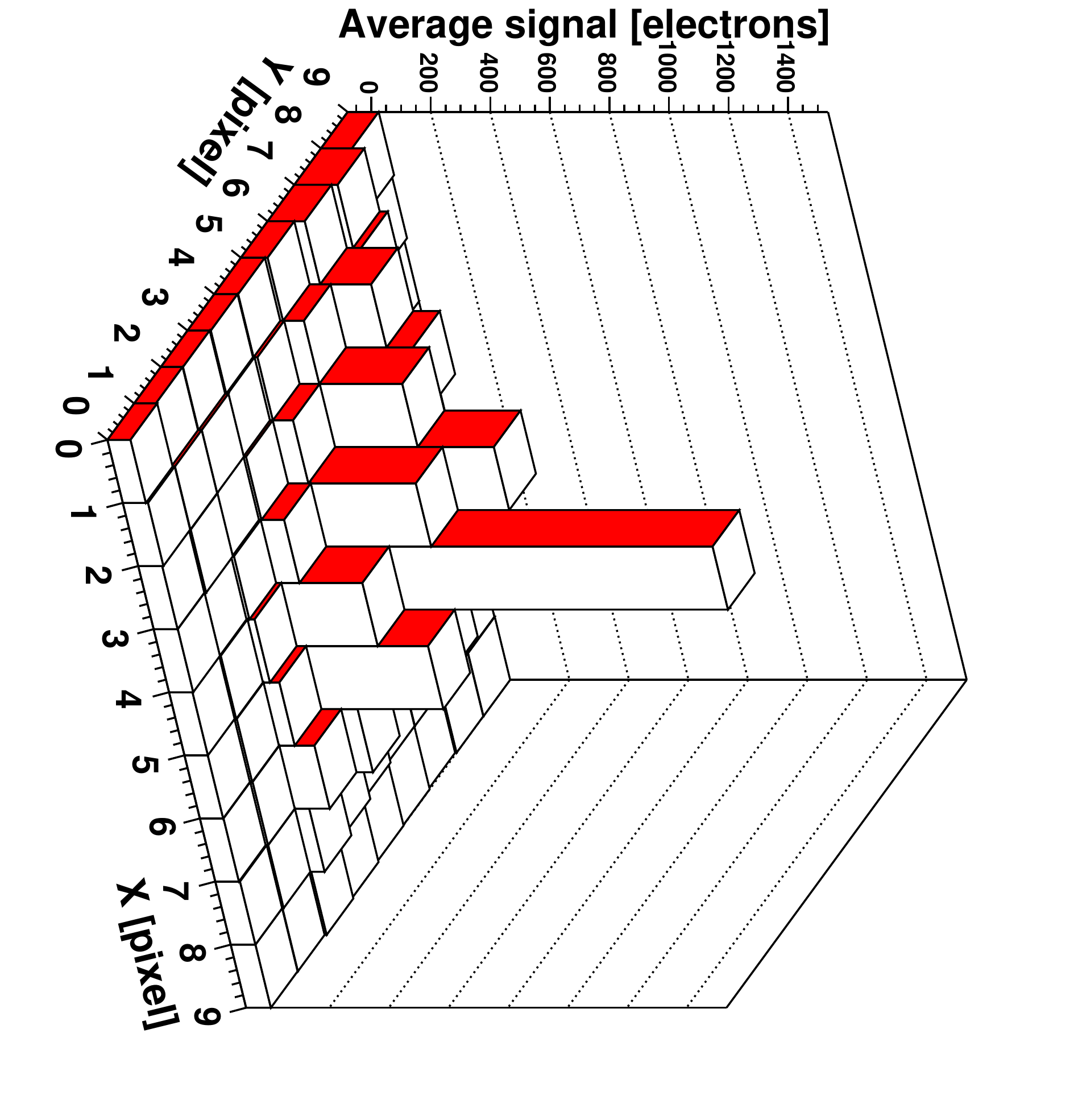}
		\label{fig:tests:MeanCl2021}
	}
	\caption[Spatial distribution of charge in an averaged cluster in the MIMOSA-5 for two different incident angle]{Spatial distribution of charge in an averaged cluster in the MIMOSA-5 for two different incident angles: (a) $\theta = 4.3^{\circ}$ and $\phi = -44.0^{\circ}$, (b) $\theta = 78.1^{\circ}$ and $\phi = -37.8^{\circ}$.}
	\label{fig:tests:MeanCluster}
	\end{center}
\end{figure}

\begin{figure}[!h] 
	\begin{center}
	\subfigure[]{
		\includegraphics[width=0.45\textwidth,angle=0]{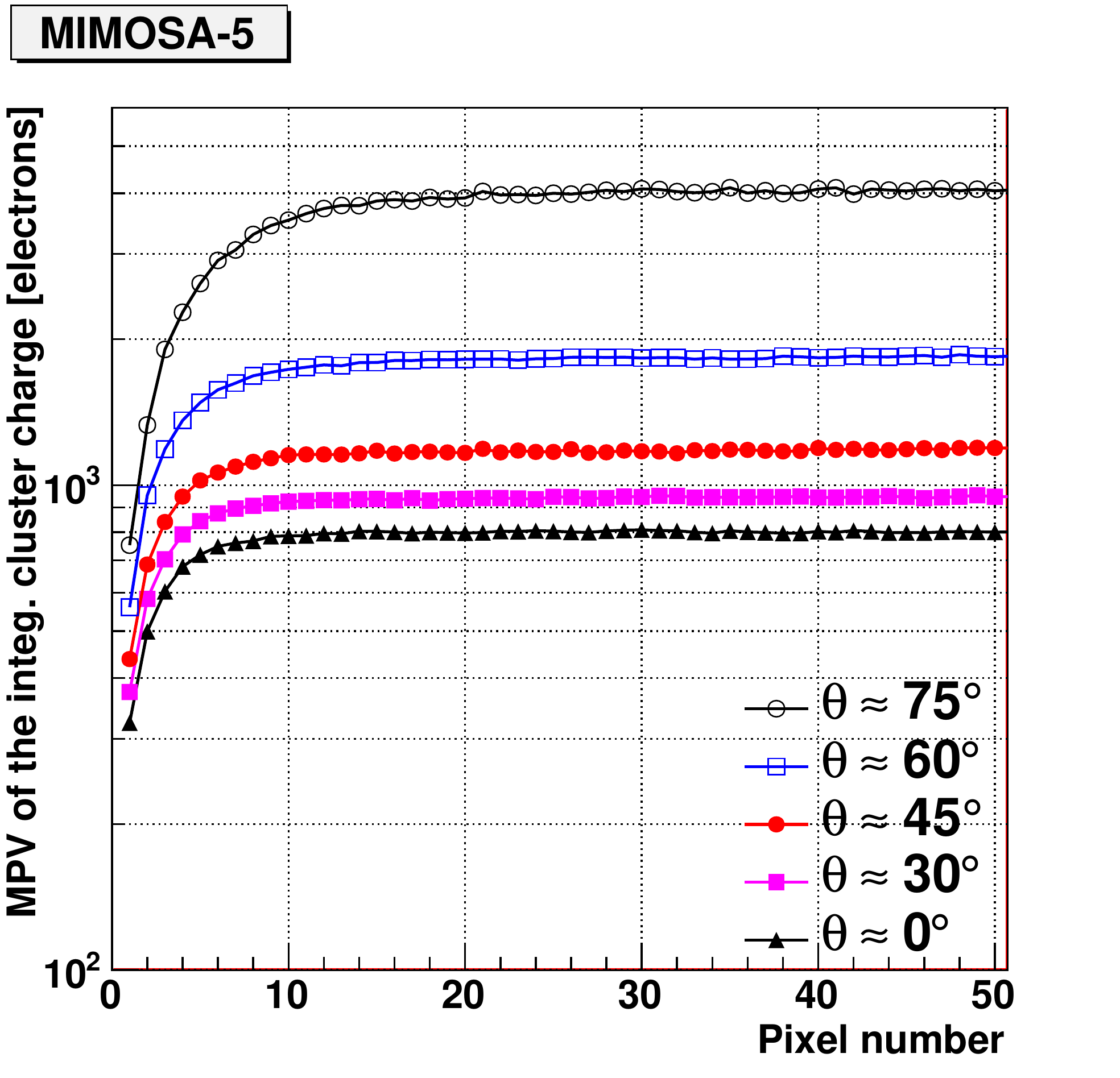}
		\label{fig:tests:inclined_multiplicity:M5}
	}
	\hspace{0.1cm}
	\subfigure[]{
		\includegraphics[width=0.45\textwidth,angle=0]{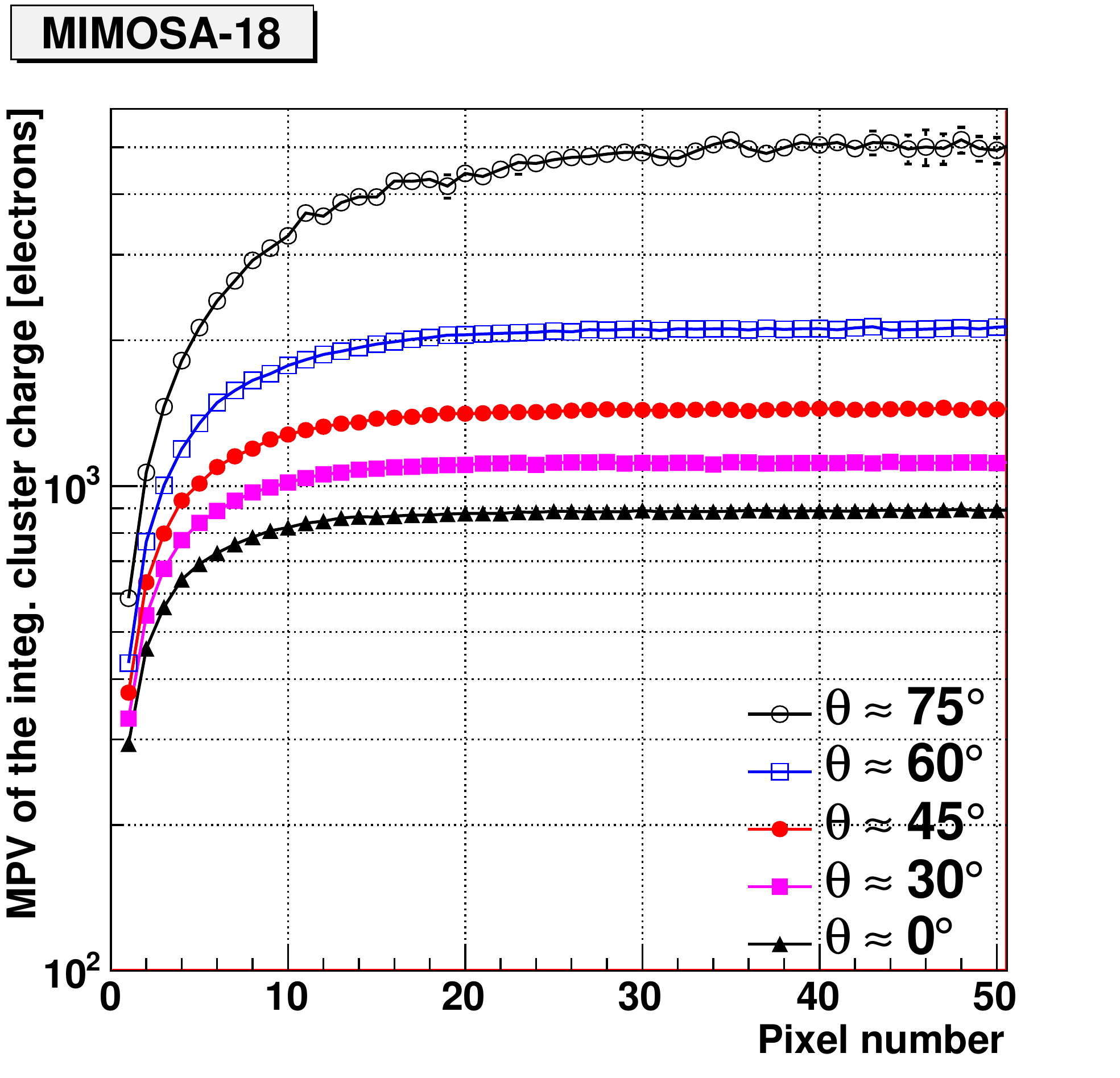}
		\label{fig:tests:inclined_multiplicity:M18}
	}
	\caption[The MPV of the integrated cluster charge as a function of the pixel number $n$ for different track inclinations]{The MPV of the integrated cluster charge as a function of the pixel number $n$ for different track inclinations in the case of (a) MIMOSA-5 and (b) MIMOSA-18 (see text).}
	\label{fig:tests:inclined_multiplicity}
	\end{center}
\end{figure}

\begin{figure}[!h] 
	\begin{center}
	\subfigure[]{
		\includegraphics[width=0.45\textwidth,angle=0]{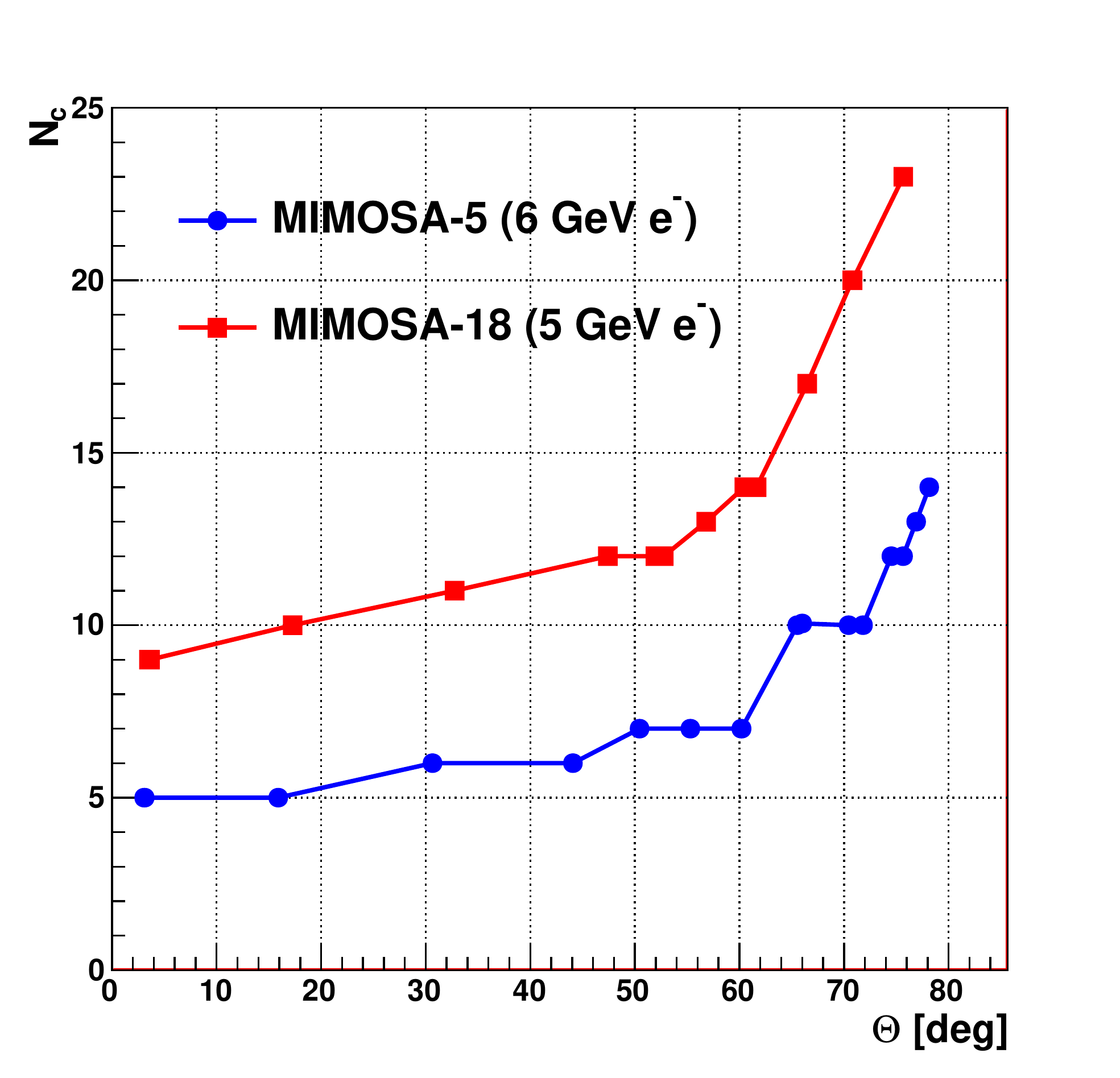}
		\label{fig:tests:inclination_scan:multiplicity}
	}
	\hspace{0.1cm}
	\subfigure[]{
		\includegraphics[width=0.45\textwidth,angle=0]{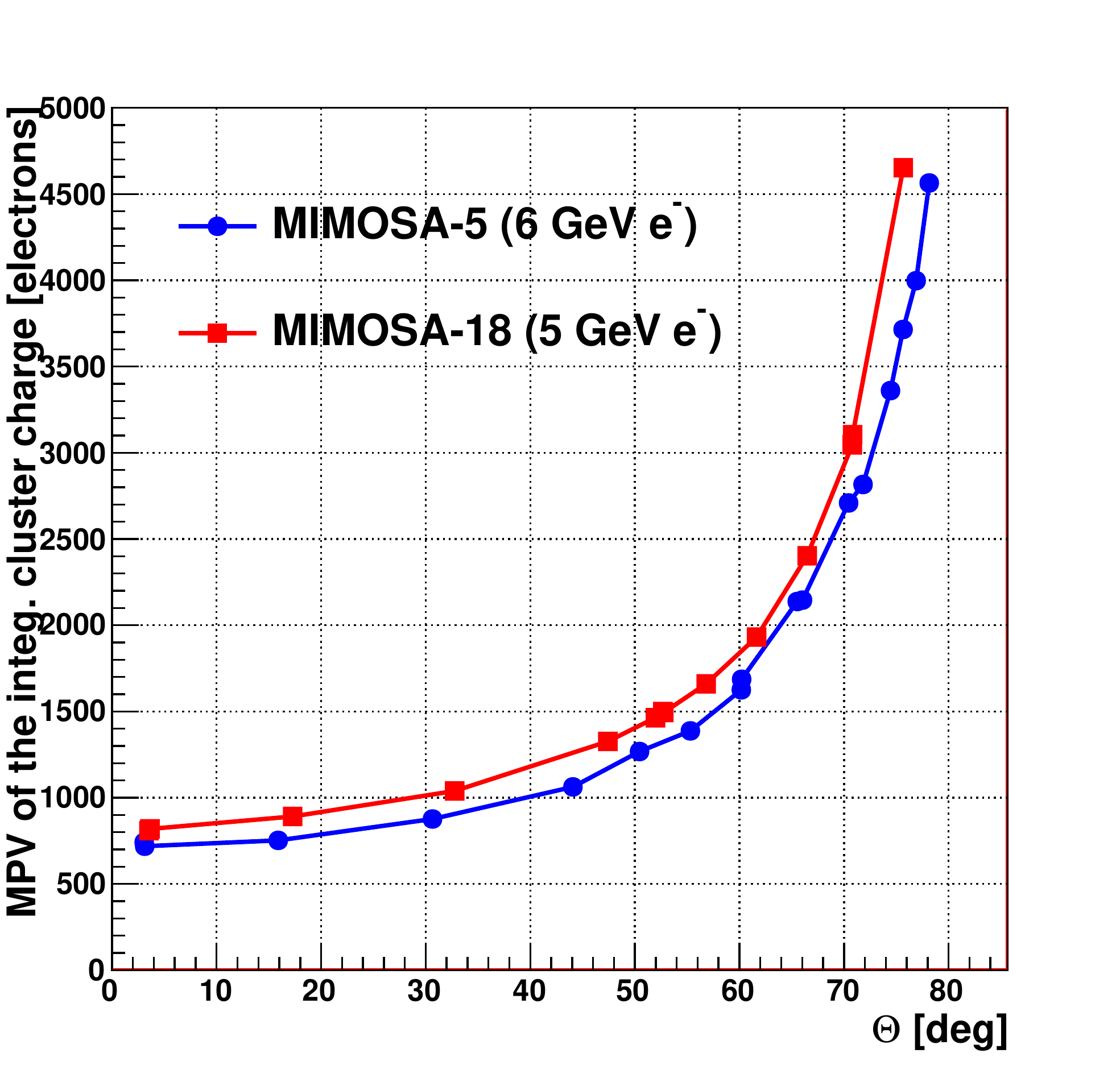}
		\label{fig:tests:inclination_scan:charge}
	}
	\caption[MAPS charge collection properties measured with the inclined tracks]{(a) The $N_{c}$ dependence on the incident angle $\theta$. The $N_{c}$ is defined as a number of pixels in which 90\% of the total cluster charge is contained; (b)~dependence of the charge MPV on the incident angle $\theta$ for clusters composed of $N_{c}$ (see text).}
	\label{fig:tests:inclination_scam}
	\end{center}
\end{figure}

\begin{figure}[!h] 
	\begin{center}
	\subfigure[Beam settings - $\phi = -45.3^{\circ}$, $\theta = 55.3^{\circ}$]{
		\includegraphics[width=0.45\textwidth,angle=90]{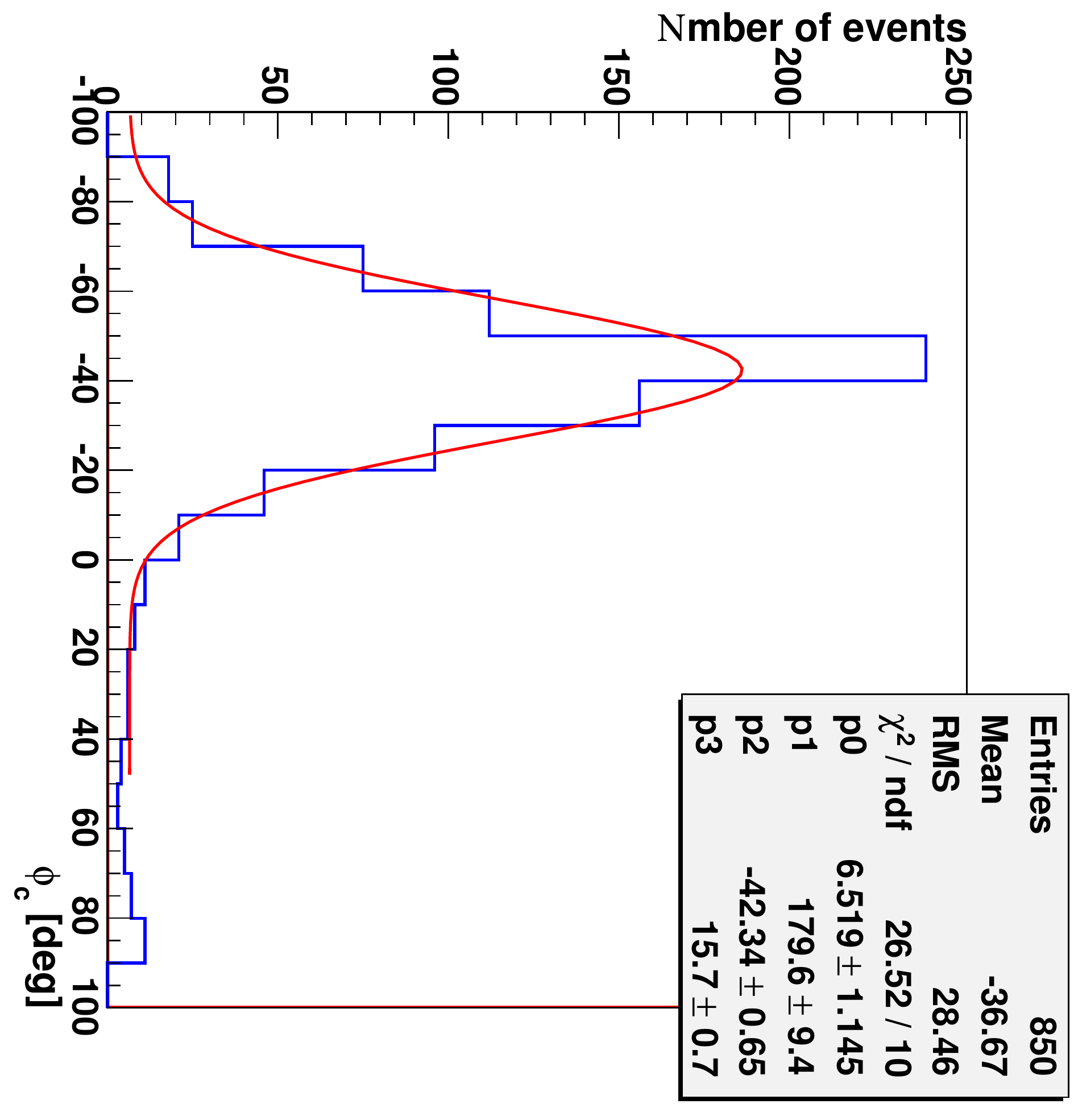}
		\label{fig:tests:PhiRecoM5:2017}
	}
	\hspace{0.1cm}
	\subfigure[Beam settings - $\phi = 0.3^{\circ}$, $\theta = 60.2^{\circ}$]{
		\includegraphics[width=0.45\textwidth,angle=90]{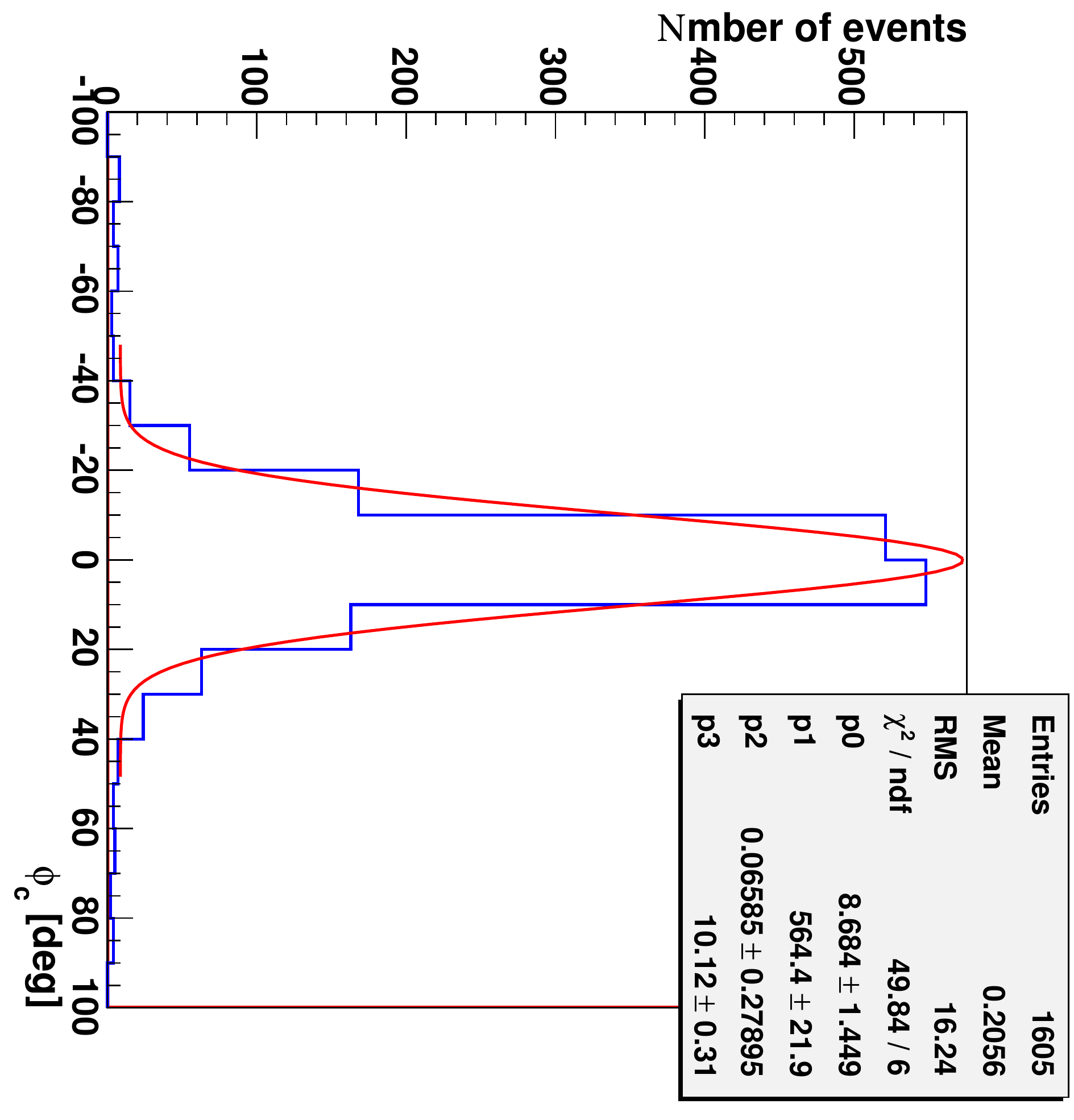}
		\label{fig:tests:PhiRecoM5:2008}
	}
	\subfigure[Beam settings - $\phi = -1.37^{\circ}$, $\theta = 70.4^{\circ}$]{
		\includegraphics[width=0.45\textwidth,angle=90]{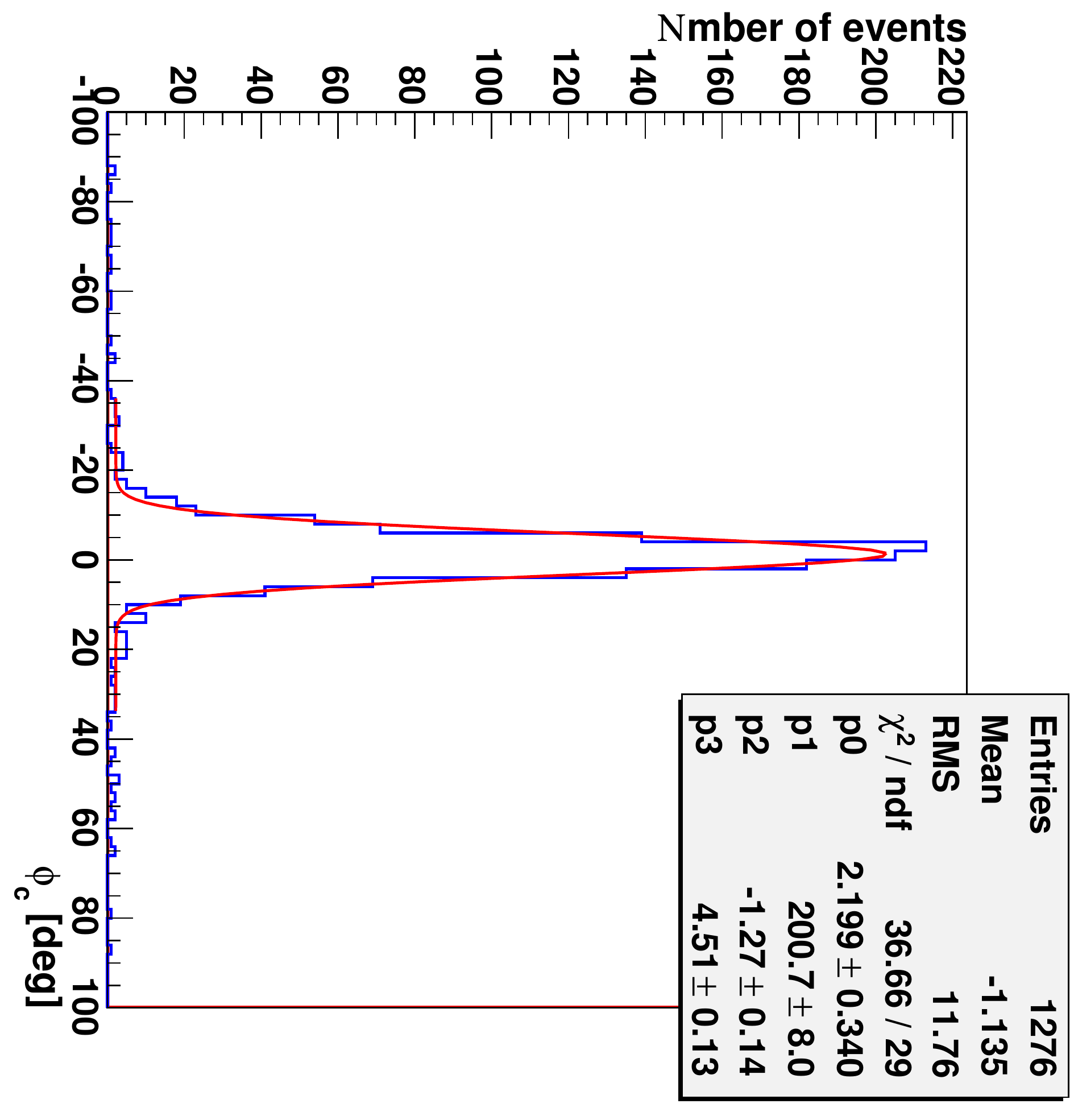}
		\label{fig:tests:PhiRecoM5:4005}
	}
	\hspace{0.1cm}
	\subfigure[Beam settings - $\phi = -37.8^{\circ}$, $\theta = 78.1^{\circ}$]{
		\includegraphics[width=0.45\textwidth,angle=90]{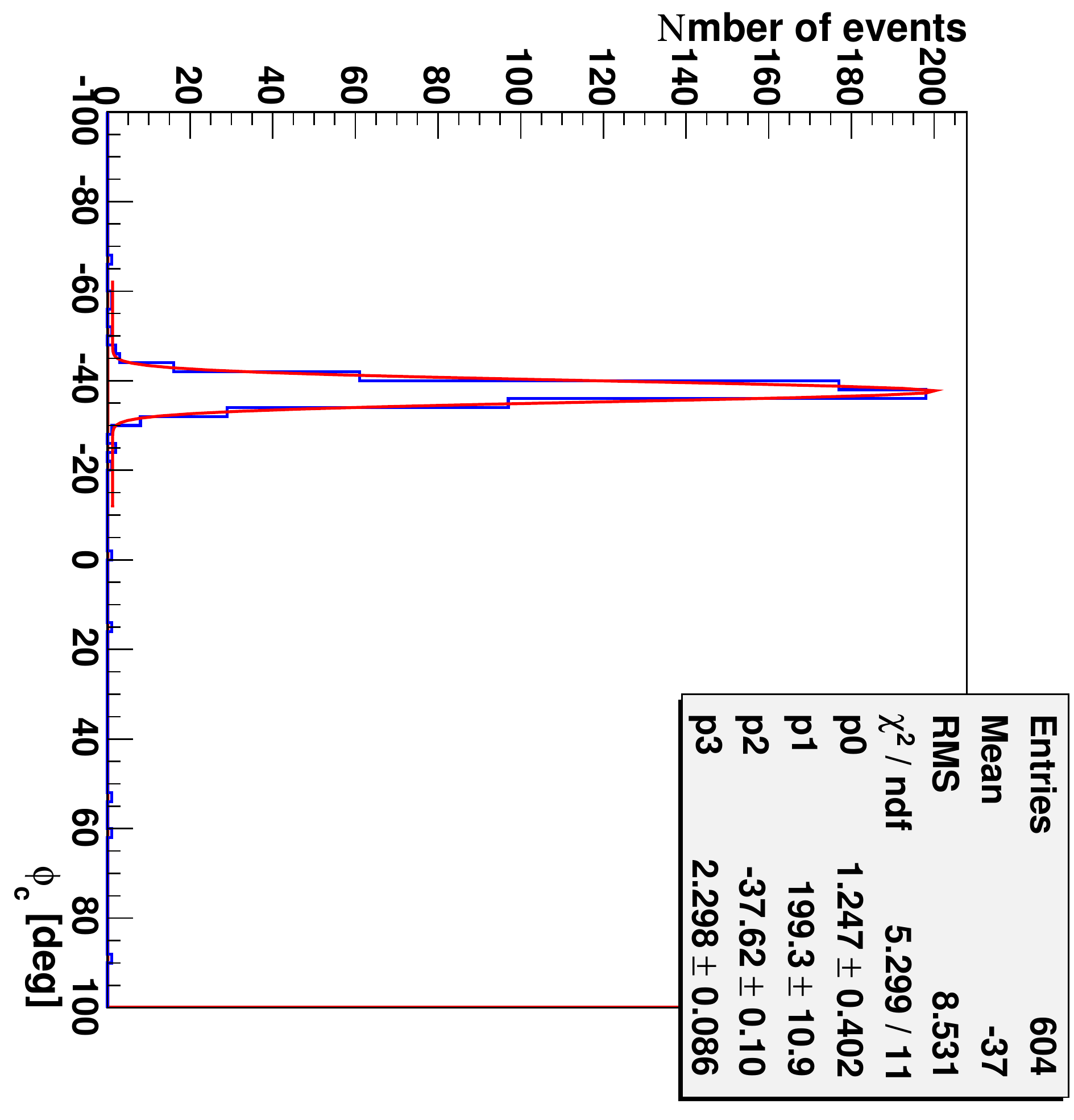}
		\label{fig:tests:PhiRecoM5:2021}
	}
	\caption[Distributions of the reconstructed $\phi_c$ angles of clusters measured in the MIMOSA-5]{Distributions of the reconstructed $\phi_c$ angles of clusters measured with 6~GeV electrons in MIMOSA-5 for different beam settings as indicated. The peak in each histogram was fitted with the Gaussian function shifted by a constant value $p0$. The resulting values of the Gauss mean and dispersion are displayed in the insets: $p2$ and $p3$, respectively.}
	\label{fig:tests:PhiRecoM5}
	\end{center}
\end{figure}

\begin{figure}[!h] 
	\begin{center}
	\subfigure[Beam settings - $\phi = 1.8^{\circ}$, $\theta = 56.8^{\circ}$]{
		\includegraphics[width=0.45\textwidth,angle=90]{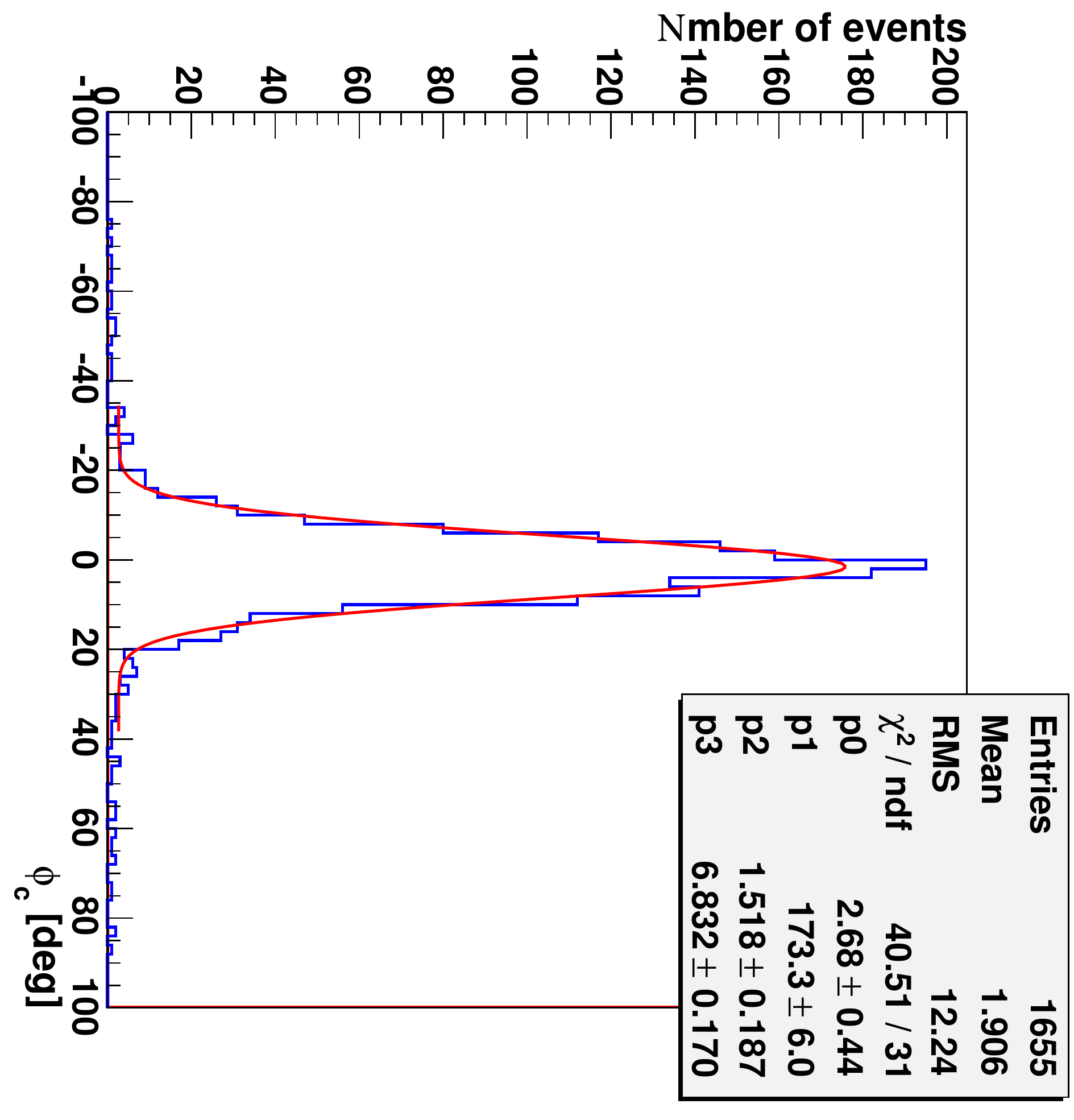}
		\label{fig:tests:PhiRecoM18:4030}
	}
	\hspace{0.1cm}
	\subfigure[Beam settings - $\phi = 1.13^{\circ}$, $\theta = 61.6^{\circ}$]{
		\includegraphics[width=0.45\textwidth,angle=90]{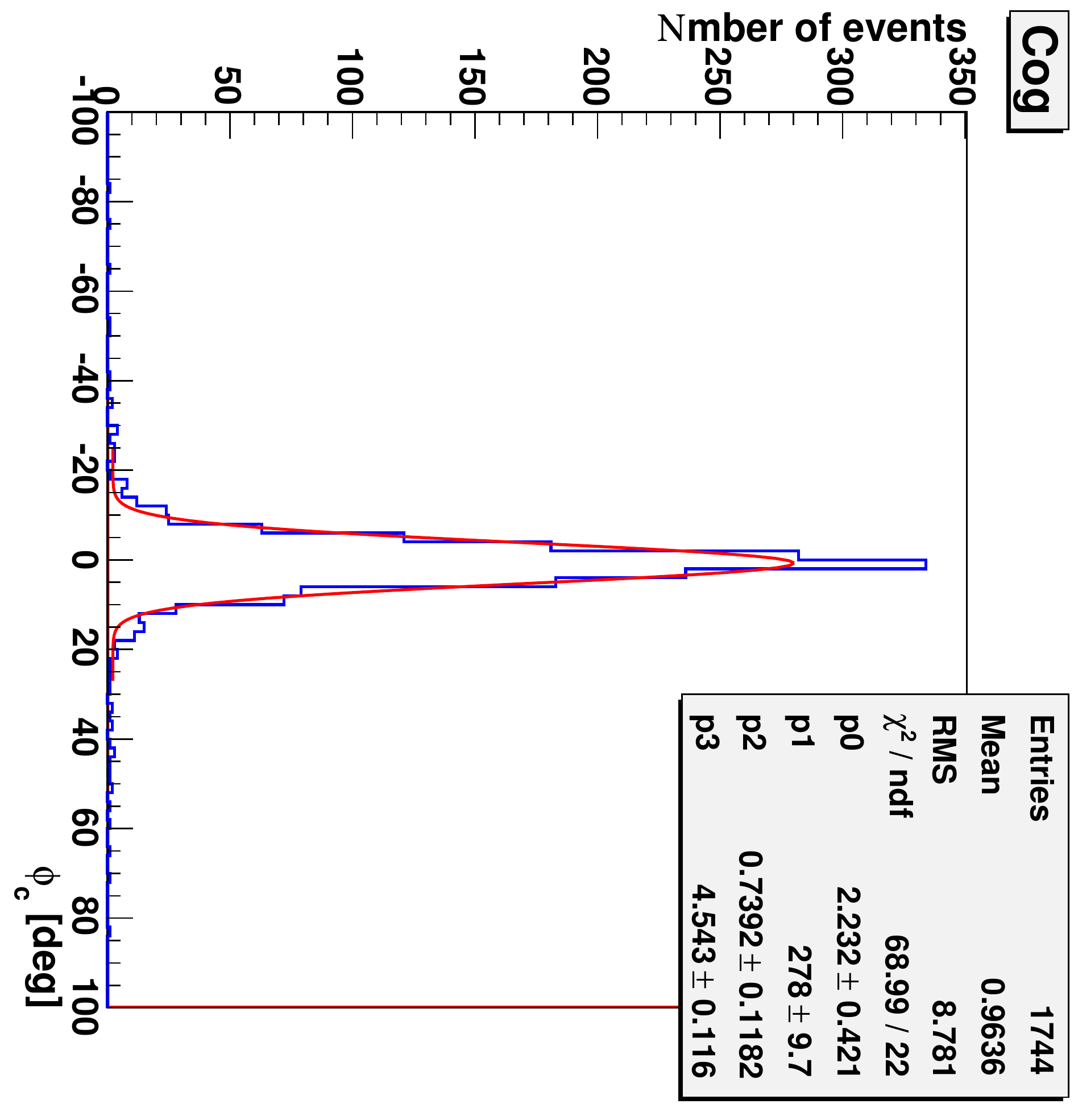}
		\label{fig:tests:PhiRecoM18:4018}
	}
	\subfigure[Beam settings - $\phi = 1.0^{\circ}$, $\theta = 70.8^{\circ}$]{
		\includegraphics[width=0.45\textwidth,angle=90]{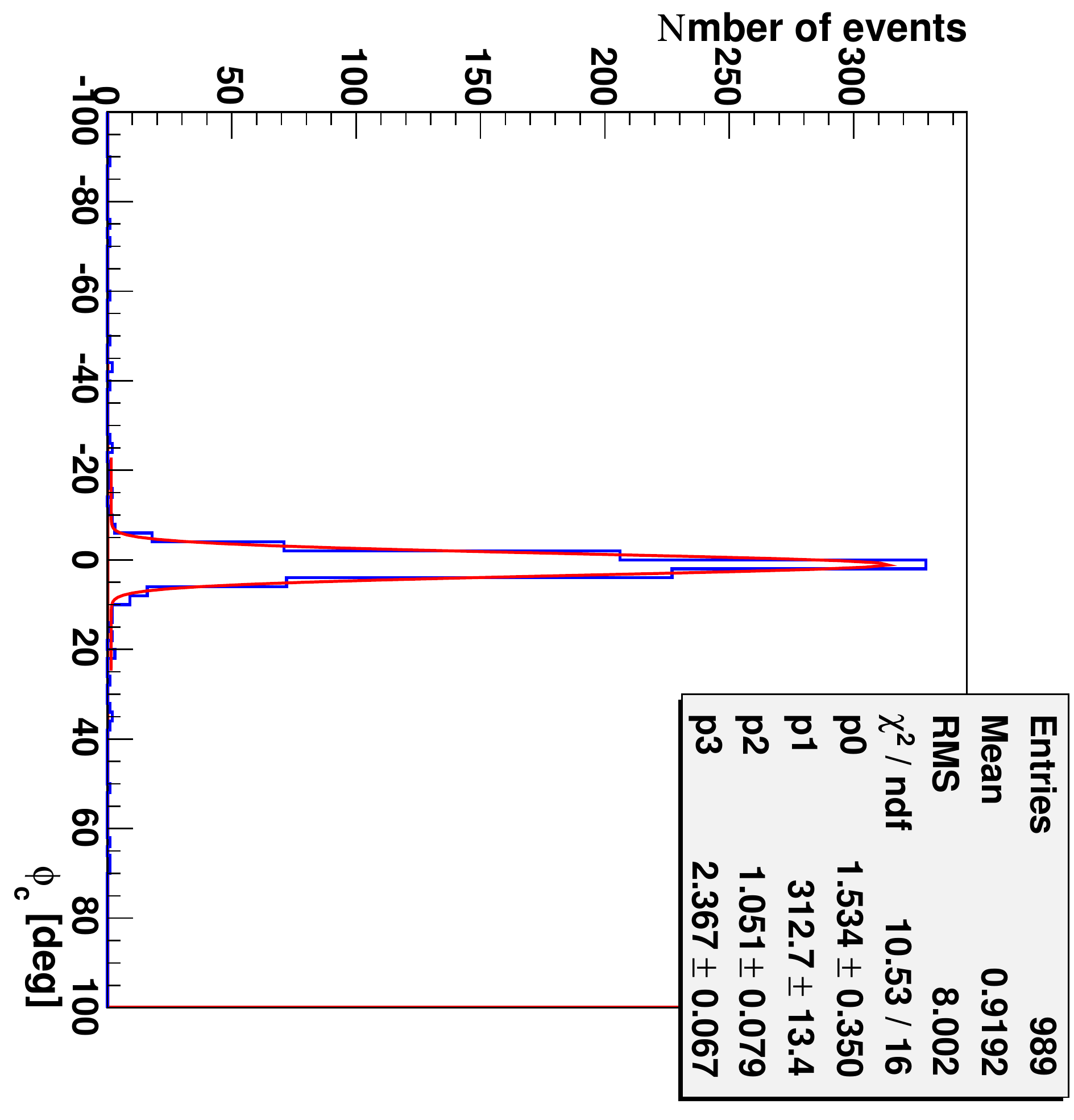}
		\label{fig:tests:PhiRecoM18:4024}
	}
	\hspace{0.1cm}
	\subfigure[Beam settings - $\phi = 0.5^{\circ}$, $\theta = 75.7^{\circ}$]{
		\includegraphics[width=0.45\textwidth,angle=90]{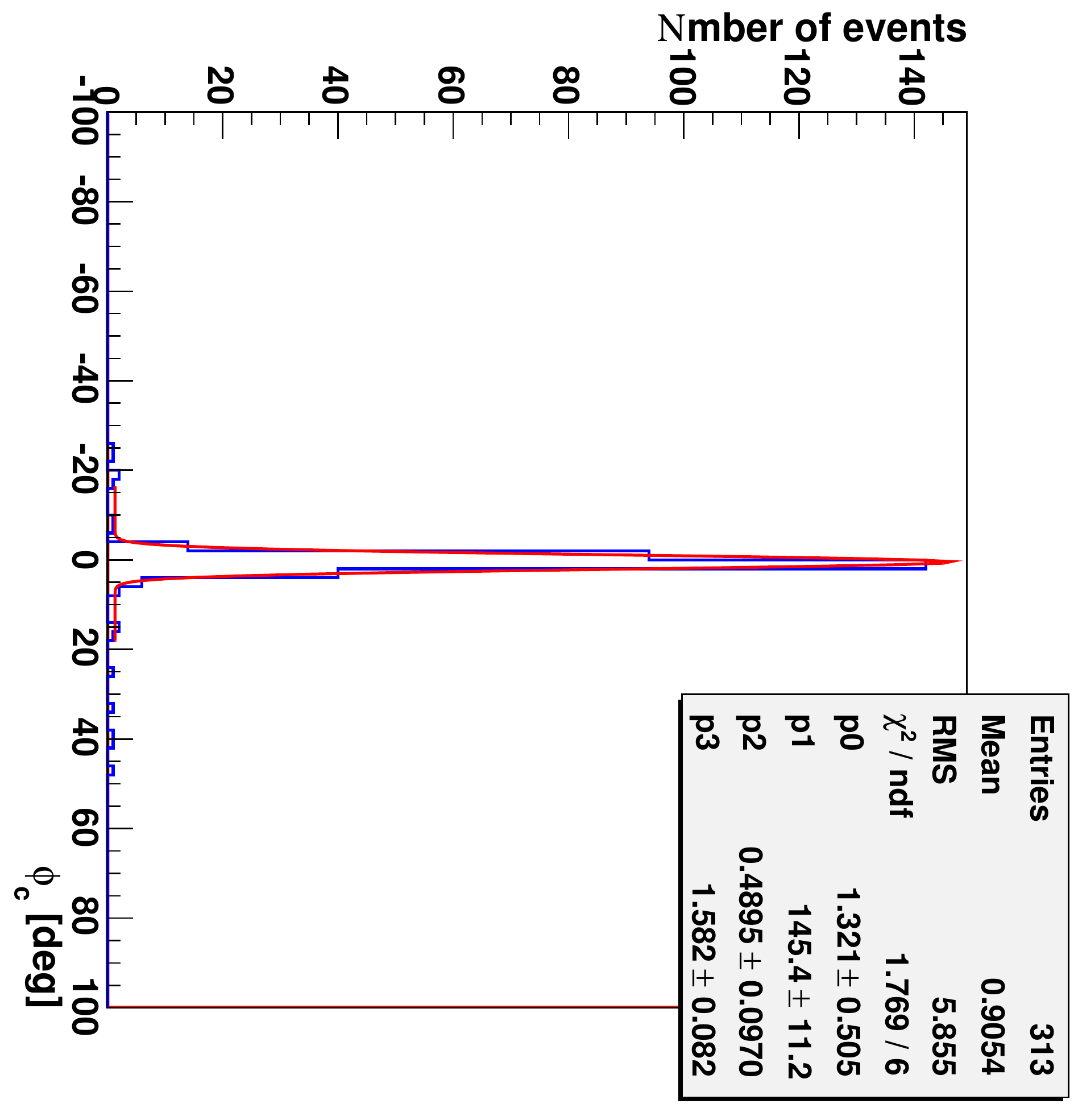}
		\label{fig:tests:PhiRecoM18:4016}
	}
	\caption[Distributions of the reconstructed $\phi_c$ angles of clusters measured in the MIMOSA-18]{Distributions of the reconstructed $\phi_c$ angles of clusters measured with 5~GeV electrons in MIMOSA-18 for different beam settings as indicated. The peak in each histogram was fitted with the Gaussian function shifted by a constant value $p0$. The resulting values of the Gauss mean and dispersion are displayed in the insets: $p2$ and $p3$, respectively.}
	\label{fig:tests:PhiRecoM18}
	\end{center}
\end{figure}

\begin{figure}[!h] 
	\begin{center}
	\includegraphics[width=0.8\textwidth,angle=90]{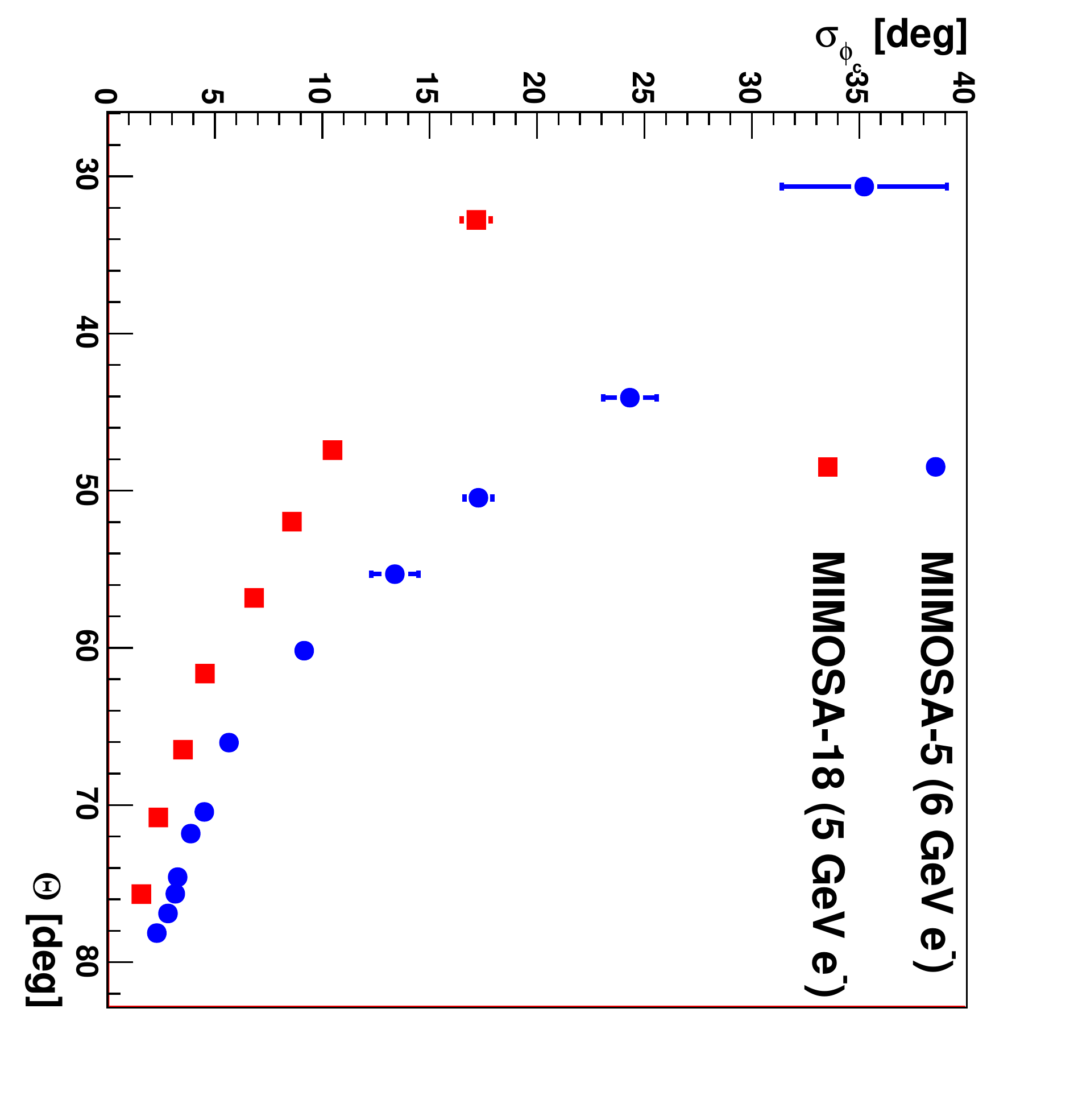}
	\caption[Dispersion of the $\phi_{c}$ distribution as a function of the incident angle $\theta$ in MIMOSA-5 and MIMOSA-18]{Dispersion of the $\phi_{c}$ distribution as a function of the incident angle $\theta$ in MIMOSA-5 and MIMOSA-18.}
	\label{fig:tests:PhiReco}
	\end{center}
\end{figure}

\begin{figure}[!h] 
	\begin{center}
	\subfigure[Charged particle tracks inclination $\theta = 55.3^{\circ}$]{
		\includegraphics[width=0.45\textwidth,angle=90]{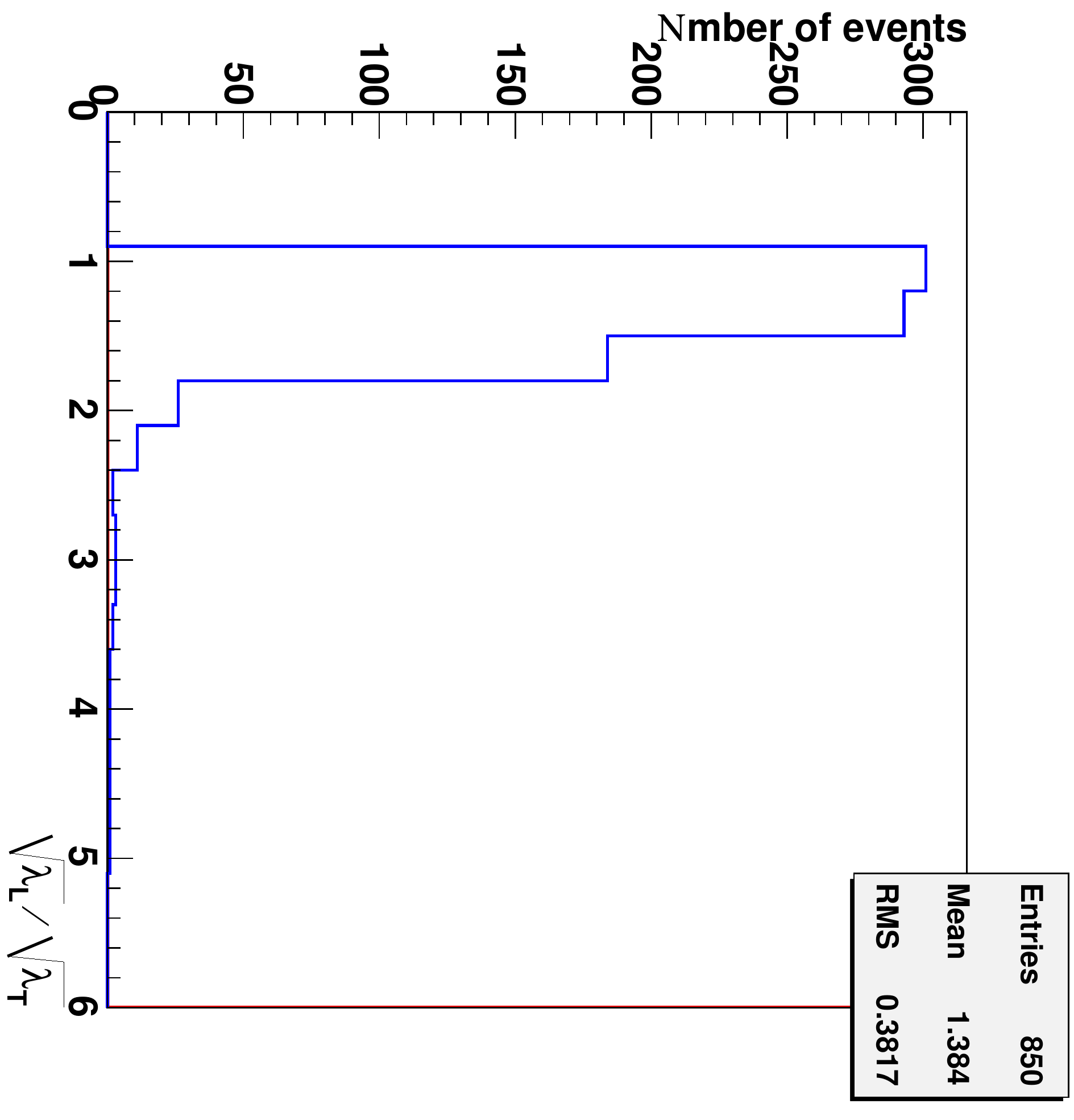}
		\label{fig:tests:LambdaRecoM5:2017}
	}
	\hspace{0.1cm}
	\subfigure[Charged particle tracks inclination $\theta = 60.2^{\circ}$]{
		\includegraphics[width=0.45\textwidth,angle=90]{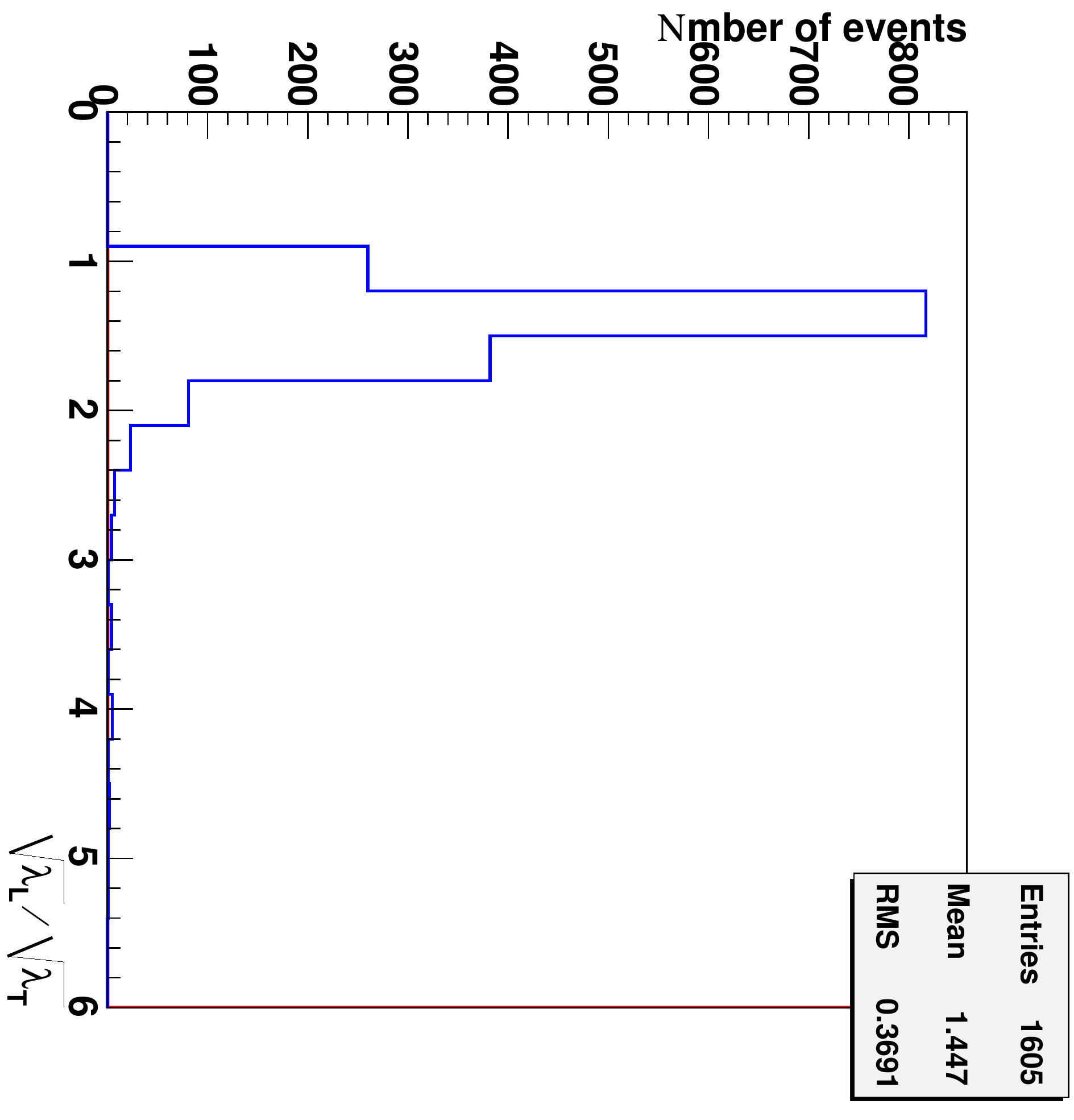}
		\label{fig:tests:LambdaRecoM5:2008}
	}
	\subfigure[Charged particle tracks inclination $\theta = 70.4^{\circ}$]{
		\includegraphics[width=0.45\textwidth,angle=90]{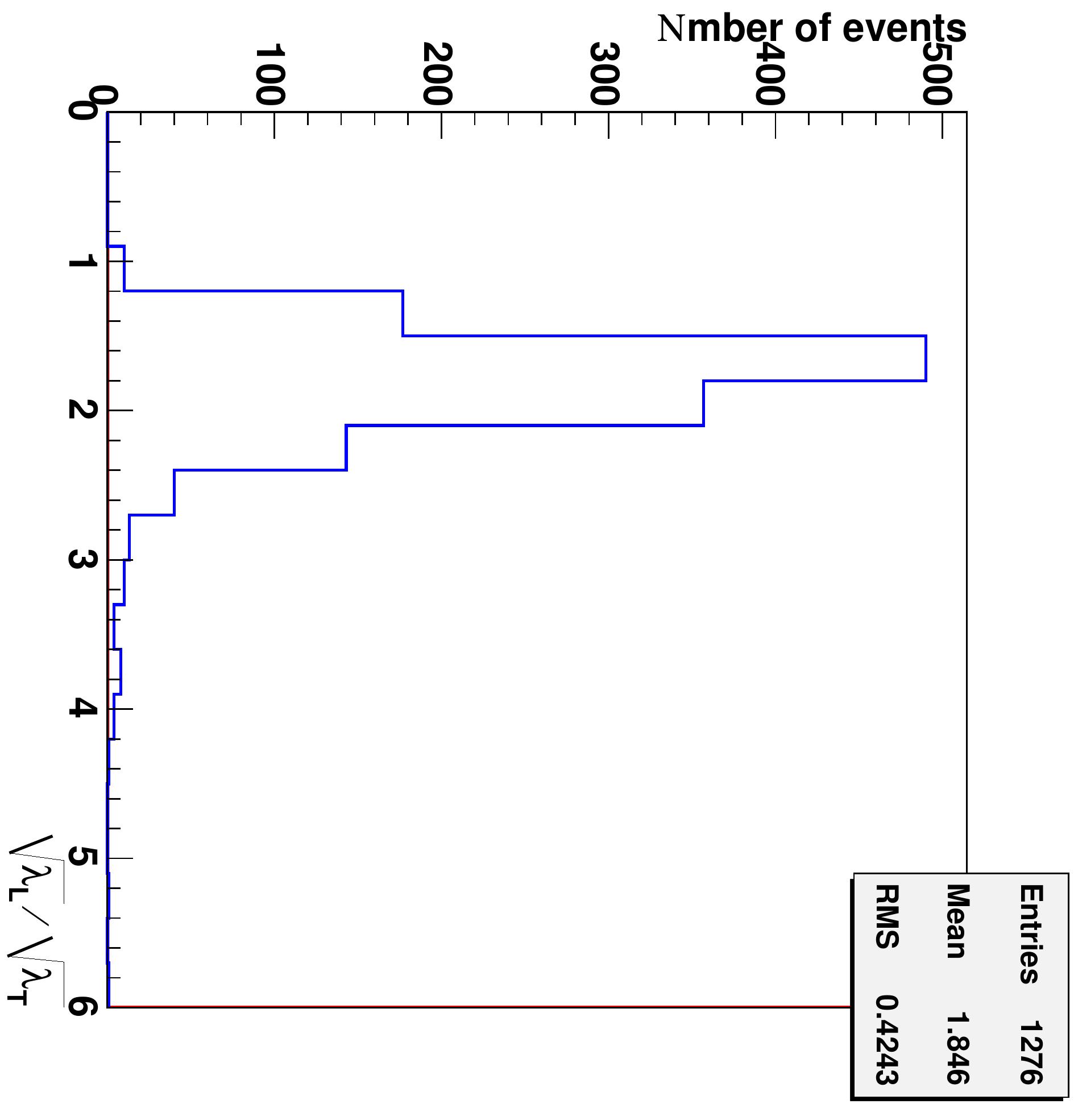}
		\label{fig:tests:LambdaRecoM5:4005}
	}
	\hspace{0.1cm}
	\subfigure[Charged particle tracks inclination $\theta = 78.1^{\circ}$]{
		\includegraphics[width=0.45\textwidth,angle=90]{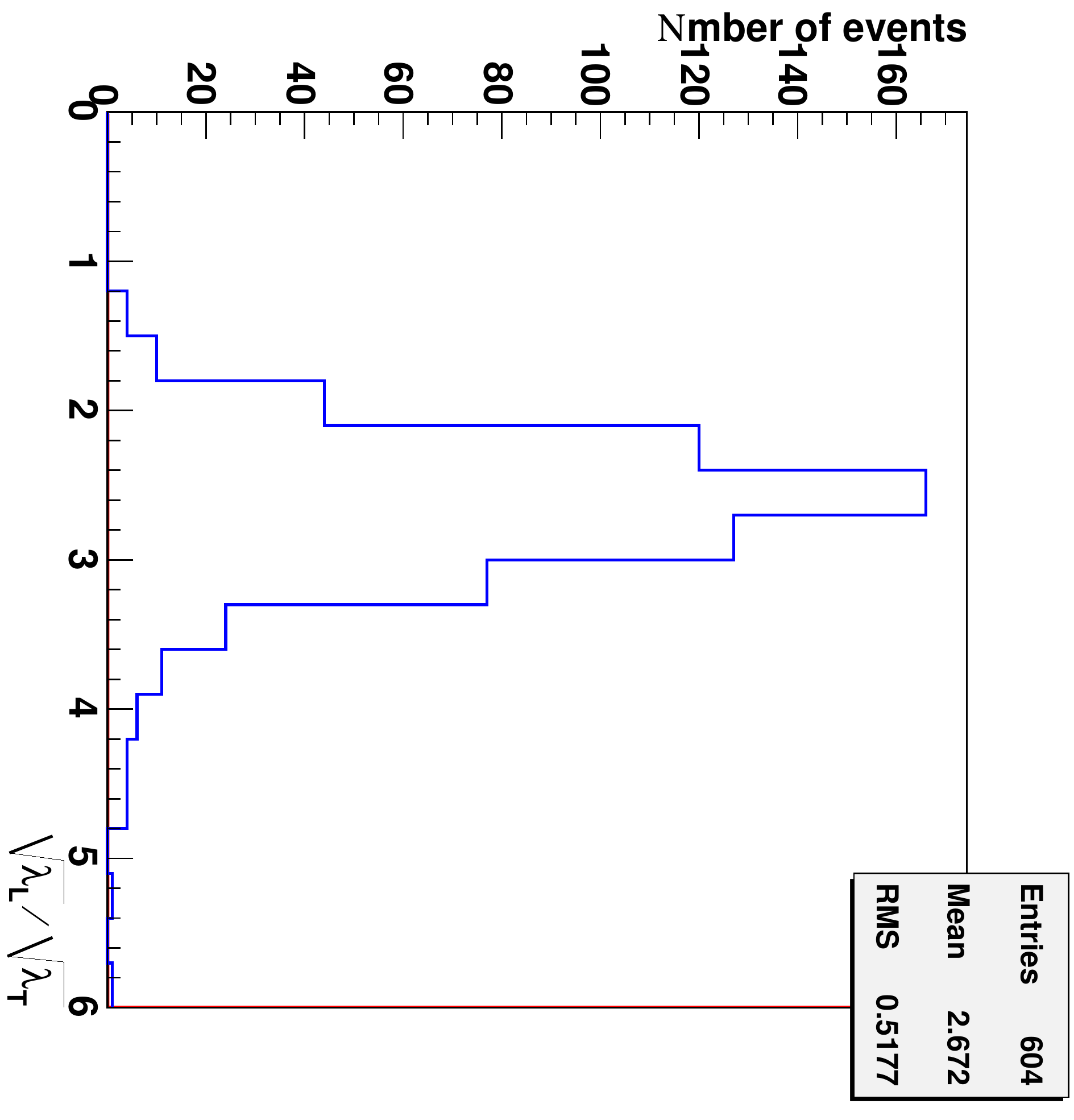}
		\label{fig:tests:LambdaRecoM5:2021}
	}
	\caption[Distributions of the $\sqrt{\lambda_{L}/\lambda_{T}}$ ratio reconstructed in clusters measured in MIMOSA-5]{Distributions of the $\sqrt{\lambda_{L}/\lambda_{T}}$ ratio reconstructed in clusters measured with 6~GeV electrons in MIMOSA-5 for different angle $\theta$.}
	\label{fig:tests:LambdaRecoM5}
	\end{center}
\end{figure}

\begin{figure}[!h] 
	\begin{center}
	\subfigure[Charged particle tracks inclination $\theta = 56.8^{\circ}$]{
		\includegraphics[width=0.45\textwidth,angle=90]{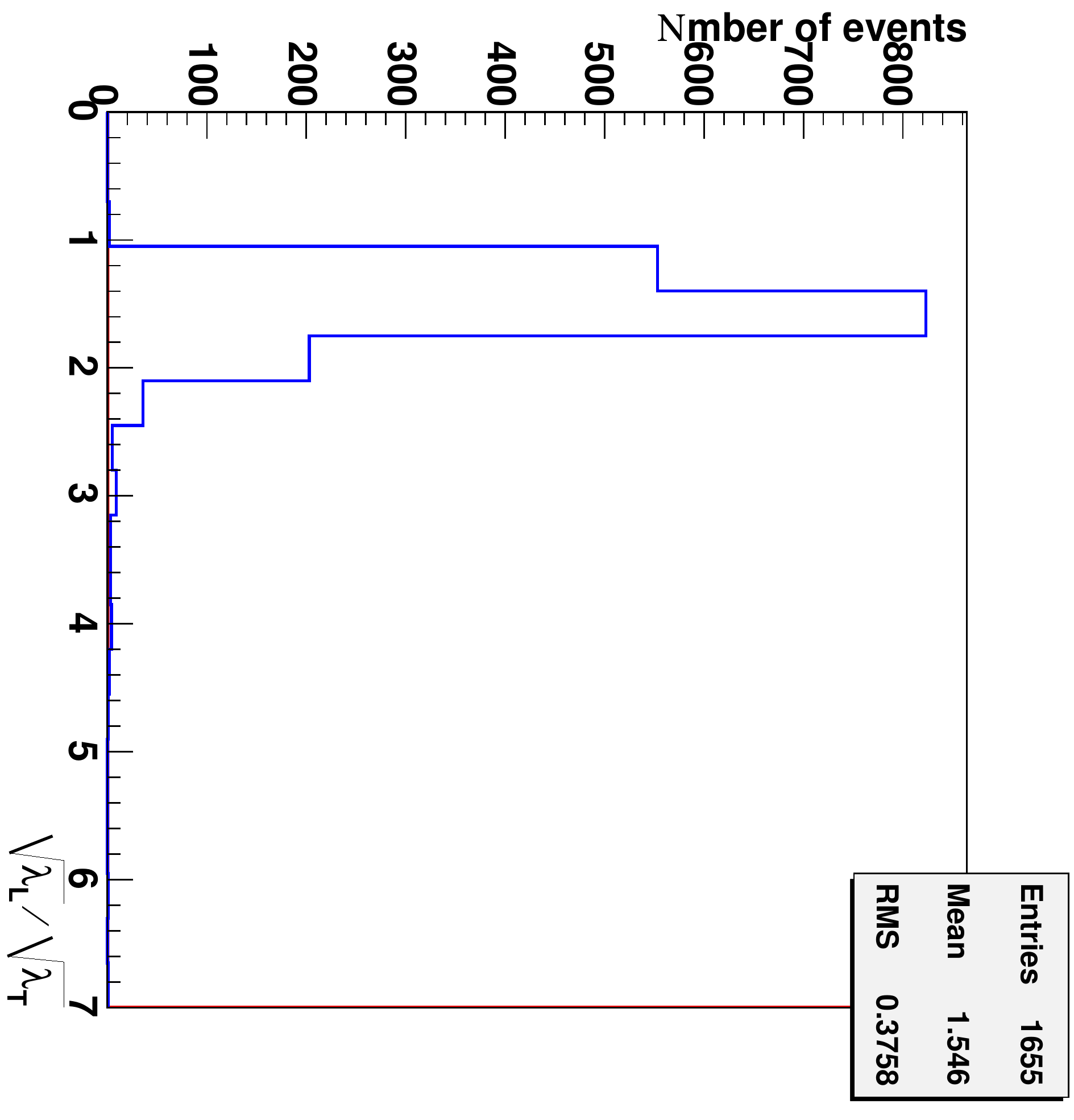}
		\label{fig:tests:LambdaRecoM18:4030}
	}
	\hspace{0.1cm}
	\subfigure[Charged particle tracks inclination $\theta = 61.6^{\circ}$]{
		\includegraphics[width=0.45\textwidth,angle=90]{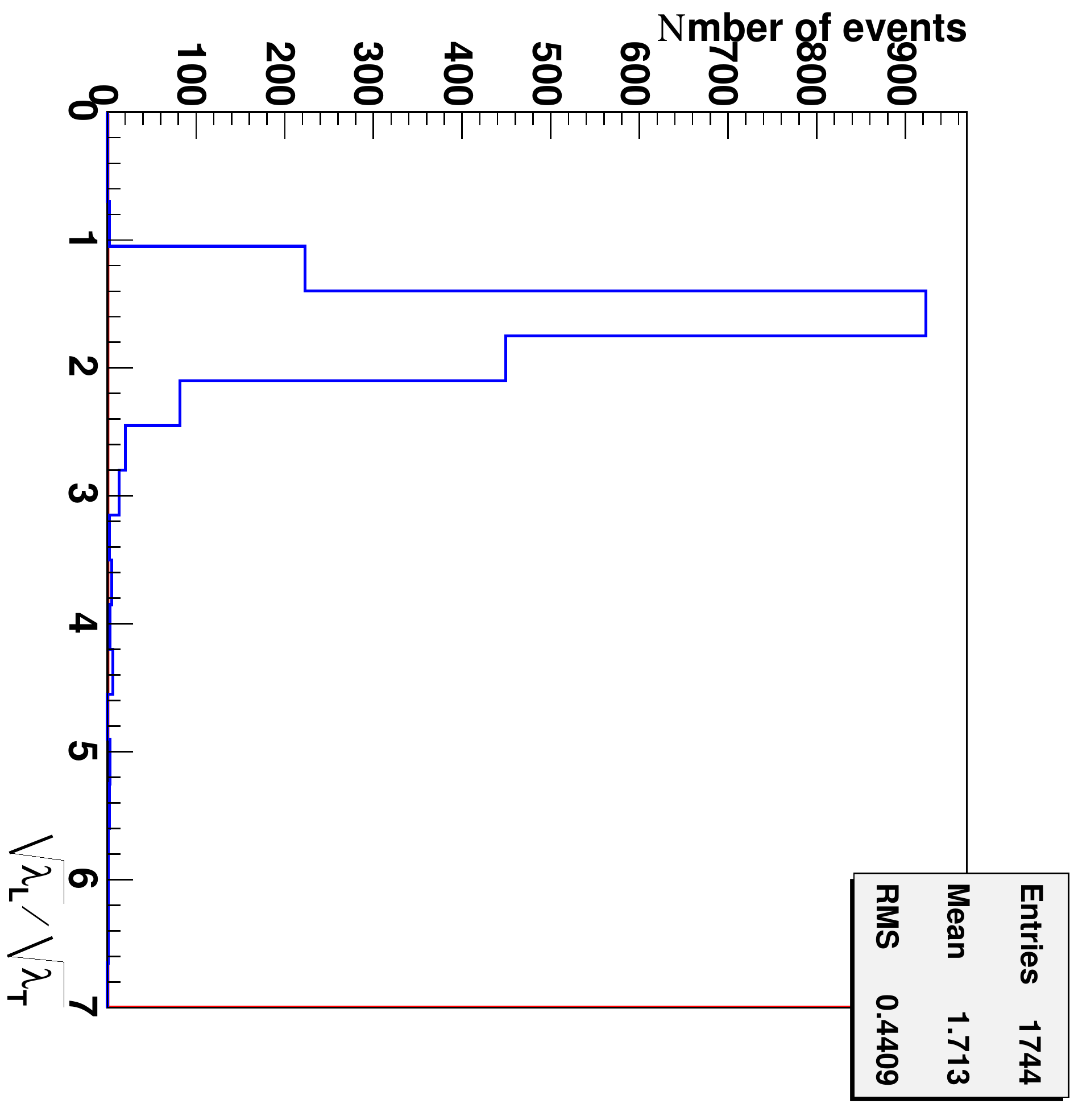}
		\label{fig:tests:LambdaRecoM18:4018}
	}
	\subfigure[Charged particle tracks inclination $\theta = 70.8^{\circ}$]{
		\includegraphics[width=0.45\textwidth,angle=90]{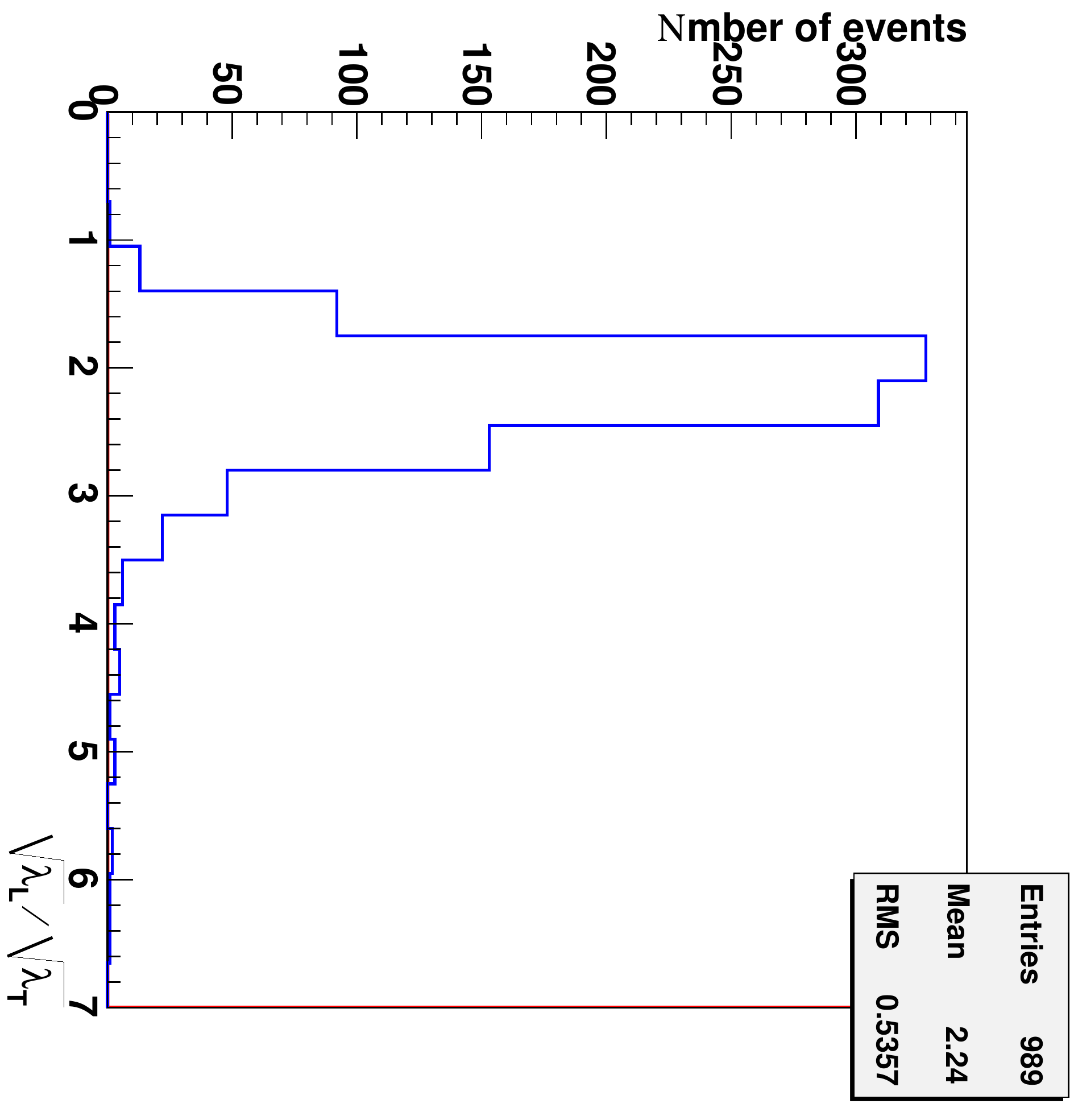}
		\label{fig:tests:LambdaRecoM18:4024}
	}
	\hspace{0.1cm}
	\subfigure[Charged particle tracks inclination $\theta = 75.8^{\circ}$]{
		\includegraphics[width=0.45\textwidth,angle=90]{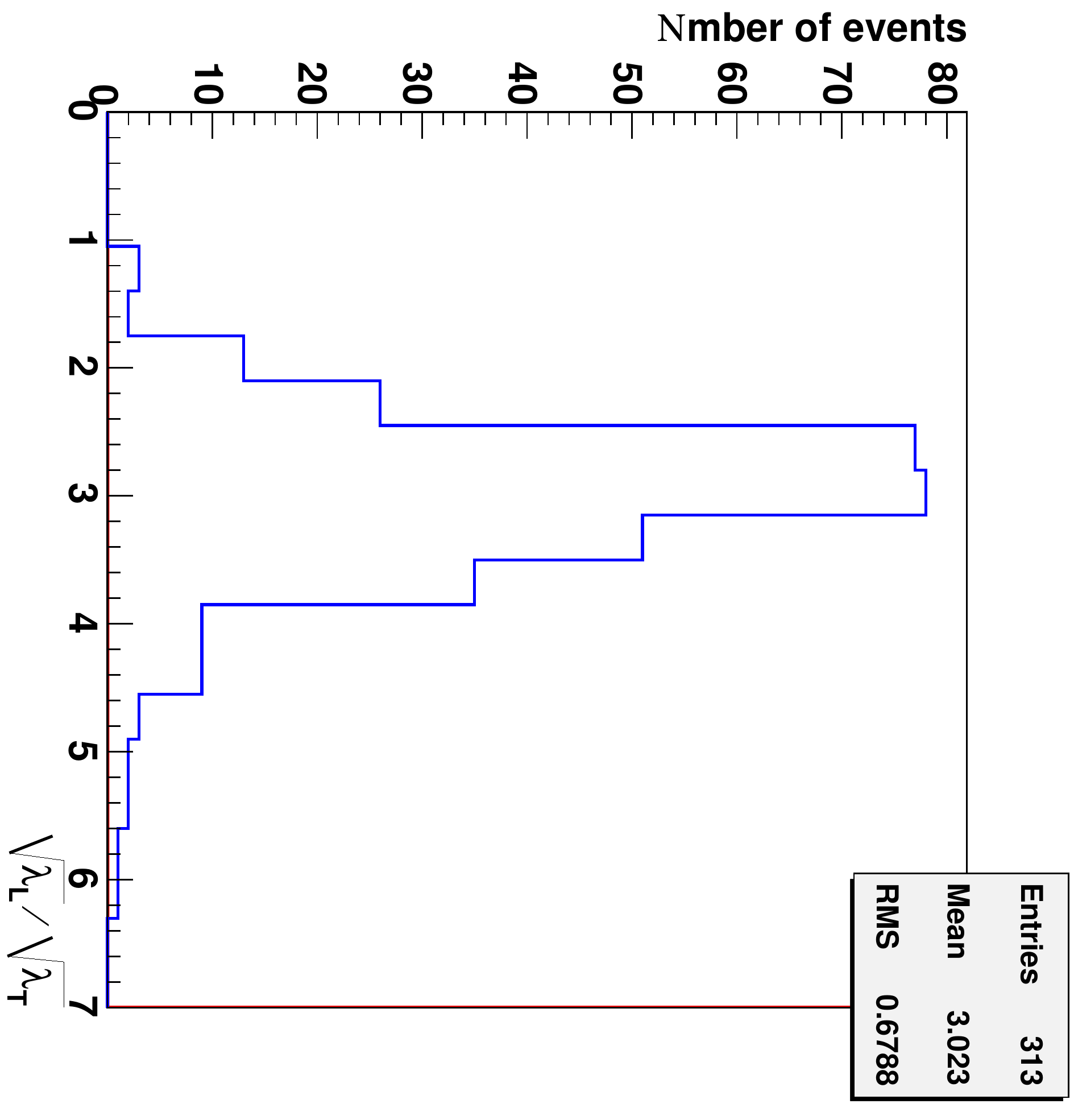}
		\label{fig:tests:LambdaReco18:4016}
	}
	\caption[Distributions of the $\sqrt{\lambda_{L}/\lambda_{T}}$ ratio reconstructed in clusters measured in MIMOSA-18]{Distributions of the $\sqrt{\lambda_{L}/\lambda_{T}}$ ratio reconstructed in clusters measured with 5~GeV electrons in MIMOSA-18 for different angle $\theta$.}
	\label{fig:tests:LambdaRecoM18}
	\end{center}
\end{figure}

\begin{figure}[!h] 
	\begin{center}
	\includegraphics[width=0.8\textwidth,angle=90]{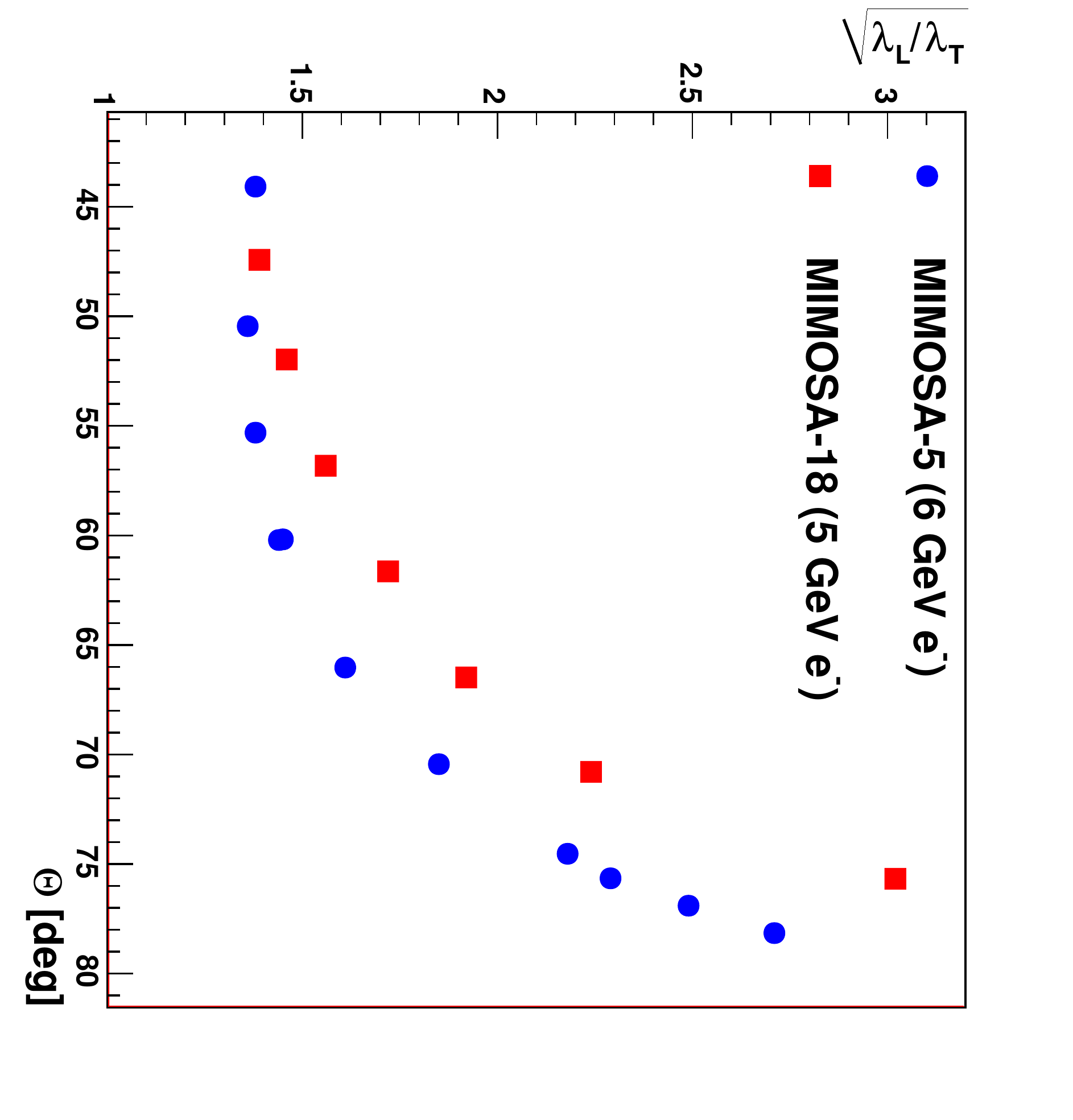}
	\caption[Measured ratio of the longitudinal and transverse dimensions of a cluster ($\sqrt{\lambda_{L}/\lambda_{T}}$) as a function of the incident angle $\theta$ in MIMOSA-5 and MIMOSA-18]{Measured ratio of the longitudinal and transverse dimensions of a cluster ($\sqrt{\lambda_{L}/\lambda_{T}}$) as a function of the incident angle $\theta$ in MIMOSA-5 and MIMOSA-18.}
	\label{fig:tests:LambdaReco}
	\end{center}
\end{figure}

\begin{figure}[!h] 
	\begin{center}
	\subfigure[]{
		\includegraphics[width=0.45\textwidth,angle=90]{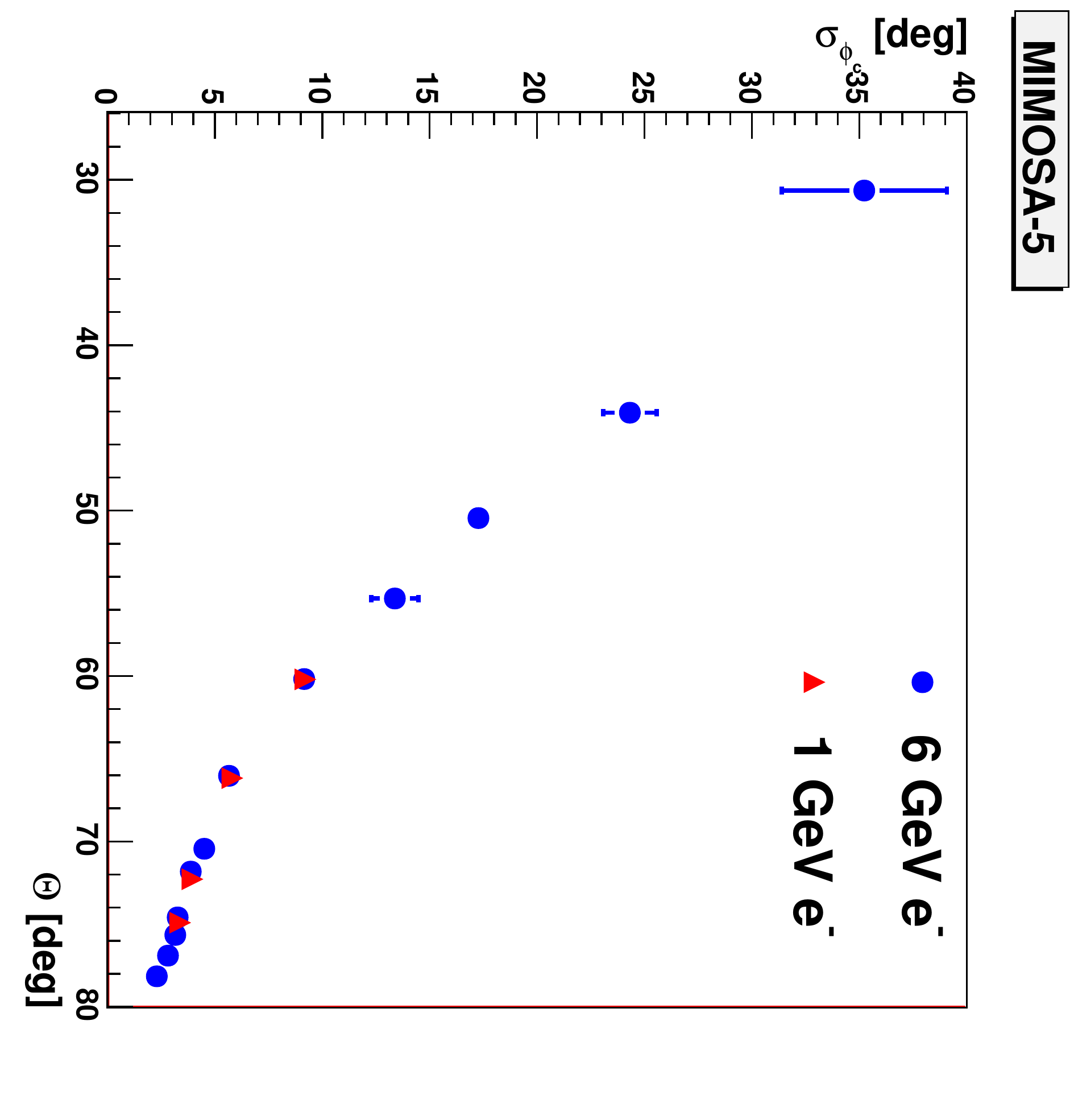}
		\label{fig:tests:PhiReco_EnergyScan:SigmaM5}
	}
	\hspace{0.1cm}
	\subfigure[]{
		\includegraphics[width=0.45\textwidth,angle=90]{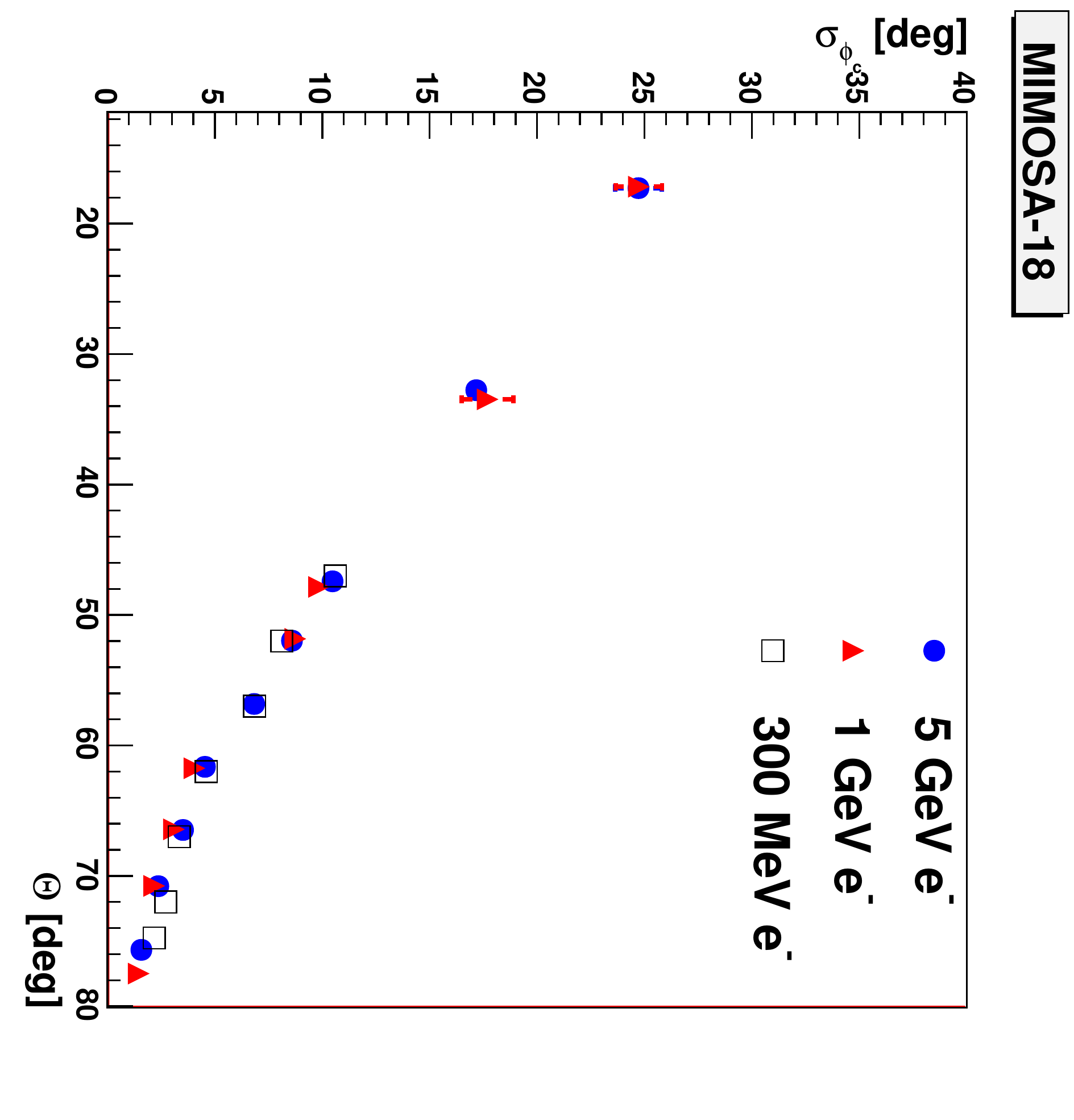}
		\label{fig:tests:PhiReco_EnergyScan:SigmaM18}
	}
	\subfigure[]{
		\includegraphics[width=0.45\textwidth,angle=90]{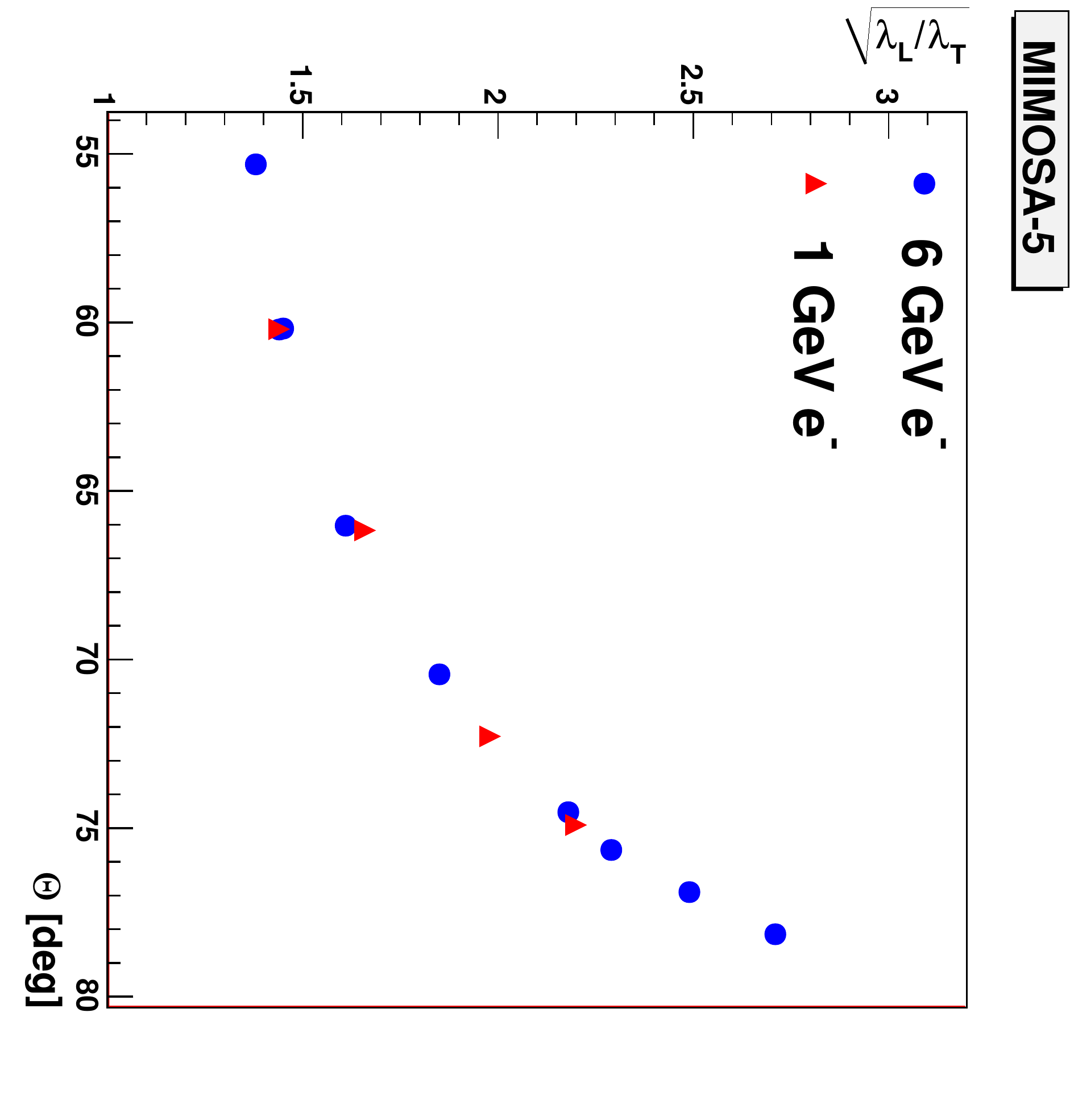}
		\label{fig:tests:Eigen_EnergyScan:M5}
	}
	\hspace{0.1cm}
	\subfigure[]{
		\includegraphics[width=0.45\textwidth,angle=90]{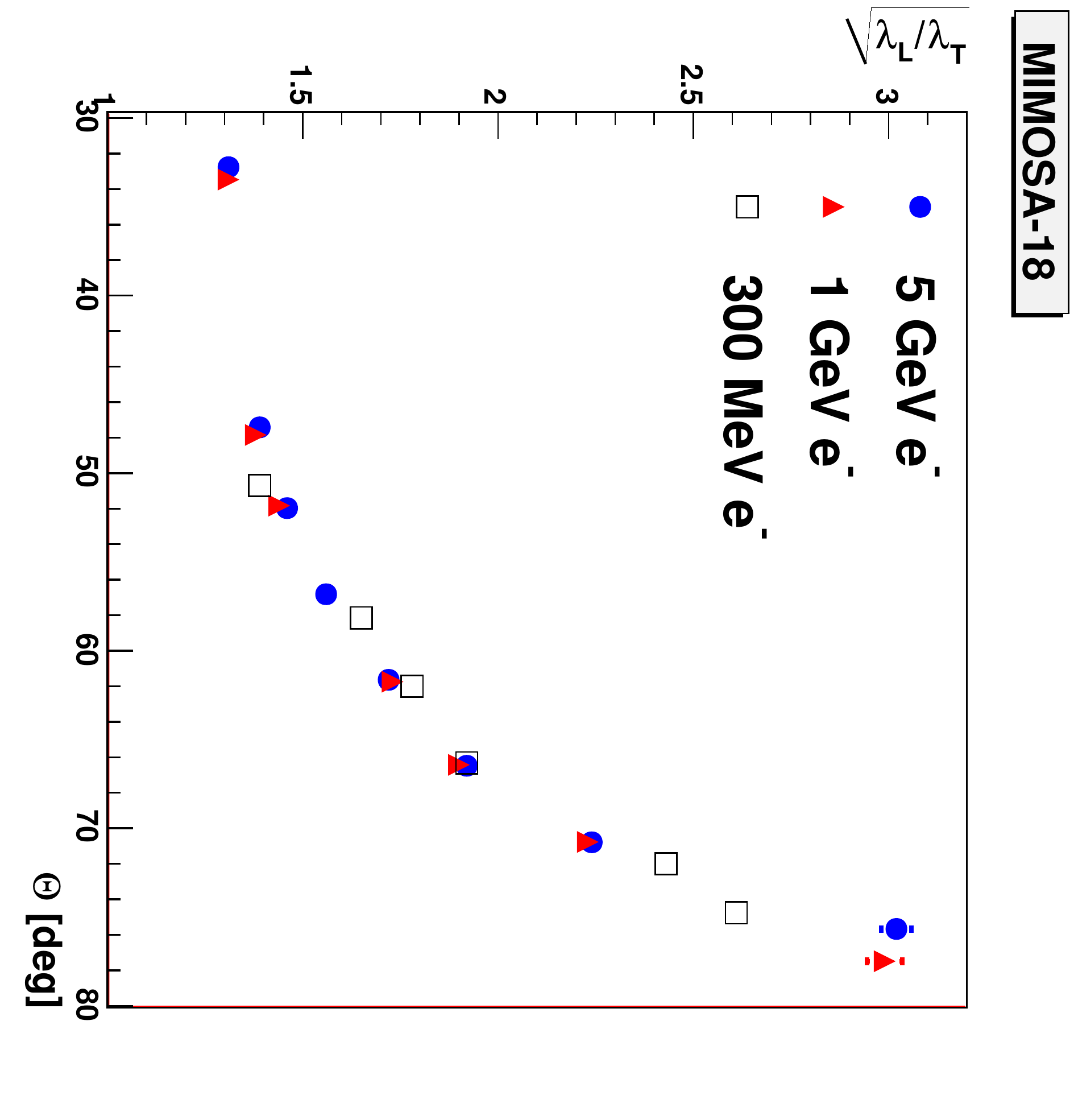}
		\label{fig:tests:Eigen_EnergyScan:M18}
	}
	\caption[Dispersion of the $\phi_{c}$ distribution and cluster elongation ($\sqrt{\lambda_{L}/\lambda_{T}}$) as a function of the incident angle $\theta$ for different beam energies]{Dispersion of the $\phi_{c}$ distribution as a function of the incident angle $\theta$ for different electron beam energies, measured in (a) MIMOSA-5 and (b) MIMOSA-18. The ratio of the longitudinal and transverse dimensions of a cluster ($\sqrt{\lambda_{L}/\lambda_{T}}$) as a function of the $\theta$ angle for different electron beam energies for (c) MIMOSA-5 and (d) MIMOSA-18.}
	\label{fig:tests:PhiReco_EnergyScan}
	\end{center}
\end{figure}

With the adjustable support it was possible to set the pixel matrix orientation with respect to the beam. Mechanics enabled rotation of the MIMOSA plane around $x$ and $y$ axes and manual setting of the angles with an accuracy of approximately $\pm$1$^{\circ}$. Precise determination of the detector angular orientation $\phi_{x}$, $\phi_{y}$ and $\phi_{z}$ was performed offline according to the alignment procedure described in section~\ref{ch:tests:exp_results:alignment_DUT}. Using the reconstructed rotation angles $\phi_{x}$, $\phi_{y}$ and $\phi_{z}$ it was possible to evaluate the $\theta$ and $\phi$ angles describing the track orientation with respect to the detector surface, as shown in fig.~\ref{fig:tests:exp:TrackOrientation}.
\\
The angle of incidence $\theta$ is defined as the angle between the track and the vector normal to the matrix surface. The angle $\phi$ describes orientation of the track projection on the surface with respect to the pixel netting. Both angles $\theta$ and $\phi$ can be expressed in terms of the rotation angles $\phi_{x}$, $\phi_{y}$ and $\phi_{z}$ which are determined from alignment:
\begin{equation}
\label{eq:exper:alignDUT_theta}
\theta = \arccos{(\cos{\phi_{x}} \cdot \cos{\phi_{y}})},
\end{equation}
\begin{equation}
\label{eq:exper:alignDUT_Phi}
\phi = \arctan{\left(\frac{\sin{\phi_{x}} \cdot \cos{\phi_{z}} + \cos{\phi_{x}} \cdot \sin{\phi_{y}} \cdot \sin{\phi_{z}}}{\sin{\phi_{x}} \cdot \sin{\phi_{z}} - \cos{\phi_{x}} \cdot \sin{\phi_{y}} \cdot \cos{\phi_{z}}}\right)}.
\end{equation}
Particles passing through the detector at low incident angles leave statistically round clusters. Since in a MAPS detector the charge is transported by diffusion, it is expected that clusters arising from sufficiently inclined tracks should be elongated in the track direction (projected to the pixel plane) while unaltered in the perpendicular direction. Shapes of individual clusters are subject to statistical fluctuations due to fluctuations of charge deposited in pixels. In fig.~\ref{fig:tests:MeanCluster} two distinct cases of average clusters, measured in the MIMOSA-5 detector, are shown: averaged clusters arising from ($a$) electron tracks incident at $\theta \approx 4^\circ$ and ($b$) tracks incident at $\theta\approx 78^\circ$.\\
The average cluster measured for the incident angle $\theta \approx 4^\circ$ fig.~\ref{fig:tests:MeanCl2012} is symmetric and the charge spread is limited to the seed and its 8 adjacent pixels, while for large tracks incidenting at $\theta\approx 78^\circ$ fig.~\ref{fig:tests:MeanCl2021} the average cluster is elongated in the track direction and is composed of a larger number of pixels. Since the length of the particle path in the epitaxial layer of the MAPS detector is inversely proportional to $\cos{\theta}$, a higher ionisation of the silicon medium is expected for larger track inclinations. Thus average charge collected in clusters from tracks of low incident angles is smaller then in the case of those arising from more inclined tracks. 
\\
Studies of cluster shapes, which are presented below, refer to 9$\times$9 and 15$\times$15 pixel clusters for the MIMOSA-5 and MIMOSA-18, respectively. The clusters of chosen sizes contain majority of pixels participating in the charge collection in the whole studied range of the $\theta$ angle.\\
The cluster charge characteristics were studied for clusters created by tracks traversing the matrices at various angles. The pixels in a cluster, this time 9$\times$9 (MIMOSA-5) and 15$\times$15 (MIMOSA-18), were ordered as described in section~\ref{ch:tests:exp_results:X-rays:charge_spreading} and the MPV of the distributions of the integrated charge for the first $n$ pixels was determined as described in section~\ref{ch:tests:exp_results:charge_collection:cluster}. The dependences of the MPV on the pixel number $n$ for different angles $\theta$ are shown in fig.~\ref{fig:tests:inclined_multiplicity:M5} and fig.~\ref{fig:tests:inclined_multiplicity:M18} for MIMOSA-5 and MIMOSA-18, respectively.
\\
The saturation level of the signal increases from approx. 800~electrons for $\theta \approx 0^\circ$ to approx. 4000~electrons for $\theta \approx 75^\circ$ and from approx. 900~electrons for $\theta \approx 0^\circ$ to approx. 5000~electrons for $\theta \approx 75^\circ$ for MIMOSA-5 and MIMOSA-18, respectively. The onset of saturation (the value of the pixel number $n$ when the dependence flattens out) also depends on the $\theta$ angle as can be seen in these figures. These features are understood in terms of increasing track length inside the epitaxial layer with increasing $\theta$ angle: the larger is this angle, the more pixels participate in charge collection.\\
Further studies of cluster shapes described below are restricted to clusters composed of the first $n \equiv N_{c}$ pixels containing 90\% of the total cluster charge. The $N_{c}$ dependence on the incident angle $\theta$ is shown in fig.~\ref{fig:tests:inclination_scan:multiplicity} and the MPV of the corresponding clusters is shown in fig.~\ref{fig:tests:inclination_scan:charge}.
\\
As expected, the number of pixels contributing to a cluster increases with the angle of incidence $\theta$. Cluster multiplicities measured in MIMOSA-18 are higher than in MIMOSA-5, since the former has smaller pixels. In the case of MIMOSA-5 a step-like structure is visible in fig.~\ref{fig:tests:inclination_scan:multiplicity}. This may be explained as follows. For perpendicular tracks ($\theta\approx0^{\circ}$) only 5 pixels contain 90\% of the total charge. With increasing $\theta$ angle this value remains constant until $\theta\approx30^{\circ}$ when the increase of the track length is large enough to pass under another pixel and a step increase in $N_{c}$ is observed about this value. This effect is more pronounced at larger inclinations (higher steps). In the case of MIMOSA-18 this effect is hardly visible since it has smaller pixels of only 10~$\mu$m pitch.\\
The number of charge carriers generated in the active volume of the detector is proportional to the length of the track in the epitaxial layer. The latter is inversely proportional to $\cos{\theta}$. Thus the cluster charge increases with track inclination. Since 5~GeV and 6~GeV electrons are minimum ionising particles (MIP) and MIMOSA-5 and MIMOSA-18 detectors have epitaxial layers of the same thickness, a similar amount of generated charge in both detectors is expected. A slight shift between MPV of the collected charge in the MIMOSA-5 and the MIMOSA-18 is observed due to the higher charge collection efficiency of MIMOSA-18 than that of MIMOSA-5.\\
The following procedure was applied to the data in order to measure longitudinal and transverse dimensions of clusters. The charge distribution matrix in a given cluster was defined as:
\begin{equation}
\label{eq:exper:Matrix}
\left(\begin{array}{cc}
	\sum_{i = 1}^{N_{c}}\frac{q_{i}}{Q}\left(x_{i}-\overline{x}\right)^{2} & \sum_{i = 1}^{N_{c}}\frac{q_{i}}{Q}\left(x_{i}-\overline{x}\right)\left(y_{i}-\overline{y}\right)\\
	\sum_{i = 1}^{N_{c}}\frac{q_{i}}{Q}\left(x_{i}-\overline{x}\right)\left(y_{i}-\overline{y}\right) & \sum_{i = 1}^{N_{c}}\frac{q_{i}}{Q}\left(y_{i}-\overline{y}\right)^{2} \\
	\end{array} \right),
\end{equation}
where $Q$ is the cluster charge, $q_{i}$ is a charge of the $i$-th pixel, $x_{i}$, $y_{i}$ are its positions in the MIMOSA coordinates system and the $\overline{x}$ and $\overline{y}$ are the coordinates of the cluster in the MIMOSA system. The $\overline{x}$ and $\overline{y}$ were assumed to be coordinates of the charge-weighted centre of gravity of the cluster:
\begin{equation}
\label{eq:exper:CentreOfGravity}
\overline{x} = \sum_{i = 1}^{N_{c}}\frac{q_{i}}{Q} x_{i},~~~
\overline{y} = \sum_{i = 1}^{N_{c}}\frac{q_{i}}{Q} y_{i}.
\end{equation}
Diagonalisation of the matrix~(\ref{eq:exper:Matrix}) allowed to determine the eigenvectors, $\vec v_{L}$ and $\vec v_{T}$, which coincide with the longitudinal and transverse axes of the cluster, respectively. The corresponding square roots of the eigenvalues, $\sqrt{\lambda_{L}}$ and $\sqrt{\lambda_{T}}$, are proportional to the cluster longitudinal and transverse dimensions.\\
Given the eigenvectors $\vec v_{L}$ and $\vec v_{T}$, one may evaluate the azimuthal angle of the cluster w.r.t. the axis of the pixel matrix, $\phi_c$, for each individual cluster. Distributions of the reconstructed $\phi_c$ angles measured in the MIMOSA-5 and MIMOSA-18 detectors for different beam settings are shown in fig.~\ref{fig:tests:PhiRecoM5} and fig.~\ref{fig:tests:PhiRecoM18}, respectively. The histograms are fitted with the Gaussian functions to obtain precise determination of peak positions and widths. The mean values of the reconstructed $\phi_c$ angles are in an agreement with the actual values established from alignment, $\phi$, and the precision of the $\phi_c$ angle reconstruction improves with increasing track inclination.
\\
The accuracy of determining the axes of a cluster diminishes as the track becomes steeper and clusters less elongated, which is reflected in increasing dispersion of the $\phi_{c}$ distribution. Measurements of the dispersion $\sigma_{\phi_{c}}$ for several values of $\theta$ are shown in fig.~\ref{fig:tests:PhiReco}. The experimental points refer to measurements performed with the MIMOSA-5 and MIMOSA-18 exposed to 6~GeV and 5~GeV electron beams, respectively, for various $\phi$ values. From fig.~\ref{fig:tests:PhiReco} it is well visible that MIMOSA-18, equipped with pixels of smaller pitch and characterised by lower noise, provides much higher precision of the $\phi_{c}$ angle determination in a wider range of the $\theta$ angle, as compared to MIMOSA-5.
\\
The ratio $\sqrt{\lambda_{L}/\lambda_{T}}$ is a measure of cluster elongation. Distributions of the $\sqrt{\lambda_{L}/\lambda_{T}}$ ratio for clusters reconstructed in the MIMOSA-5 and MIMOSA-18 detectors for tracks of different inclinations are presented in fig.~\ref{fig:tests:LambdaRecoM5} and fig.~\ref{fig:tests:LambdaRecoM18}, respectively. The measured dependence of $\sqrt{\lambda_{L}/\lambda_{T}}$ ratio on the incident angle $\theta$ is shown in fig~\ref{fig:tests:LambdaReco}. It was concluded that the cluster elongation depends strongly on $\theta$. It was estimated that this observation is valid for $\theta > 55^{\circ}$ and $\theta > 45^{\circ}$ in the MIMOSA-5 and MIMOSA-18, respectively, since for tracks incident at lower angles the elongation of clusters cannot be reliably determined. This can be explained by the fact that in these clusters approximately the same numbers of pixels are involved in charge collection.
\\
The MIMOSA-5 and MIMOSA-18 prototypes were also exposed to lower energy electron beams in order to verify if the procedure of the $\phi_{c}$ angle determination is sensitive to beam energies. Measurements with 1~GeV electrons were performed at DESY while with 300~MeV electrons at the DAFNE beam test facility (Frascati). Since there was no tracking system (telescope) available at DAFNE, it was impossible to use the offline alignment procedure for precise determination of the angles $\theta$ and $\phi$. Moreover, lack of tracking resulted in a greater fraction of noise clusters in the data sample thus deteriorating precision of determination of the cluster eigenvectors and eigenvalues. The results for the MIMOSA-5 and MIMOSA-18 presented in fig.~\ref{fig:tests:PhiReco_EnergyScan:SigmaM5} and \ref{fig:tests:PhiReco_EnergyScan:SigmaM18}, respectively, imply that precision of the $\phi$ angle determination and cluster elongation expressed by the ratio $\sqrt{\lambda_{L}/\lambda_{T}}$ fig.~\ref{fig:tests:Eigen_EnergyScan:M5} and~\ref{fig:tests:Eigen_EnergyScan:M18} are not dependent on the electron beam energy in the studied range. This can be understood since electrons of those energies are minimum ionising particles and multiple scattering (energy dependent) apparently has no effect on the cluster shape.

\chapter{Simulation of the MAPS detector response to charged particles}
\label{ch:digi}

Optimisation of the vertex detector design for an experiment at the ILC must be based on the Monte Carlo (MC) studies. Simulation studies performed in the ILC community so far focused mainly on the general detector geometry. Tracking was done assuming a given value for the spatial resolution of a given pixel matrix. However, dedicated studies presented above in chapter~\ref{ch:tests:exp_results} show that many features of the detector response, like e.g. cluster multiplicity or amount of charge collected in individual pixels, depend significantly on the incident angle of the track, $\theta$. Thus any detailed Monte Carlo simulation of the VTX must include a precise description of a given matrix response to charged particles on the pixel level. An attempt of such a description is presented below. This is based on a simple model of charge diffusion in a MAPS, supplemented by a digitisation procedure to obtain simulated pixel clusters.

\section{A Simple model of charge diffusion}
\label{ch:digi:simple_model}
\begin{figure}[!h]
        \begin{center}
                \resizebox{0.65\textwidth}{!}{
                        \includegraphics[]{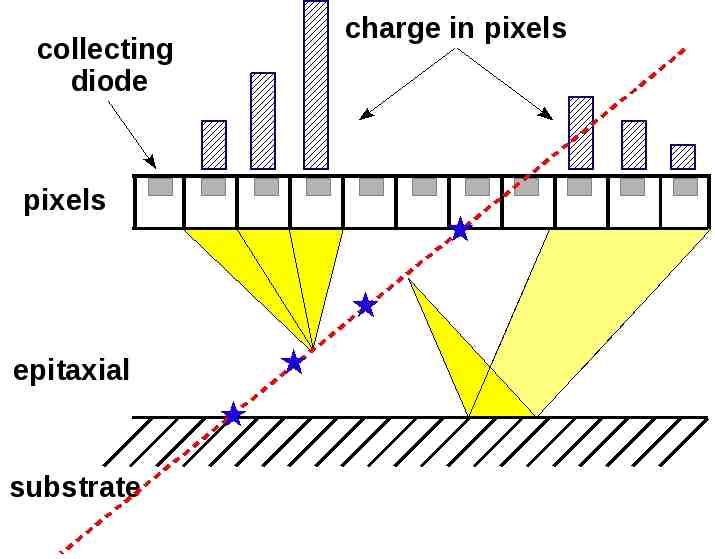}
                }
                \caption[Schematic layout of the MAPS detector]{Schematic cross-section of the MAPS detector. The dashed line marks a charged particle track and star markers separate steps in simulations (see text below).}
                \label{fig:digi:SimpleModel:MAPSCross}
        \end{center}
\end{figure}

\begin{figure}[!h]
        \begin{center}
	\subfigure[]{
		\includegraphics[width=0.4\textwidth]{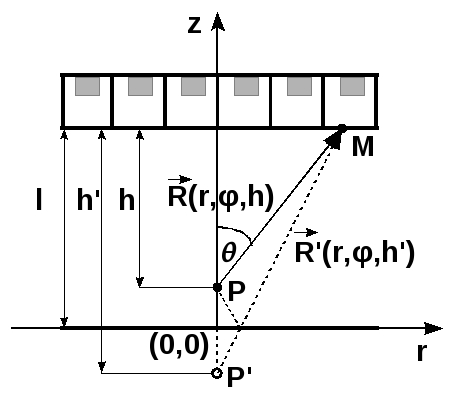}
		\label{fig:digi:SimpleModel:formula:CoorSys}
	}
	\hspace{0.1cm}
	\subfigure[]{
		\includegraphics[width=0.49\textwidth,angle=90]{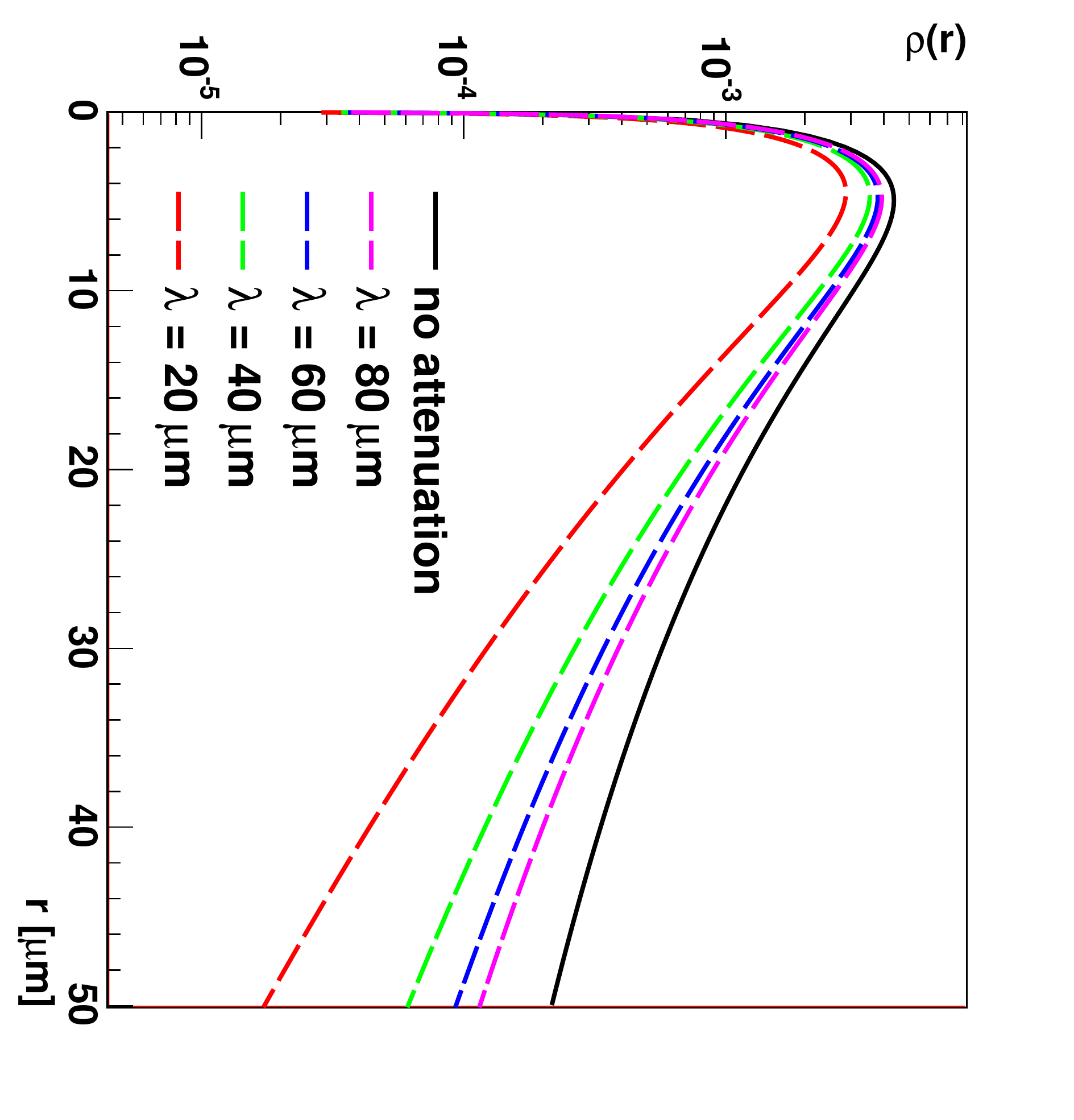}
		\label{fig:digi:SimpleModel:formula:pdf_versus_lambda}
	}
        \caption[Probability density function $\rho(r)$ for different values of the attenuation lengths $\lambda$]{(a) Illustration of variables entering formula (\ref{eq:digi:pdf_digi1}). (b) Probability density function $\rho(r)$ for different values of the attenuation lengths $\lambda$, for carriers generated at the depth of $h=7~\mu$m (thickness of the epitaxial layer was 14~$\mu$m).}
        \label{fig:digi:SimpleModel:formula}
        \end{center}
\end{figure}

\begin{figure}[!h]
        \begin{center}
	\subfigure[$h =2~\mu m$]{
		\includegraphics[width=0.45\textwidth,angle=0]{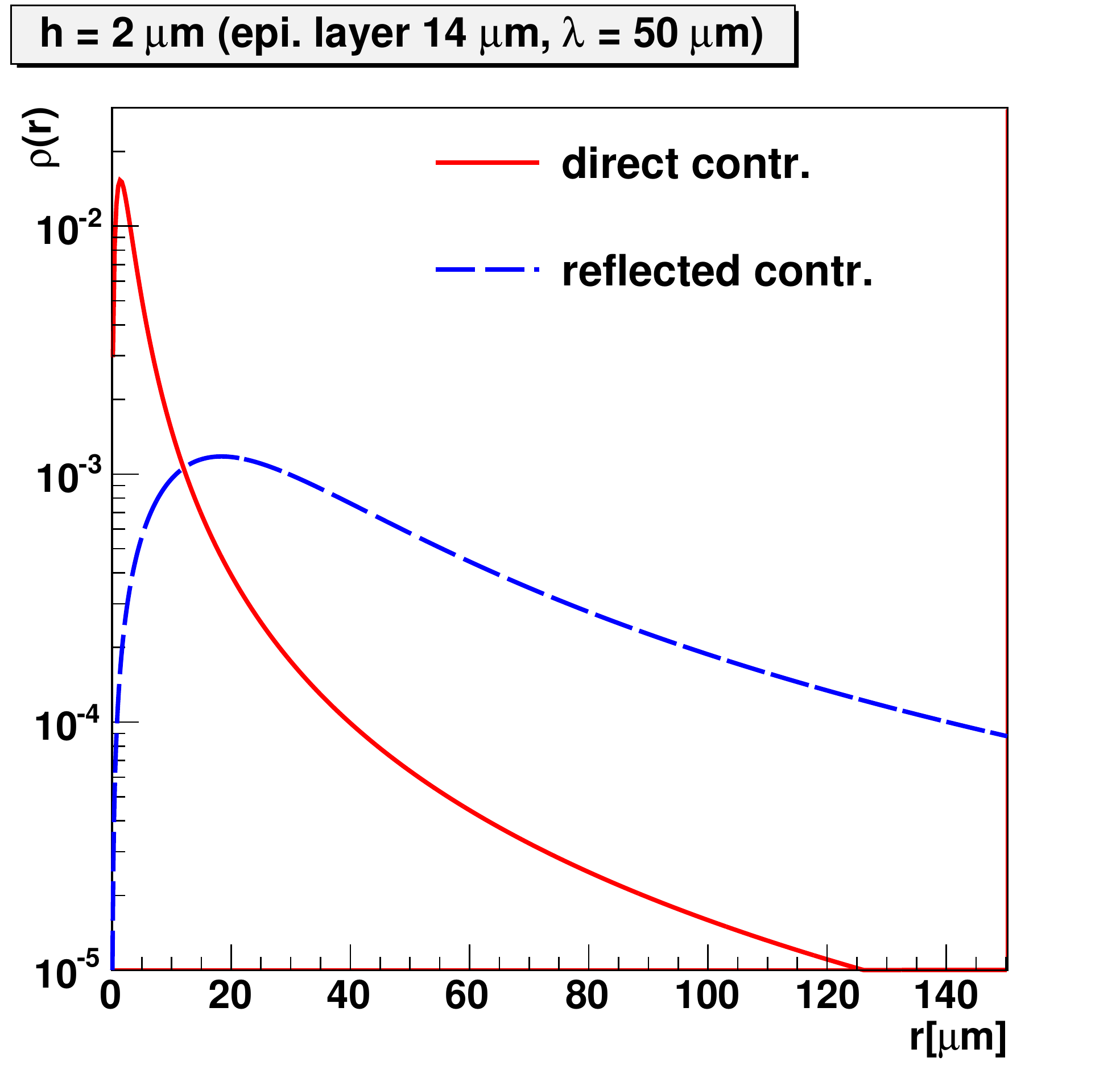}
		\label{fig:digi:SimpleModel:DepScan:2}
	}
	\hspace{0.1cm}
	\subfigure[$h = 4~\mu m$]{
		\includegraphics[width=0.45\textwidth,angle=0]{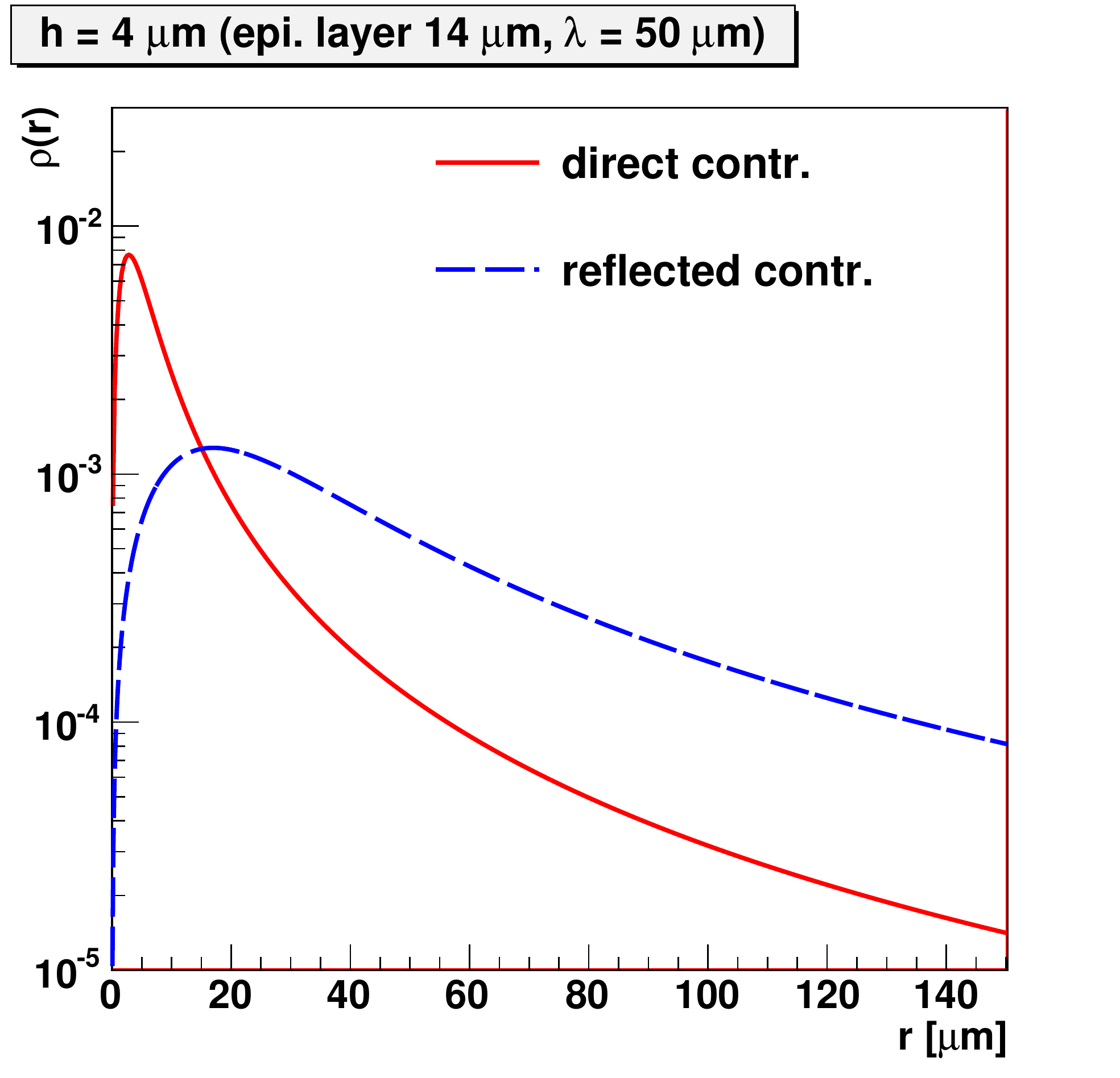}
		\label{fig:digi:SimpleModel:DepScan:4}
	}
	\subfigure[$h = 10~\mu m$]{
		\includegraphics[width=0.45\textwidth,angle=0]{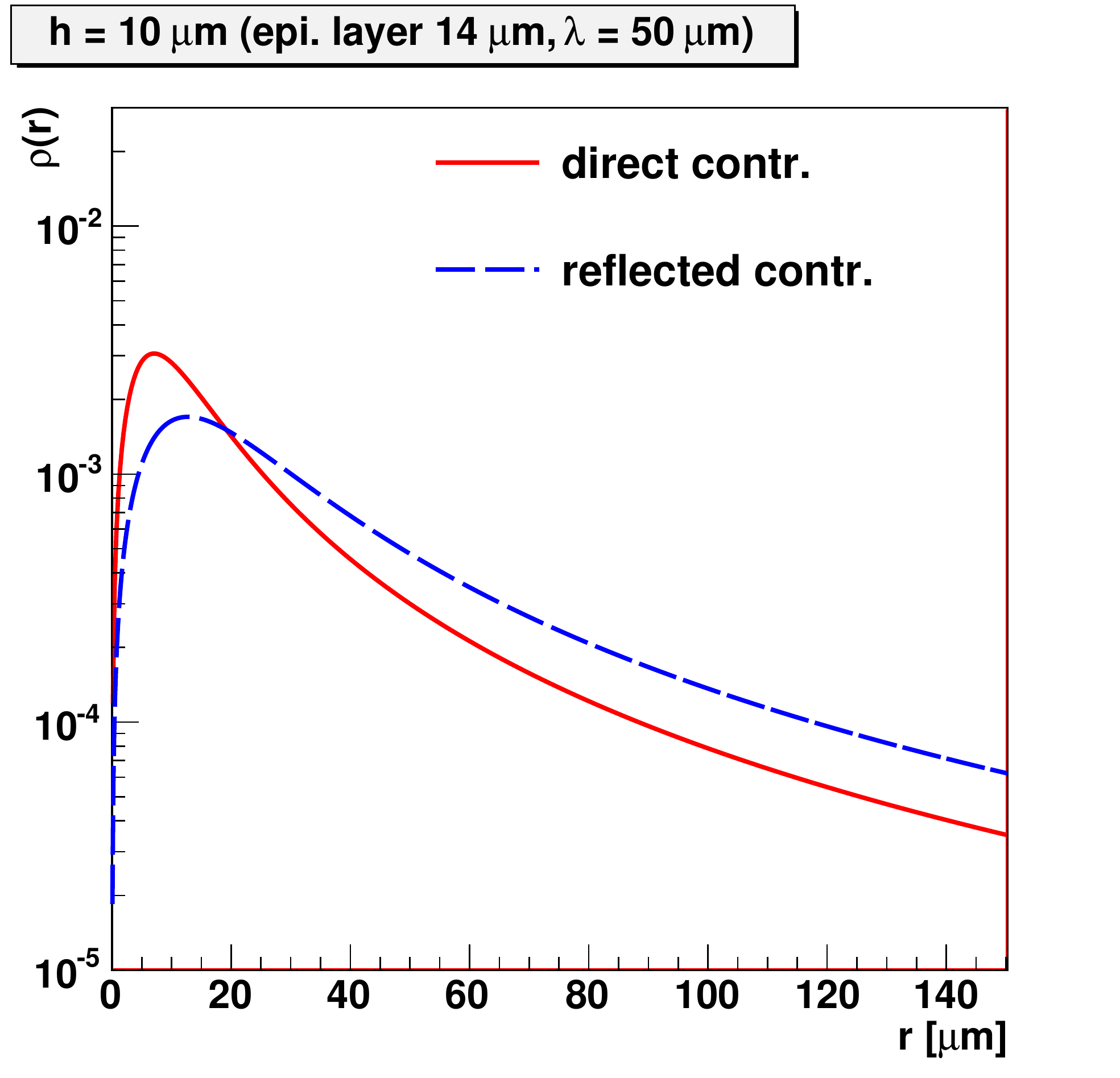}
		\label{fig:digi:SimpleModel:DepScan:10}
	}
	\hspace{0.1cm}
	\subfigure[$h = 12~\mu m$]{
		\includegraphics[width=0.45\textwidth,angle=0]{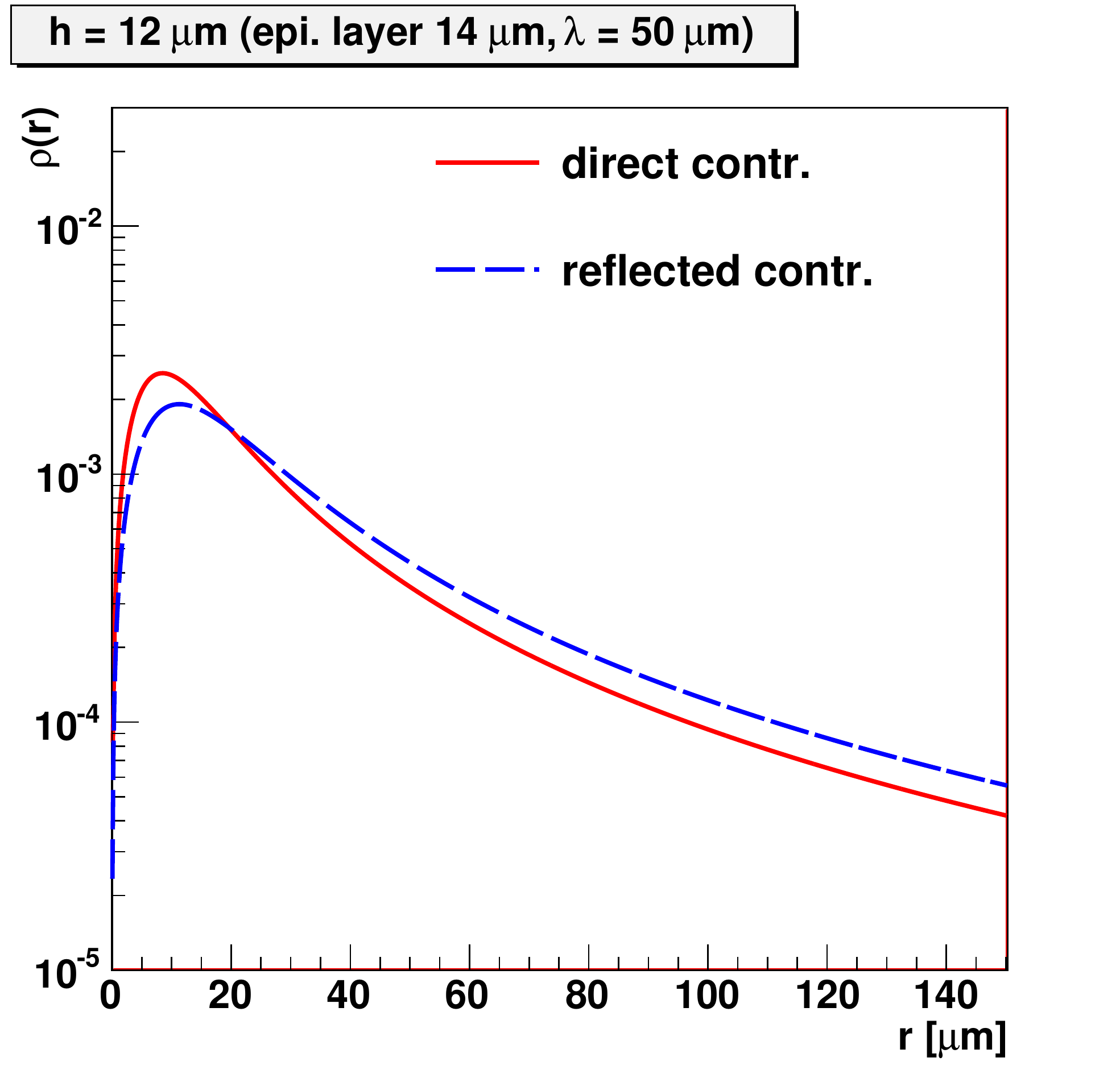}
		\label{fig:digi:SimpleModel:DepScan:12}
	}
        \caption[Probability density function $\rho(r)$ for different depths $h$, for detector with 14~$\mu$m epitaxial layer and attenuation length $\lambda$ = 50~$\mu$m]{Probability density function $\rho(r)$ for: (a) $h=2~\mu m$, (b) $h=4~\mu m$, (c) $h=10~\mu m$ and (d) $h=12~\mu m$, for detector with 14~$\mu$m epitaxial layer and attenuation length $\lambda$ = 50~$\mu$m (see text).}
        \label{fig:digi:SimpleModel:DepScan}
        \end{center}
\end{figure}

There are three main layers in a MAPS device: a substrate, an epitaxial layer and a pixel layer, as shown in fig.~\ref{fig:digi:SimpleModel:MAPSCross}. A charge collecting diode is located in the centre of each pixel. The diode is surrounded by a shallow depletion region where an electric field is present.
\\
A charged particle traversing the detector generates electron-hole pairs. Those of them which are created in the epitaxial layer, which is electric field free, diffuse isotropically and are collected in pixels. The charge generated in the highly doped substrate is lost due to the fast recombination of carriers in this region.\\
In the present model the following three assumptions have been adopted: ($i$) The charge generated in the epitaxial layer diffuses isotropically. Approximately 50\% of carriers move directly towards the collecting diodes and 50\% in the opposite direction, towards the substrate. ($ii$) After reaching the substrate the electrons are reflected towards the collecting diodes due to the potential barrier, which results from different doping concentrations of the substrate and in the epitaxial layer; the reflection angle is assumed equal to the angle of incidence. ($iii$) The charge reaching pixels is smaller that the primary ionisation because of trapping charge carriers in the silicon.\\
Basing on the above assumptions one can write a charge distribution function $\rho(\vec{R})$, describing the probability that a charge carrier, generated inside the epitaxial layer at the depth $h$ in the point $P$, reaches the point $M$ on the detector surface ($\vec{R} = \vec{PM}$), see fig.~\ref{fig:digi:SimpleModel:formula:CoorSys}:
\begin{equation}
\label{eq:digi:pdf_digi1}
	\rho(\vec{R})drd\phi = \frac{d\Omega}{4\pi}\cdot\exp{\left(-\frac{|\vec{R}|}{\lambda}\right)} = \frac{hr}{4\pi |\vec{R}|^{3}}\cdot\exp{\left(-\frac{|\vec{R}|}{\lambda}\right)}drd\phi,
\end{equation}
where $d\Omega=\sin{\theta}d\phi d\theta$ is the solid angle element and the exponential term $\exp{(-|\vec{R}|/\lambda)}$, is the probability that a charge carrier reaches its destination, M. The parameter $\lambda$ in (\ref{eq:digi:pdf_digi1}), called attenuation length, is the only free parameter of the model and has to be determined experimentally. Assuming that the reflection angle is equal to the angle of incidence, the probability density function for reflected electrons can by found by transforming $h$ to $h' = 2l - h$ and $\vec{R}$ to $\vec{R'} = (r,\phi,h')$, where $l$ is thickness of the epitaxial layer. Probability density function $\rho(r)$ (\ref{eq:digi:pdf_digi1}) for electrons generated at the depth of $h=7~\mu$m and moving directly towards the pixel layer are shown in fig.~\ref{fig:digi:SimpleModel:formula:pdf_versus_lambda} for different values of $\lambda$.
\\
The uppermost (solid) line in fig.~\ref{fig:digi:SimpleModel:formula:pdf_versus_lambda} refers to the case of no attenuation ($\lambda = \infty$) and the dashed lines corresponding to different $\lambda$. Choosing the appropriate $\lambda$ value will be done according to the criterion of agreement of simulated and measured quantities, as described in detail below in section~\ref{ch:digi:lambda_determination}.\\
The probability density function $\rho(r)$ for different values of $h$, is shown in fig.~\ref{fig:digi:SimpleModel:DepScan} for the epitaxial layer of 14~$\mu$m and $\lambda=50~\mu$m. In each figure two curves are shown: ($i$) the red, solid one corresponding to the electrons which are moving directly towards the collecting diodes and ($ii$) the blue, dashed one referring to the electrons reflected by the potential barrier at the border between the epitaxial layer and the substrate.\\
For electrons generated at small depths, fig.~\ref{fig:digi:SimpleModel:DepScan}(a)~and~(b), the contribution at low $r$ for electrons collected directly is larger than for those reflected from the substrate, while it is opposite at large $r$. This difference becomes less significant with increasing depth $h$, as shown in fig.~\ref{fig:digi:SimpleModel:DepScan}(c)~and~(d). The above observations suggest that the peripheral pixels in clusters are formed from the reflected carriers or those generated deep in the epitaxial layer.

\section{Determination of the attenuation length $\lambda$}
\label{ch:digi:lambda_determination}

\begin{figure}[!h]
        \begin{center}		
	\subfigure[MIMOSA-5]{	
		\includegraphics[width=0.45\textwidth,angle=90]{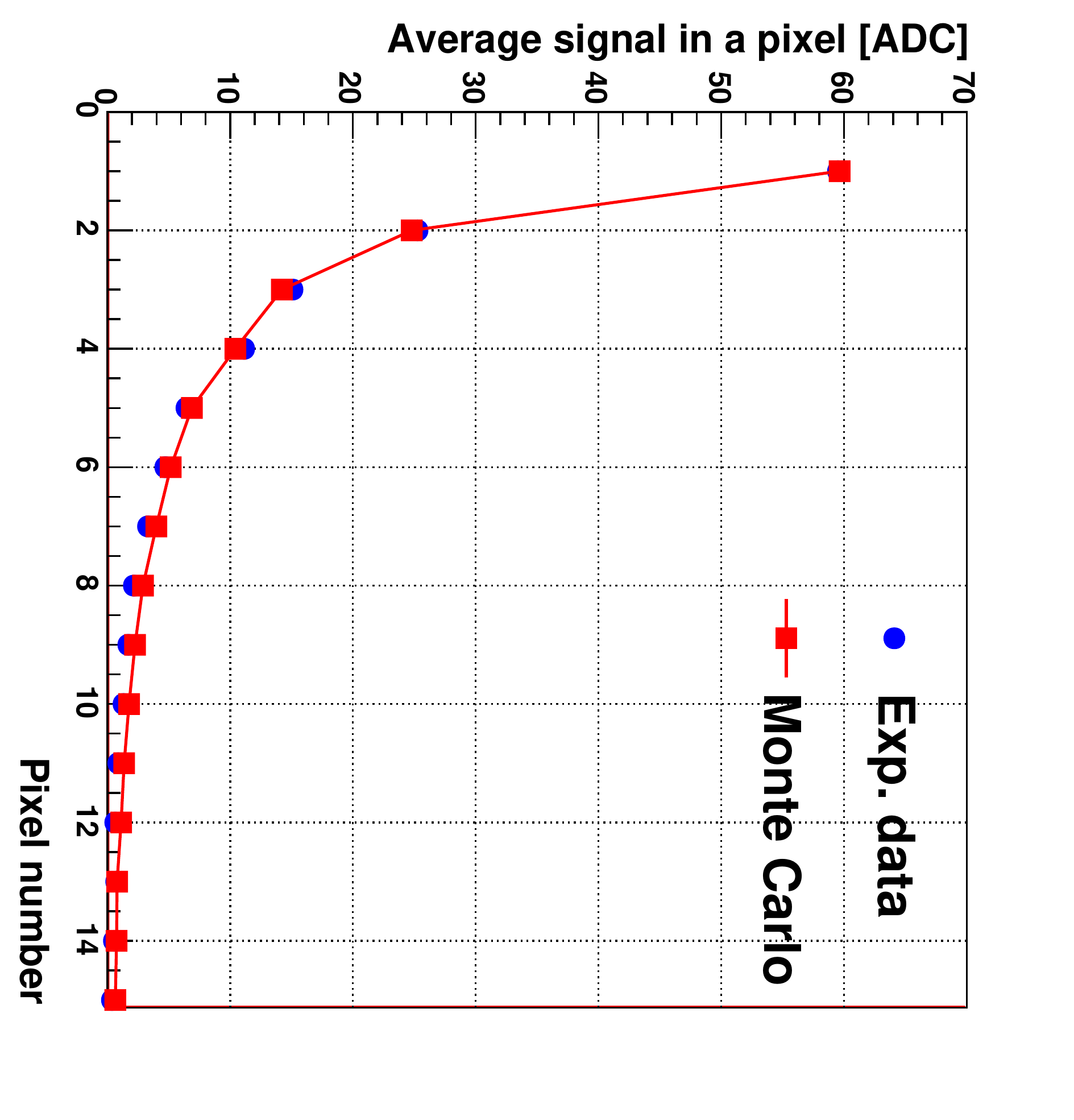}
		\label{fig:digi:Comp1:M5Comp}
	}
	\hspace{0.1cm}
	\subfigure[MIMOSA-18]{
		\includegraphics[width=0.45\textwidth,angle=90]{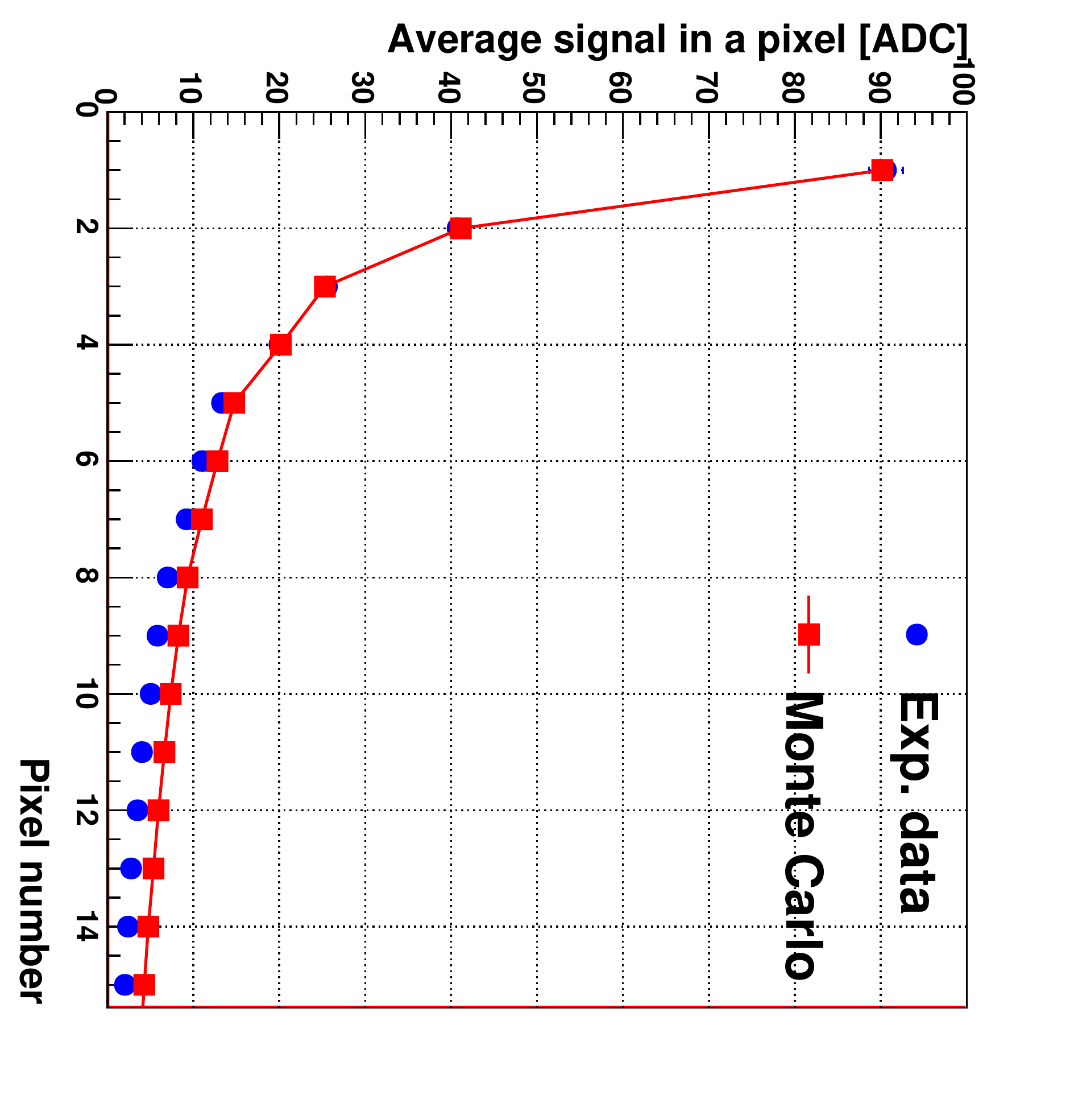}
		\label{fig:digi:Comp1:M18Comp}
	}
        \caption[Comparison of average signals collected in the consecutive pixels for measured and simulated clusters in the MIMOSA-5 and MIMOSA-18 detectors]{Comparison of average signals collected in the consecutive pixels for measured and simulated clusters in the (a) MIMOSA-5 and (b) MIMOSA-18 detectors exposed to 6.5~GeV and 5~GeV for perpendicular electron tracks ($\theta\approx0^{\circ}$). The simulations were done with (a) $\lambda=45~\mu$m, $\eta = 0.247$ and (b) $\lambda=50~\mu$m, $\eta = 0.55$. Blue dots denote the measured values and red squares show the simulation results.}
        \label{fig:digi:Comp1}
        \end{center}
\end{figure}

\begin{figure}[!h]
        \begin{center}
	\subfigure[]{
		\includegraphics[width=0.45\textwidth,angle=0]{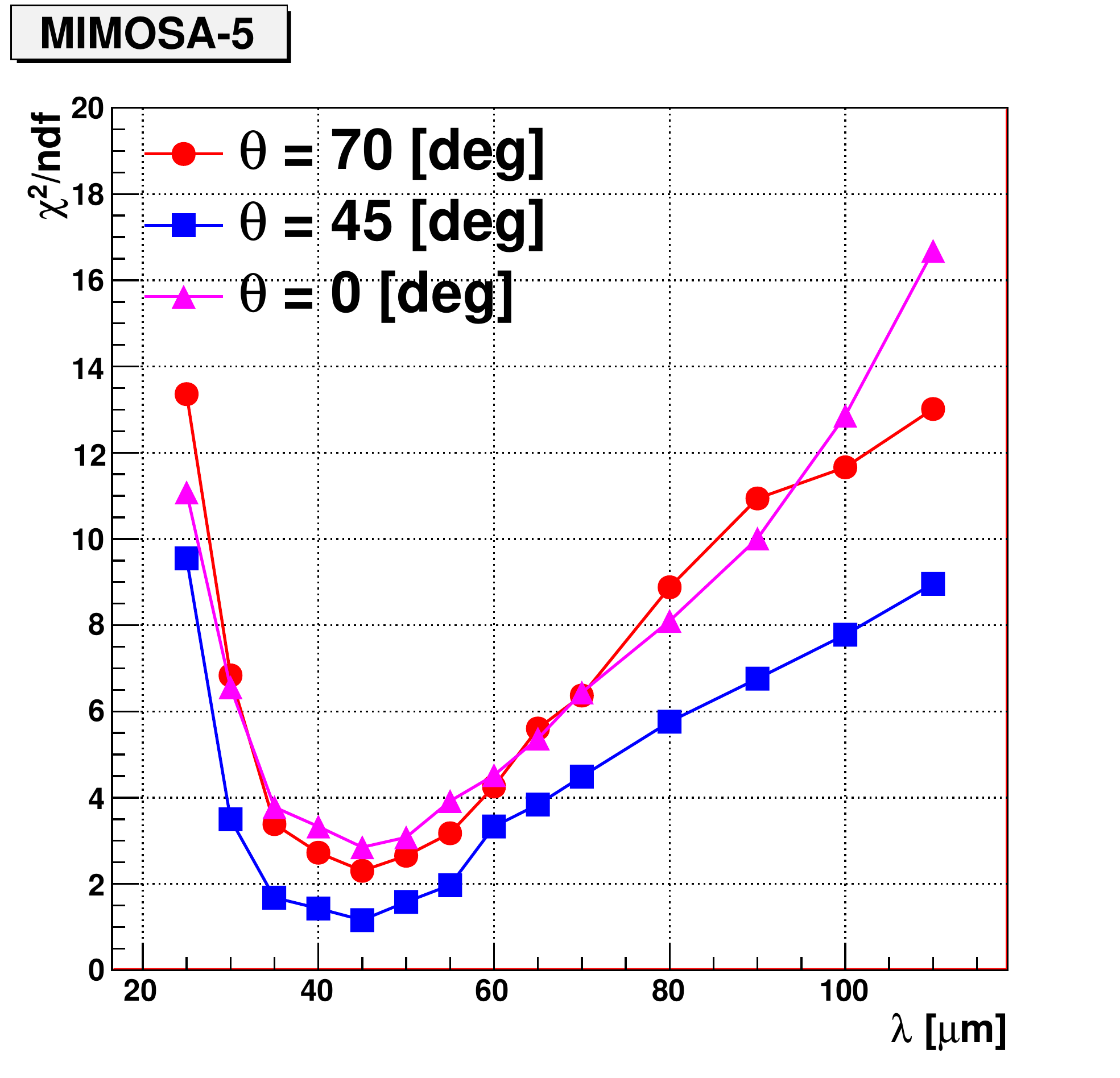}
		\label{fig:digi:CompChi2:M5}
	}
	\hspace{0.1cm}
	\subfigure[]{
		\includegraphics[width=0.45\textwidth,angle=90]{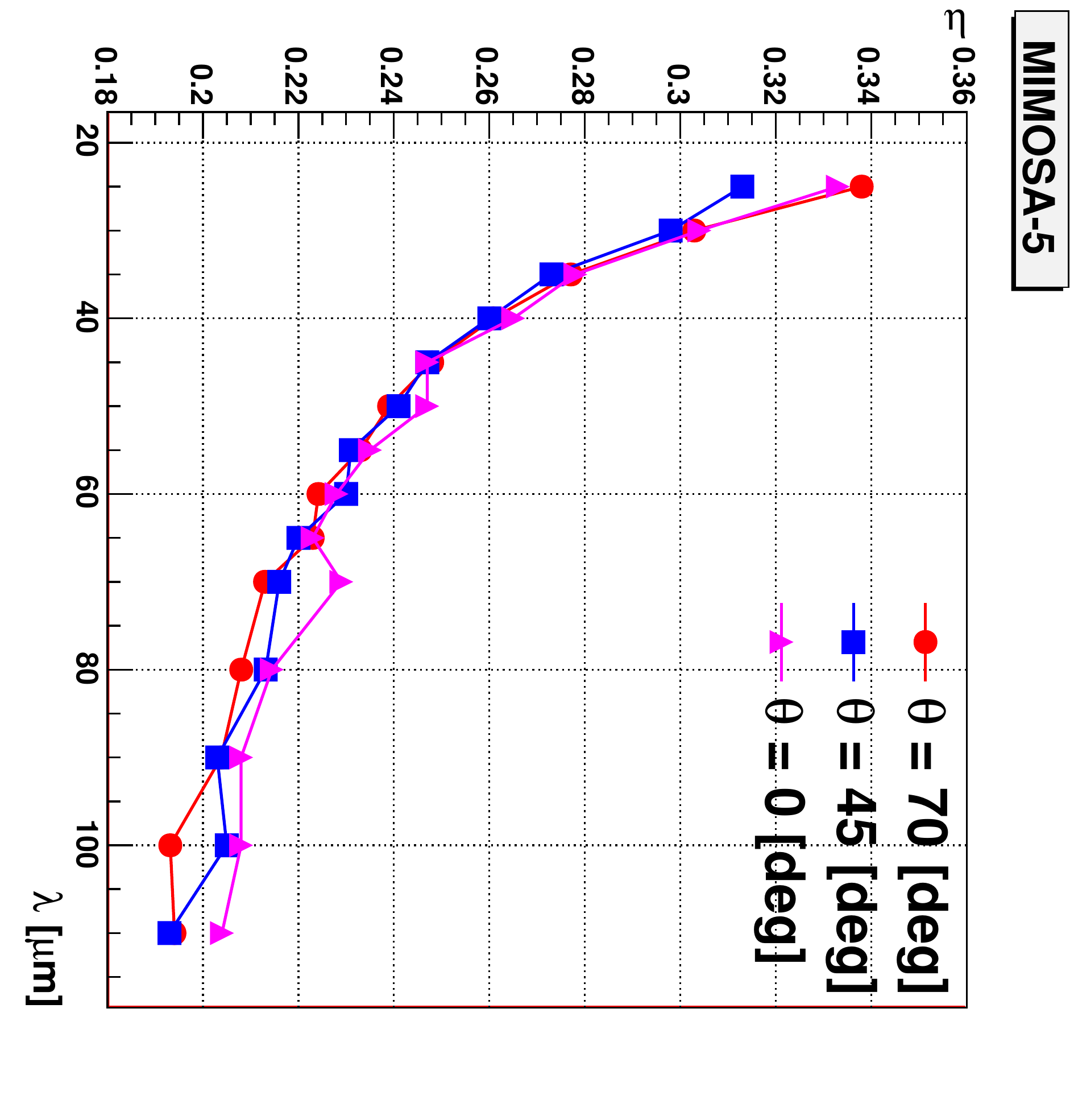}
		\label{fig:digi:CompChi2:EtaM5}
	}
        \caption[The $\chi^{2}/$ndf and conversion factor $\eta$ vs. the attenuation length $\lambda$ for the MIMOSA-5]{(a) The $\chi^{2}/$ndf and (b) conversion factor $\eta$ vs. the attenuation length $\lambda$ for the MIMOSA-5.}
        \label{fig:digi:CompChi2}
        \end{center}
\end{figure}

\begin{figure}[!h]
        \begin{center}
	\subfigure[]{
		\includegraphics[width=0.29\textwidth]{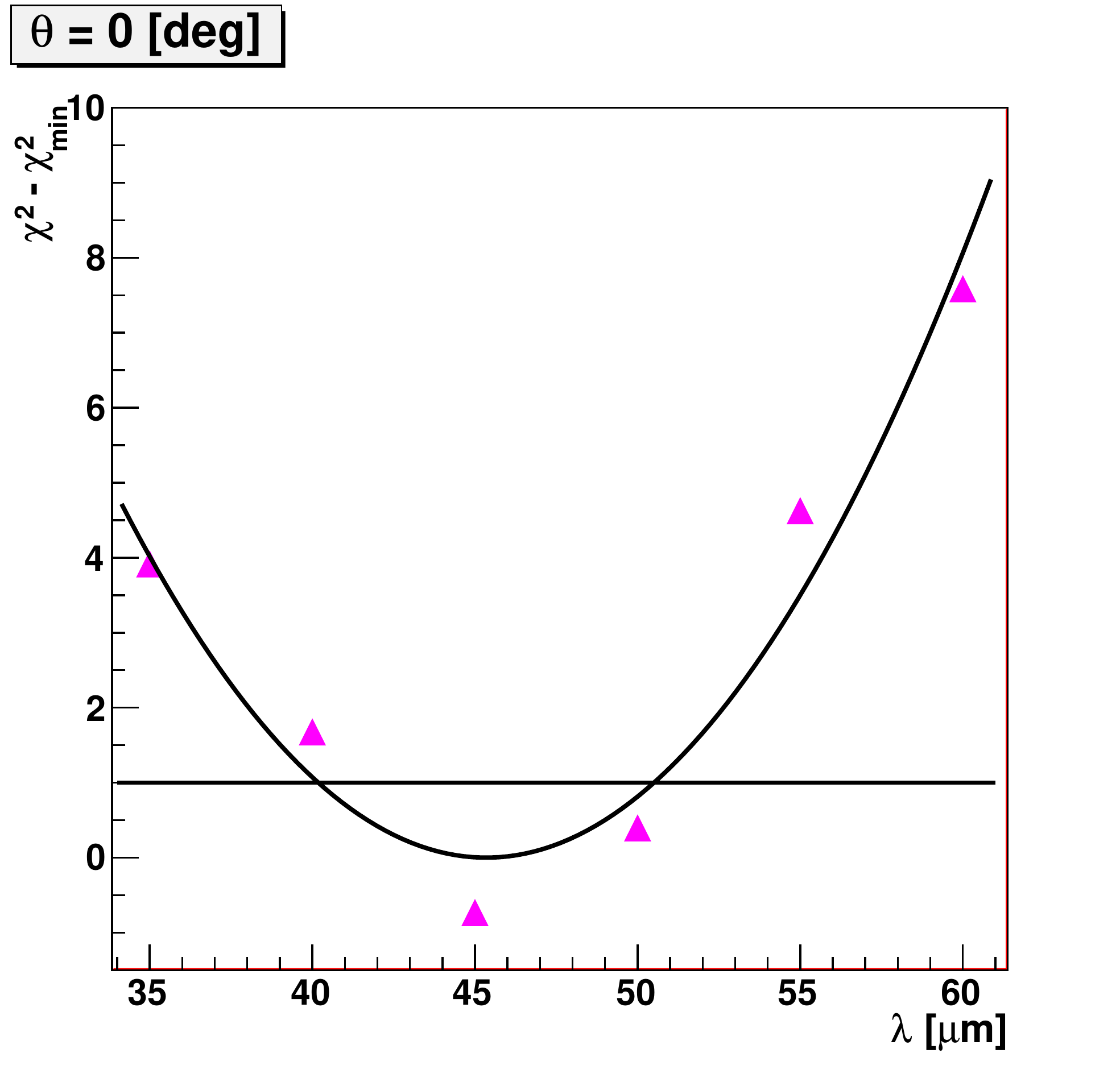}
		\label{fig:digi:test:Chi2_minimum:2012}
	}
	\hspace{0.05cm}
	\subfigure[]{
		\includegraphics[width=0.29\textwidth]{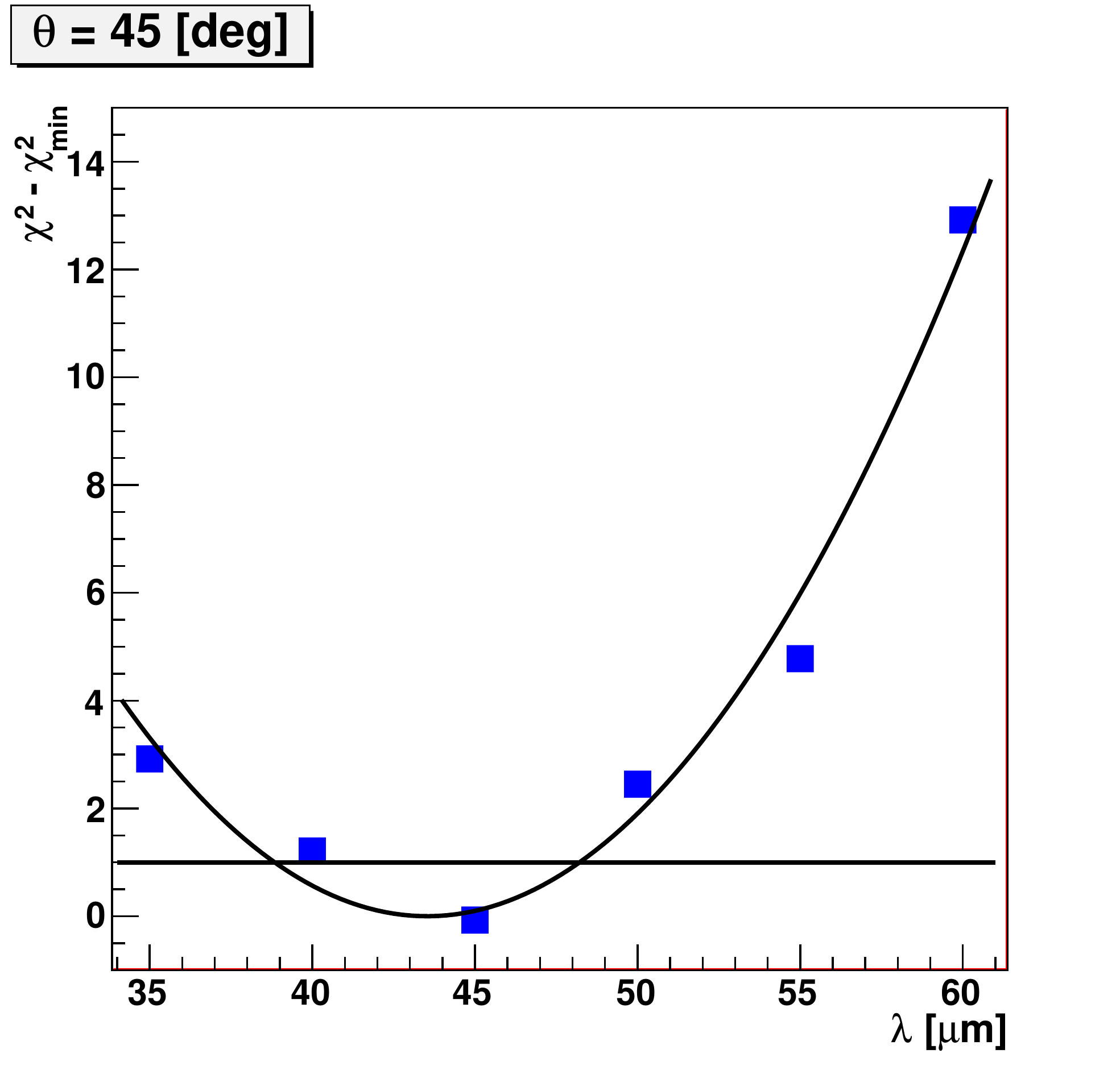}
		\label{fig:digi:Chi2_minimum:2009}
	}
	\hspace{0.05cm}
	\subfigure[]{
		\includegraphics[width=0.29\textwidth]{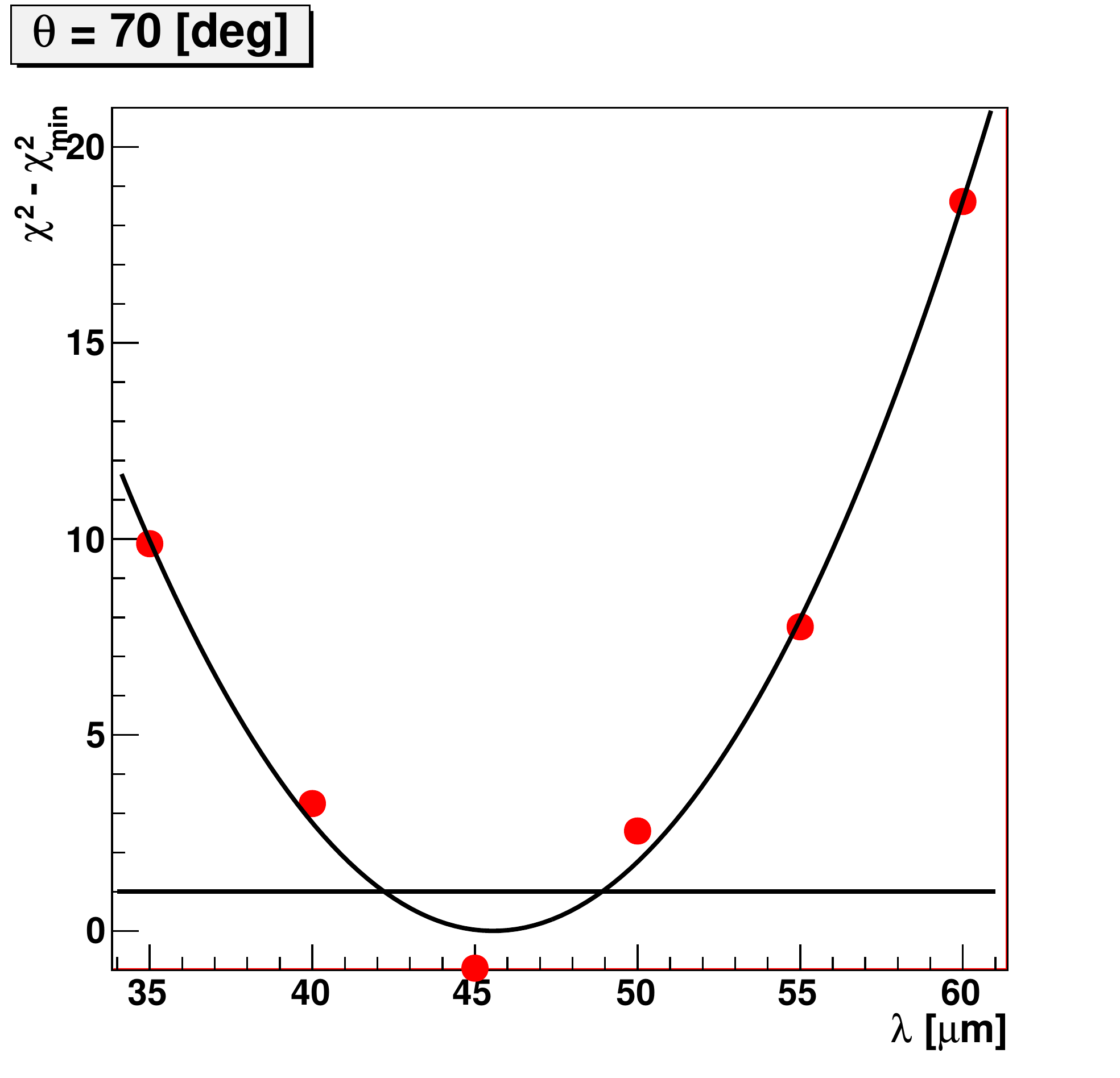}
		\label{fig:digi:Chi2_minimum:4005}
	}
        \caption[The $\chi^{2}$ dependence on the attenuation length $\lambda$ in the vicinity of minimum]{The $\chi^{2}$ dependence on the attenuation length $\lambda$ in the vicinity of the minimum. Experimental points were fitted with parabola functions to determine the positions of the minimum. The data points were shifted downwards so that the minimum of the parabola corresponds to $\chi^{2} = 0$.}
        \label{fig:digi:Chi2_minimum}
        \end{center}
\end{figure}

Passage of electrons through a MAPS detector was simulated using the Geant4 Monte Carlo package \cite{digi:Geant4}. In this package the particle trajectory is followed in steps of unequal lengths, as shown in fig.~\ref{fig:digi:SimpleModel:MAPSCross}. The length of a given step depends on the interaction processes to which the particle is subjected. The probability of each process is given by the corresponding cross section (or equally mean free interaction path). Values of steps are generated for each process according to exponential distribution of the mean free interaction path. It is then decided which of the possible processes will take place: the one with the smallest actually generated step. It is checked if the particle will traverse the step and not stop due to energy loss nor cross the detector boundary. If a selected process is an interaction or a decay, secondaries are generated after this step and subsequently followed in the same way. The tracking lasts as long as the particle has enough energy and is confined within the detector volume. The Coulomb scattering is taken into account at the end of each step and results in changing the direction of motion. In the case of e.g. $\delta$-electrons or bremsstrahlung photon emission a cut is applied to suppress generation of large numbers of soft electrons and photons. This cut is equivalent to a requirement of a minimum path length potentially passed by such secondaries, 10~$\mu$m in the present work.\\
Parametrisation of the MAPS response requires only the following information which is supplied by the Geant4 for each step:
\begin{itemize}
 \item 
 coordinates  of the start and end points of the step, expressed in the local coordinate system of the MAPS detector,
 \item
 energy loss due to ionisation.
\end{itemize}
Knowing the energy required for a single electron-hole pair creation, which is 3.6~eV in silicon, the ionisation energy loss can be expressed by a number of created electron-hole pairs. The charge carriers are assumed to be generated uniformly along the step length. Electrons originating from the epitaxial layer diffuse into pixels according to the formula (\ref{eq:digi:pdf_digi1}) presented in section~\ref{ch:digi:simple_model}. In simulations this is done in two ways.\\
In the first approach a single step is divided into a number of identical substeps. With the increasing number of substeps the precision of simulations improves at the expense of the CPU time. Consider electrons created in a given substep. It is calculated how they will be distributed over neighbouring pixels, according to the probability density function (\ref{eq:digi:pdf_digi1}). This function is integrated over a given pixel surface for both directly moving carriers as well as those reflected from the substrate. Charge from a given substep is collected by several pixels. In practice this is a group of the closest $N \times N$ pixels, where $N$ value depends on the pixel pitch, the epitaxial layer thickness and properties of the silicon. Since the number of electrons from a given substep, reaching a given pixel, $n$, is a subject to the binomial distribution, the amount of collected charge has the corresponding variance:
\begin{equation}
\label{eq:digi:sigma_number}
	\sigma_{n}^{2} = n_{0}p(1 - p),
\end{equation}
where $n_{0}$ is the number of electrons generated in a given substep and $p$ is the effective probability that they are not absorbed before reaching the pixel. This probability is given by:
\begin{equation}
\label{eq:digi:binomial_probability}
	p = \exp{\left(-\frac{|\vec{R}|}{\lambda}\right)},
\end{equation} 
where $\vec{R}$ is the position vector of the centre of the pixel seen from the centre of the substep and $\lambda$ is the effective attenuation length for electrons in silicon. Contributions for all steps are added and the pixel signals are thus obtained. The advantage of this method is the possibility to store the values of the probability density function (\ref{eq:digi:pdf_digi1}) integrals, which enables fast simulations.\\
The second approach is as follows. It is assumed that electron carriers are created along the step according to a uniform probability distribution. Each electron propagates in a direction according to isotropic diffusion (flat distribution of $\cos{\theta}$ and $\phi$ in spherical coordinates). Approximately 50\% of electrons move directly toward the collecting diodes and the other 50\% in the direction of the substrate where, due to the potential barrier, they are reflected towards collecting diodes. Since the reflection angle is assumed to be equal to the angle of incidence on the boundary, propagation of the reflected electron is equivalent to a propagation of its image located at the depth $h' = 2l - h$, where $l$ is the thickness of the epitaxial layer, with the azimuthal angle $\theta' = \pi - \theta$ and no polar angle $\phi$ dependence, see fig.~\ref{fig:digi:SimpleModel:formula:CoorSys}. Knowing the point from which the electron originates and its direction of motion, the location of the destination pixel is determined. Moreover, the electron undergoes many competing processes that can lead to its absorption on the way to this pixel. It is assumed that the probability of absorption per unit of the track length is constant, leading to the exponential dependence of the probability for electron surviving the distance $|\vec{R}|$ (\ref{eq:digi:binomial_probability}). The number of electrons surviving the distance $|\vec{R}|$ from the origin to the pixel location is obtained by generating from the (\ref{eq:digi:binomial_probability}) distribution. Electrons which survive and reach a given pixel contribute to its signal. The advantage of this approach is that it accounts for the natural source of fluctuations in the number of charge carriers, as a result of a statistical nature of the absorption process. Thus this method is expected to provide more realistic description of charge diffusion in a MAPS detector. More details are given below.\\
Apart from fluctuations which are related to the absorption process, the MAPS signal is also influenced by the noise originating from the detector leakage current and readout electronics. Moreover the analog signal in a realistic device undergoes the analog to digital conversion (ADC) which causes the data to be expressed in ADC units instead of charge units (electrons). This is described by the following formula:
\begin{equation}
\label{eq:digi:ADC}
	s_{ADC} = Integer(\eta \cdot s_{e} + N),
\end{equation}
where $s_{ADC}$ is the final pixel signal expressed in the ADC units, $\eta$ is the conversion factor expressed in [ADC/e], $s_{e}$ is the simulated signal expressed in number of electrons, $N$ is the detector noise related to the leakage current and the readout electronics. The detector noise $N$ is known from measurements.\\
The values of parameters $\lambda$ and $\eta$ in simulations have to be chosen such as to obtain the best agreement with the data. This is achieved using $\lambda = 45~\mu$m, $\eta = 0.247$ for MIMOSA-5 and for $\lambda = 50~\mu$m, $\eta = 0.55$ for MIMOSA-18 (see below for details). The comparison shown in fig.~\ref{fig:digi:Comp1} regards average values of signals collected in pixels of reconstructed clusters. For the purpose of this particular comparison the pixels were ordered according to decreasing charge (signal height). The plots present the dependence of the pixel charge versus the pixel number, where the pixel with the highest charge (seed) is the pixel number 1. The data were collected with 6.5~GeV and 5~GeV electron beams for MIMOSA-5 and MIMOSA-18, respectively, with a perpendicular orientation with respect to the detector surfaces. 
\\
In the case of the MIMOSA-5 the simulations are in an excellent agreement with measurements while for the MIMOSA-18 the simulations overestimate signals in pixels which collected little charge. This discrepancy may be due to a different readout architecture used in the MIMOSA-18 which is not accounted for in simulations. Readout in the MIMOSA-18 is based on the so-called self bias diodes for which the response to collected charge is not properly described by formula (\ref{eq:digi:ADC}). In devices equipped with self bias diodes, the pixels are continuously discharging. In order to provide a realistic description of the MIMOSA-18 response the effect of this discharge would have to be included into simulations (this is outside the scope of this work). Thus the above the following studies are restricted to the MIMOSA-5 detector.\\
The level of agreement between simulations and experimental data, fig~\ref{fig:digi:Comp1:M5Comp}, is measured using the $\chi^{2}$ defined as follows:
\begin{equation}
\label{eq:digi:chi2_PixelSignal}
	\chi^{2} = \sum_{i=1}^{N}\frac{(D^{i} - S^{i})^{2}}{(\sigma_{D}^{i})^{2} + (\sigma_{S}^{i})^{2}},
\end{equation}
where $D^{i}$ is the value in the $i$-th experimental point, $S^{i}$ is a corresponding result of a Monte Carlo simulation and $\sigma_{D}^{i}$, $\sigma_{S}^{i}$ are the corresponding uncertainties. The sum in (\ref{eq:digi:chi2_PixelSignal}) runs over the $N$ first points, where the $N$ value has been selected as follows: $N = 5$ for $\theta = 0^{\circ}$, $N = 6$ for $\theta = 45^{\circ}$ and $N = 10$ for $\theta = 70^{\circ}$ (these values are read out from fig.~\ref{fig:tests:inclination_scan:multiplicity}). The $\chi^{2}$/ndf as a function of $\lambda$ are shown in fig.~\ref{fig:digi:CompChi2:M5} for three track inclinations: $\theta = 0^{\circ}, 45^{\circ}$ and $70^{\circ}$.
\\
The minimum in $\chi^{2}$/ndf is observed for $\lambda\approx45~\mu$m for all three values of $\theta$. The precise value of $\lambda$ is obtained as follows. Each of the curves shown in fig.~\ref{fig:digi:CompChi2:M5} is fitted with a parabola using 6 points closest to the visible minimum, as shown in fig.~\ref{fig:digi:Chi2_minimum}. The value of the attenuation length $\lambda$ is assumed for each case as that corresponding to the minimum of the parabola. The following results have been obtained: $\lambda = 45.6\pm3.3~\mu$m, $43.5\pm4.7~\mu$m and $45.3\pm5.3~\mu$m for $\theta=0^{\circ}$, $45^{\circ}$ and $70^{\circ}$, respectively. A weighted average of those values amounts to $\lambda=45.0\pm2.4~\mu$m and is used in further simulations of the MIMOSA-5 detector response to charged particles for all incident angles.\\
\\
For each value of $\lambda$ there exists the best value of $\eta$. It can be seen from fig.~\ref{fig:digi:CompChi2:EtaM5} that the correlation of $\eta$ and $\lambda$ is very much the same for all three values of the angle $\theta$. The value of $\eta$ chosen for further simulations was obtained as an arithmetic average of 3 values corresponding to $\lambda = 45~\mu$m (0.247, 0.247 and 0.248 for $\theta = 0^{\circ}, 45^{\circ}$ and $70^{\circ}$, respectively), and amounts to 0.247$\pm$0.001.

\section{Comparison of simulations with measurements}
\label{ch:digi:digi_test}

In the previous section the procedure to determine the attenuation length $\lambda$ and a conversion factor $\eta$ (\ref{eq:digi:ADC}) was described. In order to test the presented model of the signal formation in the MAPS, more detailed studies of the simulation have been performed. For comparison the same cluster selection criteria were applied to the simulation and to the experimental data ($t_{s} = 4$ and $t_{n} = 0.5$).

\subsection{Charge (signal) characteristics}
\label{ch:digi:digi_test:coll_charge}

\begin{figure}[!htbp]
        \begin{center}
	\subfigure[Seed signal ($\theta=0^{\circ}$)]{
		\includegraphics[width=0.35\textwidth,angle=90]{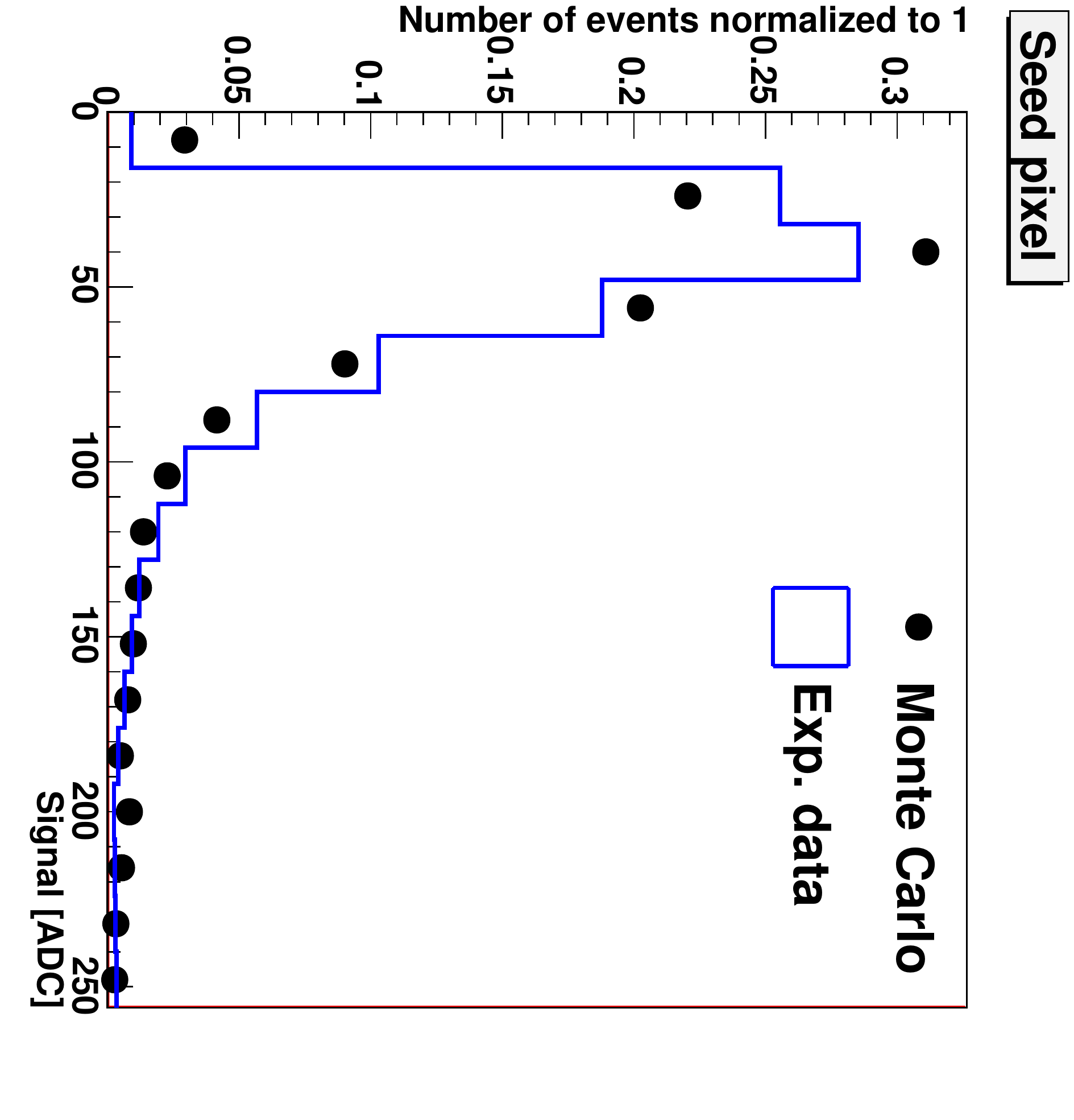}
		\label{fig:digi:test:Seed_2012}
	}
	\hspace{0.05cm}
	\subfigure[Cluster signal ($\theta=0^{\circ}$)]{
		\includegraphics[width=0.35\textwidth,angle=90]{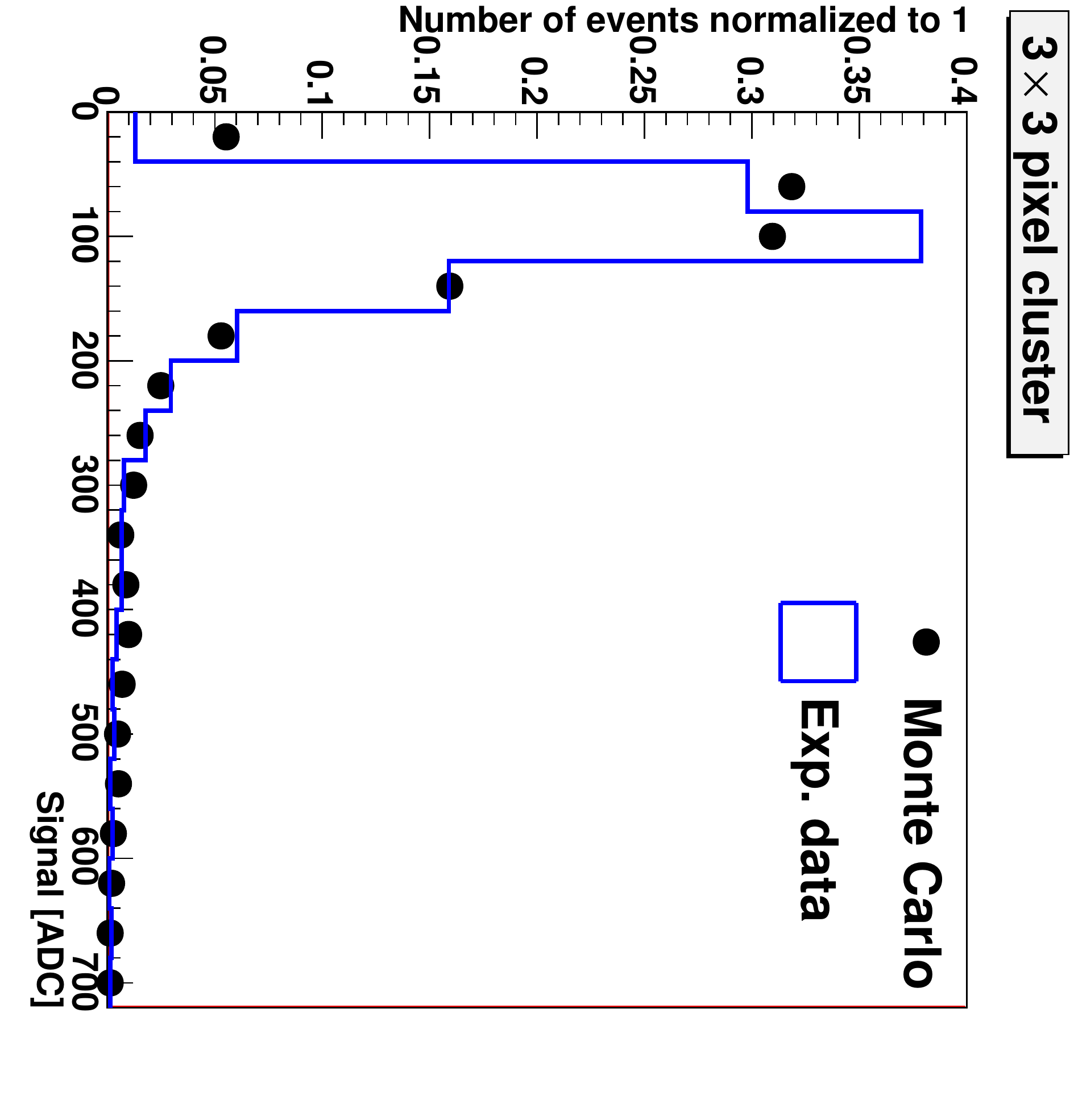}
		\label{fig:digi:test:Cluster_2012}
	}
	\subfigure[Seed signal ($\theta=45^{\circ}$)]{
		\includegraphics[width=0.35\textwidth,angle=90]{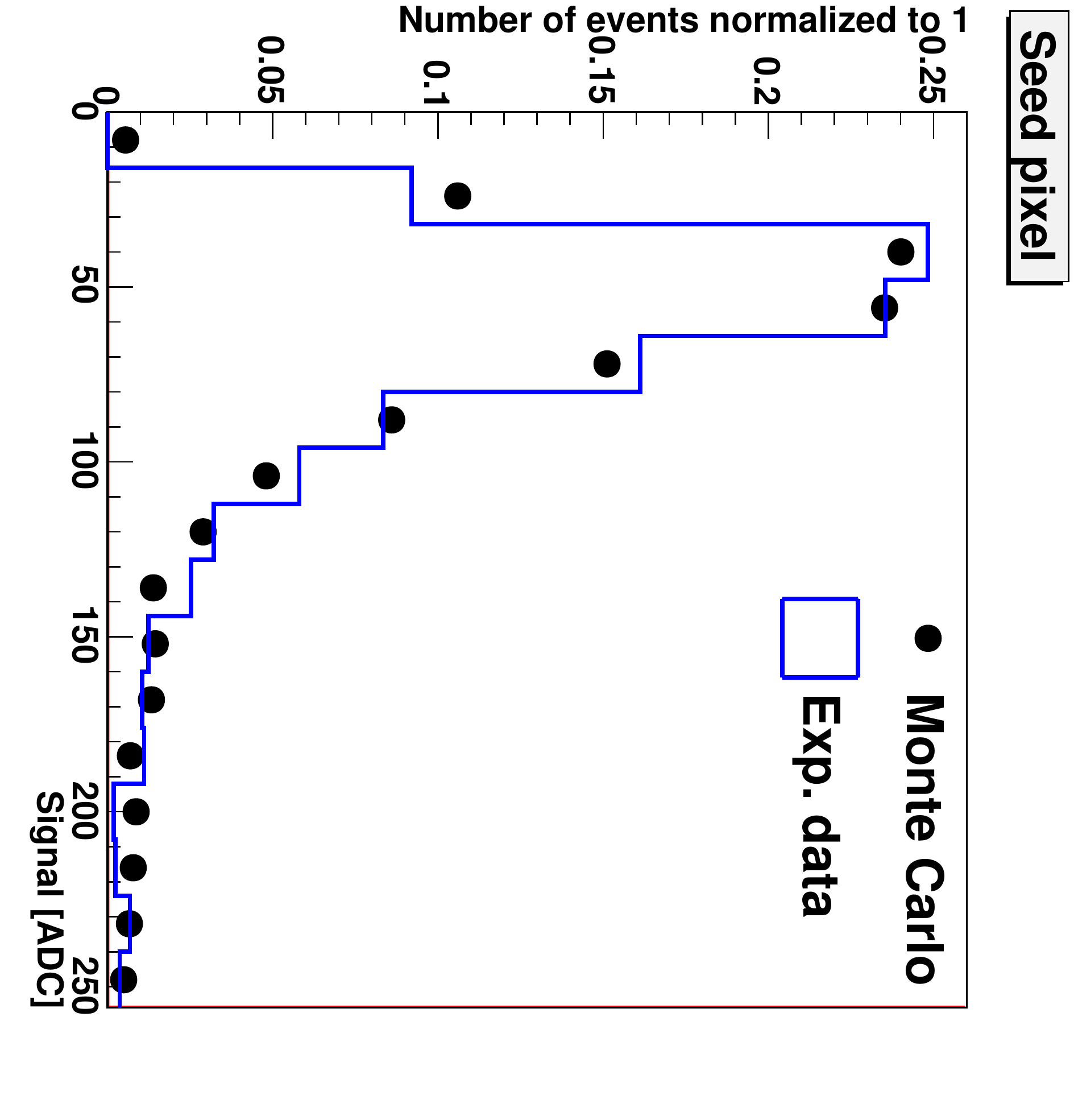}
		\label{fig:digi:test:Seed_2009}
	}
	\hspace{0.05cm}
	\subfigure[Cluster signal ($\theta=45^{\circ}$)]{
		\includegraphics[width=0.35\textwidth,angle=90]{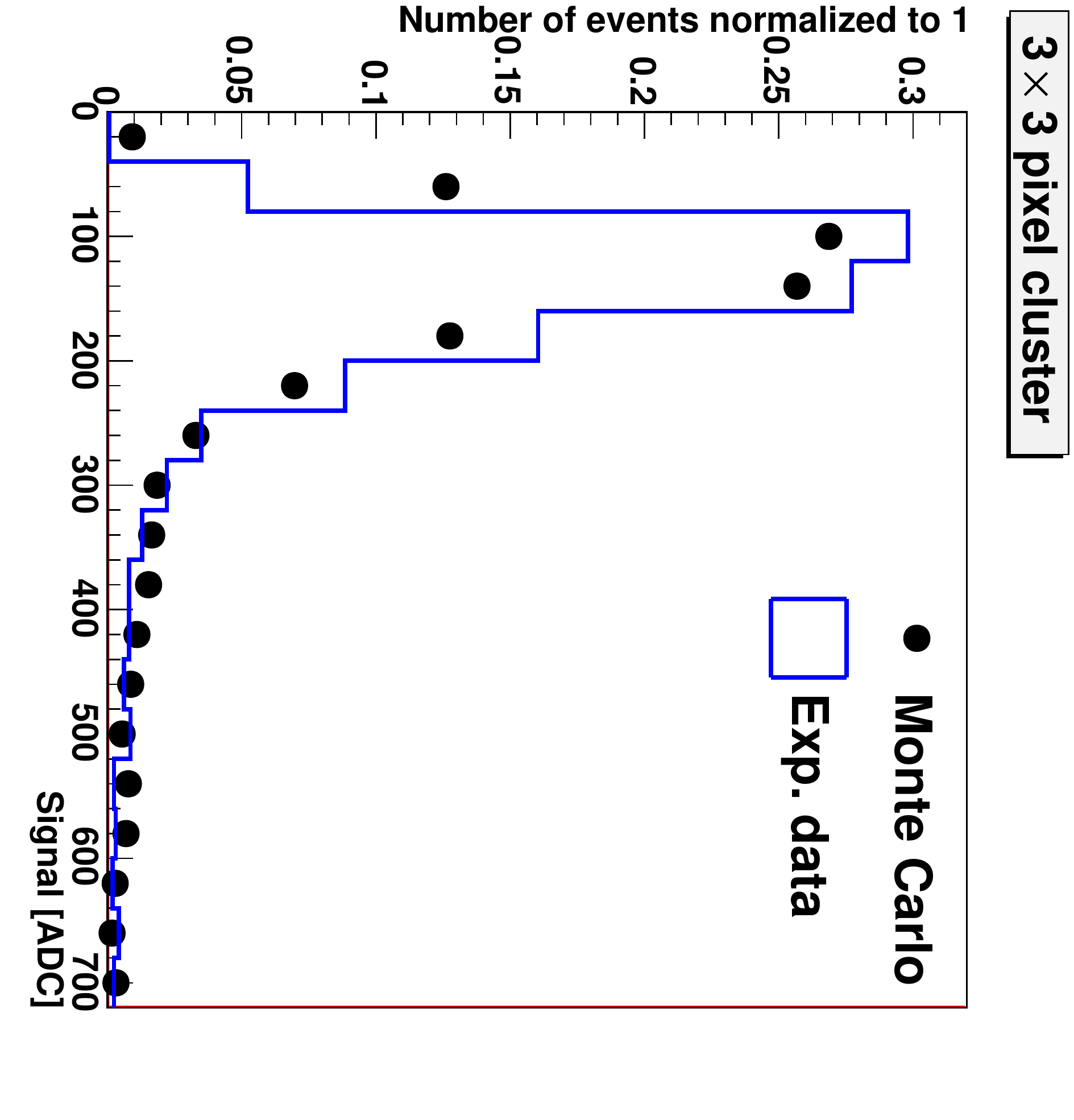}
		\label{fig:digi:test:Cluster_2009}
	}
	\subfigure[Seed signal ($\theta=70^{\circ}$)]{
		\includegraphics[width=0.35\textwidth,angle=90]{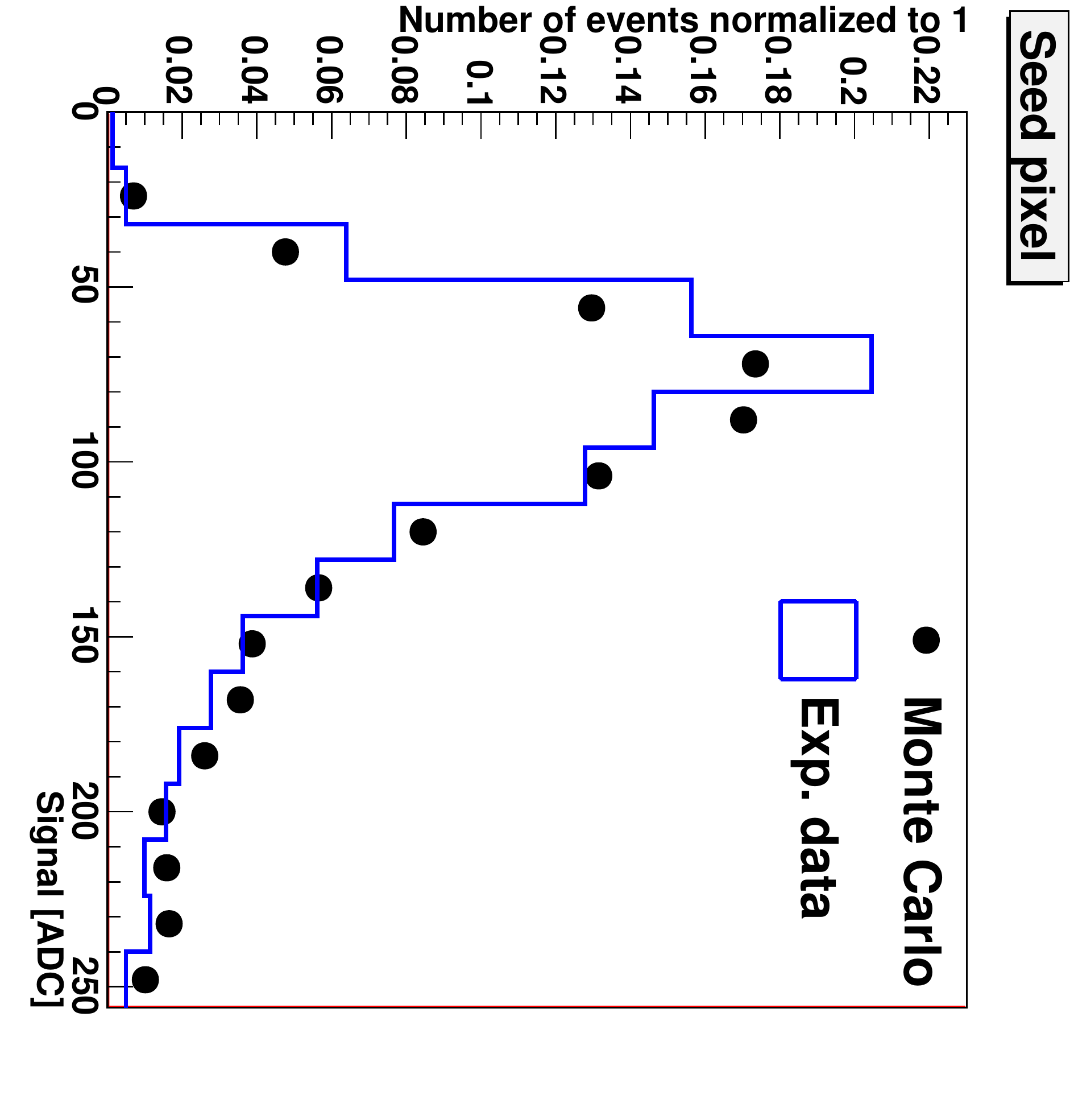}
		\label{fig:digi:test:Seed_4005}
	}
	\hspace{0.05cm}
	\subfigure[Cluster signal ($\theta=70^{\circ}$)]{
		\includegraphics[width=0.35\textwidth,angle=90]{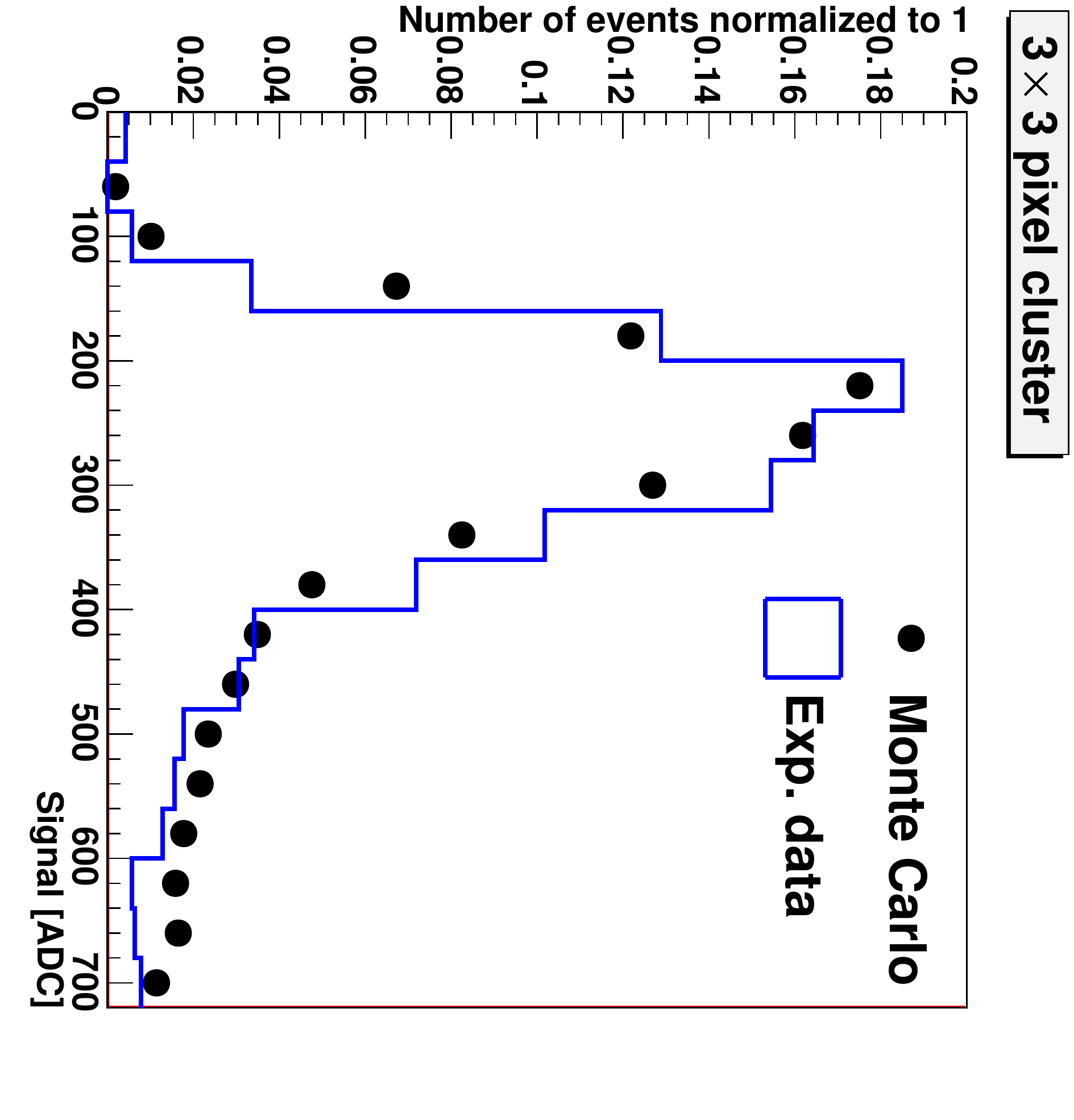}
		\label{fig:digi:test:Cluster_4005}
	}
        \caption[Comparison of the signal distribution in clusters measured and simulated in the MIMOSA-5 detector.]{Charge (signal) distributions measured in the seed pixels (a), (c), (e) and in the $3\times3$ pixel clusters (b), (d), (f) with electron tracks of 6.5~GeV which incidents detector surface at $\theta=0^{\circ}$, $45^{\circ}$ and  $70^{\circ}$, respectively. Histograms denote measured distributions while dots results of simulations.}
        \label{fig:digi:test:M5}
        \end{center}
\end{figure}

\begin{figure}[!h]
        \begin{center}
	\subfigure[Seed S/N ($\theta=0^{\circ}$)]{
		\includegraphics[width=0.29\textwidth,angle=90]{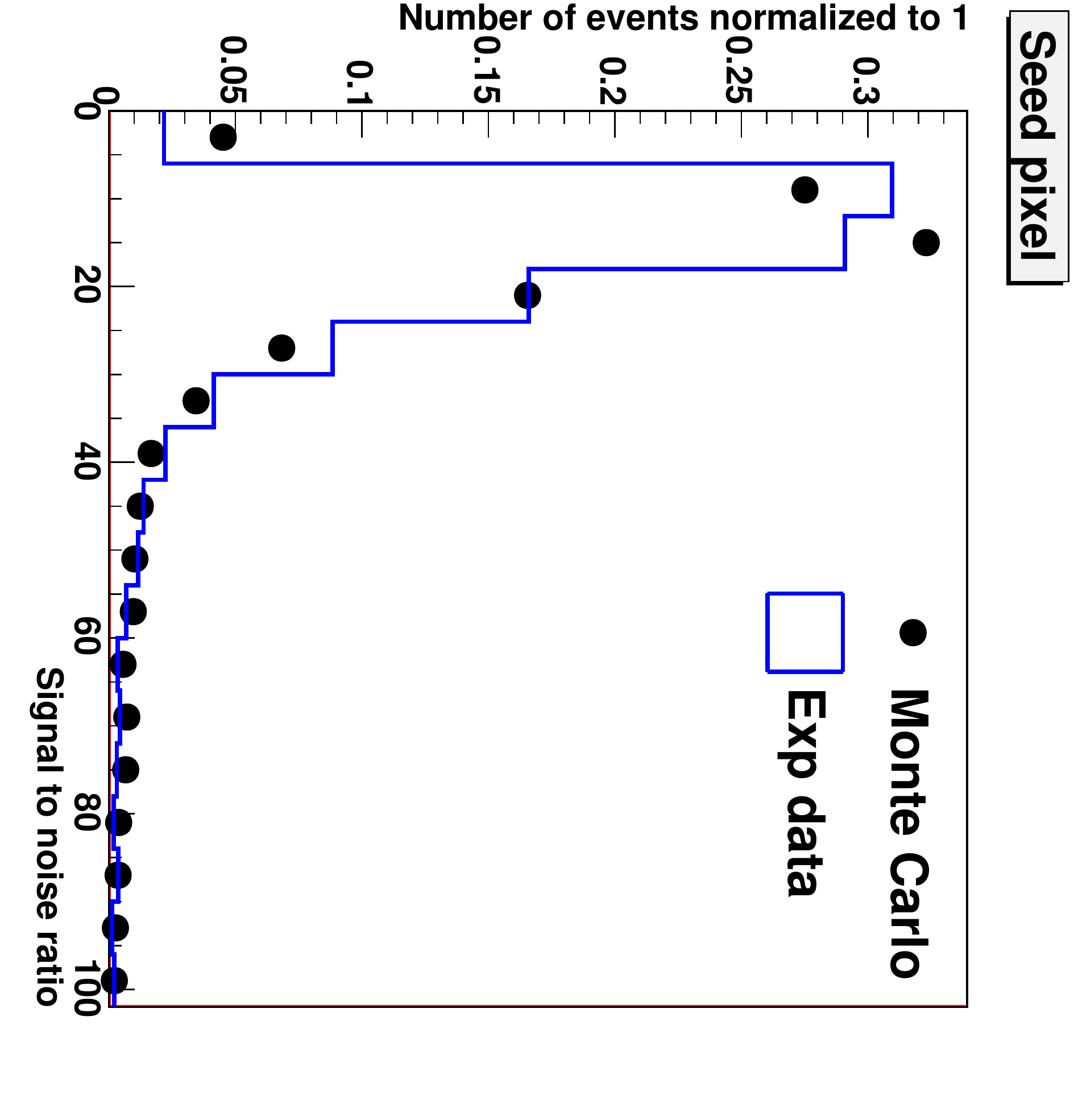}
		\label{fig:digi:test:SeedS2N_2012}
	}
	\hspace{0.05cm}
	\subfigure[Seed S/N ($\theta=45^{\circ}$)]{
		\includegraphics[width=0.29\textwidth,angle=90]{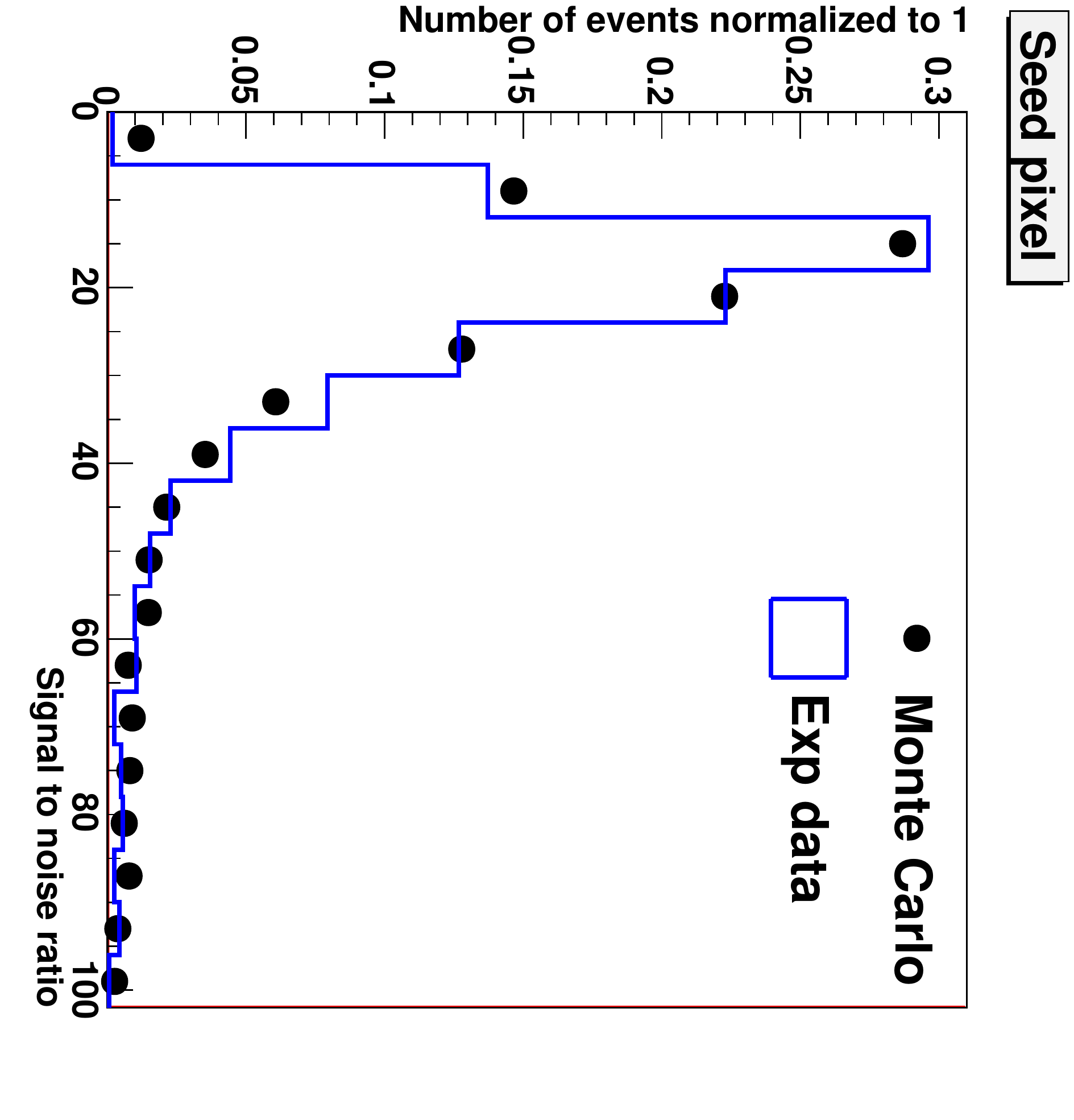}
		\label{fig:digi:test:SeedS2N_2009}
	}
	\hspace{0.05cm}
	\subfigure[Seed S/N ($\theta=70^{\circ}$)]{
		\includegraphics[width=0.29\textwidth,angle=90]{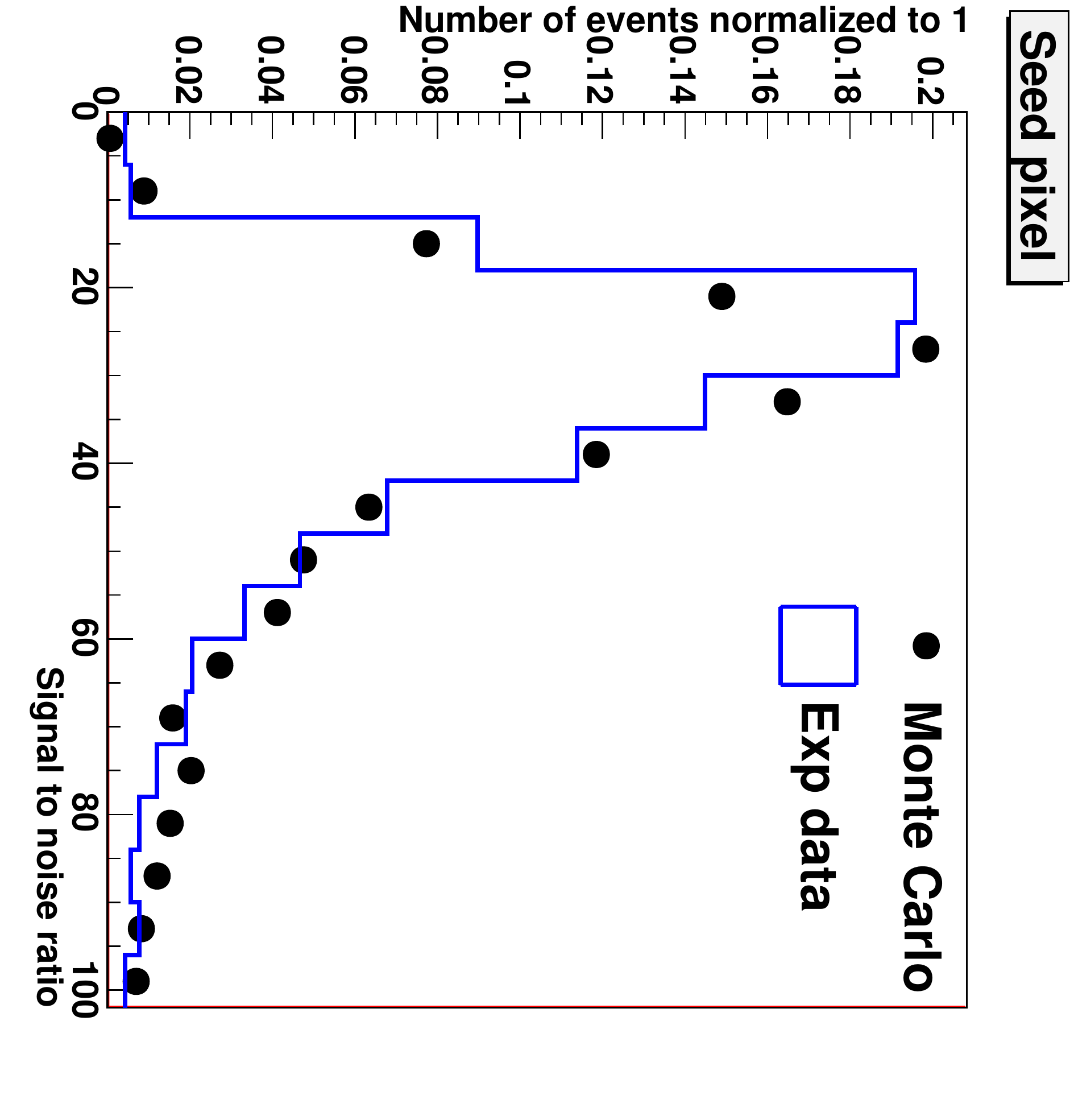}
		\label{fig:digi:test:SeedS2N_4005}
	}
        \caption[Comparison of the signal to noise distribution in clusters measured and simulated in the MIMOSA-5 detector.]{Distributions of the signal to noise ratio measured in the seed pixel for the 6.5~GeV electron tracks traversing detector volume at $\theta=0^{\circ}$, $45^{\circ}$ and  $70^{\circ}$, respectively. Histograms denote measured distributions while dots results of simulations.}
        \label{fig:digi:test:S2N}
        \end{center}
\end{figure}

The charge (signal) distributions (in terms of ADC counts) are shown in fig.~\ref{fig:digi:test:M5}. Histograms denote measured distributions while dots results of simulations. Rows correspond to tracks incident at $\theta=0^{\circ}$, $45^{\circ}$ and $70^{\circ}$. In the first column the charge distributions are shown for the seed pixels and in the second one charge distributions for $3\times3$ pixel clusters.
\\
The measured shapes of the charge distributions are of Landau type with peak positions shifting towards higher values as $\theta$ increases. Simulations reproduce this feature of the data with small disagreement in the height of the peak; the peak position is well reproduced. Those insufficiencies are discussed in more details at the end of this chapter.\\
The presented model of signal formation in the MAPS devices provides reasonably good qualitative description of the MAPS detector response to the charged particles. With this simple charge diffusion model it is possible to predict the amount of the signal collected in the MIMOSA-5 detector exposed to electron tracks at different inclinations.\\
In fig.~\ref{fig:digi:test:S2N} distributions of the signal to noise ratio in the seed pixel for $\theta=0^{\circ}$, $45^{\circ}$ and  $70^{\circ}$ are shown. Since measured distributions are very well described by simulation the simple and naive method of including detector noise and ADC conversion (\ref{eq:digi:ADC}) seems to be sufficient.

\subsection{Charge sharing among pixels}
\label{ch:digi:digi_test:charge_sharing}

\begin{figure}[!h]
        \begin{center}
	\subfigure[$\theta = 0^{\circ}$]{
		\includegraphics[width=0.29\textwidth,angle=90]{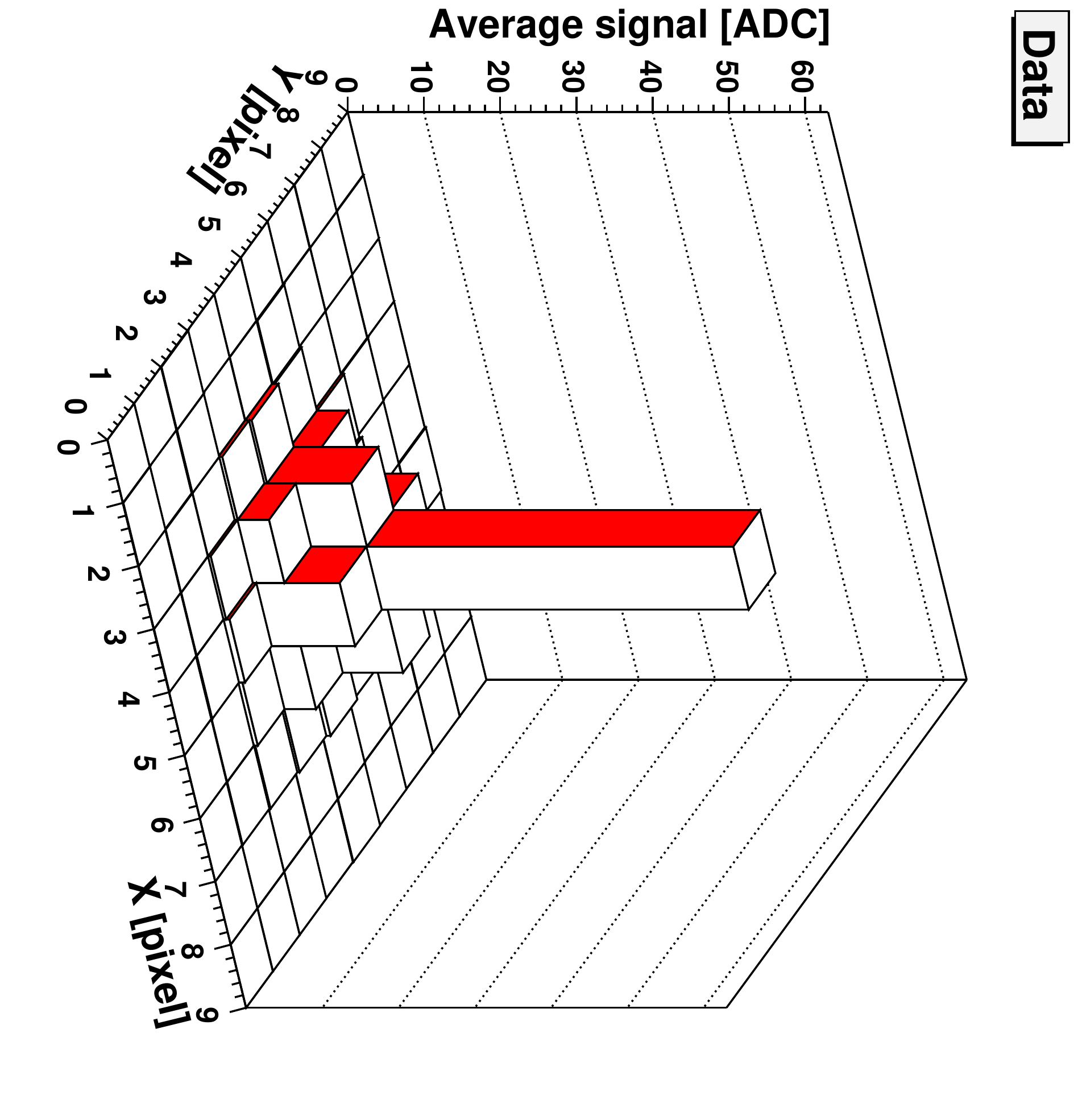}
		\label{fig:digi:MeanClust:Data_2012}
	}
	\hspace{0.05cm}
	\subfigure[$\theta = 0^{\circ}$]{
		\includegraphics[width=0.29\textwidth,angle=90]{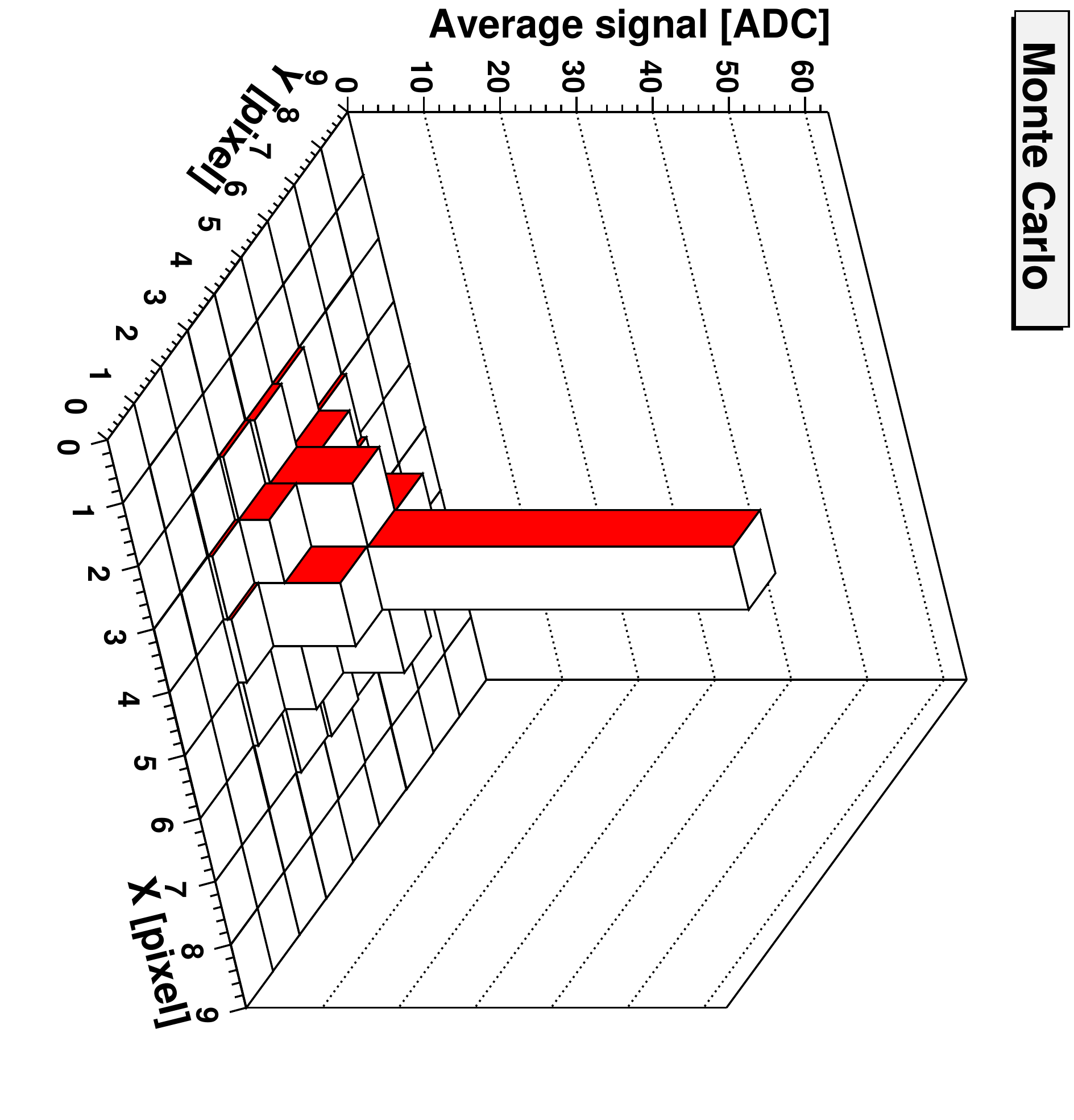}
		\label{fig:digi:MeanClust:MC_2012}
	}
	\hspace{0.05cm}
	\subfigure[$\theta = 0^{\circ}$]{
		\includegraphics[width=0.29\textwidth,angle=90]{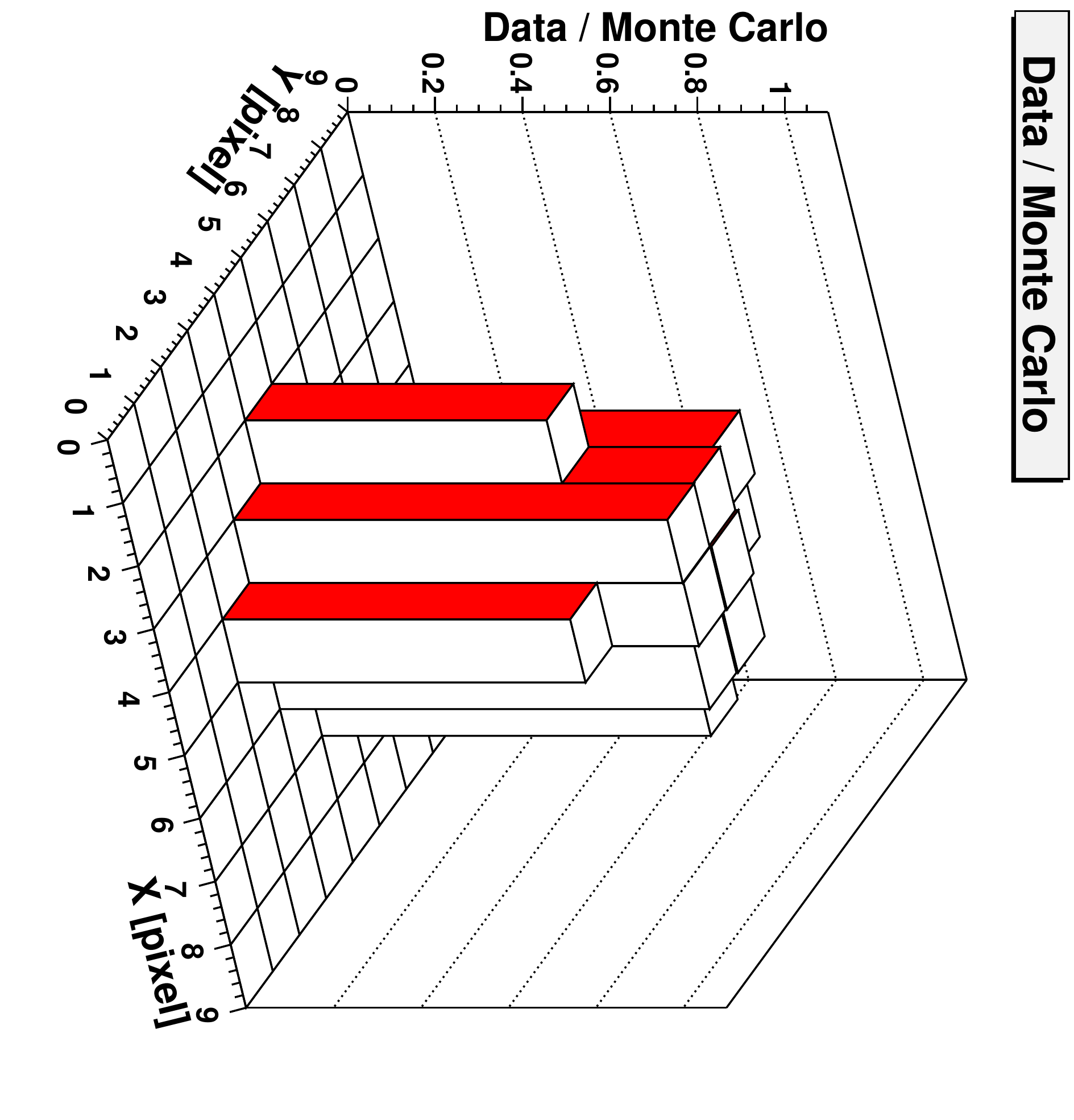}
		\label{fig:digi:MeanClust:Comp_2012}
	}
	\subfigure[$\theta = 45^{\circ}$]{
		\includegraphics[width=0.29\textwidth,angle=90]{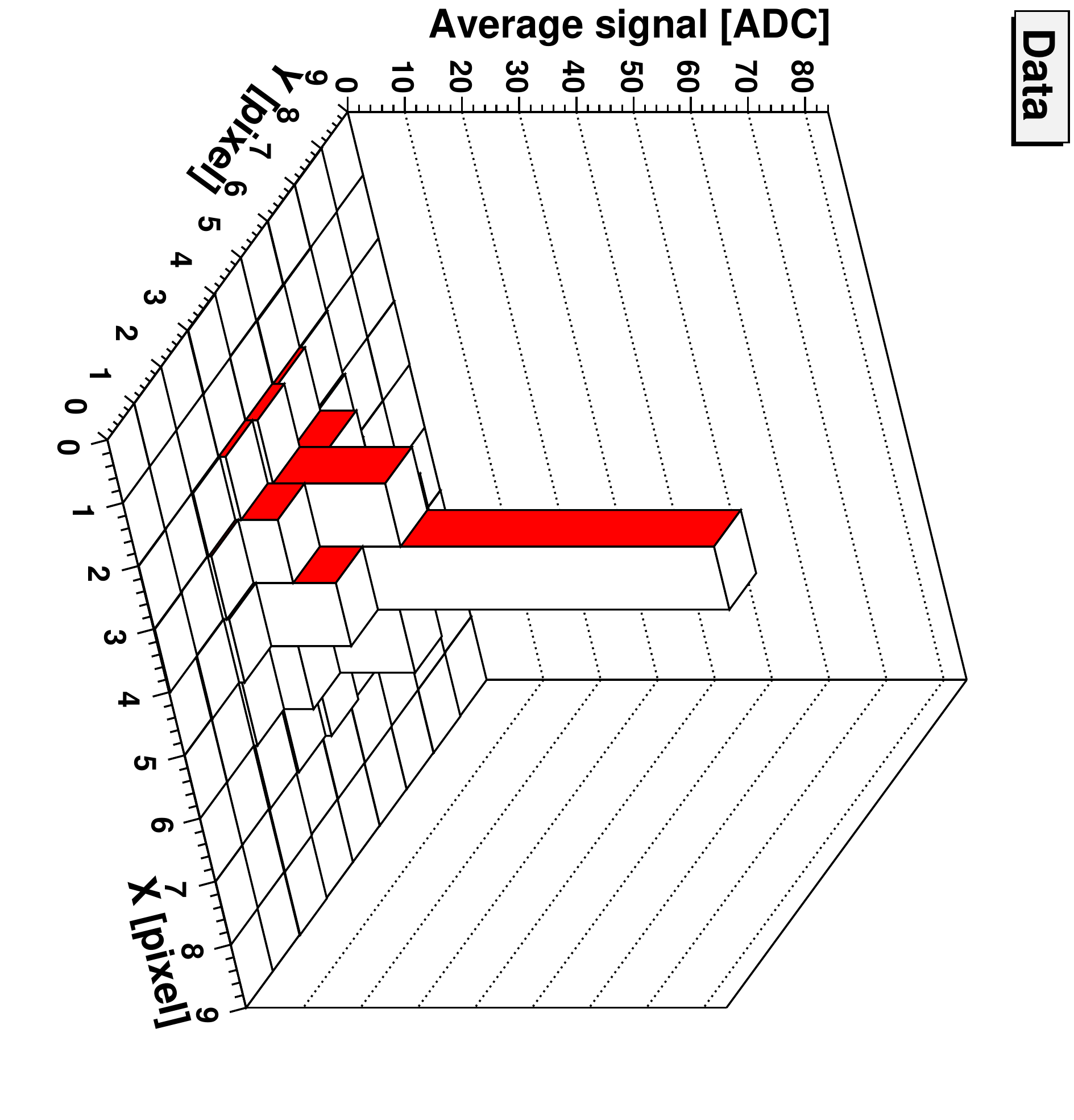}
		\label{fig:digi:MeanClust:Data_2009}
	}
	\hspace{0.05cm}
	\subfigure[$\theta = 45^{\circ}$]{
		\includegraphics[width=0.29\textwidth,angle=90]{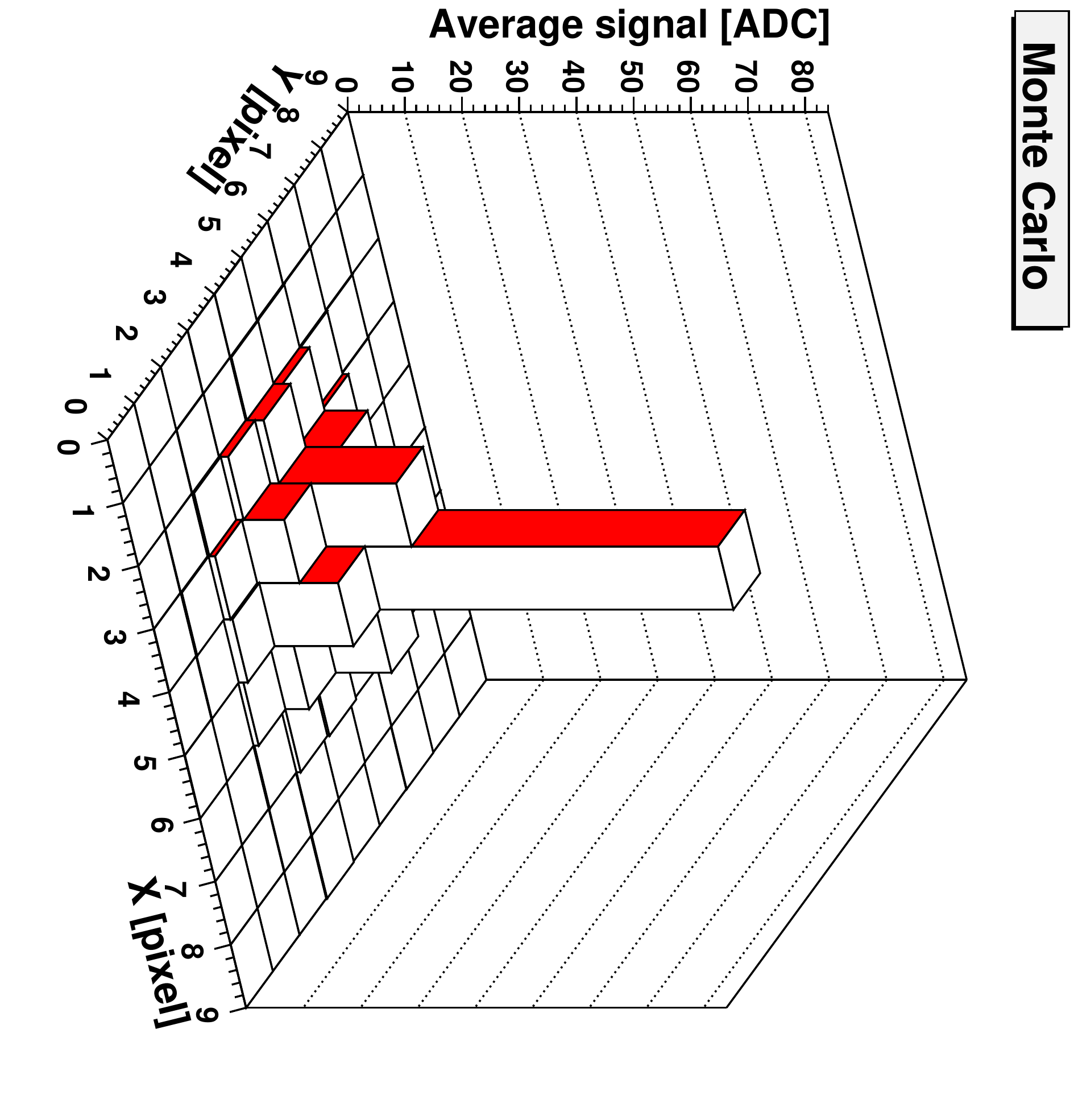}
		\label{fig:digi:MeanClust:MC_2009}
	}
	\hspace{0.05cm}
	\subfigure[$\theta = 45^{\circ}$]{
		\includegraphics[width=0.29\textwidth,angle=90]{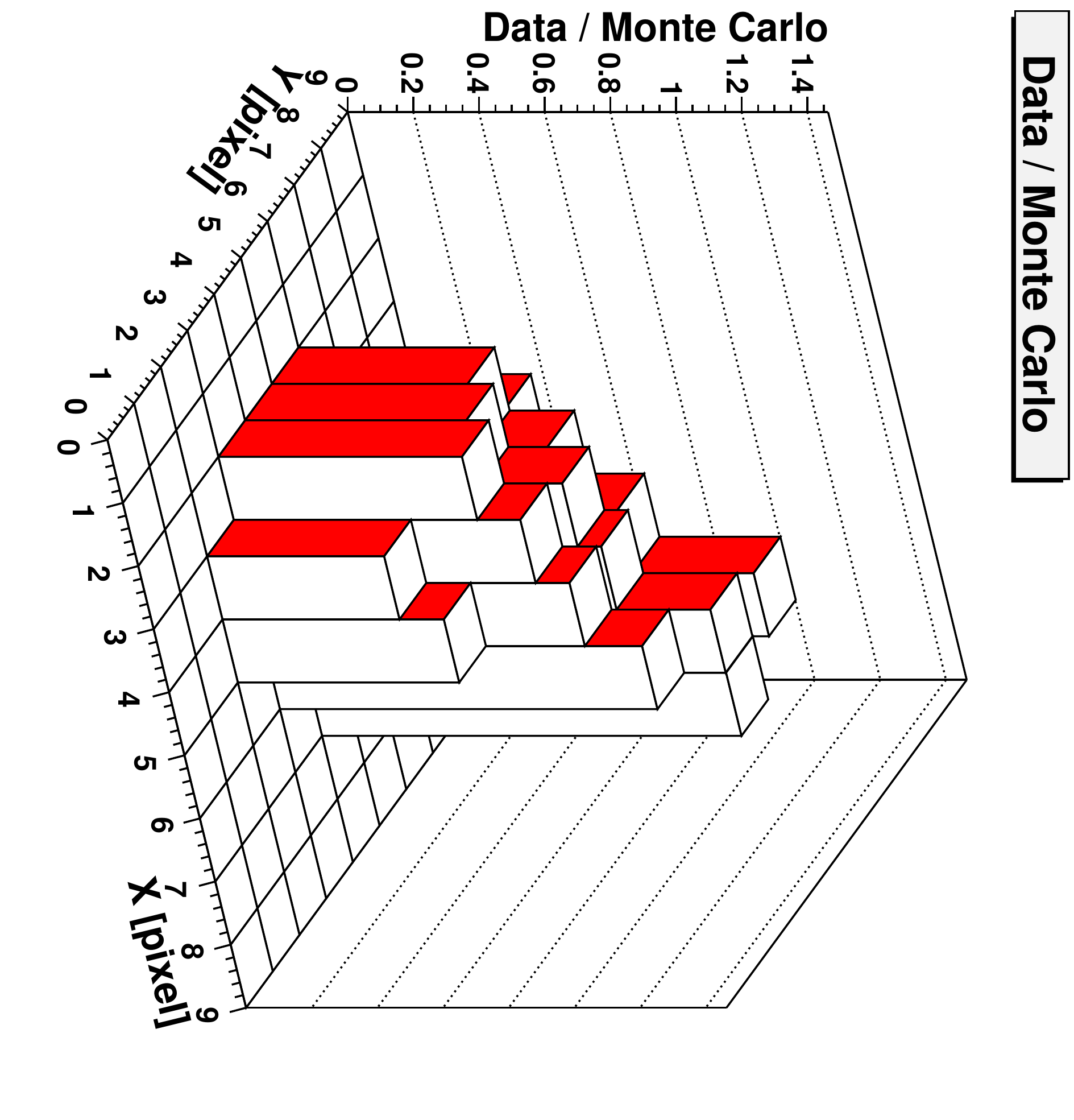}
		\label{fig:digi:MeanClust:Comp_2009}
	}
        \caption[Average clusters measured and simulated in the MIMOSA-5 detector]{Plots in fig. (a) and (d) contains average clusters measured in the MIMOSA-5 detector for incident angle $\theta$ equal $0^{\circ}$ and $45^{\circ}$, respectively. Fig. (b) and (e) refer to the corresponding simulated average clusters. In fig. (c) and (d) results of dividing measured average clusters (a) and (d) by related simulated average clusters (b) and (e) are shown.}
        \label{fig:digi:MeanClust1}
        \end{center}
\end{figure}

\begin{figure}[!h]
        \begin{center}
	\subfigure[$\theta = 60^{\circ}$]{
		\includegraphics[width=0.29\textwidth,angle=90]{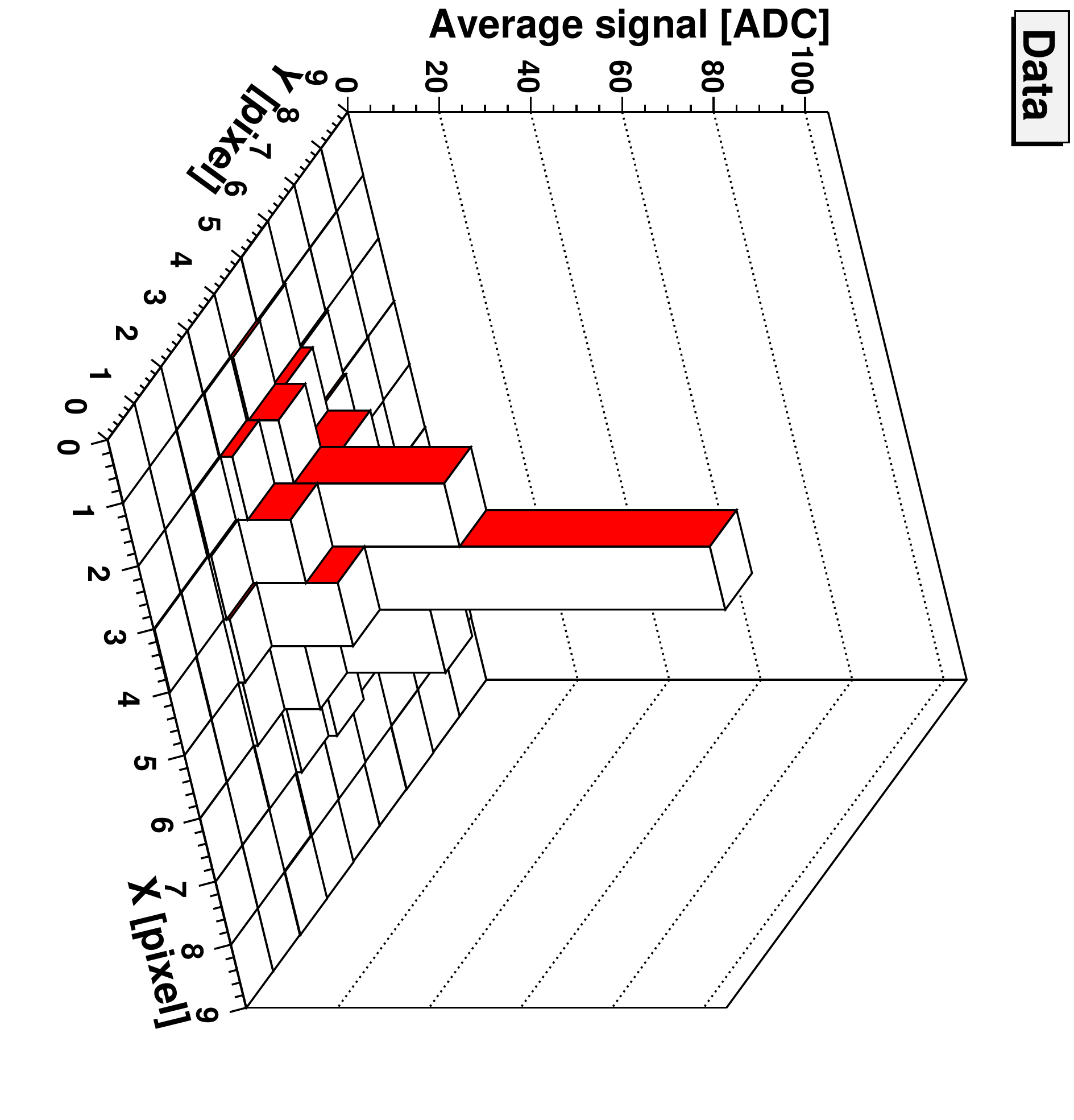}
		\label{fig:digi:MeanClust:Data_4002}
	}
	\hspace{0.05cm}
	\subfigure[$\theta = 60^{\circ}$]{
		\includegraphics[width=0.29\textwidth,angle=90]{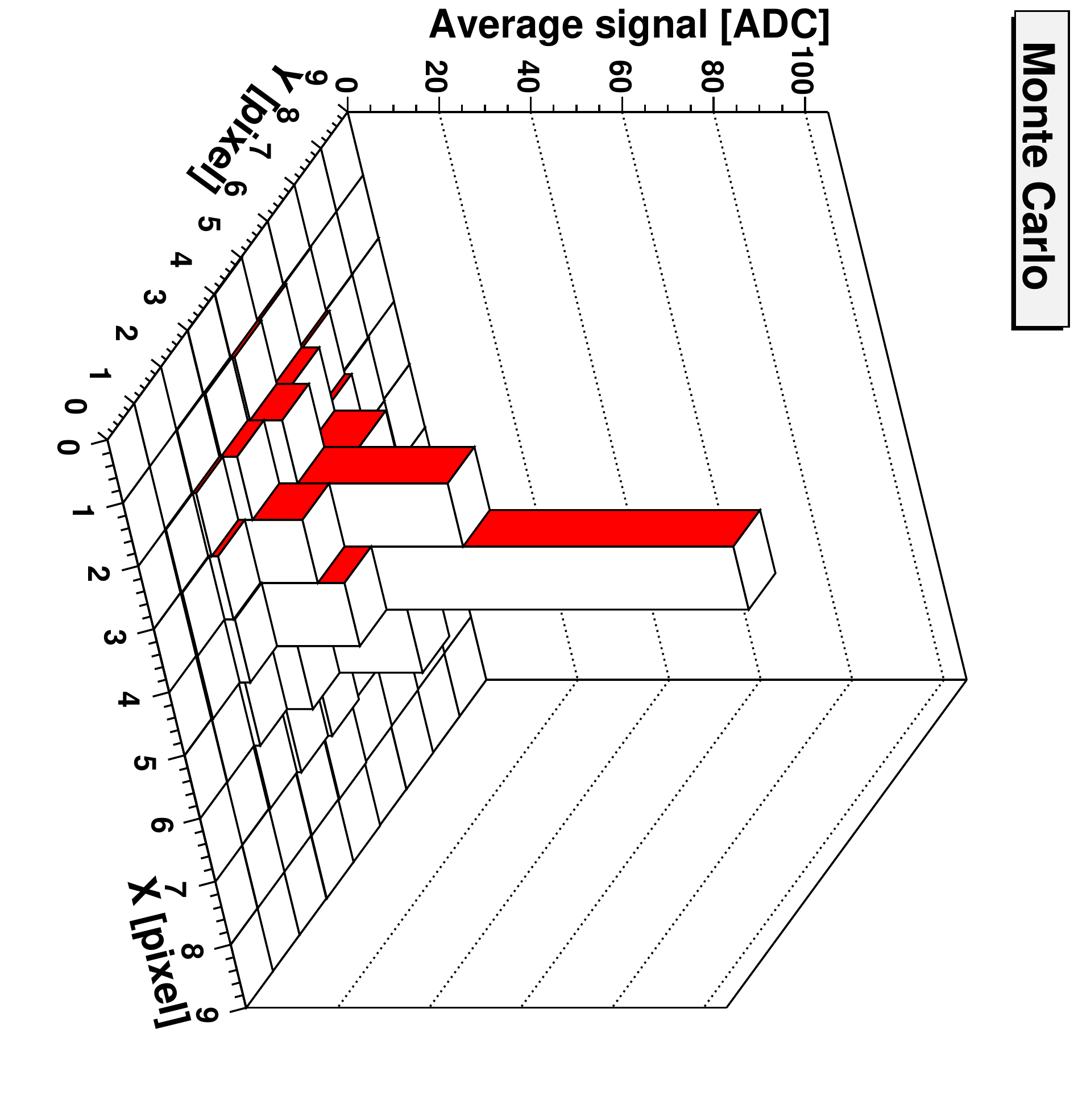}
		\label{fig:digi:MeanClust:MC_4002}
	}
	\hspace{0.05cm}
	\subfigure[$\theta = 60^{\circ}$]{
		\includegraphics[width=0.29\textwidth,angle=90]{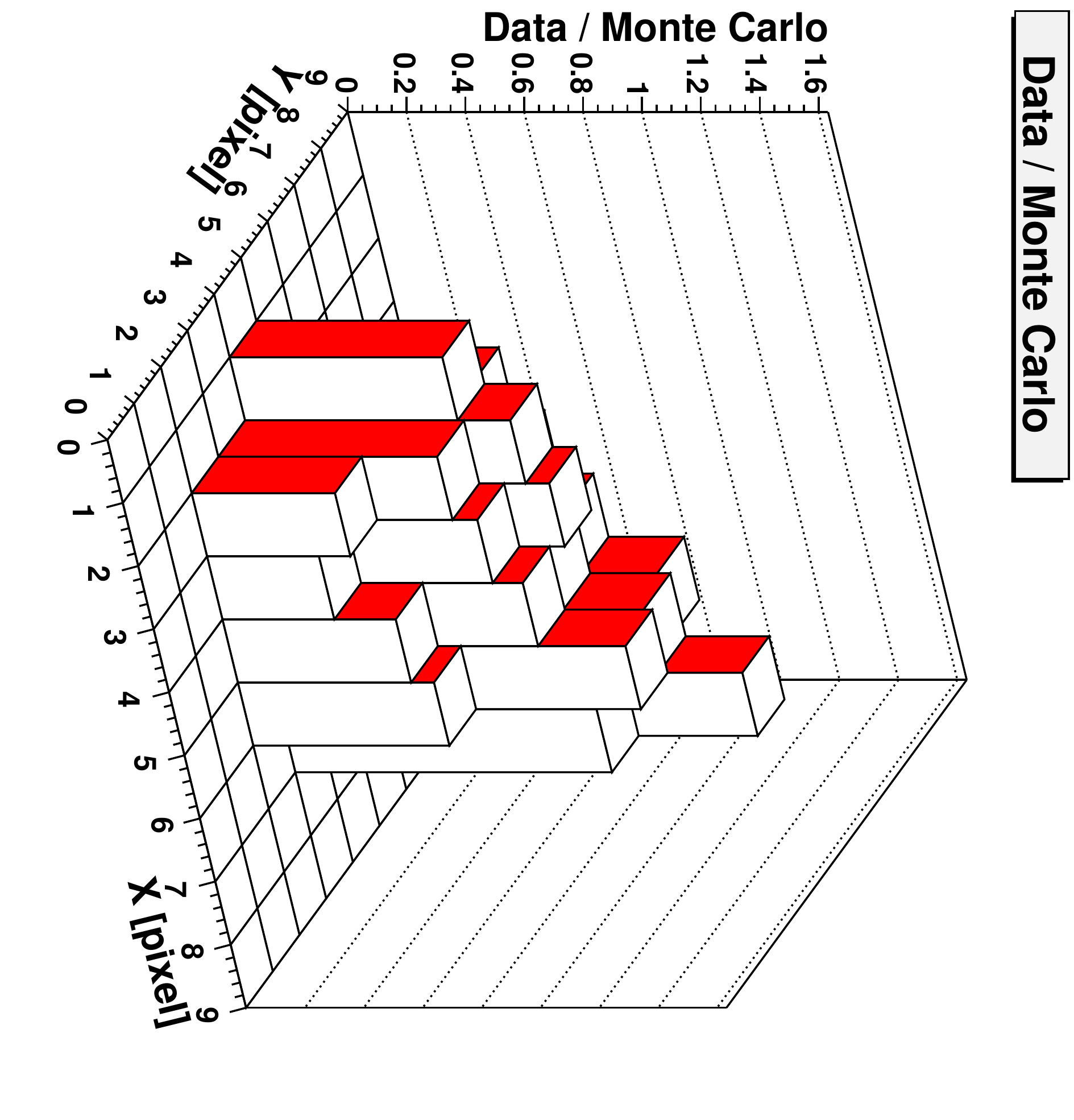}
		\label{fig:digi:MeanClust:Comp_4002}
	}
	\subfigure[$\theta = 75^{\circ}$]{
		\includegraphics[width=0.29\textwidth,angle=90]{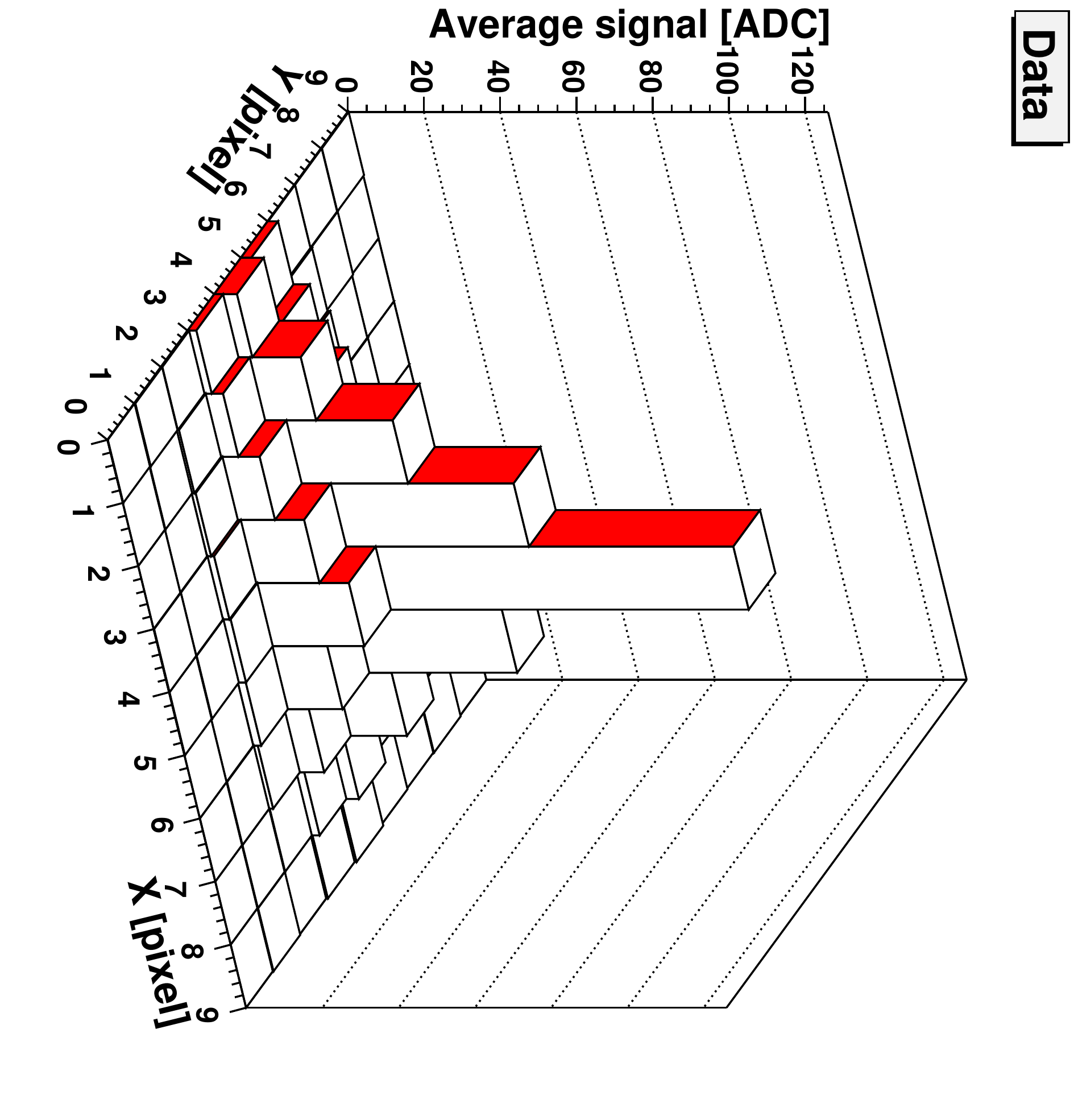}
		\label{fig:digi:MeanClust:Data_4007}
	}
	\hspace{0.05cm}
	\subfigure[$\theta = 75^{\circ}$]{
		\includegraphics[width=0.29\textwidth,angle=90]{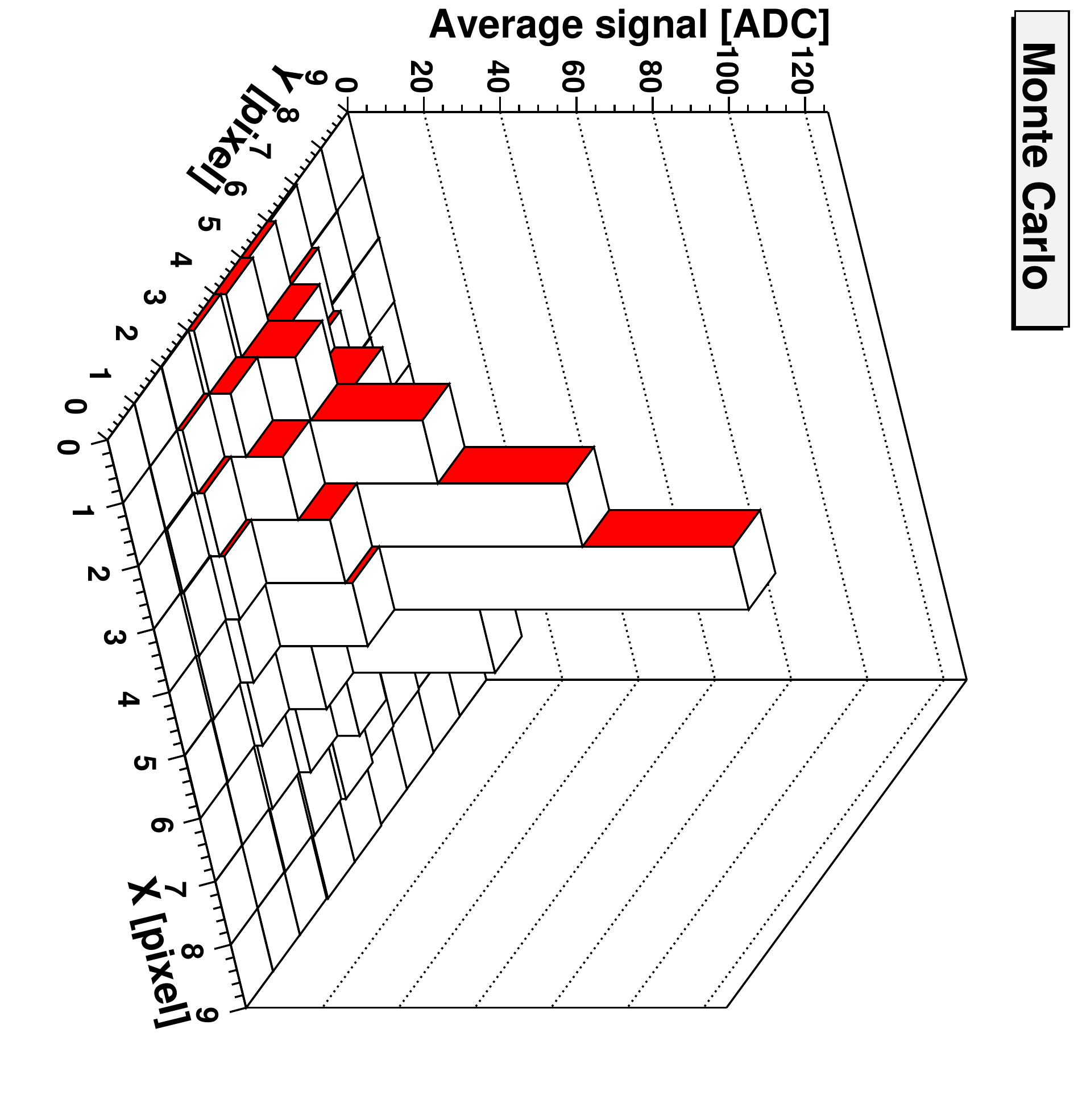}
		\label{fig:digi:MeanClust:MC_4007}
	}
	\hspace{0.05cm}
	\subfigure[$\theta = 75^{\circ}$]{
		\includegraphics[width=0.29\textwidth,angle=90]{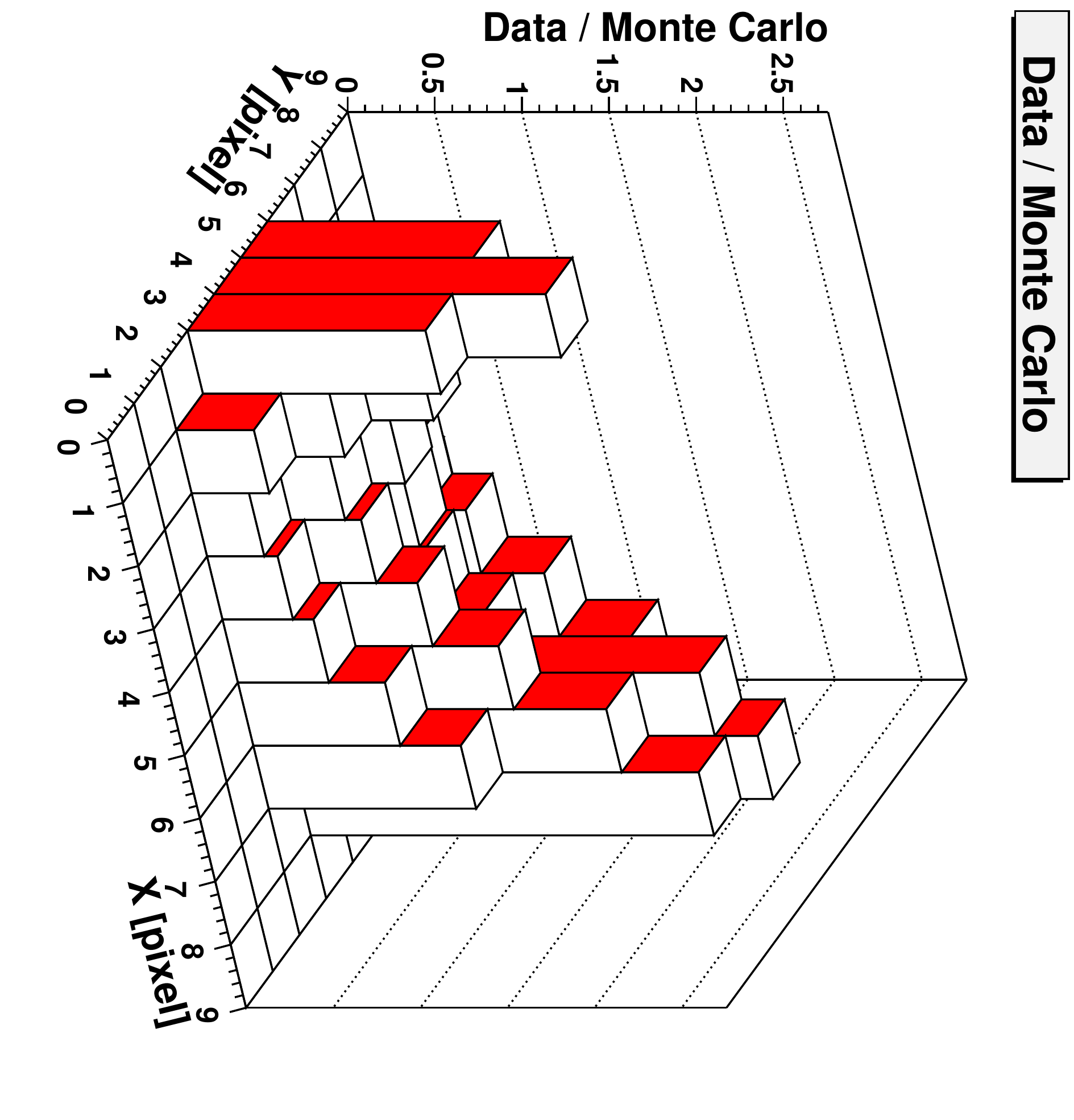}
		\label{fig:digi:MeanClust:Comp_4007}
	}
        \caption[Average clusters measured and simulated in the MIMOSA-5 detector]{Plots in fig. (a) and (d) contains average clusters measured in the MIMOSA-5 detector for incident angle $\theta$ equal $60^{\circ}$ and $75^{\circ}$, respectively. Fig. (b) and (e) refer to the corresponding simulated average clusters. In fig. (c) and (d) results of dividing measured average clusters (a) and (d) by related simulated average clusters (b) and (e) are shown.}
        \label{fig:digi:MeanClust2}
        \end{center}
\end{figure}

The algorithms used for reconstructing the hit position exploit information on the charge sharing between pixels contained in a cluster. Thus the parametrisation of the MAPS detector response should provide a correct description of charge sharing among pixels. The average clusters measured in the MIMOSA-5 for different track inclinations are compared to the simulated ones in fig.~\ref{fig:digi:MeanClust1} and ~\ref{fig:digi:MeanClust2}. In case of the average clusters measured with the inclined tracks, the charged particles enter the detector to the right of the seed pixel (larger $x$ values).
\\
The measured and simulated average cluster for tracks perpendicular to the detector surface are symmetric as shown in fig.~\ref{fig:digi:MeanClust:Data_2012} and \ref{fig:digi:MeanClust:MC_2012}. The ratio of data and Monte Carlo is close to flat as can be seen in fig.~\ref{fig:digi:MeanClust:Comp_2012}. With the increasing track inclination the measured and simulated average clusters elongate in the direction of a track and become asymmetric as shown in fig.~\ref{fig:digi:MeanClust:Data_2009} and \ref{fig:digi:MeanClust:MC_2009}. Unfortunately the proportion between average signals collected in pixels of the measured clusters is not perfectly reproduced by simulations. The discrepancy between simulation and measurements is well visible for large track inclinations (see fig.~\ref{fig:digi:MeanClust2}).
\\
Comparison of simulated and measured average clusters for inclined tracks, in particular those at large inclination angles $\theta$ like in fig.~\ref{fig:digi:MeanClust:Data_4007} and \ref{fig:digi:MeanClust:MC_4007}, indicates that simulated signals in pixels on the left side of the seed (for $0 < x < 3$) are overestimated with respect to the corresponding measured signals. For pixels placed on the opposite side of the seed (for $5 < x < 8$) the situation is different and the simulated average signals for those pixels are underestimated with respect to measurements. Thus the spatial distribution of charge in the simulated clusters exhibits higher asymmetry in the direction of the charged particle track than the spatial distribution of signal in the measured clusters.\\
Additionally for high incident angles $\theta$ the simulated clusters have smaller elongation than the measured clusters. This effect i very well visible in fig.~\ref{fig:digi:MeanClust:Comp_4007}, where the marginal pixels for $x=0$ in the measured cluster present much higher average signal than those in the simulated clusters. 

\subsection{The hit position}
\label{ch:digi:digi_test:hit_position}

\begin{figure}[!htbp]
        \begin{center}
	\subfigure[$\theta = 0^{\circ}$]{
		\includegraphics[width=0.45\textwidth,angle=0]{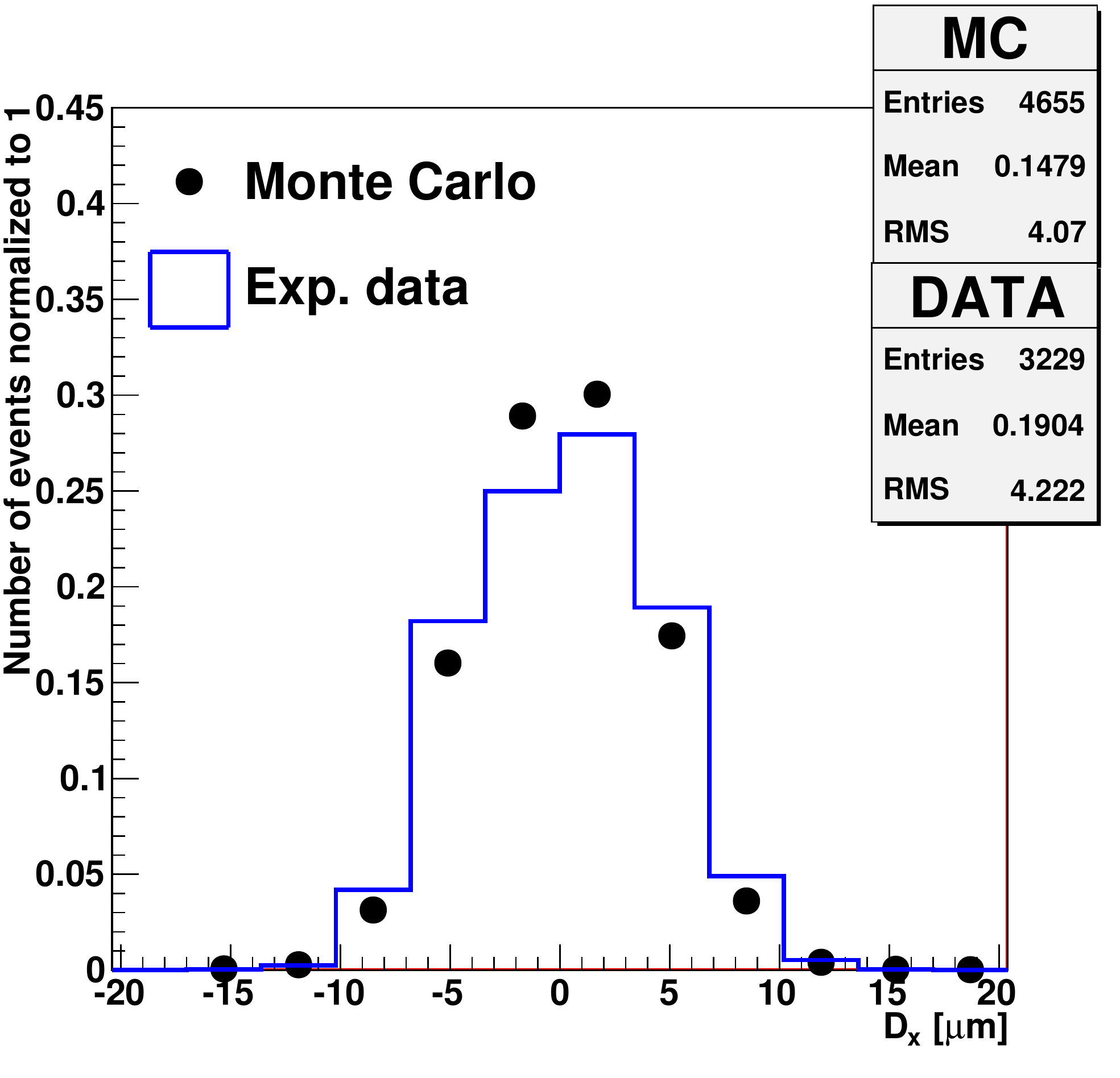}
		\label{fig:digi:CompCog:2012}
	}
	\hspace{0.1cm}
	\subfigure[$\theta = 45^{\circ}$]{
		\includegraphics[width=0.45\textwidth,angle=0]{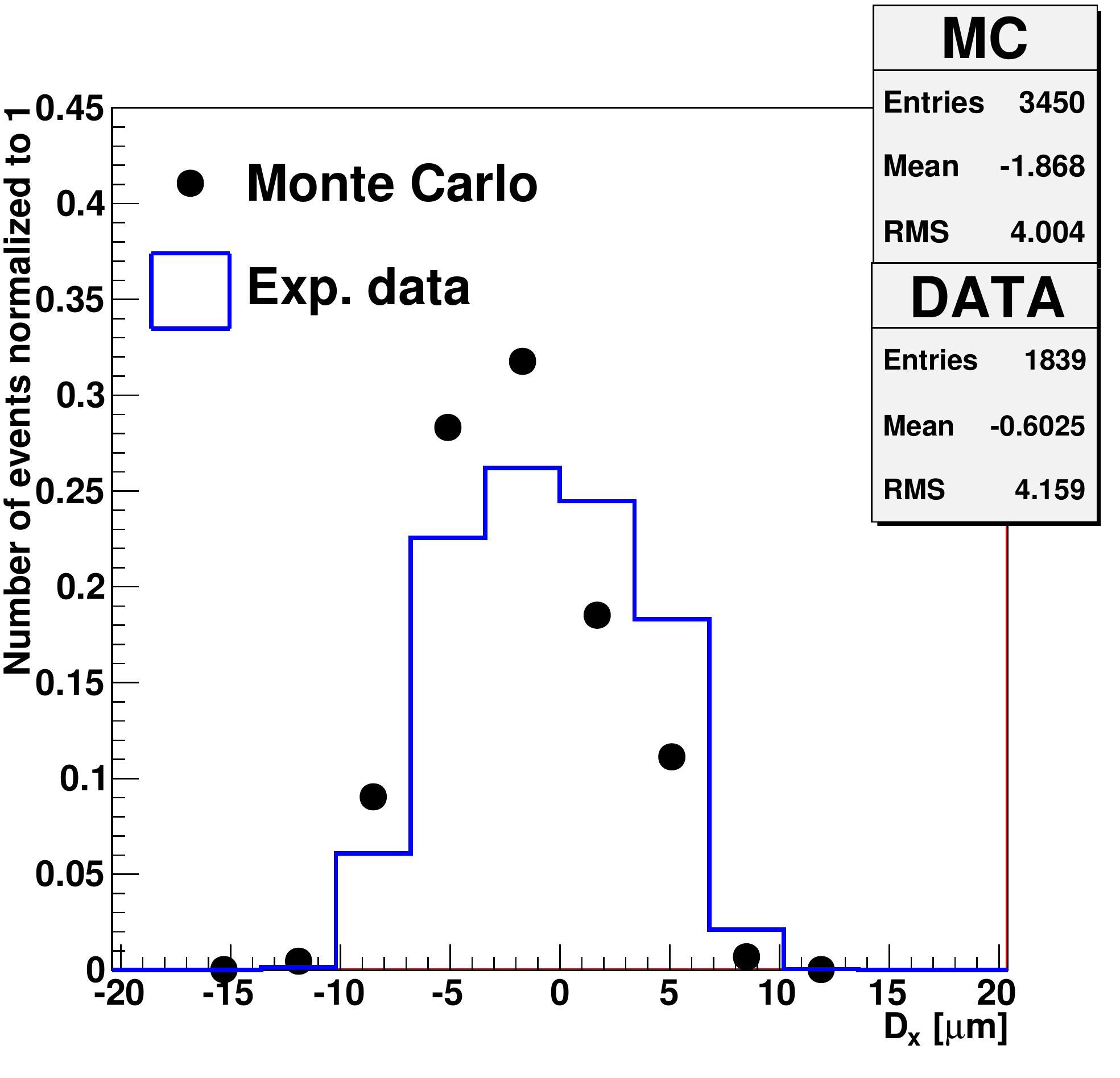}
		\label{fig:digi:CompCog:2009}
	}
	\subfigure[$\theta = 60^{\circ}$]{
		\includegraphics[width=0.45\textwidth,angle=0]{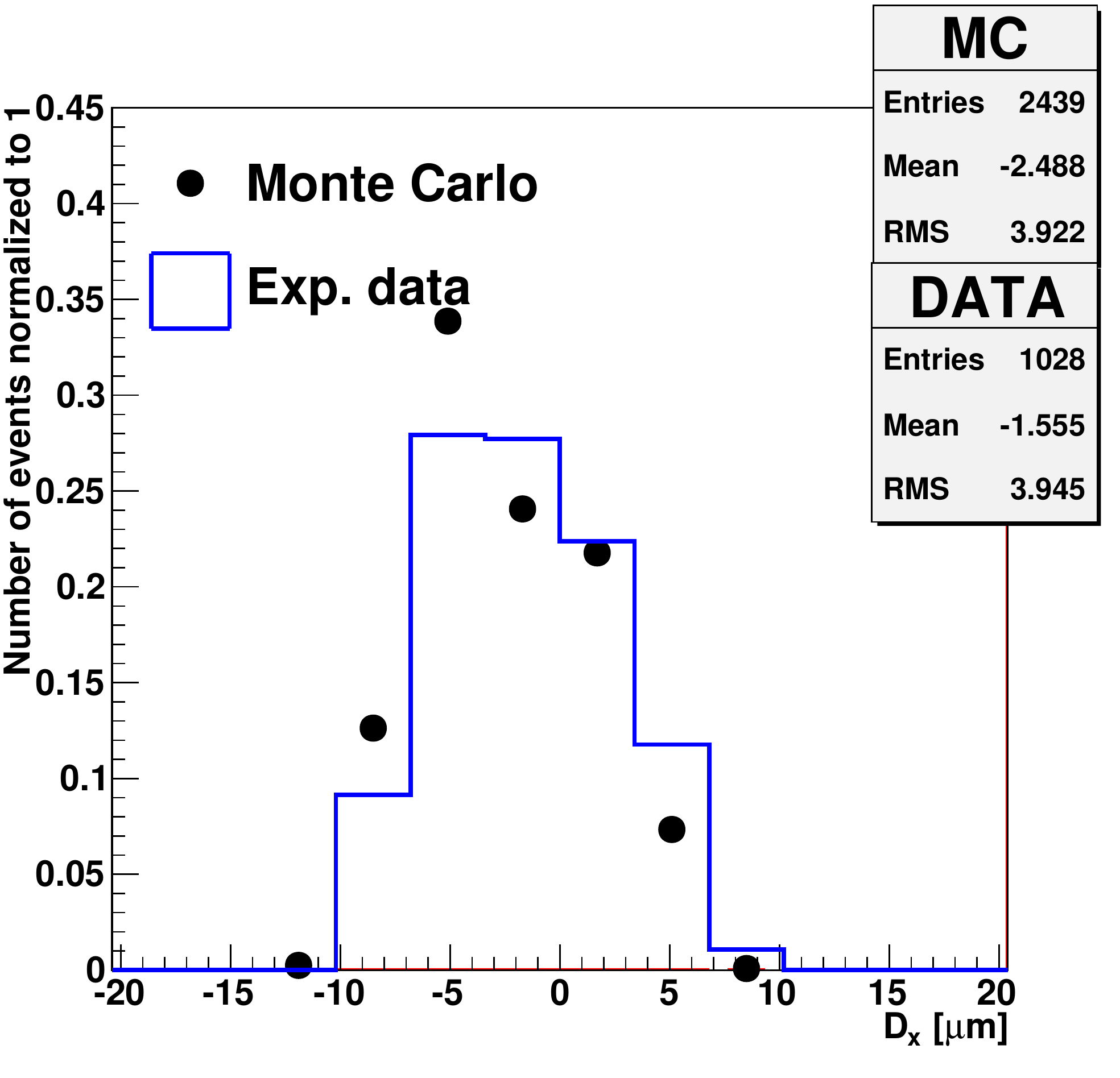}
		\label{fig:digi:CompCog:4002}
	}
	\hspace{0.1cm}
	\subfigure[$\theta = 75^{\circ}$]{
		\includegraphics[width=0.45\textwidth,angle=0]{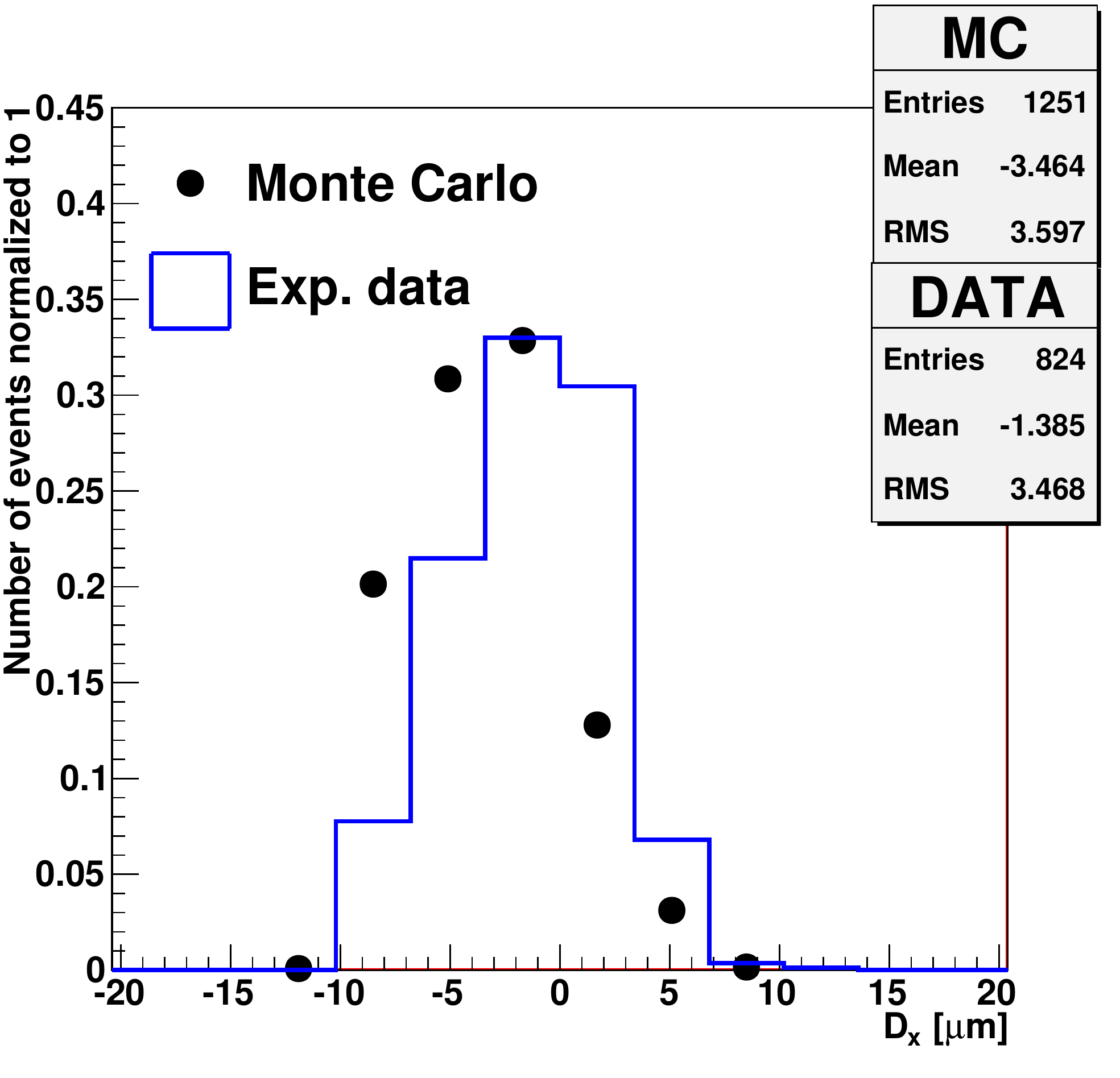}
		\label{fig:digi:CompCog:4007}
	}
        \caption[Distribution of difference $D_{x}$ between cluster position in the $x$ direction reconstructed according to a charge weighted centre of gravity ($x_{CoG}$) and a $x$ position of corresponding seed pixel ($x_{seed}$)]{Distribution of difference $D_{x}$ between cluster position in the $x$ direction reconstructed according to a charge weighted centre of gravity ($x_{CoG}$) and $x$ position of corresponding seed pixel ($x_{seed}$). The $x$ direction is along the charged particle track and a cluster centre of gravity is calculated exploiting information on the charge spatial distribution in the $3\times3$ pixel cluster formed around the seed. The black dots correspond to simulation results while blue histograms represent experimental data. Fig. (a), (b), (c) and (d) refer to different tack inclinations of $\theta = 0^{\circ}$, $\theta = 45^{\circ}$, $\theta = 60^{\circ}$ and $\theta = 75^{\circ}$, respectively.}
        \label{fig:digi:CompCog}
        \end{center}
\end{figure}

\begin{figure}[!h]
        \begin{center}
	\includegraphics[width=0.7\textwidth]{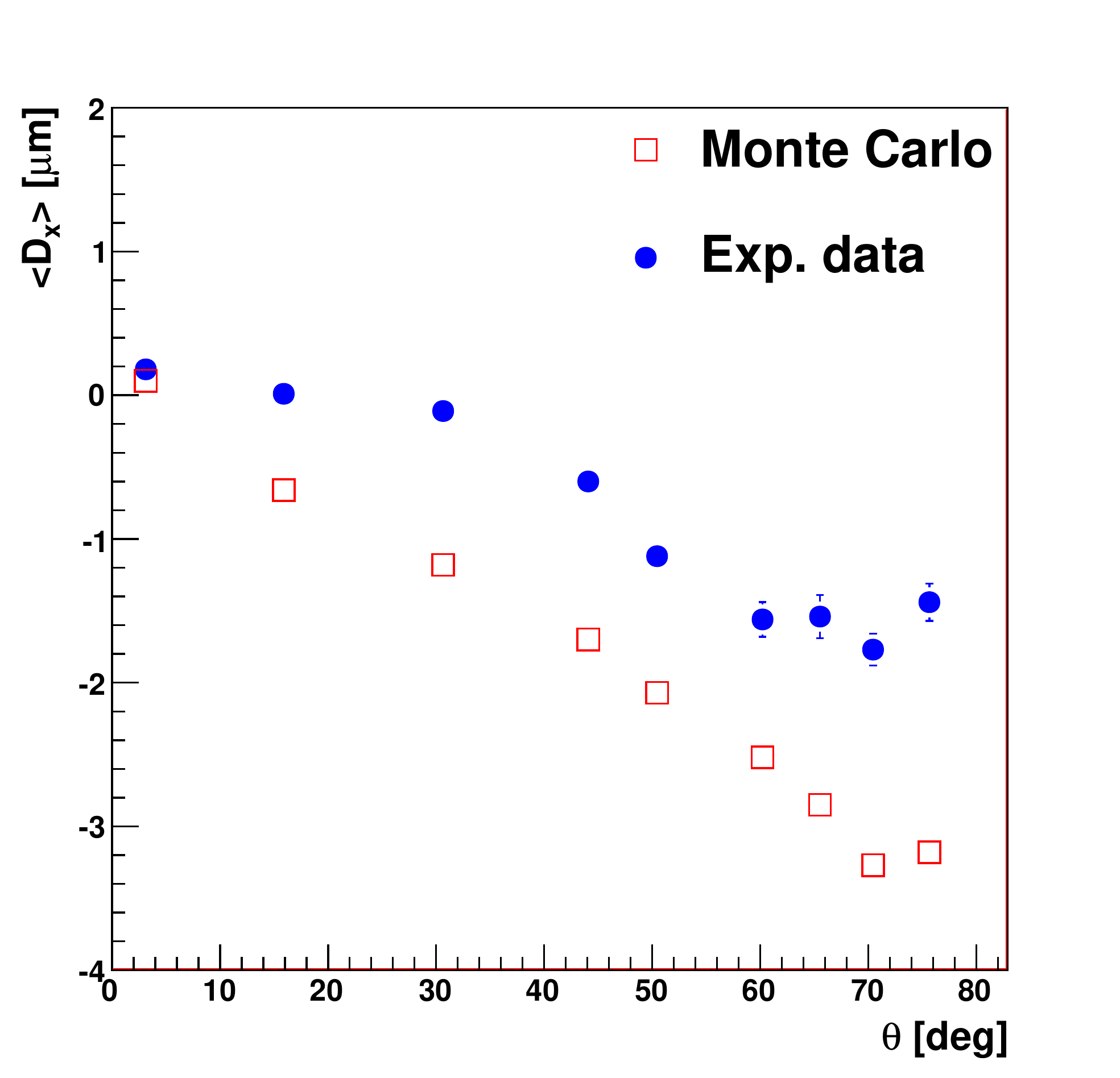}
        \caption[Distance between positions of the seed pixel and the cluster CoG as a function of the incident angle $\theta$]{Distance between positions of the seed pixel and the cluster charge weighted centre of gravity ($CoG$) as a function of the incident angle $\theta$.}
        \label{fig:digi:CompMeanCog}
        \end{center}
\end{figure}

Given a measured cluster with its charge distribution in pixels, a natural question arises where lies the entry point of the particle track - the ``hit''. One of the ways to answer this question is to use simulations and compare them with the measurements. It is natural to expect that the hit should be located within or in the vicinity of the seed pixel. Although the simple simulation algorithm presented above seems somewhat imperfect, it may serve as a guide to learn information on the hit position. For this purpose $3\times3$ pixel clusters have been used for all values of the angle $\theta$. For each measured and simulated cluster, coordinates of the charge weighted centre of gravity ($CoG$) were calculated. In further discussion only the $x$ coordinate (along the particle track projected to the pixel plane) was considered. A variable $D_{x}$ was introduced as: $D_{x} = x_{CoG} - x_{seed}$, where $x_{seed}$ is the $x$ coordinate of the centre of the seed. Distributions of the variable $D_{x}$ for different values of the angle $\theta$ are shown in fig.~\ref{fig:digi:CompCog}.
\\
The agreement of simulations with the data is very good for perpendicular tracks fig.~\ref{fig:digi:CompCog:2012}. Recall that it is so for charge distribution in pixels as shown in fig.~\ref{fig:digi:MeanClust:Comp_2012}. The agreement is worse for larger values of $\theta$. Note that the mean value of the parameter $D_{x}$ decreases with increasing $\theta$, in other words the $CoG$ moves away from the seed centre in the direction of along the track projection. This trend is reproduced by simulation as shown in fig.~\ref{fig:digi:CompMeanCog}. 

\subsection{Cluster multiplicity}
\label{ch:digi:digi_test:cluster_multiplicity}

\begin{figure}[!h]
        \begin{center}
	\subfigure[]{
		\includegraphics[width=0.45\textwidth,angle=90]{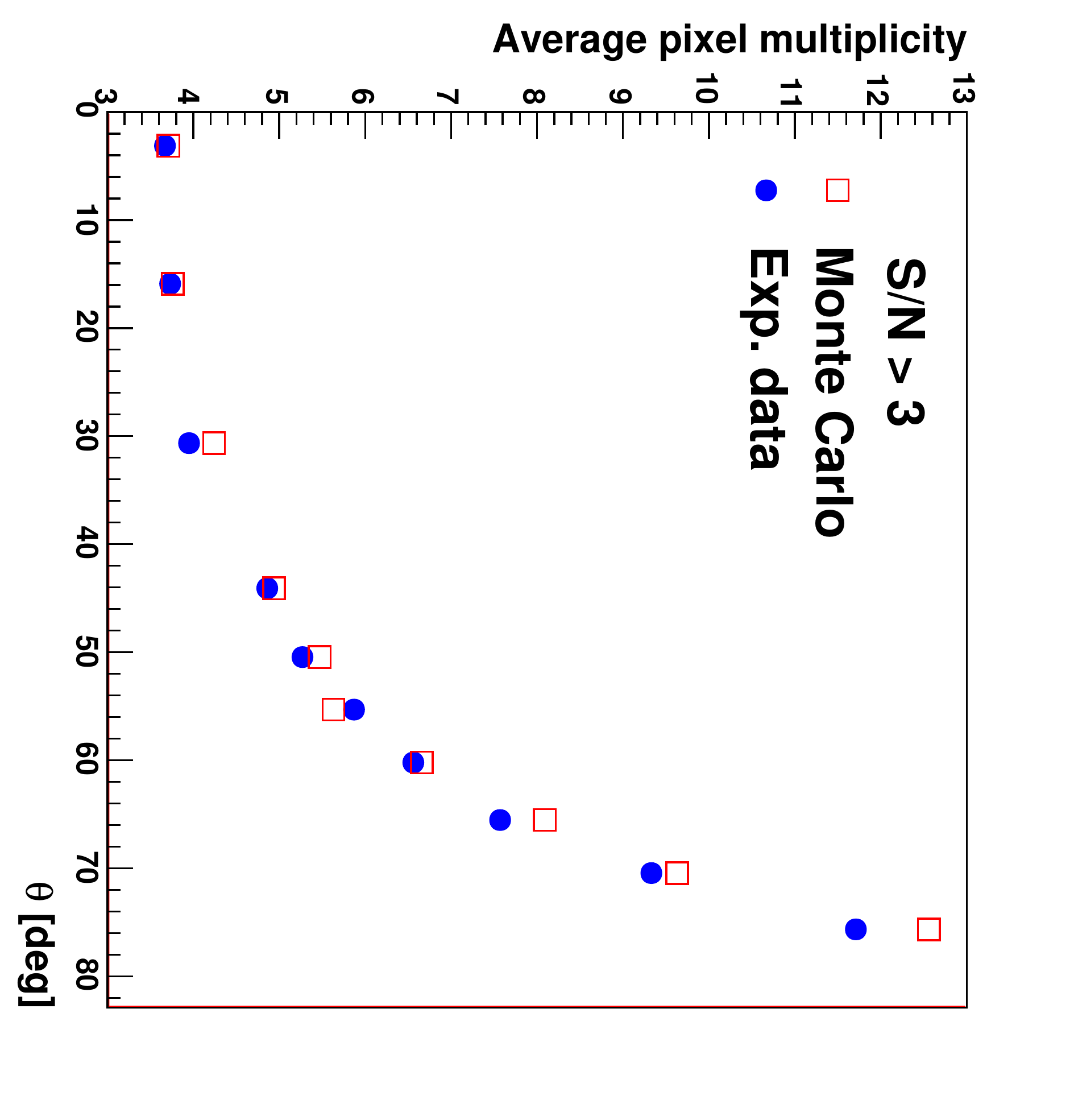}
		\label{fig:digi:Multiplicity_M5:3}
	}
	\hspace{0.1cm}
	\subfigure[]{
		\includegraphics[width=0.45\textwidth,angle=90]{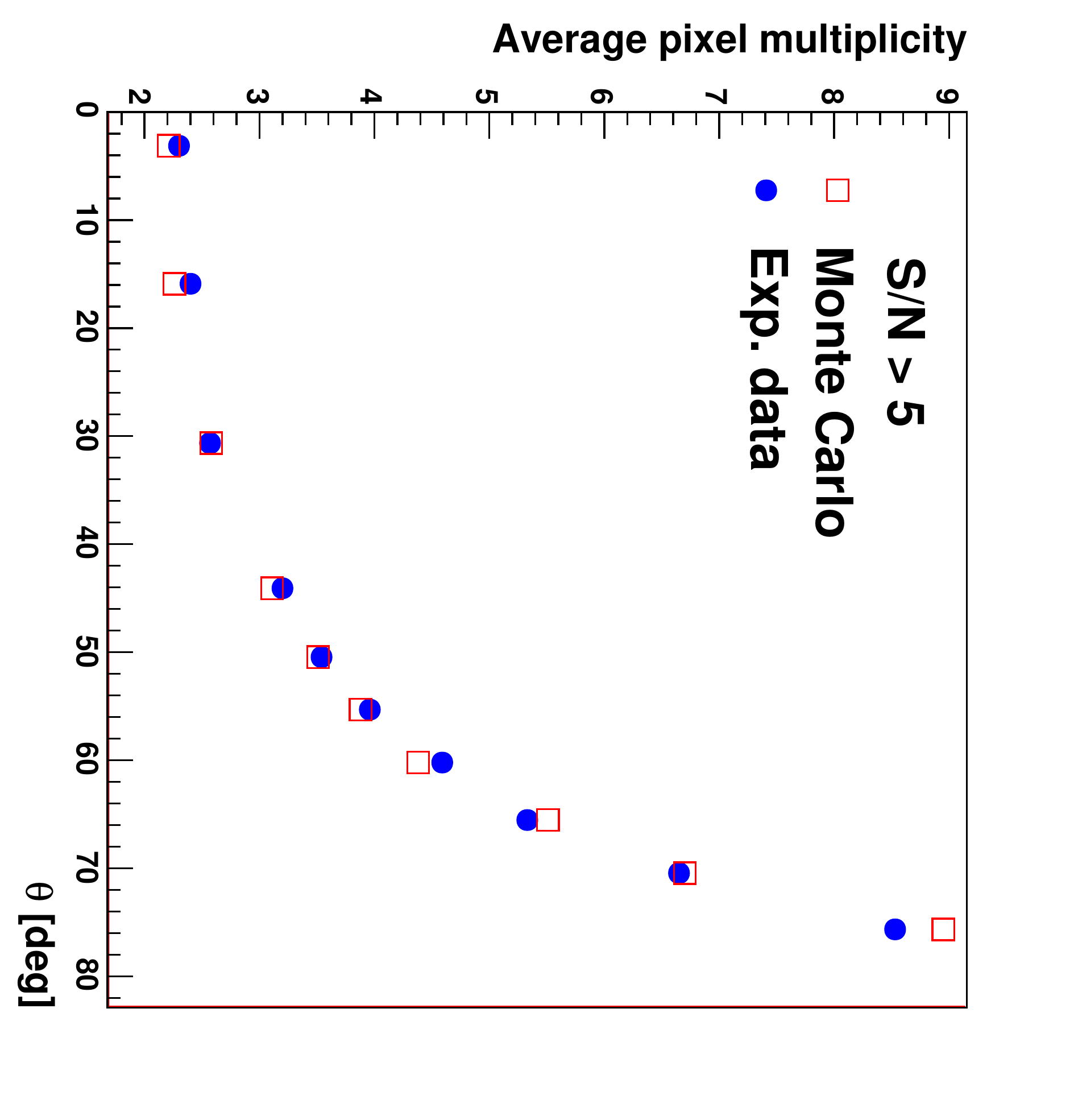}
		\label{fig:digi:Multiplicity_M5:5}
	}
        \caption[MIMOSA-5 cluster multiplicity dependence on track inclination $\theta$ - comparison of measurements and simulations]{MIMOSA-5 cluster multiplicity dependence on track inclination $\theta$ - comparison of measurements and simulations for (a) $S/N > 3$; (b) $S/N > 5$.}
        \label{fig:digi:Multiplicity_M5}
        \end{center}
\end{figure}

Cluster multiplicity is defined as the number of pixels forming a given cluster under a condition that the $S/N$ ratio in each pixel exceeds a present threshold value. Average cluster multiplicity in the MIMOSA-5 pixel matrix as function of the angle $\theta$ are shown in fig.~\ref{fig:digi:Multiplicity_M5:3} and \ref{fig:digi:Multiplicity_M5:5} for two threshold values of the $S/N$ ratio - 3 and 5, respectively. It can be seen that the cluster multiplicity increases with the cluster elongation (i.e. with the increasing angle $\theta$). The results of simulations reproduces the measured trend very satisfactorily for both $S/N$ cuts.

\subsection{Cluster shapes}
\label{ch:digi:digi_test:cluster_shape}

\begin{figure}[!h]
        \begin{center}
	\subfigure[]{
		\includegraphics[width=0.45\textwidth,angle=90]{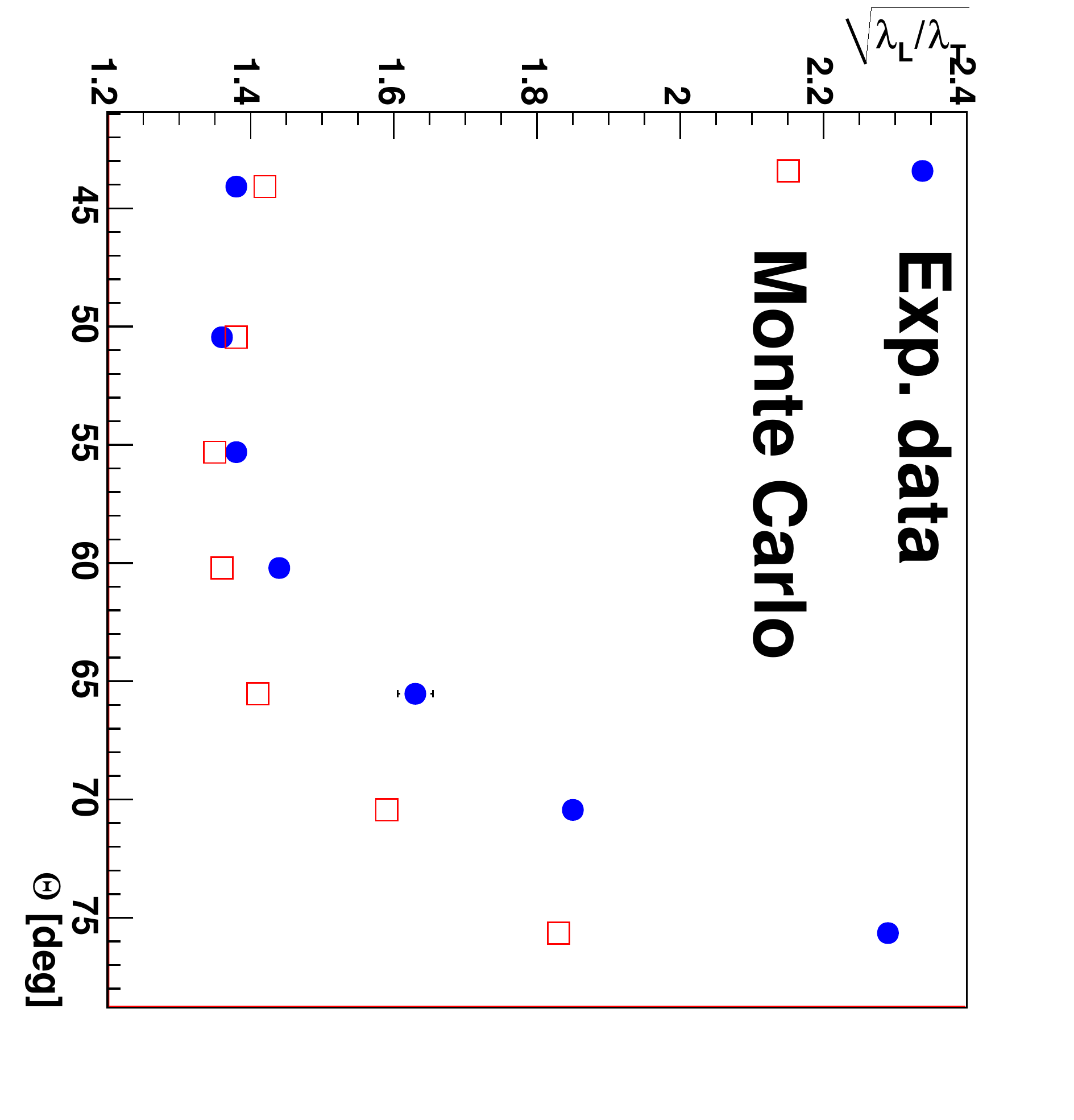}
		\label{fig:digi:cl_shape:Elongation}
	}
	\hspace{0.1cm}
	\subfigure[]{
		\includegraphics[width=0.45\textwidth,angle=90]{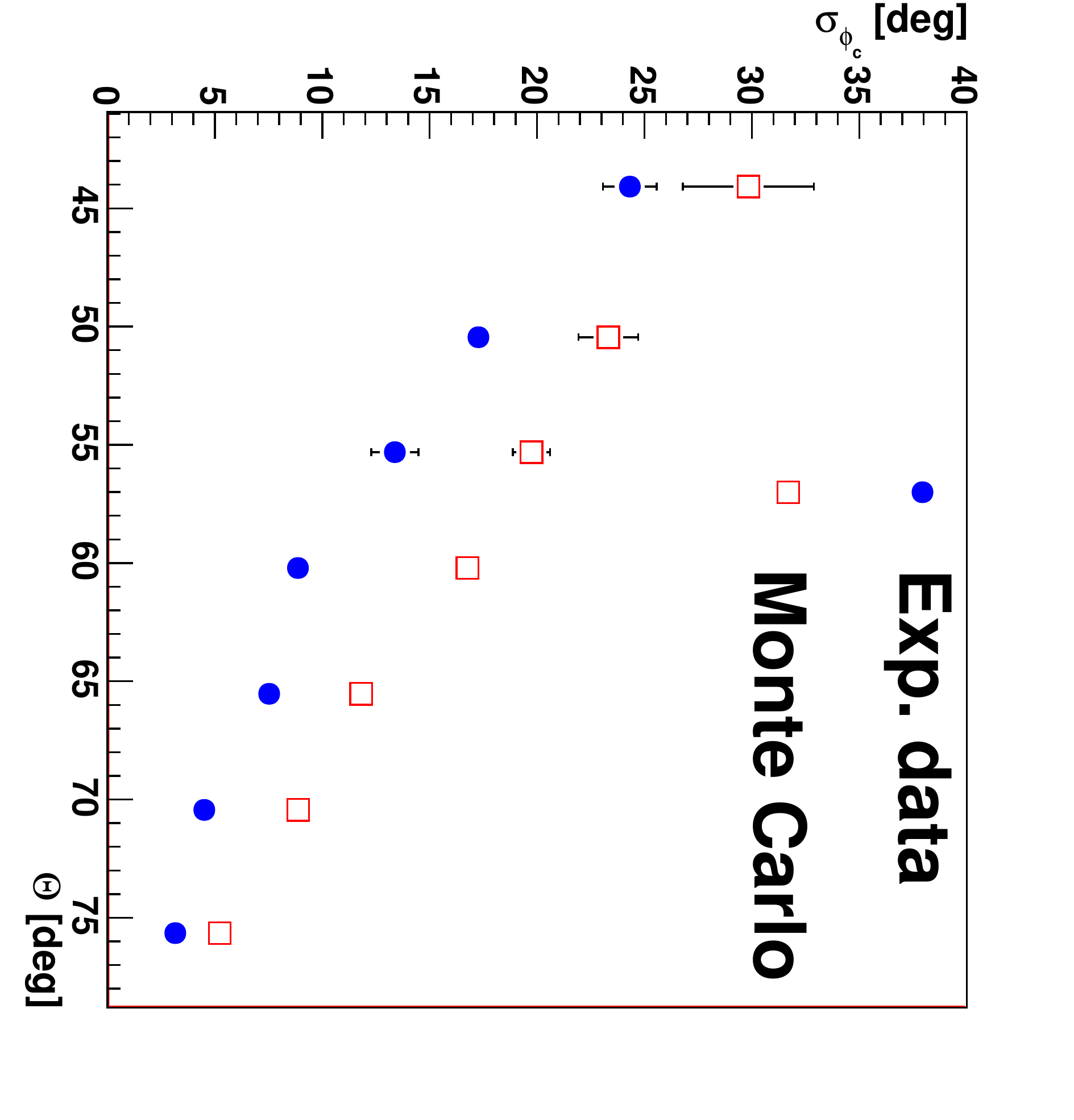}
		\label{fig:digi:cl_shape:SigmaPhi}
	}
        \caption[Superposition of results of the cluster shape reconstruction procedure applied to experimental and corresponding simulation samples of data]{Comparison simulations (open squares) and measurements (full dots) for 6~GeV electron tracks. (a) ratio of the longitudinal and transverse dimensions of a cluster as a function of the angle $\theta$. (b) dependence of the $\phi_{c}$ angle width on the track inclination.}
        \label{fig:digi:cl_shape}
        \end{center}
\end{figure}

In section ~\ref{ch:tests:exp_results:cluster_shape} it was demonstrated that cluster axes may be determined when the cluster is sufficiently elongated, what in case of MIMOSA-5 corresponds to $\theta > 55^{\circ}$, as shown in fig.~\ref{fig:tests:LambdaReco}. The procedure of determining cluster axes, described in section~\ref{ch:tests:exp_results:cluster_shape}, was applied to a sample of simulated clusters for 6~GeV electron tracks. The results are presented in fig.~\ref{fig:digi:cl_shape} and compared with the corresponding measurements. It follows from these comparisons that the $\theta$ dependence of the cluster elongation, $\sqrt{\lambda_{L}/\lambda_{T}}$, has a similar trend in simulations although the results lie systematically below the data points as shown in fig.~\ref{fig:digi:cl_shape:Elongation}. The simulations of the $\phi_{c}$ angle uncertainty, $\sigma_{\phi_{c}}$, as a function of the angle $\theta$ also show a similar trend as data but systematically overestimate the measured values, as presented in fig.~\ref{fig:digi:cl_shape:SigmaPhi}. One would expect these two features of simulations to be consistent: the lower elongation, the larger uncertainty of $\phi_{c}$ is expected.

\section{Simplifications in the parametrisation of MAPS detector response}
\label{ch:digi:simplifications}

The observed discrepancies between simulation and the experiment result possibly from the number of simplifications made in the parametrisation of the MAPS detector response:
\begin{itemize}
 \item 
The geometrical description of a MAPS detector is much simplified and consists only of three layers: the substrate, the epitaxial layer and the pixel layer. It does not contain a description of the depletion region that surrounds each collecting diode, where the electric field is present and which influences trajectories of the charge carriers in the vicinity. Moreover, a uniform pixel response to charge is assumed while the charge collection efficiency in a real MAPS device is lower for marginal regions of a pixel than for its central part.
 \item
The model of charge carrier reflection at the potential barrier between the epitaxial layer and the substrate is very simple. It assumes the equality of the angles of incidence and reflection. It also assumes that reflection is 100\% efficient and no transmission to the substrate occurs.
 \item
Electrons generated in the substrate are assumed to be totally absorbed inside the substrate, however it is known that a small fraction of them can reach the epitaxial layer and contribute to the total collected charge.
 \item
The detector and readout electronics noise which influences mostly pixels with low collected charge, requires more accurate description in the digitisation procedure.
\end{itemize}
To improve the parametrisation of MAPS detectors response, improvements of the above items should be implemented.\\
The discrepancies can be also partially attributed the the fact that the Geant4 does not provide accurate description of ionisation losses in thin silicon detectors. Geant4 exploits the Landau theory \cite{digi:Landau} for simulating ionisation energy losses. The most probable value of the simulated energy loss distribution is in agreement with the experimental data however its width is broader than expected \cite{digi:HBichsel}.

\chapter{Summary and conclusions}
\label{ch:conc}

The apparatus of the future experiment at the ILC would require  precise tracking and in particular an excellent pixel vertex detector needed for efficient heavy flavour identification. The present thesis regarded studies of MAPS pixel matrices that might be candidates for building such a vertex detector. \\
Two MAPS detectors were examined: MIMOSA-5 and MIMOSA-18, with pixels of 17~$\mu$m and 10~$\mu$m pitch. Calibration (ADC-to-charge) of each detector was performed using X-ray photons of 5.9~keV and 6.49~keV from a $^{55}$Fe source. Detector efficiency was studied using 5~GeV and 6~GeV electron beams. Cluster properties were studied for different track inclinations using 1, 5 and 6~GeV electron test beams at DESY and 300~MeV electron test beam at DAFNE/Frascati.\\
In summary, the following results were obtained.
\begin{enumerate}
\item
The pedestal level measured in the MIMOSA-5 amounts to approx. 80~electrons while in the case of the MIMOSA-18 it is approx. 0~electrons. The latter is due to self biased diodes in the readout of the MIMOSA-18. The noise level measured in the MIMOSA-5 is approx. 2.7 times higher than the one of the MIMOSA-18. It amounts to approx. 30~electrons for MIMOSA-5 and approx. 11~electrons for MIMOSA-18. 
\item
The ADC-to-electron conversion gains for MIMOSA-5 and MIMOSA-18 amounts to 9.70$\pm$0.04~e/ADC and 6.02$\pm$0.04~e/ADC. Despite the undepleted epitaxial layer, both of the tested MIMOSA detectors exhibit high charge collection efficiency amounting to 90\% for MIMOSA-5 and 97\% for MIMOSA-18.
\item
Efficiencies of the studied matrices were determined using the beam telescope. The value for a given detector varies with the method of cluster reconstruction, in particular depends on adopted values for the signal-to-noise ratio of the seed ($t_{s}$) and the signal-to-noise of the seed neighbours ($t_{n}$). Since the MIMOSA-5 exhibits approx 2.7 times higher noise level than MIMOSA-18, the result for the detection efficiency in the MIMOSA-5 is much more sensitive to the above cuts than that measured in MIMOSA-18. The efficiency of the MIMOSA-18 exceeds 99\% and almost does not depend on the above cuts in the studied range ($5 < t_{s} < 10$ and $3 < t_{n} < 5$), while the efficiency of the MIMOSA-5 ranges from 97\% for $t_{s} = 6$ and $t_{n} = 1$ to 99\% for $t_{s} = 4$ and $t_{n} \leqslant 1$.\\
Due to multiple Coulomb scattering, which dominates track position measurement at the available electron beam energies, determination of the single point resolution was not feasible. However the measurements confirmed that the spatial resolution of both tested detectors was better than the spatial resolution of the telescope used for electron tracks reconstruction ($\sigma_{TELE} \sim 8~\mu$m).
\item
For electron tracks perpendicular to the detector surface 99\% of the total generated charge is confined to 15 and 17 pixels in the MIMOSA-5 and MIMOSA-18, respectively. The most probable value (MPV) of the collected charge in 15 and 17 pixel clusters in the MIMOSA-5 and MIMOSA-18 amounts to $\sim790$ electrons and $\sim876$ electrons, respectively. The MPV of the signal-to-noise ratio for the seed pixels in the MIMOSA-5 and MIMOSA-18 are $10.9$ and $25.5$, respectively.
\item
The tests performed with inclined electron tracks have shown that the number of pixels participating in the charge collection and the amount of the charge generated inside the active volume of the detector increases with the angle of incidence, $\theta$. The MPV of the collected charge ranges from $\sim$800~electrons for $\theta \approx 0^\circ$ to $\sim$4000~electrons for $\theta \approx 75^\circ$ and from $\sim$900~electrons for $\theta \approx 0^\circ$ to $\sim$5000~electrons for $\theta \approx 75^\circ$, for MIMOSA-5 and MIMOSA-18, respectively. In the case of the MIMOSA-5, 90\% of the total charge generated by the ionising particle incident at $\theta \approx 0^{\circ}$ is collected on average in clusters composed of 5 pixels while in the case of MIMOSA-18 in clusters composed of 9 pixels. For particle tracks traversing the detector at $\theta \approx 75^{\circ}$, 90\% of the total charge is collected in clusters composed of 14 pixels and 23 pixels, for MIMOSA-5 and MIMOSA-18, respectively.
\item
A method of evaluation cluster elongation and its orientation with respect to the pixels netting was developed and tested. The method is based on diagonalisation of the two dimensional charge distribution matrix and computing eigenvectors coinciding with the longitudinal and transverse axes of clusters. Precision of the azimuthal angle reconstruction increases with increasing incident angle $\theta$ and it does not depend on the beam energy in the studied energy range. Clusters become sufficiently elongated for $\theta > 55^{\circ}$ and for $\theta > 45^{\circ}$ in the MIMOSA-5 and MIMOSA-18, respectively and the effect grows rapidly with increasing angle of incidence $\theta$.
\item
The response of a MAPS detector to charged particles was simulated using a simple model of charge diffusion. This model is based on the assumptions that charge carriers in the MAPS detectors diffuse isotropically and they may be reflected from the substrate with an angle equal to the angle of incidence. This has been the first attempt to provide an effective description of signal formation in MAPS detectors. Due to its simplicity, the algorithm is fast which makes this approach attractive for detailed Monte Carlo studies of the ILC vertex detector performances.\\
The simulation provides a good qualitative description of the MAPS detector response to charged particles. This includes basic cluster characteristics for perpendicular and inclined tracks and in particular the cluster multiplicity in the real device is very well described. Certain discrepancies of the model predictions and the measurements suggest that the parametrisation may be to simple and still requires refinements.
\end{enumerate}
The above results regarding cluster shapes have been published \cite{conc:clsuter_shapes}.\\\\
The following main conclusions can be drawn on the basis of the studies presented in this thesis:
\begin{enumerate}
\item
Pixel matrices of the MIMOSA series are suitable devices that could be used in building a vertex detector at the ILC. A big progress has been made regarding the noise level reduction which is visible from comparing MIMOSA-5 and MIMOSA-18. MIMOSA-18 is already detector characterised by low noise, high signal-to-noise ratio, small pixel pitch, high track detection efficiency and high single point resolution.
\item
It is possible to reconstruct the cluster size, its possible elongation and its orientation with respect to the incident beam direction using both MIMOSA-5 and MIMOSA-18 detectors, i.e. with 17~$\mu$m and 10~$\mu$m pixel pitch. Since MIMOSA-18 has smaller pixels it allows much more precise reconstruction of cluster properties. The final decision on the pixel size to be used in the VTX at the ILC will have to be taken after considering the readout rate. The information on cluster elongation and its orientation on the pixel matrix can be exploited for distinguishing between hits originating from final state particles of $e^{+}e^{-}$ interactions (hadrons) and hits due to the beamsstrahlung related $e^{+}e^{-}$ pairs. The presented method for reconstructing cluster elongation and its orientation with respect to the pixel netting can be also applied to other pixel technologies like hybrid pixel detectors, CCDs or DEPFET sensors.
\item
A simple, effective model of charge diffusion presented in this thesis is quite adequate for describing basic characteristics of the signal formation in a MAPS detector. It provides a fast simulation tool of a MAPS which has been implemented in a full Monte Carlo simulation of the vertex detector.
\end{enumerate}
Presently intense studies of pixel matrices are conducted at IReS (Strasbourg), with the purpose of improving the properties of MAPS detectors towards a better version of the MIMOSA series. One of the open issues regarding the vertex detector is the strategy of the readout. Charged particles leave clusters containing several pixels. One solution is to read out all pixels for further offline analysis. At contrast one might consider reading out only the seed in each cluster. Another approach is to read out only those pixels which have collected charge above a certain threshold level. Each approach to readout results in different amount of information available for reconstruction the coordinates of the traversing track. This in turn is related to the accuracy of the measurement. The model of signal formation in the MAPS devices, presented in this thesis, may be used for the purpose of these studies. In particular it may serve to verify the detector occupancy and precision of track reconstruction depending on the adopted readout strategy.\\
The studies described in this thesis are still continuing since amount of the collected data depassed the capabilities of their quick analysis. In particular, the issue of relating the readout strategy and accuracy of vertex reconstruction is under consideration. The subject of radiation damages induced by the beam related background in silicon is currently under intense studies by the author. Those activities contribute to the part of a common effort to develop a final design of a vertex detector for the International Linear Collider (ILC Technical Design Report to be prepared in 2012).

\chapter*{Acknowledgements}
\addcontentsline{toc}{chapter}{Acknowledgements}

First of all, I would like to thank my supervisor Professor Jacek Ciborowski for his continued support. His experience and wisdom made this thesis better.\\
Special thanks go to Dr. Marek Adamus who shared with me his knowledge regarding the work with hardware. His support during tests of MAPS devices was invaluable.\\
I would also like to thank Dr. Grzegorz Grzelak, Dr. Pawe{\l} {\L}u\.zniak, Dr. Piotr Nie\.zurawski and Professor Aleksander Filip \.Zarnecki for lots of discussions regarding data analysis and high energy physics.\\ 
I am very grateful to the DESY and Frascati Directorates for the possibility of using the test beam facilities. In particular I would like to thank Dr. M.I. Gregor and Dr. U. Koetz from DESY who gave me many useful advices. My gratitude goes also to Dr. A. Besson, Dr. G. Claus, Dr. R. De Masi, Dr. W. Duli\'nski, Dr. M. Goffe and Dr. M. Winter from the IReS (Strasbourg) for numerous discussions and lending the MIMOSA prototypes.\\
Finally, I would like to thank my family, in particular my wife Ma{\l}gorzata, daughter Aniela and parents Aleksandra and Janusz for their love and support.

\bibliographystyle{unsrt}
\bibliography{PhD_MAPS}

\end{document}